\documentclass[12pt]{article}%
\usepackage{amsfonts}
\usepackage{amsmath}
\usepackage{amssymb}
\usepackage{sectsty}
\usepackage[round,authoryear]{natbib}
\usepackage[left=0.6in, right=0.6in, top=1in, bottom=1in]{geometry}
\usepackage[sc,center,compact,noindentafter,medium]{titlesec}
\usepackage[onehalfspacing,nodisplayskipstretch]{setspace}
\usepackage{graphicx}
\usepackage{paralist}
\usepackage[colorlinks,bookmarksnumbered]{hyperref}%
\usepackage{hyperref}

\usepackage[dvipsnames,svgnames,x11names,hyperref]{xcolor}
\hypersetup{colorlinks=true,urlcolor=blue,citecolor=black,linkcolor=purple}    

\usepackage{multirow}

\usepackage[utf8]{inputenc}
\usepackage{mathtools}
\usepackage{amsthm}
\usepackage{float}
\usepackage{subcaption}
\usepackage{verbatim}
\usepackage{enumitem}
\usepackage{geometry}
\usepackage{physics}
\usepackage{breqn}
\usepackage{soul} 
\allowdisplaybreaks

\providecommand{\U}[1]{\protect\rule{.1in}{.1in}}
\newtheorem{theorem}{\normalfont\scshape Theorem}[section]

\newtheorem{lemma}{\normalfont\scshape Lemma}[section]
\newtheorem{assumption}{\normalfont\scshape Assumption}

\newtheorem{proposition}{\normalfont\scshape Proposition}[section]
\newtheorem{definition}{\normalfont\scshape Definition}[section]
\expandafter
\def \expandafter \normalsize \expandafter{\normalsize \setlength \abovedisplayskip{10pt plus 2pt minus 7pt}}
\expandafter
\def \expandafter \normalsize \expandafter{\normalsize \setlength \abovedisplayshortskip{0pt plus 2pt}}
\expandafter
\def \expandafter \normalsize \expandafter{\normalsize \setlength \belowdisplayskip{10pt plus 2pt minus 7pt}}
\expandafter
\def \expandafter \normalsize \expandafter{\normalsize \setlength \belowdisplayshortskip{5pt plus 2pt minus 3pt}}

\numberwithin{equation}{section}
\pltopsep=\medskipamount
\DeclareMathOperator*{\plim}{plim}
\DeclareMathOperator*{\Avar}{Avar}

\usepackage{pdflscape}

\graphicspath{{figures/}}

\begin{document}

	\title{\textsc{Factor Network Autoregressions}
		\thanks{We thank for useful comments and suggestions participants at 
		the 3rd Italian Workshop of Econometrics and Empirical Economics (Rimini, 20-21 January 2022),
		the Workshop on Dimensionality Reduction and Inference in High-Dimensional Time Series (Maastricht, 13-14 June 2022), 
		the Waseda Workshop (Tokyo, 7 February 2023), the 10th Italian  Congress of Econometrics and Empirical Economics (Cagliari, 26-28 May 2023),
		the 13th Workshop in Time Series Econometrics (Zaragoza, 16-17 March 2023), 
		the Barcelona Summer Forum (Barcelona, 12-13 June, 2023),
		the ECARES@30 Workshop (Bruxelles 30 June, 2023),
		the Joint Statistical Meetings (Toronto, 5-10 August 2023),	
		as well as seminar participants at ITAM (24 March 2023) and Lund (28 March 2023).
		All authors gratefully acknowledge financial support from the Italian Ministry of Education, University and Research (PRIN 2017, Grant 2017TA7TYC). Graziano Moramarco gratefully acknowledges funding from the European Union and the Italian Ministry of Education, University and Research (ESF REACT-EU, PON Ricerca e Innovazione 2014-2020 D.M. 1062/2021, CUP J41B21012140007).}}
	    		\author{Matteo Barigozzi\\{\small University of Bologna, Italy}
    		\and Giuseppe Cavaliere\\{\small University of Bologna, Italy}\\{\small \& Exeter Business School, UK}
    		\and Graziano Moramarco\\{\small University of Bologna, Italy}}
	\date{}
	\maketitle

	\begin{abstract}
	
	\noindent 
	
	We propose a factor network autoregressive (FNAR) model for time series with complex network structures. The coefficients of the model reflect many different types of connections between economic agents (``multilayer network"), which are summarized into a smaller number of network matrices (``network factors") through a novel tensor-based principal component approach. 
	We provide consistency and asymptotic normality results for the estimation of the factors, their loadings, and the coefficients of the FNAR,  as the number of layers, nodes and time points diverges to infinity.
	Our approach combines two different dimension-reduction techniques and can be applied to high-dimensional datasets.
	Simulation results show the goodness of our estimators in finite samples.
	In an empirical application, we use the FNAR to investigate the cross-country interdependence of GDP growth rates based on a variety of international trade and financial linkages. The model provides a rich characterization of macroeconomic network effects as well as  good  forecasts of GDP growth rates. 
	
	\end{abstract}

	
	\def\spacingset#1{\renewcommand{\baselinestretch}{#1}\small\normalsize}
	\spacingset{1}
	\thispagestyle{empty}
	\newpage\spacingset{1.7}%
	
	
	\section{Introduction}

	Network models are key tools for analyzing interconnected systems and are increasingly used in many areas of research. Typical applications involve social networks, where a high number of agents are connected along several dimensions (e.g., \citealp{zhuetal17}), or economic networks, which are usually applied to evaluate spillover effects between agents, geographical regions and economic sectors (e.g., \citealp{acemogluetal15}; \citealp{dieboldyilmaz14}; \citealp{billioetal12}). 
	From an empirical perspective, the analysis of networks calls for the development of novel statistical techniques that are able to account for complex interactions and handle large datasets, possibly recorded over time. 
	
	Several approaches have been proposed to integrate network structures into well-established time series models; the network autoregression (NAR) by \citet{zhuetal17} or the global vector autoregressive (GVAR) model by \citet{pesaranetal04} are two leading examples. 
	These models are generally designed to deal with one type of connection at a time between the agents or ``nodes" of a network, e.g., import/export flows. 
	They characterize networks in the form of a single \textit{adjacency matrix}; i.e., a matrix whose $ij$-th element represents a link between node $i$ and node $j$.

	The assumption of a single adjacency matrix, however, appears in general restrictive. In fact, modern economies or social networks are complex systems, in which agents interact through many different channels. For instance, countries are simultaneously linked not only by international trade flows, but also by financial linkages, multinational firms' activities, migration flows, etc.
	Accordingly,  models of \textit{multilayer networks}, i.e., networks where nodes are linked through multiple types of connections (\citealp{kivelaet14}), have important applications in many different areas, such as international economics and financial (systemic) risk assessment.
	
	To state the problem we consider in this paper, let $y_t$ be an $N$-dimensional vector of stationary time series, e.g., GDP growth rates for different countries. To model the dynamics of $y_t$ when $N$ is large, we could consider a VAR model in which the dynamic relations between the $N$ elements of $y_t$ are mediated by $m$ time-varying networks, denoted as $W_{k,t}$, $k=1,\ldots,m$, e.g., measuring trade flows of different commodities or various financial linkages. All together these networks, which are $N\times N$ matrices, form the $m$ layers of a multilayer network. We could then consider a multilayer NAR:
	\begin{align}
		y_t &= b_1 {\scriptsize{W_{1,t-1}}}  y_{t-1}+ \ldots +b_m {\scriptsize{W_{m,t-1}}}  y_{t-1}+\varrho y_{t-1}+ a + \zeta_t, \quad t=1,\ldots, T, \label{eq:zhunar}
	\end{align}
	where  $b_k$, $k=1,\ldots, m$, $\varrho$, and $a$ are unknown scalars, and $\zeta_{t}:=(\zeta_{1t},\cdots, \zeta_{Nt})'$ is an $N$-dimensional vector of errors. 
	This model generalizes the NAR by \citet{zhuetal17} to the case of multilayer networks. However, in empirical applications, the number $m$ of layers can be very large (e.g., for trade data one may get to hundreds or even thousands of product categories); hence, least-squares estimation of \eqref{eq:zhunar} can become quickly infeasible.
	
	In this paper, we assume that the observed layers $W_{k,t}$, $k=1,\ldots,m$, of the multilayer network have a factor structure, where the common  factors are unobserved networks, each represented by an $N\times N$ matrix, capturing the commonality between the layers. We call these latent factors \textit{network factors}; once retrieved, we can use them in \eqref{eq:zhunar} to model the dynamics of $y_t$ in place of all the $m$ layers, thus reducing the dimensionality of the problem. We call the resulting model a factor network autoregression (FNAR). To estimate the FNAR we first develop a novel, tensor-based, principal component analysis approach, which allows to estimate the network factors. We then use these network factors to estimate an $N$-dimensional FNAR.

	Our model exploits two different types of dimension-reduction techniques. 
	First, by assuming, as in \citet{zhuetal17}, that the network effects are homogeneous across nodes (or across a finite number of groups of nodes), the number of unknown parameters in our FNAR is at most linear in the number of networks considered, instead of being proportional to $N^2$ as is the case for ordinary vector autoregressions. Second, since the FNAR is based on the extraction of just few network factors common across the $m$ network layers, the number of parameters does not grow with the number of layers $m$, but with the number of network factors which we assume to be finite and independent of $N$ and $m$.
	
	In terms of theory, we first provide sufficient conditions such that the FNAR admits a (weakly) stationary and causal solution. Second, we provide sufficient conditions such that our estimators 
	of the network factors and of the FNAR coefficients are consistent and asymptotically normal as the sample size $T$, the number of layers $m$, and the number of nodes $N$, diverge. Crucially, the estimator of the FNAR coefficients we propose is consistent even when the FNAR errors admit cross-sectional correlation as well as serial correlation, due to the presence of node specific factors. Finally, the number of network factors can be determined by an eigenvalue ratio approach, by adapting to the present context those recently proposed by \citet{hanetal20} and \citet{helitrapani2022}.
	
	In an empirical application, we use the FNAR model to study the dynamics of GDP growth rates in a multilayer network of $N=24$ countries with $m=25$ different layers, reflecting international trade flows for different categories of goods and services, and a variety of cross-border financial linkages. In a first step we retrieve 6 network factors common across the $m$ layers, which, once identified, have a clear economic meaning. These are then used to construct the regressors which are employed in a FNAR to study and predict GDP growth rates for all countries.  A similar idea was explored by \citet{chen2022modeling} where, however, the tensor data of trade flows is compressed also along the node (country) dimension, thus making the interpretation of the underlying factors less clear.  
	
	The paper is organized as follows. Section \ref{sec:lit} gives a summary of the related literature. Section~\ref{sec:model} illustrates the model.  Section~\ref{sec:estimation} presents the estimation approach. Section~\ref{sec:theory} introduces the assumptions and the asymptotic properties of the estimators. 
	Section~\ref{sec:simulations} evaluates the finite sample performance of our estimators by means of Monte Carlo simulations.
	Section~\ref{sec:application} presents the empirical application and Section~\ref{sec:conclusions} concludes. All proofs are in the online Supplementary Material, which also contains additional possible estimators, further simulation results and details on the empirical application.

	\medskip
	\noindent \textsc{Notation}. The \textit{order} of a tensor is the number of dimensions, also known as \textit{modes}. The \textit{fibers} of a given mode are defined by fixing every index but one. We deal only with order-3 tensors. 
	The mode-$q$ \textit{matricization}, or unfolding,  of a tensor $ \mathcal{T}$, denoted as $\text{mat}_{(q)} (\mathcal{T})$ or, equivalently, $ \mathcal{T}_{(q)} $, is a matrix having as columns its mode-$q$ fibers.
	For a generic order-3 tensor $\mathcal{T}$ of dimensions $d_1 \times d_2 \times d_3$ with generic element $\mathcal T_{ijk}$, the mode-1 matricization is a $d_1 \times (d_2 d_3)$ matrix having as columns the $d_1$-dimensional vectors $\mathcal T_{\cdot jk}$,  the mode-2 matricization is a $d_2 \times (d_1 d_3)$ matrix having as columns the $d_2$-dimensional vectors $\mathcal T_{i\cdot k}$, and the mode-3 matricization is a $d_3 \times (d_1 d_2)$ matrix having as columns the $d_3$-dimensional vectors $\mathcal T_{ij\cdot}$. 
	
	Finally, 
	we denote the mode-$q$ multiplication of tensor $\mathcal{T}$ by a matrix $X$ of size $p \times d_q$ as  $\mathcal{T} \times_q X$, which is a tensor of size $p$ in the $q$-th dimension and the same size as $\mathcal{T}$ in the other dimensions. 
	
	\section{Related literature}\label{sec:lit}
	
	Our modelling approach is especially related to the network autoregression (NAR) model by \citet{zhuetal17},  the community network autoregression (CNAR) model by \cite{chenetal2020cnar} and the group network autoregression (GNAR) model by \citet{zhu2022simultaneous}. Differently from these papers, where just one network is considered, our main contribution is to integrate a large multilayer network into a VAR model by representing multilayer networks as a superposition of common network factors. 
	
	Our methodology for extracting the network factors is related to the recent statistical literature on factor analysis of tensor time series building on the tensor Tucker decomposition, which assumes the existence of low-rank tensor factors  (\citealp{chenyangzhang22}; \citealp{hanetal2020iter,hanetal20}; \citealp{helitrapani2022}; \citealp{zhangetal2022}; \citealp{chen2024rank}). This approach differs from the canonical polyadic (CP) decomposition which always assumes the factors to be vectors \citep{chang2023modelling,hanetal2021}. 
	
	Differently from all the above cited works, since our aim is to reduce the dimensionality of a multilayer network, our setting is a special case of the Tucker decomposition, where the factors are low-rank only along the layers' mode, while they retain all information along the nodes' two modes, i.e., they are $N\times N$ matrices which we can directly interpret as network factors. 
	Moreover, differently from some of the above cited works, we allow for serial dependence in the idiosyncratic tensor, reflecting the assumption, commonly made in the literature on factor modeling of economic data, that factors account for the cross-sectional (for us, cross-layer) variation of the data (see, e.g., \citealp{bai2003inferential}, in the vector case). 
	In Appendix F.5 we provide a simulation-based comparison between our estimates of the network factor loadings and those obtained  by means of the approaches by \citet{chenyangzhang22} where no serial idiosyncratic correlation is allowed for.

	Our paper is also related to a growing literature that uses tensor decomposition methods to estimate the coefficients of high-dimensional time series models. 
	First,
	\citet{wangetal21}  apply tensor decomposition to the order-3 tensor whose slices are the matrices of (unknown) VAR coefficients at different lags, which makes this approach an alternative to ours. However, unlike our approach, \citet{wangetal21} do not exploit data on observed networks to estimate the VAR.
	We refer to Appendix G.2.4 for an empirical comparison between this approach and our FNAR.
	Second, \citet{wangetal24} consider an autoregressive model for tensor-valued time series and use a Tucker decomposition to estimate the autoregressive coefficients, while \citet{chang2023modelling} model matrix-valued time series using a tensor  CP decomposition.
	Both these last two approaches are more distant from ours, since we do not aim to estimate an AR model for a multilayer network itself, but we aim to use a multilayer network to estimate a vector autoregression in the same spirit as \citet{zhuetal17}.

	Finally, we also contribute to two further strands of literature. First, to the literature on factor and factor-augmented models (e.g., \citealp{stockandwatson02}, \citealp{baiandng2006}) by developing a new framework where factors are matrices rather than vectors, and enter a factor-augmented autoregression by multiplying (weighting) the lagged vector of endogenous variables, rather than being included directly as regressors. Second, to two important streams of the literature on network econometrics, that is: (i) works that investigate the properties of observed networks, such as production networks (see, e.g., \citealp{acemogluetal12}), trading networks in financial and interbank markets (\citealp{denbeeetal21}) and social networks (\citealp{zhuetal17}), and (ii) works concerned with the estimation of network links from the data (e.g., \citealp{dieboldyilmaz14}, \citealp{billioetal12}, \citealp{barigozzibrownlees19}). Our approach combines these two streams of research: on the one hand, we use data on a large number of observed economic networks; on the other hand, we estimate unobserved common network factors driving them.

	
	\section{Model}\label{sec:model}
	
	We assume to observe $m$ networks, each represented by a matrix $W_{k,t}$, $k=1,\ldots, m$, of dimension $N\times N$. The networks have no self-loops, so the diagonal elements of the matrices are zero, and are weighted and {directed}, so in general the matrices have real entries and are not symmetric.  In the present high-dimensional setting, it is convenient to assume that the weights are normalized in such a way that the elements in each row of $W_{k,t}$ sum to $N$. This can be done without loss of generality (see Section \ref{sec:estimation} for specific comments on this aspect).

	We then assume that each observed network can be written as a linear combination of $r$ network factors, $F_{k,t}$, $k=1,\ldots, r$, each of them being an $N\times N$ matrix and with $r\ll m$,  plus a network $\mathcal E_{t\cdot\cdot k}$ idiosyncratic to the $k$-th layer, which is an $N\times N$  matrix. Specifically, we assume
	\begin{equation}
		W_{k,t} = u_{k1} F_{1,t}+\ldots + u_{kr} F_{r,t}+ \mathcal E_{t\cdot\cdot k}, \qquad k=1,\ldots, m, \; t=1,\ldots, T,\label{eq:netfacscalar}
	\end{equation}
	where $u_{kh}$, $h=1,\ldots, r$, are the scalar factor loadings for network $k$.
	
	If the factors are sufficiently ``pervasive'', we can think of replacing the multilayer NAR in \eqref{eq:zhunar} with a FNAR model, which is given by:
	\begin{align}
		y_t &= \beta_1 {\scriptsize\frac{F_{1,t-1}}{N}}  y_{t-1}+ \ldots +\beta_r {\scriptsize\frac{F_{r,t-1}}{N}}  y_{t-1}+\rho y_{t-1}+ \alpha+ \nu_t,  \quad t=1,\ldots, T,\label{eq:famnvar}
	\end{align}
	where  $\beta_k$, $k=1,\ldots, r$, $\rho$, and $\alpha$ are unknown scalars, and $\nu_{t}:=(\nu_{1t},\cdots, \nu_{Nt})'$ is the $N$-dimensional vector of FNAR errors.
	Notice that the network factors are rescaled by $N$ in agreement with our normalization assumption on the observed networks.
	
	Furthermore, in order to allow also for factors common across the $N$ nodes, we assume each element of the FNAR errors $\nu_{it}$, $i=1,\ldots, N$, to have a factor structure:
	\begin{align}\label{eq:moderror}
		\nu_{it} &= \lambda_{i1} G_{1t} +\ldots + \lambda_{iq} G_{qt} + \epsilon_{it}, \qquad i=1,\ldots, N, \; t=1,\ldots, T,
	\end{align}
	where $G_{kt}$, $k=1,\ldots,q$, are node factors, $\lambda_{ik}$, $k=1,\ldots, q$, are the scalar factor loadings for node $i$, and $\epsilon_{it}$ is the node idiosyncratic component.
	
	We call the term $\sum_{k=1}^{r} \beta_k N^{-1} F_{k,t-1}y_{t-1}$ the {\it network effect}, where the coefficients capture the strength of dynamic network effects between nodes, exerted through different types of relationships, which are summarized by means of few network factors. We call the term $\rho y_{t-1}$ the {\it momentum effect}, which captures the direct dynamic interaction of a node with itself. Last, we call the term $\alpha$ the {\it nodal effect}. This can be generalized to include a random effect by adding a term $Z_{t-1}\delta$, where $\delta$ is a $K$-dimensional parameter vector, common to all nodes, and $Z_{t}$ is a $N\times K$ matrix of node-specific exogenous variables. Because of the factor structure in the FNAR errors, the network nodes are correlated with each other not only through the network relationships, but also through the common factors which characterize the cross-sectional dependence at a global level.

	By comparing the FNAR in \eqref{eq:famnvar} with the multilayer NAR in \eqref{eq:zhunar} and a standard VAR(1) for $y_t$ we see that, while the latter requires estimating $N^2+N$ parameters, the multilayer NAR requires estimating $m+2$ parameters, and the FNAR further reduces the number of parameters to $r+2$. Hence, the FNAR model provides two forms of dimension reduction, both along the layer direction, by means of the network factors in \eqref{eq:netfacscalar} and along the node direction by means of the network mediated interactions in \eqref{eq:famnvar}.

	\section{Estimation}\label{sec:estimation}
	
	Estimation of the FNAR model requires two distinct steps. The first one involves estimation of the $r$ network factors.
	The second step involves fitting the FNAR equation \eqref{eq:famnvar}.
	Furthermore, it is necessary to determine the number of network and node factors.
	
	\noindent\textsc{Network factors.}	
	We start by discussing estimation of the network factors. To this aim, it is convenient to collect the  $m$ weight matrices or layers of the multinetwork into a 
	\textit{weight tensor} of order 3, which we denote as $\mathcal{W}_t$ and has size $N \times N \times  m$. In our notation the $m$ network matrices are the \textit{frontal slices} of the tensor. 
	The factor network structure in \eqref{eq:netfacscalar} is then rewritten as:
	\begin{equation}\label{eq:W_decomp}
		\mathcal{W}_t = \mathcal{F}_t \times_3 U + \mathcal{E}_t, \quad t=1,\ldots, T,
	\end{equation}
	where $\mathcal{F}_t $ is a $N \times N \times r$ tensor containing, as frontal slices, the $r$ \textit{network factors} $F_{k,t}$, $k=1,\ldots, r$, each of dimensions $N\times N$, which are common across all layers, $U$ is a $m \times r$ matrix of factor loadings, with entries $u_{kh}$, $k=1,\ldots, m$, $h=1,\ldots, r$, determining how each layer of the original network loads on the network factors, and $\mathcal{E}_t$ is an $N \times N \times m$ tensor  containing, as frontal slices, the {idiosyncratic networks}, $\mathcal E_{t\cdot\cdot k}$, $k=1,\ldots, m$. The elements of $\mathcal{E}_t$ are allowed to be
	(i) weakly correlated in all three modes, and (ii) autocorrelated over time (see Section~\ref{sec:theory} for the specific assumptions). 
	Finally, from \eqref{eq:W_decomp} the mode-3 matricization of $\mathcal W_t$, denoted as $\mathcal{W}_{(3)t}$, is a $m\times N^2$ matrix such that:
	\begin{equation}\label{eq:W_decomp3}
		\mathcal{W}_{(3)t} = U \mathcal{F}_{(3)t} + \mathcal{E}_{(3)t} \quad t=1,\ldots, T.
	\end{equation}
	where $\mathcal{F}_{(3)t}$ is $r\times N^2$ and $\mathcal{E}_{(3)t}$ is $m\times N^2$.	This expression resembles a conventional factor model, with the major difference that each of the $r$ factors is no longer a scalar but a vector of size $N^2$, containing up to $N(N-1)$ non-zero elements (as there are no self-interactions in the network).

	To estimate  $\mathcal{F}_t$, or, equivalently, $\mathcal{F}_{(3)t}$, first, we compute the sample $m\times m$ outer product of $\mathcal{W}_{(3)t}$:
	\begin{align}\label{eq:gamma}
		\widehat{\Gamma}^{\mathcal{W}} &:= \frac{1}{T} \sum_{t=1}^T \mathcal{W}_{(3)t} \mathcal{W}_{(3)t}'. 
	\end{align} 
	Second, letting  $\widehat{V}^{\mathcal{W}} $ be the $m \times r$ matrix whose $j$-th column is the normalized eigenvector corresponding to the $j$-th largest eigenvalue, $ \widehat{\mu}_j^{\mathcal{W}}$, of $  \widehat{\Gamma}^{\mathcal{W}}$, and $\widehat{M}^{\mathcal{W}}$ be the $r \times r$ diagonal matrix with  $ \widehat{\mu}_j^{\mathcal{W}}$ as its $j$-th diagonal entry,
	we estimate  the $m\times r$ loadings matrix $U$ as:
	\begin{equation}\label{eq:est_loadings}
		\widehat{U} :=  \frac 1N\widehat{V}^{\mathcal{W}}( \widehat{M}^{\mathcal{W}})^{1/2}.
	\end{equation} 
	Third, we estimate the mode-3 matricization of the network factors $\mathcal{F}_{(3)t}$ as the principal components (PCs) of $\mathcal{W}_{(3)t}$, i.e., by linear projection of $\mathcal W_{(3)t}$ onto $\widehat U$:
	$
	\widehat{\mathcal{F}}_{(3)t} := (\widehat{U}'\widehat{U})^{-1}\widehat{U}'\mathcal{W}_{(3)t}= N ( \widehat{M}^{\mathcal{W}} )^{-1/2}\widehat{V}^{\mathcal{W}'} \mathcal W_{(3)t},
	$
	which is an $r\times N^2$ matrix; once folded into an order-3 tensor, it gives the estimated $N\times N\times r$ tensor of network factors:
	\begin{equation}\label{eq:est_factors}
		\widehat{\mathcal{F}}_{t} := \mathcal{W}_{t} \times_3 [  (\widehat{U}'\widehat{U})^{-1}\widehat{U}' ]=
		\mathcal{W}_{t} \times_3 [ N  ( \widehat{M}^{\mathcal{W}} )^{-1/2} \widehat{V}^{\mathcal{W} '} ],	 
	\end{equation}
	having as layers $\widehat F_{k,t}:=\text{mat}_{(1)}( \widehat{\mathcal{F}}_{\cdot ,\cdot ,k, t} )$, $k=1,\ldots, r$, which are the $N\times N$ matrices of estimated network factors.
	
	Like in ordinary principal component analysis (PCA), the key intuition is that by exploiting the cross-sectional variation we can estimate the space spanned by the factors. In this case, the relevant cross-sectional dimension is the dimension $m$ of the layers in a network; i.e., the third dimension of the weight tensor $\mathcal{W}_{t}$. The rescaling by $N$ in estimating the loadings reflects the fact that no dimension reduction is applied to the first and second modes of $\mathcal W_t$.
	
	Notice that the estimated loadings and factor tensor are such that they satisfy the identifying conditions:
	$m^{-1}\widehat{U}'\widehat U=m^{-1}N^{-2} \widehat{M}^{\mathcal{W}}$,
	which is a diagonal matrix by construction, and
	\begin{equation}\label{eq:F3ON}
		\frac 1{N^2T}\sum_{t=1}^T \widehat{\mathcal{F}}_{(3)t}\widehat{\mathcal{F}}_{(3)t}'= 
		( \widehat{M}^{\mathcal{W}})^{-1/2}
		\widehat{V}^{\mathcal{W} '}
		\widehat{\Gamma}^{\mathcal{W}} 
		\widehat{V}^{\mathcal{W} }
		(\widehat{M}^{\mathcal{W}} )^{-1/2} = I_r,
	\end{equation}
	so that the $r$ rows of $\widehat{\mathcal{F}}_{(3)t}$ are the classical normalized PCs of the $m$ rows of ${\mathcal{W}}_{(3)t}$.
	
	There are two main differences between our approach and the one proposed by \citet{chenyangzhang22} for estimating common tensor factors from tensor times series admitting a Tucker decomposition. First, we estimate factors using only contemporaneous sample second moments, thus allowing the idiosyncratic tensor to be also autocorrelated.  
	The second difference is that we extract factors along a single dimension of the tensor, namely the dimension of the network layers. 
	This implies that the extracted factors still have a network interpretation; i.e., they are common network factors. Indeed, by construction, the estimated factor matrices $\widehat F_{k,t} $,  $k=1,\dots, r$, have zeros along the main diagonal, as the observed network matrices, and have a scale fixed by means of \eqref{eq:F3ON}.

	To preserve the network properties of the factors, we do not standardize nor demean the elements in $\mathcal{W}_{t} $ along the time dimension. Indeed, our variables are all expressed in the same unit of measurement and standardization would eliminate the fundamental interpretation of $\mathcal{W}_{t}$ as a network, as centering would affect the zero diagonal entries. The same strategy is adopted by \citet[Remark 11]{chenyangzhang22}.
	
	Our estimators in \eqref{eq:est_loadings} and \eqref{eq:est_factors} are defined consistently with the assumption that the observed networks have rows summing to $N$. This implies that the estimated network factors in \eqref{eq:est_factors} have variance growing with $N$, as shown in \eqref{eq:F3ON}, and they must be rescaled before being used in the FNAR defined in \eqref{eq:famnvar} to ensure that the scale of the estimated network coefficients $\beta_k$ does not depend on $N$.
	Clearly, we could equivalently work with row-normalized observed networks and then no rescaling by $N$ would be needed anywhere, although this would imply that, as the number of nodes $N$ grows, the entries of $\mathcal W_t$ would have to become smaller and smaller.

	\noindent\textsc{FNAR coefficients.} 
	To describe the estimation of the FNAR, it is convenient to introduce some further notation.
	Hereafter, with reference to \eqref{eq:famnvar}, 
	let $y := (y_1,\cdots ,y_T)' = (y_1,\cdots ,y_N)$ be the $T\times N$ matrix of observed data  (with $y_t$, $t=1,\ldots, T$, being $N$-dimensional, and $y_i$, $i=1,\ldots, N$, being $T$-dimensional), and define also 
	the $r$-dimensional vector $\beta:=(\beta_1,\cdots,\beta_r)'$ and the $(r+2)$-dimensional  \linebreak vector  $\theta := (\beta', \rho, \alpha )'$ of the FNAR coefficients.
	Moreover, with reference to the node factor structure in \eqref{eq:moderror}, we let 
	$\epsilon:=(\epsilon_1,\cdots ,\epsilon_T)'=(\epsilon_1,\cdots ,\epsilon_N)$  be the $T\times N$ matrix of node idiosyncratic components (with $\epsilon_t$, $t=1,\ldots, T$, being $N$-dimensional, and $\epsilon_i$, $i=1,\ldots, N$, being $T$-dimensional), and we let \linebreak $G_t:=(G_{1t},\cdots, G_{qt})'$ and $\Lambda_i=(\lambda_{i1},\cdots,\lambda_{iq})'$, so that
	$G:=(G_1,\cdots ,G_T)'$  is the $T\times q$ matrix of node factors, and $\Lambda:=(\Lambda_1,\cdots ,\Lambda_N)'$  is the $N\times q$ matrix of loadings.  
	Finally, define the order-3 tensor $\mathcal X$ of dimensions $T\times N\times (r+2)$ having in each of the first $r$ layers one of the $r$ matrices \linebreak
	$
	\mathsf F_k:=\left(N^{-1} F_{k,0} y_{0},\cdots, N^{-1} F_{k,T-1} y_{T-1}\right)^\prime
	$,  $k=1,\ldots, r$, each of size $T\times N$,
	in layer $(r+1)$ the $T\times N$ matrix  \linebreak$y_{(-1)}:=(y_0,\cdots,y_{T-1})^\prime$, and in layer $(r+2)$ a $T\times N$ matrix of ones, denoted as $1_{T\times N}$. It follows that\linebreak $\mathcal X_{(1)}:=\text{mat}_{(1)}(\mathcal X)$ is a $T\times N(r+2)$ matrix having as $t$-th row $\text{vec}(X_t)^\prime$ and $\mathcal X_{(2)}:=\text{mat}_{(2)}(\mathcal X)$ is a $N\times T(r+2)$\linebreak matrix having as $i$-th row $\text{vec}(X_i)^\prime$, where we define
	$X_t := (\mathsf F_{1,t-1,\cdot}',\cdots, \mathsf F_{r,t-1,\cdot}', y_{t-1}, \iota_N)$
	and \linebreak $X_i:=(\mathsf F_{1,\cdot,i},\cdots, \mathsf F_{r,\cdot,i}, y_{i}, \iota_T)$, with $\iota_N$ and $\iota_T$ being $N$- and $T$-dimensional vectors of ones, respectively, with $\mathsf F_{k,t-1,\cdot}$ and $\mathsf F_{k,\cdot,i}$ being the $t$th row and $i$th column of $\mathsf F_{k}$, respectively. 
	
	According to the above notation, the FNAR in \eqref{eq:famnvar}, jointly with the node factor structure in its errors given in \eqref{eq:moderror}, can be equivalently rewritten as:
	\begin{equation}\label{eq:mnvar_matrix_form}
		y_t = X_t \theta + \Lambda G_t+\epsilon_t, \;\; t=1,\ldots,T,\;\;\text{ or }\;\; y_i = X_i \theta +G\Lambda_i+\epsilon_i, \;\; i=1,\ldots,N,
	\end{equation}
	which, by stacking all $T$ or $N$ equations, can also be written as:
	\begin{align}
		y &= \text{mat}_{(1)}(\mathcal X\times_3\theta')+G\Lambda' + \epsilon = \mathcal X_{(1)} (\theta\otimes I_N) +G\Lambda'+\epsilon= (\theta'\otimes I_T)\mathcal X_{(2)}' +G\Lambda'+\epsilon.\label{eq:mnvar_tensor_form}
	\end{align}
	
	Since both the network factors (contained in $\mathcal X$) and the node factors are unobserved, estimation of \eqref{eq:mnvar_tensor_form} is infeasible. 
	To make estimation feasible,  we start by substituting $\mathcal X$ with $\widehat{\mathcal X}$, which, in turn, is obtained by replacing the network factors $F_{k,t}$ with their estimates $\widehat F_{k,t}$, $k=1,\ldots, r$, given by \eqref{eq:est_factors}. 
	
	An infeasible OLS estimator of $\theta$ would then be obtained by applying the Frisch-Waugh theorem to partial out the effect of either $G$ or $\Lambda$:
	\begin{align}
		\widehat \theta^{\text{\tiny FW,G}} \otimes I_N = \left(\widehat{\mathcal X}_{(1)}' M_G \widehat{\mathcal X}_{(1)} \right)^{-1}\left(\widehat{\mathcal X}_{(1)}'M_G\, y \right)\;\text{ or }\;
		\widehat \theta^{\text{\tiny FW,$\Lambda$}} \otimes I_T = \left(\widehat{\mathcal X}_{(2)}' M_\Lambda \widehat{\mathcal X}_{(2)} \right)^{-1}\left(\widehat{\mathcal X}_{(2)}'M_\Lambda\, y' \right),\nonumber
	\end{align}
	where  $M_G:= I_T-G(G'G)^{-1}G'$ and $M_\Lambda:=I_N-\Lambda(\Lambda'\Lambda)^{-1}\Lambda'$ are the $T\times T$ and $N\times N$ linear projectors onto the spaces orthogonal to the factors and loadings spaces, respectively. 
	Likewise,  given  $\theta$, the estimators of $\Lambda$ and $G$ would be the usual PC estimators applied to the FNAR residuals $\widehat{\nu} := y- \text{mat}_{(1)}(\widehat{\mathcal X}\times_3\theta')$. 
	
	Following this reasoning, two equivalent estimators of $\theta$ are given by:
	\begin{equation}%
		\begin{array}
			[c]{c}%
			\widehat{\theta}^{\dag}:=\left(  \sum_{i=1}^{N}\widehat{X}_{i}^{\prime
			}M_{\widehat{G}^{\dag}}\widehat{X}_{i}\right)  ^{-1}\left(  \sum_{i=1}%
			^{N}\widehat{X}_{i}^{\prime}M_{\widehat{G}^{\dag}}y_{i}\right),  \\
			\widehat{\theta}^{\ast}:=\left(  \sum_{t=1}^{T}\widehat{X}_{t}^{\prime
			}M_{\widehat{\Lambda}^{\ast}}\widehat{X}_{t}\right)  ^{-1}\left(  \sum
			_{t=1}^{T}\widehat{X}_{t}^{\prime}M_{\widehat{\Lambda}^{\ast}}y_{t}\right)  ,
		\end{array}
		\label{eq:thetastar}%
	\end{equation}
	where, letting $\widehat \nu^\dag:=y-\text{mat}_{(1)}(\widehat{\mathcal X}\times_3\widehat{\theta}^{\dag'})$ and  $\widehat \nu^*:=y-\text{mat}_{(1)}(\widehat{\mathcal X}\times_3\widehat{\theta}^{*'})$, we defined (see also Appendix A)
	\begin{align}\label{eq:lambdastar}
		\widehat G^\dag := \widehat V^{\widehat\nu^\dag}\sqrt T, 	
		\quad
		\widehat \Lambda^*:= \widehat V^{\widehat\nu^*}(\widehat M^{\widehat\nu^*})^{1/2},
	\end{align}
	with $\widehat V^{\widehat\nu^\dag}$ being the $T\times q$ matrix of normalized eigenvectors of $N^{-1}\widehat \nu^{\dag}\widehat \nu^{\dag'}$, and
	$\widehat M^{\widehat\nu^*}$ being the $q\times q$ diagonal matrix of eigenvalues of $T^{-1}\widehat \nu^{*'}\widehat \nu^{*}$ with corresponding normalized eigenvectors given by the columns of the $N\times q$ matrix $\widehat V^{\widehat\nu^*}$. 
	By iterating between \eqref{eq:thetastar} and \eqref{eq:lambdastar}, we  solve such minimization and compute the two final estimators of $\theta$.  As explained at the end of Section \ref{sec:APFF}, the two are asymptotically equivalent and they might differ numerically just because of the  iterative approaches used to compute them.
	
	Finally, notice that, obviously, once we have computed $\widehat{\theta}^\dag$ we can also estimate the loadings $\Lambda$ by linear projection of $\widehat{G}^\dag$ onto $\widehat \nu^{\dag}$, and once we have computed $\widehat{\theta}^*$ we can also estimate the node factors $G$ by linear projection of $\widehat{\Lambda}^*$ onto $\widehat \nu^{*}$. These estimates, however, are not needed for estimating $\theta$.
	
	The outlined estimators are robust to the presence of autocorrelated node factors $G_t$, and, as a consequence, the FNAR errors $\nu_t$ are allowed to be both cross-sectionally and serially correlated. The adopted algorithm is similar to the one proposed by \citet{chenetal2020cnar}, who in turn adapted the approach by \citet{bai2009panel} to the NAR setting.  Under the assumption of no autocorrelation in the node factors, the OLS and GLS estimators are also valid estimators of $\theta$ (see Appendices B and C, respectively).

	\noindent\textsc{Number of factors.}
	Letting $\widehat{\mu}_j^{\mathcal W}$, $j=1,\ldots, m$, be the $j$-th largest eigenvalue of $\widehat{\Gamma}^{\mathcal W}$ defined in \eqref{eq:gamma}, we estimate the number of network factors, $r$,
	by means of the eigenvalue ratio criterion:
	\begin{equation}\label{eq:rhat}
		\widehat r:=\arg\!\!\!\!\!\!\!\max_{j=1,\ldots, r_{\max}}(\widehat{\mu}_j^{\mathcal W}/\widehat{\mu}_{j+1}^{\mathcal W}),
	\end{equation}
	where $r_{\max}$ is a predefined maximum number of network factors such that $r_{\max}<\min\{m,T,N^2\}$. This is the criterion proposed by \citet{helitrapani2022}, which generalizes the approach proposed by \citet{hanetal20} to the tensor factor model with autocorrelated idiosyncratic components. 
	
	Likewise,  letting $\widehat{\mu}_j^{\widehat \nu}$, $j=1,\ldots, N$, be the $j$-th largest eigenvalue of the sample covariance matrix of the FNAR residuals, 
	the number of node factors, $q$, is determined by means of the criterion:
	\begin{equation}\label{eq:qhat}
		\widehat q:=\arg\!\!\!\!\!\!\!\max_{j=1,\ldots, q_{\max}}(\widehat{\mu}_j^{\widehat \nu}/\widehat{\mu}_{j+1}^{\widehat \nu}),
	\end{equation}
	where $q_{\max}$ is a predefined maximum number of node factors such that $q_{\max}<\min\{T,N\}$; see \citet{ahn2013eigenvalue}.
	In practice, since the FNAR residuals $\widehat{\nu}_t$ depend on the chosen value of $q$, we can adopt an iterative procedure, which starts by over-estimating $q$ in the first stage, identical to the one described by \citet[Section C.3 in the Supplementary Material]{bai2009panel}.	
	
	Alternative approaches to estimating $r$ and $q$ are possible; see, e.g., the information criterion by \citet{baing02},   or the randomized test by \citet{trapani2018randomized} which is based on the divergence rates of the eigenvalues. Importantly, the latter could also be used to test for no factors.
	
	\section{Theory}\label{sec:theory}
	\subsection{Assumptions}

	In the following, we allow for the number of layers $m$ to grow to infinity. Therefore, all our assumptions are stated for the infinite sequence of $N\times N$ networks 
	$W_{i,t}:=\text{mat}_{(1)}( \mathcal{W}_{\cdot ,\cdot ,i, t} )$ with $i\in\mathbb N$. Equivalently, we could state the assumptions for $i=1,\ldots, m$ with $m\in\mathbb N$. Moreover, all assumptions are stated contemplating the possibility that also the number of nodes $N$ and the sample size $T$ grow to infinity.

	\begin{assumption}[Common component of multilayer network]\label{as:common_component}\
		\vspace{-10pt}
		\begin{enumerate}[label=(\roman*)]
			\item  $\lim_{m \rightarrow \infty} \norm{m^{-1} U'U - \Sigma_U} = 0$
			where $\Sigma_U$ is $r \times r$ finite and positive definite, and, for all $k \in \mathbb{N}$, $\norm{U_{k\cdot}}  \leq M_U$
			for some finite $M_U$
			independent of $k$.
			\label{as:common_component_i}
			\vspace{-10pt}
			\item 
			For all $t \in \mathbb{Z}$ and all $N \in \mathbb{N}$, 
			$ \Gamma^{\mathcal F}:=\mathbb{E} [ \mathcal{F}_{(3)t} \mathcal{F}_{(3) t}'  ] $ is $r \times r$ positive definite, and such that 
			$\norm{N^{-2} \Gamma^{\mathcal F}}\le M_{\mathcal F}$ for some finite $M_{\mathcal F}$
			independent of $N$.
			\label{as:common_component_ii}
			\vspace{-10pt}
			\item For all $N\in\mathbb N$ and all $t \in \mathbb{Z}$,
			$\mathbb E\left[\norm{N^{-1}\mathcal{F}_{(3)t}}^4\right]\le K_{\mathcal F}$ for some finite $K_{\mathcal{F}}$ independent of $t$ and $N$.
			\label{as:common_component_4}
			\vspace{-10pt}
			\item  
			For all $i,j=1,\dots,r$ and all $T, N\in\mathbb N$,
			$
			\mathbb{E} \left[  
			\abs{ \frac{1}{\sqrt{T} N } \sum_{t=1}^{T} 
				\left\{ 
				\mathcal{F}_{(3)t i \cdot} \mathcal{F}_{(3)t j \cdot}^\prime 
				- 
				\mathbb{E} \left[ \mathcal{F}_{(3)t i \cdot} \mathcal{F}_{(3)t j \cdot}^\prime \right]
				\right\}  	
			}^2 
			\right]
			\leq C_{\mathcal{F}}
			$
			for some finite $C_{\mathcal{F}}$ independent of $i,j,T$, and $N$.
			\label{as:common_component_iii}
			\vspace{-10pt}
			\item There exists an integer $\overline{M}$ such that for all $m>\overline{M}$, $r$ is a finite positive integer, independent of $m$.
			\vspace{-10pt}
			\label{as:common_component_iv}
			\item For all $h\in\mathbb N$, all $s\in\mathbb Z$, and all $l=1,\ldots,r$,
			$\mathbb E\left[\left\vert 
			\mathcal{F}_{(3)slh}\right\vert\right] \le C_{\mathcal F}^\prime$, for some finite $C_{\mathcal{F}}^\prime$ independent of $h,s,$ and $l$.
			\label{as:common_component_v}
		\end{enumerate}
	\end{assumption}

	Assumptions \ref{as:common_component}\ref{as:common_component_i} and  \ref{as:common_component}\ref{as:common_component_ii} imply that we consider only pervasive, or strong, factors. In other words, the network factors are loaded by most or all the network layers. Notice that the factor tensor has dimension $N\times N\times r$, hence the rescaling introduced in Assumption \ref{as:common_component}\ref{as:common_component_ii}, which accounts for its first two modes having diverging dimensions.
	Furthermore, under Assumption  \ref{as:common_component}\ref{as:common_component_ii}, the 2nd order moments of the process $\{N^{-1}\text{vec}(\mathcal F_{(3)t}), t\in\mathbb Z\}$ are finite and independent of time, for all $N\in\mathbb N$. 
	
	Assumptions  \ref{as:common_component}\ref{as:common_component_4} and \ref{as:common_component}\ref{as:common_component_iii} imply that, given the factors, we can consistently estimate $\Gamma^{\mathcal F} := \mathbb{E} [ \mathcal{F}_{(3)t} \mathcal{F}_{(3)t}' ]$, as proved in Lemma E.5(i).  
	Therefore, given the loadings, we can also consistently estimate $\Gamma^\chi:= U\mathbb E [ \mathcal{F}_{(3)t} \mathcal{F}_{(3)t}' ] U' $. 
	Assumption  \ref{as:common_component}\ref{as:common_component_iv} simply states that the number of factors is finite for all $m,N\in\mathbb N$ and that in order to find such factors we need $m$ to be large enough. Finally, Assumption  \ref{as:common_component}\ref{as:common_component_v} is very mild as it simply requires each element of the network factors to have finite first moment.

	\begin{assumption}[Idiosyncratic component of multilayer network]\label{as:idiosyncratic_component}\
		\vspace{-10pt}
		\begin{enumerate}[label=(\roman*)]
			\item For all $m, N \in \mathbb{N}$ and all $ t \in \mathbb{Z}$, $\mathbb{E}\left[ \mathcal{E}_{(3)t} \right]= 0_{m \times N^2}$ and $\Gamma^{\mathcal{E}} := \mathbb{E} [\mathcal{E}_{(3)t} \mathcal{E}_{(3)t}']$ is $m \times m$ positive definite.		
			\label{as:idiosyncratic_component_i}
			\vspace{-10pt}
			\item 
			For all $N \in \mathbb{N}$, all $t,s \in \mathbb{Z}$, and all $i,j=1,\ldots N^2$, 
			$
			{N^{-2}}
			\sum_{h=1}^{N^2} \sum_{k=1}^{N^2}
			\abs{
				\mathbb{E} \left[ \mathcal{E}_{(3) t i h} \mathcal{E}_{(3) s j k} \right]	 
			}
			\leq \rho_{\mathcal{E}}^{\abs{t-s}} M_{ij}  
			$
			and, for all $N \in \mathbb{N}$, all $t,s \in \mathbb{Z}$, and all $i,j,k=1, \dots, N^2$,
			$
			{N^{-2}}
			\sum_{h=1}^{N^2}
			\abs{
				\mathbb{E} \left[ \mathcal{E}_{(3) t i h} \mathcal{E}_{(3) s j k} \right]	 
			}
			\leq \rho_{\mathcal{E}}^{\abs{t-s}} M_{ij}  
			$
			for 
			some finite $\rho_{\mathcal{E}}$ and $M_{ij}$ independent of $t,s,k$ and $N$ such that
			$0 \leq \rho_{\mathcal{E}} <1 $, $\sum_{i=1, i\ne j}^m M_{ij}  \leq  M_{\mathcal{E}}$ and  $\sum_{j=1,j\ne i}^m M_{ij}  \leq  M_{\mathcal{E}}$, for some finite $ M_{\mathcal{E}}$ independent of $i,j$ and $m$.
			\label{as:idiosyncratic_component_ii}
			\vspace{-10pt}
			\item For all $i,j\in\mathbb N$ and all $t \in \mathbb{Z}$,
			$\mathbb E[\vert \mathcal{E}_{(3)t ij}\vert^4]\le K_{\mathcal E}$ for some finite $K_{\mathcal{E}}$ independent of $i,j,t$.
			\label{as:idiosyncratic_component_4}
			\vspace{-10pt}
			\item 
			For all $m,T, N \in \mathbb{N}$ and all $j=1,\ldots, N^2$ and all $s=1,\ldots, T$,
			\begin{equation*}
				\mathbb{E} \left[  \abs{ \frac{1}{\sqrt{mT} N^{2} } \sum_{i=1}^{m} \sum_{t=1}^{T} \sum_{h=1}^{N^2} \sum_{k=1}^{N^2} 
					\left\{ 
					\mathcal{E}_{(3)t i h} \mathcal{E}_{(3)t j k}
					-
					\mathbb{E} \left[  \mathcal{E}_{(3)t i h} \mathcal{E}_{(3)t j k} \right]
					\right\} }^2 \right]
				\leq C_{\mathcal{E}}
			\end{equation*}		
			and
			\begin{equation*}
				\mathbb{E} \left[  \abs{ \frac{1}{\sqrt{mT} N^{2} } \sum_{i=1}^{m} \sum_{t=1}^{T} \sum_{h=1}^{N^2} \sum_{k=1}^{N^2} 
					\left\{ 
					\mathcal{E}_{(3)t i h} \mathcal{E}_{(3)s i k} 
					-
					\mathbb{E} \left[  \mathcal{E}_{(3)t i h} \mathcal{E}_{(3)s i k} \right]
					\right\} }^2 \right]
				\leq C_{\mathcal{E}}
			\end{equation*}		
			for some finite $C_{\mathcal{E}}$ independent of $j, s, m, T, N$.
			\label{as:idiosyncratic_component_iv}
			\vspace{-10pt}
			\item 
			For all $m,T, N \in \mathbb{N}$ and all $j=1,\ldots, N^2$ and all $s=1,\ldots, T$,
			\begin{equation*}
				\mathbb{E} \left[  \abs{ \frac{1}{\sqrt{mT} N} \sum_{i=1}^{m} \sum_{t=1}^{T} \sum_{h=1}^{N^2}
					\left\{ 
					\mathcal{E}_{(3)t i h} \mathcal{E}_{(3)t j h}
					-
					\mathbb{E} \left[  \mathcal{E}_{(3)t i h} \mathcal{E}_{(3)t j h} \right]
					\right\} }^2 \right]
				\leq C_{\mathcal{E}}
			\end{equation*}		
			and
			\begin{equation*}
				\mathbb{E} \left[  \abs{ \frac{1}{\sqrt{mT} N} \sum_{i=1}^{m} \sum_{t=1}^{T} \sum_{h=1}^{N^2}
					\left\{ 
					\mathcal{E}_{(3)t i h} \mathcal{E}_{(3)s i h} 
					-
					\mathbb{E} \left[  \mathcal{E}_{(3)t i h } \mathcal{E}_{(3)s i h} \right]
					\right\} }^2 \right]
				\leq C_{\mathcal{E}}
			\end{equation*}		
			for some finite $C_{\mathcal{E}}$ independent of $j, s, m, T, N$.
			\label{as:idiosyncratic_component_v}
		\end{enumerate}
	\end{assumption}
	
	This assumption controls the serial and cross-sectional dependence of the entries of the idiosyncratic tensor. In particular, according to Assumption \ref{as:idiosyncratic_component}\ref{as:idiosyncratic_component_ii}, the covariance across time and layers is controlled in a standard way; hence, the classical summability conditions hold  (see Lemma E.1 and \citealp[Assumption C]{bai2003inferential}).  The covariances across the elements of the $N^2$-dimensional vector $\mathcal E_{(3)t j\cdot}$ are of order $N^4$ for any given $t=1,\ldots, T$ and $j=1,\ldots, m$, and we require to scale their sum by $N^2$, thus assuming a standard summability condition.  
	Assumptions  \ref{as:idiosyncratic_component}\ref{as:idiosyncratic_component_4}, \ref{as:idiosyncratic_component}\ref{as:idiosyncratic_component_iv}, and \ref{as:idiosyncratic_component}\ref{as:idiosyncratic_component_v} imply that, 
	given the idiosyncratic tensor, we can consistently estimate $m^{-1}N^{-2}\Gamma^{\mathcal E}$, for any $m,N\in\mathbb N$, as proved in Lemma E.5(ii).

	\begin{assumption}[Moment conditions - part 1]\label{as:MB_Indep}
		For all $i,j\in\mathbb N$, all $k=1,\ldots,r$, and all $t\in\mathbb Z$,  $\mathbb E[\mathcal{F}_{(3)tkj}\mathcal{E}_{(3)tij}]=0$,
		and, for all $m,N,T\in\mathbb N$ and all $t=1,\ldots, T$,
		\begin{align}
			&\mathbb{E} \left[\frac 1{m N^{2}} \sum_{i=1}^{m} \norm{ \frac 1{\sqrt T} \sum_{t=1}^{T}  \mathcal{F}_{(3)t} \mathcal{E}_{(3)t i\cdot}^\prime  }_F^2 \right] \leq C_{\mathcal{F}\mathcal{E}},\nonumber\\
			&\mathbb{E}\left[\frac 1{ N^{4} }\left\Vert\frac 1{\sqrt{mT}}\sum_{i=1}^{m}  \sum_{s=1}^{T}\mathcal{F}_{(3)s}\left\{\mathcal{E}_{(3)s i \cdot}^\prime \mathcal{E}_{(3)t i \cdot}-\mathbb E[\mathcal{E}_{(3)s i \cdot}^\prime \mathcal{E}_{(3)t i \cdot}]\right\}  \right\Vert_F^2\right]\le C_{\mathcal{F}\mathcal{E}}^\prime,\nonumber
		\end{align}
		for some finite $C_{\mathcal{F}\mathcal{E}}$ and  $C_{\mathcal{F}\mathcal{E}}^\prime$ independent of $t$, $m, N$, and $T$.
	\end{assumption}
	
	Uncorrelatedness of 	the processes $\{\mathcal F_{(3)t}\}$ and $\{\mathcal E_{(3)t}\}$ is a natural assumption, while the moment conditions controlling higher-order dependence are standard and are direct extensions of what typically assumed in the vector case  \citep[Assumptions D and F1]{bai2003inferential}.
	
	Define the $m\times m$ matrix $  \Gamma^{\mathcal W}:=\mathbb E[\mathcal W_{(3)t}\mathcal W_{(3)t}']$ and let $\Gamma^\chi$ be as previously defined. Denote the eigenvalues of $\Gamma^\chi$ as $\mu_j^\chi$, $j=1,\ldots, r$, in decreasing order. Then, as established by Lemmas E.1(i) and  E.1(iv), we have  that  $\mu_j^\chi\asymp mN^2$ and $\norm{N^{-2}\Gamma^{\mathcal{E}}}$ is finite for all $N\in\mathbb N$.	
	As a consequence, by Assumption \ref{as:MB_Indep} and Weyl's inequality, the matrix $\Gamma^{\mathcal W}=\Gamma^\chi+\Gamma^{\mathcal E}$ is characterized by an eigengap between the $r$-th and the $(r+1)$-th largest eigenvalues which widens as $m\to\infty$.
	This property allows us to identify the number of network factors $r$ and it is the rationale for the eigenvalue ratio criterion for estimating $r$ defined in \eqref{eq:rhat}. This also means that the network factor model in \eqref{eq:W_decomp3} is always identified as long as $m\to\infty$.

	In general, the network factors and their loadings are not separately identified unless we impose further restrictions. To this end we make the following assumption.		
	\begin{assumption}[Identification of network factors and loadings]\label{as:MB_Assumption_9}\
		\vspace{-10pt}
		\begin{enumerate}[label=(\roman*)]
			\item For all $m \in \mathbb{N}$, $m^{-1}U'U$ is diagonal with distinct entries. \label{as:MB_Assumption_9_i}
			\vspace{-10pt}
			\item For all $N, T \in \mathbb{N}$, ${N^{-2}T^{-1}}\sum_{t=1}^{T} \mathcal{F}_{(3)t} \mathcal{F}_{(3)t}' = I_r$. \label{as:MB_Assumption_9_ii}
		\end{enumerate}	
	\end{assumption}
	
	Under Assumption \ref{as:MB_Assumption_9} the columns of the loadings matrix $U$ and the layers of the tensor factor $\mathcal F_t$ are identified up to a sign multiplication. 
	This identification scheme is a classical one adopted for example by \citet{bai2009panel} in the vector factor model case.		
	\begin{assumption}[CLTs]\label{as:CLT}\
		\vspace{-10pt}
		\begin{enumerate}[label=(\roman*)]
			\item For any given $i=1,\ldots, m$, 
			as $N,T\to\infty$, 
			$
			\frac{1}{N\sqrt {T}}  \sum_{t=1}^{T} \mathcal{F}_{(3)t}  \mathcal{E}_{(3)ti\cdot}' \overset{d}{\to}\mathcal N(0_r,\Phi_i),
			$
			where
			$
			\Phi_i:=\lim_{N,T\to\infty}  \mathbb E\left[\left (\frac 1{N\sqrt {T }}\sum_{t=1}^{T} \mathcal{F}_{(3)t} \mathcal{E}_{(3)t i \cdot}' \right)
			\left(\frac 1{N\sqrt { T }}\sum_{t=1}^{T} \mathcal{F}_{(3)t} \mathcal{E}_{(3)t i \cdot}' \right)'
			\right].
			$
			\label{as:CLT_i}
			\vspace{-10pt}
			\item For any given $t=1,\ldots, T$ and $j=1,\ldots, N^2$, as $m\to\infty$, 
			$
			\frac{   1}{\sqrt{m}}\sum_{i=1}^m u_i \mathcal E_{(3)tij}\overset{d}{\to}\mathcal N(0_{r},\Pi_{tj}),
			$
			where
			$\Pi_{tj}:=\lim_{m\to\infty}\mathbb E\left[\left(
			\frac{1}{\sqrt{m}}\sum_{i=1}^m u_i \mathcal {E}_{(3)t i j}
			\right)
			\left(
			\frac{1}{\sqrt{m}}\sum_{i=1}^m u_i \mathcal {E}_{(3)t i j}
			\right)'
			\right]$ 
			and $u_i'$ is the $i$-th row of $U$.
			\label{as:CLT_iii} 
		\end{enumerate}	
	\end{assumption}
	
	Assumption \ref{as:CLT}\ref{as:CLT_i} is standard in the vector factor model case; i.e., when $N=1$, where it is satisfied for example by strong-mixing processes \citep[Assumption F4]{bai2003inferential}. 
	In Assumption 
	\ref{as:CLT}\ref{as:CLT_iii}	
	we directly assume a cross-sectional CLT, which is a standard approach in the vector factor model case \citep[Assumption F3]{bai2003inferential}. 
	
	The FNAR errors $\nu_t$ follow a factor model given in \eqref{eq:moderror}, characterized by the following assumption.
	
	\begin{assumption}[FNAR errors]\label{as:errors_nu}\
		\vspace{-10pt}
		\begin{enumerate}[label=(\roman*)]
			\item 
			$
			\lim_{N \to \infty}
			\norm{ N^{-1} \Lambda' \Lambda - \Sigma_{\Lambda} } = 0
			$
			where $\Sigma_{\Lambda}$ is $q \times q$ finite and positive definite, and, for all $i \in \mathbb{N}$, 
			$\norm{\Lambda_{i\cdot}} \leq M_{\Lambda}$
			for some finite $M_{\Lambda}$ independent of $i$.
			\label{as:errors_nu_common_i}
			\vspace{-10pt}
			\item For all $t\in\mathbb Z$, 
			$\mathbb{E}[G_t] = 0_q $, 
			$\Gamma^{G} := \mathbb{E}\left[G_t G_t' \right]$ is $q \times q$ positive definite,
			and such that
			$\norm{\Gamma^{G}} \leq M_G$
			for some
			finite $ M_G$ independent of $t$.
			\label{as:errors_nu_common_ii}
			\vspace{-10pt}			
			\item  For all $t\in\mathbb Z$, $\mathbb E[\norm{G_t}^4]\le K_G$ for some finite $K_G$ independent of $t$.
			\label{as:errors_nu_common_iii}
			\vspace{-10pt}
			\item 
			For all $i,j = 1, \dots, q$ and all $T \in \mathbb{N}$, 
			$
			\mathbb{E} \left[  \abs{ \frac{1}{\sqrt{T}} \sum_{t=1}^{T} 
				\left\{ 
				G_{it} G_{jt}
				-
				\mathbb{E} \left[ G_{it} G_{jt} \right]
				\right\} }^2 \right]
			\leq C_{G}
			$
			for some finite $ C_{G}$ independent of $i,j$, and $T$. \label{as:errors_nu_common_iia}
			\vspace{-10pt}
			\item 
			There exists an integer $\underline{N}$ such that for all $N > \underline{N}$, $q$ is a finite positive integer, independent of $N$.
			\label{as:errors_nu_common_iv}
			\vspace{-10pt}
			\item For all $i \in \mathbb{N}$ and all $t\in\mathbb Z$, 
			$\mathbb{E}[\epsilon_{it}] = 0$,
			$\mathbb E[\epsilon_{it}^2]= \sigma_i^2$ such that $\sigma_i^2 \ge \underline M_{\epsilon}$ and $\sigma_i^2\le \overline M_{\epsilon}$ for some finite $\underline M_{\epsilon}$ and 
			$\overline M_{\epsilon}$ independent of $i$ and $t$. \label{as:errors_nu_idio_i}
			\vspace{-10pt}
			\item 
			For all $i,j \in \mathbb{N}$ and all $t,s \in \mathbb{Z}$, $\mathbb{E} \left[ \epsilon_{ it} \epsilon_{js} \right]=0$ if $i\ne j$ and 
			$\mathbb{E} \left[ \epsilon_{ it} \epsilon_{is} \right]=0$ if $t\ne s$. \label{as:errors_nu_idio_ii}
			\vspace{-10pt}
			\item 
			For all $i \in \mathbb N$ and all $t \in \mathbb{Z}$, 
			$\mathbb{E}[\epsilon_{it}^4] \leq K_{\epsilon}$
			for some finite $K_{\epsilon}$ independent of $i$ and $t$. \label{as:errors_nu_idio_iii}
			\vspace{-10pt}
			\item 
			For all $N, T \in \mathbb{N}$, 
			$
			\mathbb{E} \left[  \abs{ \frac{1}{\sqrt{NT}} \sum_{i=1}^{N} \sum_{t=1}^{T} 
				\left\{ 
				\epsilon_{it}^2
				-
				\mathbb{E} \left[ \epsilon_{it}^2 \right]
				\right\} }^2 \right]
			\leq C_{\epsilon}
			$
			for some finite $C_{\epsilon}$ independent of $N$ and $T$.\label{as:errors_nu_idio_iv}
		\end{enumerate}
	\end{assumption}	
	
	This assumption is similar to the usual set of assumptions for the vector factor model \citep{bai2003inferential}. The comments to these assumptions are analogous to those previously made for the network factors and are therefore omitted. 
	Concerning the idiosyncratic components, we follow \citet{chenetal2020cnar} and assume zero correlations both in time and across units. Given that we are considering a factor structure for the FNAR errors, this assumption is not very restrictive, as most of the correlations are likely to be already captured by the lagged terms in the FNAR and by the common factors $G_t$. Nevertheless, it is possible to develop the following asymptotic theory by allowing for the usual kind of weak cross- and autocorrelations between the components of $\epsilon_t$.  
	
	\begin{assumption}[Independence of network and node factors]\label{as:MB_Assumption_9_nu_common_idio}
		For all $N\in\mathbb N$, the processes $\{\mathcal F_t,\,t\in\mathbb Z\}$,  $\{\epsilon_{t},\, t \in \mathbb{Z}\}$, 
		and $\{G_{t},\, t \in \mathbb{Z}\}$ are
		mutually independent.
	\end{assumption}
	
	Assumption \ref{as:MB_Assumption_9_nu_common_idio} is taken from 
	\citet[Assumption D]{bai2009panel} 
	and is made just to simplify the proof. 
	In principle it could be relaxed to allow for weak dependence between $\{G_t\}$ and $\{\epsilon_{t}\}$ 
	by means of a condition similar to those required in Assumption \ref{as:MB_Indep}, which, in turn, derives from \citet[Assumption D]{bai2003inferential}.
	
	Because of Assumptions \ref{as:errors_nu}\ref{as:errors_nu_common_i}, \ref{as:errors_nu}\ref{as:errors_nu_common_ii}, 
	\ref{as:errors_nu}\ref{as:errors_nu_idio_i},
	\ref{as:errors_nu}\ref{as:errors_nu_idio_ii},
	and \ref{as:MB_Assumption_9_nu_common_idio}, the FNAR errors have covariance matrix $V=  \Lambda\Gamma^G\Lambda' + S$, with $S=\mathbb{E} \left[ \epsilon_{t} \epsilon_{t}' \right]$, which is positive definite for all $N\in\mathbb N$. This also implies that $V^{-1}$ is finite for all $N\in\mathbb N$. Moreover, $V$ has the usual eigengap property; i.e., its largest $q$ eigenvalues diverge at rate $N$, while the remaining $N-q$ stay bounded for all $N\in\mathbb N$. This implies that the factor model in \eqref{eq:moderror}, and therefore the number of factors $q$, is always identified as $N\to\infty$. This is the rationale for the eigenvalue ratio criterion for estimating $q$ defined in \eqref{eq:qhat}.

	\begin{assumption}[Identification of node factors and loadings]\label{as:errors_nu_id}\
		\vspace{-10pt}
		\begin{enumerate}[label=(\roman*)]
			\item For all $N \in \mathbb{N}$, ${N}^{-1}\Lambda'\Lambda$ is diagonal with distinct entries. \label{as:errors_nu_id_i}
			\vspace{-10pt}
			\item For all $T \in \mathbb{N}$, ${T}^{-1}\sum_{t=1}^{T} G_tG_t' = I_q$. \label{as:errors_nu_id_ii}
		\end{enumerate}	
	\end{assumption}
	
	Under Assumption \ref{as:errors_nu_id} the columns of the loadings matrix $\Lambda$ and the factors $G_t$ are identified up to a sign multiplication. 	
	
	Turning to the FNAR defined in \eqref{eq:famnvar}, we make the following assumption.
	
	\begin{assumption}[Stability of FNAR]\label{as:stability}\
		\vspace{-10pt}
		\begin{enumerate}[label=(\roman*)]
			\item For all $t\in\mathbb Z$ and all $N\in\mathbb N$, 
			$\det(I_N-\rho I_N- N^{-1}\sum_{j=1}^r\beta_j\mathbb E[F_{j,t}])\ne 0$.
			\label{as:stability_i}
			\vspace{-10pt}
			\item  For all $t\in\mathbb Z$ and all $N\in\mathbb N$, 
			$\det ({\rho^2 I_{N^2}+ {N^{-2}}\sum_{j=1}^r \beta_j^2 \mathbb E [ F_{j,t}\otimes F_{j,t}]}- zI_{N^2}) = 0$ has roots $z_j^*\in\mathbb C$, $j=1,\ldots,N^2$, such that $|z_j^*|\le C_S$ for some finite $C_S<1$ independent of $j,t$, and $N$.
			\label{as:stability_ii}
		\end{enumerate}
	\end{assumption}
	
	This assumption is a generalization to the case of random multivariate AR models of the usual stability conditions for a VAR.
	As shown below it implies, together with Assumptions \ref{as:common_component}\ref{as:common_component_ii} and \ref{as:MB_Assumption_9_nu_common_idio}, that the FNAR has a stationary solution for all $N\in\mathbb N$.  Notice that Assumption \ref{as:stability}\ref{as:stability_i} is stated for the general case in which $\mathbb E[F_{j,t}]\ne 0$, otherwise the condition needed to ensure the existence of the mean is simply $|\rho|<1$.
	
	\begin{assumption}[Moment conditions - part 2]
		\label{as:X_nu_v} For all $m,N,T\in\mathbb N$,
		\begin{align}
			&\mathbb E\left[\norm{\frac 1{\sqrt {mT}N^{2}}\sum_{t=1}^T\sum_{i=1}^m u_i \mathcal E_{(3)ti\cdot} (y_{t-1}\otimes X_t)}^2 \right]\le\mathfrak K_1,\nonumber\\
			&\mathbb E\left[\norm{\frac 1{\sqrt {mT}N^{2}}\sum_{t=1}^T\sum_{i=1}^m u_i \mathcal E_{(3)ti\cdot} (y_{t-1}\otimes \nu_t)}^2 \right]\le\mathfrak K_2,\nonumber
		\end{align} 
		for some finite $\mathfrak K_1$ and $\mathfrak K_2$ independent of $m,N$, and $T$.
	\end{assumption}
	
	To get an intuition of this assumption, consider the $m\times (r+2)$ matrix process $\{\mathcal E_{(3)t} (y_{t-1}\otimes X_t)\}$. We are saying that this process is weakly correlated along the time dimension, which is a standard requirement, but it is also weakly correlated across its $m$ rows. The latter requirement is fulfilled by the idiosyncratic terms $\mathcal E_{(3)t}$ via Assumption \ref{as:idiosyncratic_component}\ref{as:idiosyncratic_component_ii}, and here is extended to the case in which $\mathcal E_{(3)t}$ is multiplied by $y_{t-1}\otimes X_t$ which is weakly dependent of $\mathcal E_{(3)t}$ because of Assumption \ref{as:MB_Indep}.

	\begin{assumption}[CLT for FNAR]\label{as:Z}	
		Let $Z_i:={
			M_{ G}X_i-\frac 1N\sum_{k=1}^N(\Lambda_i'\left(\frac{\Lambda'\Lambda} N)^{-1}\Lambda_k \right)M_{ G}X_k
		}$ such that $Z_i :=(Z_{i1} \cdots Z_{iT})'$ is $T\times (r+2)$
		and $W_t:={
			M_{ \Lambda}X_t-\frac 1T\sum_{s=1}^T(G_t'G_s )M_{ \Lambda}X_s
		}$ such that $W_t :=(W_{1t} \cdots W_{Nt})'$ is $N\times (r+2)$.
		Then, as $N,T\to\infty$,
		\begin{enumerate}[label=(\roman*)]
			\item $
			\frac{1}{\sqrt{NT}} \sum_{i=1}^{N} Z_i' \epsilon_{i} \overset{d}{\to} N(0,D_1)
			$, 
			where $D_1 := \lim_{N \to \infty}  \frac{1}{N} \sum_{i=1}^N \mathbb{E} \left[ Z_{it}Z_{it}'\right] \sigma_i^2$  is an $(r+2)\times (r+2)$ positive definite matrix.
			\label{as:Z_i}
			\vspace{-10pt}
			\item $\frac{1}{NT} \sum_{i=1}^{N} Z_i' Z_{i} \overset{p}{\to}\Sigma_{ZZ}$,
			where 
			$\Sigma_{ZZ}:=\lim_{N\to\infty}\frac 1N \sum_{i=1}^{N} \mathbb E[Z_{it} Z_{it}']$ is an $(r+2)\times (r+2)$ positive definite matrix.
			\label{as:Z_ii}
			\vspace{-10pt}
			\item $
			\frac{1}{\sqrt{NT}} \sum_{t=1}^{T} W_t' \epsilon_{t} \overset{d}{\to} N(0,D_2)
			$, 
			where $D_2 := \lim_{N \to \infty}  \frac{1}{N} \sum_{i=1}^N \mathbb{E} \left[ W_{it}W_{it}'\right] \sigma_i^2$  is an $(r+2)\times (r+2)$ positive definite matrix.
			\label{as:Z_iii}
			\vspace{-10pt}
			\item $\frac{1}{NT} \sum_{t=1}^{T} W_t' W_{t} \overset{p}{\to}\Sigma_{WW}$,
			where 
			$\Sigma_{WW}:=\lim_{N\to\infty}\frac 1N \sum_{i=1}^{N} \mathbb E[W_{it} W_{it}']$ is an $(r+2)\times (r+2)$ positive definite matrix.
			\label{as:Z_iv}
		\end{enumerate}
		
	\end{assumption}
	
	Notice that we do not rule out autocorrelation in the node factors, as only the node specific idiosyncratic components are required to have no correlation for the above CLTs to hold, and this is ensured by Assumption \ref{as:errors_nu}\ref{as:errors_nu_idio_ii}. These assumptions are similar to the conditions in \citet[Assumptions A and E]{bai2009panel}.

	\subsection{Stationarity}
	In order to develop the theory for the FNAR, we first discuss under which conditions equation \eqref{eq:famnvar} admits a stationary causal solution. Given the difficulty of the problem we limit ourselves to consider weakly stationary solutions. This poses two issues. First, the FNAR is defined for an $N$-dimensional vector $y_t$ where we allow $N\to\infty$. Second, the FNAR is an autoregressive model with stochastic time-varying coefficients. Regarding the former issue, we adopt the definition proposed by \citet{zhuetal17}. 
	
	\begin{definition}\label{def:station}
		Let $\{y_t\}$ be an $N$-dimensional stochastic process with $N\in\mathbb N$. Let $W:=\{\omega:=(\omega_1\cdots\omega_N)' \in\mathbb R^N\,:\, \sum_{i=1}^N \abs{\omega_i}<\infty , N\in\mathbb N\}$. Then, $\{y_t\}$ is weakly stationary if for all $N\in\mathbb N$ and any given $\omega\in W$, $y_t^\omega:=\lim_{N\to\infty} \omega'y_t$ exists almost surely and $\{y_t^w\}$ is weakly stationary and causal.
	\end{definition}	
	
	Turning to the second problem, we have the following result.
	
	\begin{proposition}[Stationarity of FNAR]\label{prop:station}
		Under Assumptions \ref{as:common_component}\ref{as:common_component_ii}, \ref{as:errors_nu}\ref{as:errors_nu_idio_i}, \ref{as:MB_Assumption_9_nu_common_idio}, and \ref{as:stability}, for all $N\in\mathbb N$, the FNAR has a unique weakly stationary and causal solution. 		
	\end{proposition}		
	
	In general, one might object that when $N\to\infty$ a meaningful concept of stationarity cannot be stated, as no causal solution to the FNAR can exist since Assumption \ref{as:stability} will break down, see, e.g., the remark by \citet{zhou2020network}, in a similar context. What we mean by Definition \ref{def:station} and Proposition \ref{prop:station} is that we are implicitly assuming that there exists a space where the causal solution is well-defined even when $N \to\infty$. This is the same approach adopted by \citet{zhuetal17}. An interesting implication of this definition is that, under our assumptions, we can ensure that any finite linear combination of the elements of $\{y_t\}$ satisfies a finite dimensional FNAR with a causal solution.

	\subsection{Asymptotic properties of network factors and FNAR coefficients}\label{sec:APFF}
	Consistency and asymptotic normality of the estimated network factor loadings are given next.
	
	\begin{theorem}[Consistency and asymptotic normality of loadings]\label{theorem:CLT_loadings}\
		\vspace{-10pt}
		\begin{enumerate}[label=(\roman*)]
			\item Under Assumptions \ref{as:common_component}-\ref{as:MB_Assumption_9}, 
			as $m,N,T \to \infty$,
			\[
			\norm{ \frac{\widehat{U} - U J}{\sqrt{m}} } 
			= O_p \left(
			\max \left( \frac{1}{N \sqrt{T}}, \frac{1}{ m  } \right)  	\right),
			\]
			where $J$ is a $r \times r$ diagonal matrix whose diagonal entries are equal to $\pm 1$.
			\label{theorem:CLT_loadings_i}
			\vspace{-10pt}
			\item Under Assumptions \ref{as:common_component}-\ref{as:CLT}, for any given $i=1,\ldots, m$, 
			as $m,N,T\to\infty$, 	if $N\sqrt T/m \to 0$,	
			\[
			N\sqrt T \left(\widehat u_i'-u_i' J \right) \overset{d}{\to}\mathcal N\left(0_r, \Phi_i\right),
			\]
			where $\widehat u_i'$ and $u_i'$ are the $i$-th rows of $\widehat U$ and $U$, respectively,  $\Phi_i$ is defined in Assumption \ref{as:CLT}\ref{as:CLT_i}, and $J$ is 			defined in part \ref{theorem:CLT_loadings_i}.
			\label{theorem:CLT_loadings_ii}
		\end{enumerate}
	\end{theorem}
	
	Theorem \ref{theorem:CLT_loadings} shows that, when applying PCA to a given mode of the tensor $\mathcal W_t$, the dimensions of all other modes contribute to a faster convergence rate, hence allowing for more degrees of freedom. This is an advantage with respect to the vector case, since even for moderately small values of $T$ we can still have good estimates of the loadings matrix and therefore of the network factors. In particular, 
	we see that the estimated loadings vector $\widehat u_i$ has a consistency rate $\min(m,N\sqrt T)$ and is asymptotically normal if $N\sqrt T/m\to 0$. This is the generalization to the multilayer network case (i.e., to order-3 tensors) of the usual vector case, which corresponds to setting $N=1$ (see \citealp[Theorem 2]{bai2003inferential}). 
	
	Next we prove consistency and asymptotic normality of the estimated network factors.
	
	\begin{theorem}[Consistency and asymptotic normality of network factors]\label{theorem:CLT_factors}\
		\vspace{-10pt}
		\begin{enumerate}[label=(\roman*)]
			\item Under Assumptions \ref{as:common_component}-\ref{as:MB_Assumption_9},
			for any given $t=1,\ldots, T$, as $m,N,T \to \infty$,
			\[
			\norm{ \frac{\widehat{\mathcal{F}}_{(3)t} - J \mathcal{F}_{(3)t}}{N} } = 
			O_p \left(\max 
			\left(
			\frac 1{N^{2} T},
			\frac{1}{\sqrt{m} } 
			\right) \right), 
			\]
			where 
			$J$ is defined in Theorem \ref{theorem:CLT_loadings}\ref{theorem:CLT_loadings_i}. 
			\label{theorem:CLT_factors_i}
			\vspace{-10pt}
			\item Under Assumptions \ref{as:common_component}-\ref{as:CLT}, for any given $t=1,\ldots, T$ and  $j=1,\ldots,N^2$, as $m,N,T\to\infty$, 
			if $\sqrt m/(N^2T)\to 0$,	
			\[
			\sqrt{m}\left(\widehat {\mathcal F}_{(3)t\cdot j}-J {\mathcal F}_{(3)t\cdot j} 
			\right)\overset{d}{\to}\mathcal N\left(0_{r}, \Sigma_U^{-1}\Pi_{tj}\Sigma_U^{-1} \right),
			\]
			where $\Pi_{tj}$ is defined in Assumption \ref{as:CLT}\ref{as:CLT_iii} and $J$ is defined in Theorem \ref{theorem:CLT_loadings}\ref{theorem:CLT_loadings_i}.
			\label{theorem:CLT_factors_iii}
		\end{enumerate}
	\end{theorem}
	
	Theorem \ref{theorem:CLT_factors}\ref{theorem:CLT_factors_i} proves consistency of the whole network factor tensor.  
	Theorem \ref{theorem:CLT_factors}\ref{theorem:CLT_factors_iii} proves asymptotic normality of any given column of $\widehat {\mathcal F}_{(3)t}$, which is equivalent to asymptotic normality of any of the $N^2$ entries of each of the $r$ layers of the multilayer network factor $\widehat {\mathcal F}_t$. 
	This is the natural generalization to the multinetwork case of the usual vector case, i.e., when $N=1$ (see \citealp[Theorem 1]{bai2003inferential}).

	
	We then turn to the asymptotic properties of the estimated FNAR coefficients.
	Hereafter, let 
	\begin{equation}\label{eq:barJ}
		\bar J :=\left[\begin{array}{cc}
			J &\!\!\!\!0_{r\times 2}\\ [-3pt]
			0_{2\times r}&\!\!\!\! I_2	
		\end{array}
		\right],
	\end{equation} 
	with $J$ as in Theorem \ref{theorem:CLT_loadings}\ref{theorem:CLT_loadings_i}.
	Then, we analyze the properties of the estimators $\widehat{\theta}^\dag$ and $\widehat{\theta}^*$ by noticing that
	\begin{equation}%
		\begin{array}
			[c]{c}%
			\widehat{\theta}^{\dag}-\bar{J}\theta=\left(  \frac{1}{NT}\sum_{i=1}%
			^{N}\widehat{X}_{i}^{\prime}M_{\widehat{G}^{\dag}}\widehat{X}_{i}\right)
			^{-1}\left(\frac{1}{NT}  \sum_{i=1}^{N}\widehat{X}_{i}^{\prime}M_{\widehat
				{G}^{\dag}}(G\Lambda_{i}+\epsilon_{i})+\frac{1}{NT}\sum_{i=1}^{N}\widehat{X}_{i}^{\prime
			}M_{\widehat{G}^{\dag}}u_{i}\right)  \\
			\widehat{\theta}^{\ast}-\bar{J}\theta=\left(  \frac{1}{NT}\sum_{t=1}%
			^{T}\widehat{X}_{t}^{\prime}M_{\widehat{\Lambda}^{\ast}}\widehat{X}%
			_{t}\right)  ^{-1}\left(\frac{1}{NT}  \sum_{t=1}^{T}\widehat{X}_{t}^{\prime
			}M_{\widehat{\Lambda}^{\ast}}(\Lambda G_{t}+\epsilon_{t})+\frac{1}{NT}\sum_{t=1}%
			^{T}\widehat{X}_{t}^{\prime}M_{\widehat{\Lambda}^{\ast}}u_{t}\right)
		\end{array}
		\label{eq:thetastarAE}%
	\end{equation}
	where $u_i:= (X_i\bar J-\widehat X_i)\bar J\theta$ and $u_t:= (X_t\bar J-\widehat X_t)\bar J\theta$, and recall that $\theta := (\beta', \rho, \alpha )'$.

	The following theorem holds. 
	
	\begin{theorem}[CLT for FNAR coefficients estimated by iterative OLS]\label{th:CLT_FNAR_BAI}
		Under Assumptions \ref{as:common_component}-\ref{as:X_nu_v} and \ref{as:Z}, and if ${ \sqrt {NT}}/ m \to 0$ and ${ {N}}/ {m} \to 0$, as $m,N,T\to\infty$,  then:
		\begin{enumerate}[label=(\roman*)]
			\item if $T/N\to 0$ and $\sqrt N/T\to 0$, we have 
			\begin{equation}
				\sqrt {NT}(\widehat{\theta}^{\dag}-\bar J\theta)
				\overset{d}{\to}
				\mathcal{N} ( 0_{r+2},
				\Sigma_{ZZ}^{-1} D_1\Sigma_{ZZ}^{-1}
				),\nonumber
			\end{equation}
			where $D_1$ and $\Sigma_{ZZ}$ are defined in Assumptions \ref{as:Z}\ref{as:Z_i} and \ref{as:Z}\ref{as:Z_ii}, and $\bar J$  is defined in \eqref{eq:barJ};
			\label{th:CLT_FNAR_BAI_i}
			\vspace{-10pt}
			\item if $\sqrt T/N\to 0$ and $\sqrt N/T\to 0$ and $\sigma_i^2=\sigma^2$ for all $i=1,\ldots, N$, we have 
			\begin{equation}
				\sqrt {NT}(\widehat{\theta}^{\dag}-\bar J\theta)
				\overset{d}{\to}
				\mathcal{N} ( 0_{r+2},\sigma^2
				\Sigma_{ZZ}^{-1} 
				),\nonumber
			\end{equation}
			where $\Sigma_{ZZ}$ is defined in Assumption \ref{as:Z}\ref{as:Z_ii}, and $\bar J$  is defined in \eqref{eq:barJ};
			\label{th:CLT_FNAR_BAI_ii}
			\vspace{-10pt}
			\item if $T/N\to 0$ and $\sqrt N/T\to 0$, we have 
			\begin{equation}
				\sqrt {NT}(\widehat{\theta}^{*}-\bar J\theta)
				\overset{d}{\to}
				\mathcal{N} ( 0_{r+2},
				\Sigma_{WW}^{-1} D_2\Sigma_{WW}^{-1}
				),\nonumber
			\end{equation}
			where $D_2$ and $\Sigma_{WW}$ are defined in Assumptions \ref{as:Z}\ref{as:Z_iii} and \ref{as:Z}\ref{as:Z_iv}, and $\bar J$  is defined in \eqref{eq:barJ};
			\label{th:CLT_FNAR_BAI_iii}
			\vspace{-10pt}
			\item if $\sqrt T/N\to 0$ and $\sqrt N/T\to 0$ and $\sigma_i^2=\sigma^2$ for all $i=1,\ldots, N$, we have 
			\begin{equation}
				\sqrt {NT}(\widehat{\theta}^{*}-\bar J\theta)
				\overset{d}{\to}
				\mathcal{N} ( 0_{r+2},\sigma^2
				\Sigma_{WW}^{-1} 
				),\nonumber
			\end{equation}
			where $\Sigma_{WW}$ is defined in Assumption \ref{as:Z}\ref{as:Z_iv}, and $\bar J$  is defined in \eqref{eq:barJ};
			\label{th:CLT_FNAR_BAI_iv}	
		\end{enumerate}
	\end{theorem}
	
	Parts \ref{th:CLT_FNAR_BAI_i} and \ref{th:CLT_FNAR_BAI_ii} extend Theorem 2 in \citet{bai2009panel} to the FNAR case. The interesting cases are parts \ref{th:CLT_FNAR_BAI_i}  and \ref{th:CLT_FNAR_BAI_iii}, where we do not impose homoskedastic idiosyncratic components in the FNAR errors. 	Notice that the network coefficients, $\beta_j$, $j=1,\ldots, r$, which are the first $r$ elements of $\theta$, are consistently estimated only up to a sign, due to the indeterminacy in the identification of the network factors.
	
	Estimators of the asymptotic variance-covariance matrix of $\widehat \theta^\dag$ and $\widehat\theta^*$ under the assumptions in parts \ref{th:CLT_FNAR_BAI_i}  and \ref{th:CLT_FNAR_BAI_iii}  are (see also \citealp{bai2009panel}):
	\begin{align}
		\widehat{\Avar}\left[ \sqrt{NT} (\widehat{\theta}^{\dag} -\bar J\theta) \right]
		&=
		\left(\frac{1}{NT} \sum_{i=1}^{N} \widehat{Z}_{i}' \widehat{Z}_{i} \right)^{-1}
		\left(\frac{1}{NT} \sum_{i=1}^{N} \widehat{Z}_{i}' \widehat{Z}_{i} \frac 1T \sum_{t=1}^T \widehat{\epsilon}_{it}^2 \right)
		\left(\frac{1}{NT} \sum_{i=1}^{N} \widehat{Z}_{i}' \widehat{Z}_{i} \right)^{-1},\nonumber\\
		\widehat{\Avar}\left[ \sqrt{NT} (\widehat{\theta}^{*} -\bar J\theta) \right]
		&=
		\left(\frac{1}{NT} \sum_{t=1}^{T}\widehat{W}_{t}' \widehat{W}_{t} \right)^{-1}
		\left(\frac{1}{NT}  \sum_{t=1}^{T} \widehat{W}_{t}' \widehat{W}_{t} \frac 1T \sum_{t=1}^T \widehat{\epsilon}_{it}^2 \right)
		\left(\frac{1}{NT} \sum_{t=1}^{T}\widehat{W}_{t}' \widehat{W}_{t} \right)^{-1}, \label{eq:AVARtheta}
	\end{align}	
	where
	$
	\widehat Z_{i} := 
	M_{\widehat G^\dag}\widehat{X}_i-\frac 1N\sum_{k=1}^N \left(\widehat\Lambda_i^{\dag'}\left(\frac{\widehat{\Lambda}^{\dag'}\widehat\Lambda^\dag} N\right)^{-1}\widehat\Lambda_k^\dag\right)M_{\widehat G^\dag}\widehat X_k
	,
	$
	and
	$
	\widehat W_t:=
	M_{\widehat \Lambda^*}\widehat X_t-\frac 1T\sum_{s=1}^T (\widehat G_t^{*'}\widehat G_s^* )M_{\widehat \Lambda^*}\widehat X_s
	.	
	$

	Four important comments about this result follow.
	First, the proof is based on showing that if ${ \sqrt {NT}}/ m \to 0$ and $N/m\to 0$, as $m,N,T\to\infty$, then the network factors can be treated as observed; i.e., the generated regressors bias is asymptotically negligible (see Proposition C.1). 
	Under these conditions, and if also $T/N\to 0$ and $\sqrt N/T\to 0$, the iterative estimators are $\sqrt{NT}$-consistent. 
	
	Second, if the node factors are not autocorrelated we can also compare the iterative estimators with the OLS and the GLS estimators studied in Appendix B and C, respectively. In this case, the GLS is also $\sqrt{NT}$-consistent since, similarly to the iterative estimator, it rescales $X_t$ by the FNAR error covariance matrix $V$, which is $O(N)$ by Assumption \ref{as:errors_nu}. For the same reason the OLS estimator is just $\sqrt T$-consistent since it does not control for the FNAR error covariance, and it would be $\sqrt{NT}$-consistent only if $V$ were a diagonal matrix; i.e., when no node factor is present, as assumed by \citet{zhuetal17}. The same results on OLS and GLS are obtained by \citet{chenetal2020cnar} for the case of observed networks.

	Third, the two estimators are asymptotically equivalent and thus also equally efficient. To see this notice that $\widehat{\theta}^\dag$ and $\widehat{\theta}^*$ are such that they solve
	\begin{align}
		\widehat{\theta}^{\dag}&=\arg\min_{\theta}\frac 1{NT}\sum_{i=1}^N (y_i-X_i\theta-G\Lambda_i)^\prime (y_i-X_i\theta-G\Lambda_i) =\frac 1{NT}\sum_{i=1}^N\sum_{t=1}^T (y_{it}-X_{it}\theta -\Lambda_i^\prime G_t)^2,\label{eq:lossdag}\\
		\widehat{\theta}^{*}&=\arg\min_{\theta}\frac 1{NT}\sum_{t=1}^T (y_t-X_t\theta-\Lambda G_t)^\prime (y_t-X_t\theta-\Lambda G_t)=\frac 1{NT}\sum_{t=1}^T\sum_{i=1}^N (y_{it}-X_{it}\theta -\Lambda_i^\prime G_t)^2,\label{eq:lossstar}
	\end{align}
	so the two losses are identical and must have the same minimum. Once we fix the identification constraints as in Assumption \ref{as:errors_nu_id}, the only difference is then about the implementation of the minimizations. For $\widehat{\theta}^{\dag}$, by replacing $\Lambda_i= (G^\prime G)^{-1} G^\prime(y_i-X_i\theta)=T^{-1}G^\prime(y_i-X_i\theta)$ in \eqref{eq:lossdag}, we can solve for $G$ and $\theta$ only. For $\widehat{\theta}^{*}$, by replacing $G_t=(\Lambda^\prime\Lambda)^{-1}\Lambda'(y_t-X_t\theta)$ in \eqref{eq:lossstar}, we can solve for $\Lambda$ and $\theta$ only. These two procedures lead to the two solutions given in \eqref{eq:thetastar}. However, since in practice the solutions are obtained by iteration, the two estimates might not coincide exactly, although they will be very similar. And, as expected, the estimated standard errors will coincide (see the results in Section \ref{sec:application} and Appendix F.4).
	
	Fourth, and last, we should view Theorem \ref{th:CLT_FNAR_BAI} as giving the asymptotic distribution of the theoretical estimator minimizing \eqref{eq:lossdag} or \eqref{eq:lossstar}. This is the same point of view adopted by \citet{bai2009panel}. In practice, it might be important to investigate how the initialization of the algorithm affects such convergence. In the simpler case of a panel regression having errors with a factor structure, \citet[Theorem 3]{jiang2021recursive} show that any initial estimator could still lead to a consistent iterated estimator, depending on the structure of the regressors which can be quite general. However, the estimator computed in practice might have a slower convergence rate if the initial estimator is not consistent. We do not explore this aspect further here, but we limit ourselves to notice that in our numerical exercises of Sections \ref{sec:simulations} and \ref{sec:application}, convergence is always achieved in few steps and the iterated estimator works well even in presence of weak serial correlation of the node factors.

	
	\section{Monte Carlo simulations}\label{sec:simulations}

	To evaluate the finite sample performance of the proposed estimators, we generate artificial time series of $y_t$ and $\mathcal{W}_t$, for $t=1, \dots, T$, according to the model equations \eqref{eq:netfacscalar}, \eqref{eq:famnvar}, and \eqref{eq:moderror}. 
	We fix $r=1$ and $q=1$; i.e., one network factor and one node-specific factor. We also fix		
	the values of FNAR parameters $\beta=0.5$, $\rho=0.3$, and $\alpha=0.2$. We consider $N \in \{10,20,50,100,200\}$ nodes, $m \in \{20,50,100\}$ layers, and $T \in \{50,100\}$ time periods. 
	Also, for each value of the pair ($N ,m$), we randomly generate the entries of the loading	vectors $U$ and $\Lambda$ once (and independently)
	from $\mathcal{N}(1,1)$, and then we keep them fixed across Monte Carlo (MC) iterations (see \citealp{chenetal2020cnar}). All other quantities are generated at each MC iteration. All results are based on 500 iterations. Full details on the data generating process are in Appendix  F.1.

	\begin{table}[H]
		\centering
		\scriptsize{
			\caption{Monte Carlo RMSEs - $T=50$, case II: dependent $\mathcal E_t$}	\label{tab:MC_RMSE_Case2_T50}
			\scalebox{0.9}[0.8]{
				\begin{tabular}{ll | cc|cc|cc|cc|cc }
					\hline \hline
					&& \multicolumn{2}{|c|}{$N = 10$} & \multicolumn{2}{c|}{$N = 20$} & \multicolumn{2}{c|}{$N = 50$}& \multicolumn{2}{c|}{$N = 100$} & \multicolumn{2}{c}{$N = 200$} \\
					&               & RMSE          & ReRMSE     & RMSE          & ReRMSE     & RMSE          & ReRMSE     & RMSE          & ReRMSE  & RMSE          & ReRMSE     \\
					\hline
					\multirow{5}{*}{$m = 20$}  & $\beta$     & 0.084 & 16.9\% & 0.078 & 15.6\% & 0.077 & 15.4\% & 0.075 & 15.0\% & 0.078 & 15.6\% \\
					& $\rho$        & 0.045 & 14.9\% & 0.031 & 10.2\% & 0.021 & 7.1\%  & 0.015 & 5.0\%  & 0.011 & 3.6\%  \\
					& $\alpha$      & 0.084 & 41.8\% & 0.043 & 21.6\% & 0.029 & 14.4\% & 0.020 & 10.1\% & 0.014 & 7.2\%  \\
					& $\mathcal{F}$ & 0.206 & 21.7\% & 0.211 & 21.7\% & 0.214 & 21.7\% & 0.216 & 21.7\% & 0.216 & 21.7\% \\
					& $U$           & 0.085 & 3.9\%  & 0.079 & 3.6\%  & 0.078 & 3.5\%  & 0.078 & 3.5\%  & 0.078 & 3.5\%  \\
					\hline
					\multirow{5}{*}{$m = 50$}  & $\beta$     & 0.090 & 17.9\% & 0.078 & 15.6\% & 0.073 & 14.6\% & 0.080 & 16.0\% & 0.077 & 15.4\% \\
					& $\rho$        & 0.047 & 15.8\% & 0.032 & 10.7\% & 0.020 & 6.7\%  & 0.014 & 4.8\%  & 0.010 & 3.5\%  \\
					& $\alpha$      & 0.068 & 34.1\% & 0.045 & 22.4\% & 0.030 & 15.2\% & 0.019 & 9.4\%  & 0.015 & 7.4\%  \\
					& $\mathcal{F}$ & 0.132 & 13.9\% & 0.135 & 13.9\% & 0.137 & 13.9\% & 0.138 & 13.9\% & 0.138 & 13.9\% \\
					& $U$           & 0.037 & 2.1\%  & 0.029 & 1.6\%  & 0.027 & 1.5\%  & 0.026 & 1.5\%  & 0.026 & 1.5\%  \\
					\hline
					\multirow{5}{*}{$m = 100$} & $\beta$     & 0.086 & 17.1\% & 0.081 & 16.1\% & 0.078 & 15.6\% & 0.078 & 15.7\% & 0.078 & 15.7\% \\
					& $\rho$        & 0.047 & 15.6\% & 0.031 & 10.3\% & 0.021 & 7.0\%  & 0.015 & 5.0\%  & 0.011 & 3.6\%  \\
					& $\alpha$      & 0.087 & 43.6\% & 0.044 & 22.1\% & 0.027 & 13.6\% & 0.019 & 9.6\%  & 0.014 & 7.0\%  \\
					& $\mathcal{F}$ & 0.093 & 9.8\%  & 0.096 & 9.8\%  & 0.097 & 9.8\%  & 0.098 & 9.8\%  & 0.098 & 9.8\%  \\
					& $U$           & 0.026 & 1.6\%  & 0.016 & 1.0\%  & 0.013 & 0.8\%  & 0.012 & 0.7\%  & 0.012 & 0.7\%  \\
					\hline \hline
				\end{tabular}   	
			}
		}
	\end{table}
	
	\vspace{-5pt}
	
	\begin{table}[H]
		\centering
		\scriptsize{
			\caption{Monte Carlo RMSEs - $T=100$, case II: dependent $\mathcal E_t$}	\label{tab:MC_RMSE_Case2_T100}
			\scalebox{0.9}[0.8]{
				\begin{tabular}{ll | cc|cc|cc|cc|cc }
					\hline \hline
					&& \multicolumn{2}{|c|}{$N = 10$} & \multicolumn{2}{c|}{$N = 20$} & \multicolumn{2}{c|}{$N = 50$}& \multicolumn{2}{c|}{$N = 100$} & \multicolumn{2}{c}{$N = 200$} \\
					&               & RMSE          & ReRMSE     & RMSE          & ReRMSE     & RMSE          & ReRMSE     & RMSE          & ReRMSE   & RMSE      & ReRMSE     \\
					\hline
					\multirow{5}{*}{$m = 20$}  & $\beta$     & 0.056 & 11.1\% & 0.056 & 11.2\% & 0.055 & 11.0\% & 0.053 & 10.6\% & 0.056 & 11.2\% \\
					& $\rho$        & 0.031 & 10.2\% & 0.022 & 7.2\%  & 0.014 & 4.6\%  & 0.010 & 3.3\%  & 0.008 & 2.5\%  \\
					& $\alpha$      & 0.056 & 28.0\% & 0.029 & 14.5\% & 0.020 & 10.1\% & 0.014 & 7.2\%  & 0.010 & 5.0\%  \\
					& $\mathcal{F}$ & 0.206 & 21.7\% & 0.211 & 21.7\% & 0.214 & 21.7\% & 0.216 & 21.7\% & 0.216 & 21.7\% \\
					& $U$           & 0.081 & 3.7\%  & 0.078 & 3.6\%  & 0.078 & 3.5\%  & 0.078 & 3.5\%  & 0.078 & 3.5\%  \\
					\hline
					\multirow{5}{*}{$m = 50$}  & $\beta$     & 0.064 & 12.9\% & 0.055 & 11.0\% & 0.054 & 10.9\% & 0.053 & 10.7\% & 0.056 & 11.2\% \\
					& $\rho$        & 0.031 & 10.3\% & 0.021 & 7.0\%  & 0.014 & 4.7\%  & 0.010 & 3.3\%  & 0.008 & 2.5\%  \\
					& $\alpha$      & 0.048 & 24.0\% & 0.032 & 15.8\% & 0.019 & 9.6\%  & 0.014 & 6.9\%  & 0.010 & 5.2\%  \\
					& $\mathcal{F}$ & 0.132 & 13.9\% & 0.135 & 13.9\% & 0.137 & 13.9\% & 0.138 & 13.9\% & 0.138 & 13.9\% \\
					& $U$           & 0.032 & 1.8\%  & 0.027 & 1.5\%  & 0.026 & 1.5\%  & 0.026 & 1.5\%  & 0.026 & 1.5\%  \\
					\hline
					\multirow{5}{*}{$m = 100$} & $\beta$     & 0.060 & 12.1\% & 0.056 & 11.3\% & 0.054 & 10.8\% & 0.055 & 10.9\% & 0.055 & 11.0\% \\
					& $\rho$        & 0.030 & 10.0\% & 0.022 & 7.4\%  & 0.014 & 4.5\%  & 0.010 & 3.4\%  & 0.007 & 2.5\%  \\
					& $\alpha$      & 0.059 & 29.4\% & 0.030 & 15.1\% & 0.018 & 8.8\%  & 0.014 & 7.1\%  & 0.010 & 4.9\%  \\
					& $\mathcal{F}$ & 0.093 & 9.8\%  & 0.096 & 9.8\%  & 0.097 & 9.8\%  & 0.098 & 9.8\%  & 0.098 & 9.8\%  \\
					& $U$           & 0.020 & 1.3\%  & 0.014 & 0.9\%  & 0.012 & 0.8\%  & 0.012 & 0.7\%  & 0.012 & 0.7\% \\
					\hline \hline
				\end{tabular}   	
			}
		}
	\end{table}

	Tables \ref{tab:MC_RMSE_Case2_T50}-\ref{tab:MC_RMSE_Case2_T100} report the {RMSE} and Relative RMSE ({ReRMSE}) of the estimates.
	Here we report results under case II, which corresponds to idiosyncratic terms $\mathcal E_t$ having serial and cross-layer correlation, and using the iterative estimator $\widehat{\theta}^\dag$ in \eqref{eq:thetastar}. Additional results for case I of uncorrelated idiosyncratic terms are in Appendix  F.2. 
	As predicted by the theory, the accuracy of estimates for $\beta$, $\rho$ and $\alpha$ improves with both $N$ and $T$, and the RMSE of network factors and loadings decreases when the number of layers $m$ increases. Furthermore, as shown in Appendix F.3, the MC distributions of the estimated network coefficient are all strongly centered around the true value $\beta=0.5$ and become narrower as $T$ and $N$ increase.
	
	Last, in Appendix F.5 we compare our estimates of the loadings $U$ with those obtained using the TOPUP and TIPUP estimation methods proposed by  \citet{chenyangzhang22}. As expected our approach improves over those estimators in presence of serial idiosyncratic correlation.

	
	\section{Empirical application}\label{sec:application}
	
	In this section, we present an application of the FNAR for studying cross-country macroeconomic interdependence determined by global trade flows and cross-border financial relationships.
	
	\noindent\textsc{Data.} 	
	For a sample of $N=24$ countries, we use $m=25$ networks constructed using bilateral import/export flows for different good (layers 1--9) and services (10--19) categories, bilateral financial positions for different types of financial claims (20--23) and cross-border mergers and acquisitions classified by sector of economic activity (24--25). 	The list of countries and network layers, including details on how the networks are built, are given in Appendix G.1.

	\noindent\textsc{Network factors.}	Due to data limitations in the time series of financial positions, we collect data for the networks at the annual frequency from 2001 to 2019, so the factor analysis is conducted on a sample of length $T_1=19$.\footnote{We have few missing values over this period. In these cases, we use the previous year's value or the closest available year's value.} Although this is a short time span, we recall that in tensor PCA the effective sample size when estimating the loadings space is $N^{2}T_1$ 
	(see Theorem \ref{theorem:CLT_loadings}). 	
	
	We then extract the common network factors from the 25 layers of the network, and we set $\widehat r=6$ network factors, as in \citet{chenyangzhang22}.		
	From Figures G.9-G.12 in Appendix G.2 we can interpret the six network factors as follows. The first network factor conveys approximately the average country weights across all layers of the network. In particular, the factor values are very close to the average weights (scaled by a constant), and the loading coefficients are almost the same for all layers. The countries with the largest factor weights for the US are its major economic partners: Canada, UK, Mexico, China, Japan and Germany. 
	
	The second factor captures a difference between financial relationships and trade in goods. The factor loadings for financial layers have opposite sign (positive) compared to the loadings for trade-in-goods layers (negative). Recall that factors are identified up to a sign. 
	The largest positive weights are assigned to economies having relatively large financial sectors with global reach: UK, US, and Hong Kong.  In the case of the US connections, a large positive weight is assigned to the UK, whose tight economic links with the US are mostly concentrated in the financial sector, and large negative weights are assigned to Canada, Mexico, and China, i.e., the US biggest trade partners.
	
	The third factor distinguishes between equity and debt relationships, being the only factor where equity, on the one hand, and debt, on the other hand, show loadings with opposite signs. The fourth factor is strongly associated with M\&A relationships.
	The fifth factor is mainly driven by agricultural/extractive goods (positive weights, especially for vegetable fuels, oils, fats, and waxes). It also loads on trade in manufacturing goods (negative weights). Positive weights are assigned to countries with strong trade links with the US in non-manufacturing sectors, such as Canada, Saudi Arabia, and Italy, while negative weights are associated with large manufacturing partners, like China. This factor also distinguishes between stocks of portfolio holdings and flows associated with M\&A deals and banking. Finally, the sixth factor captures a distinction between goods-sector M\&A integration and services-sector integration.

	Next, in line with conventional PCA, we evaluate the fraction of variance in network layers explained by each factor, denoted as $v^{(k)}$, $k=1,\ldots, 6$ (computed as in Appendix G.2). We have $v^{(1)}=0.68$, $v^{(2)}=0.07$, $v^{(3)}=0.03$, $v^{(4)}=0.03$, $v^{(5)}=0.02$, and $v^{(6)}=0.02$. Thus, overall the 6 factors explain about 85\% of the total variance of $\mathcal W$. However, the importance of different factors varies greatly across countries; see Table G.11 in Appendix G.2.

	\noindent\textsc{FNAR coefficients.}	
	The endogenous vector $y_t$, $t=1,\ldots, T$,  collects (quarterly) real GDP growth rates for all $N$ considered countries and for the sample 2001Q1-2019Q4; i.e., $T_2=76$. To address heterogeneity of nodal and momentum effects, we split the countries into two groups: (1) advanced economies ($N_1=15$), and (2) emerging economies ($N_2=9$) and the vector $y_t$ is partitioned accordingly as $y_t=(y_t^{(1)'}; y_t^{(2)'})'$.		
	We consider the following FNAR, for $t=1,\ldots, T$,
	\begin{align}
		y_t =&\,\sum_{j=1}^r \beta_j \frac{\widetilde F_{j,t-1}}N y_{t-1}
		+\rho^{(1)} \left(\begin{array}{c}
			y^{(1)}_{t-1}\\
			0_{N_2}
		\end{array}
		\right) 
		+\rho^{(2)} \left(\begin{array}{c}
			0_{N_1}\\
			y^{(2)}_{t-1}
		\end{array}
		\right)+\alpha^{(1)} \left(\begin{array}{c}
			\iota_{N_1}\\
			0_{N_2}
		\end{array}
		\right) 
		+\alpha^{(2)} \left(\begin{array}{c}
			0_{N_1}\\
			\iota_{N_2}
		\end{array}
		\right) 
		+\nu_t, \label{eq:FNARGDP}
	\end{align}
	where $\widetilde F_{j,t} = \widehat F_{j,\tau}$ for $4(\tau-1)+1\le t \le 4\tau$, $\tau=1,\ldots, T_1$. In other words, the network factors $F_{k,t}$, $t=1,\ldots, T_1$, which are computed on a yearly basis, are treated as constant throughout all quarters of a given year. 
	Hereafter, we let $\theta:=(\beta', \rho^{(1)},\rho^{(2)},\alpha^{(1)}, \alpha^{(2)})'$. 
	
	By means of the criterion defined in \eqref{eq:qhat}, we find evidence of one common node factor, i.e., $\widehat q=1$. We then estimate the model by GLS as described in Appendix C. Last, we consider the iterative estimators $\widehat\theta^\dag$ or $\widehat\theta^*$ defined in \eqref{eq:thetastar} and we initialize the algorithm by using the GLS estimator and the estimated node loadings, $\widehat \Lambda$, and factor, $\widehat G_t$, computed by PCA on the GLS residuals as described in Appendix A. 
	Since these residuals do not display significant autocorrelation, we are confident that the GLS estimator is $\sqrt{NT}$-consistent and, based on the results of \citet{jiang2021recursive} we conjecture that Theorem \ref{th:CLT_FNAR_BAI} holds for our iterated estimators. Convergence is reached in 8 or 4 iterations for $\widehat\theta^\dag$ or $\widehat\theta^*$, respectively.
	
	Table \ref{tab:theta} reports the estimated coefficients and their standard errors with significance reported according to the usual $Z$-test. The coefficients on $N^{-1}\widehat F_{1,t-1}y_{t-1}$ and $N^{-1}\widehat F_{5,t-1}y_{t-1}$ are always strongly significant, while there is mixed evidence regarding the coefficients on $N^{-1}\widehat F_{2,t-1}y_{t-1}$, $N^{-1}\widehat F_{4,t-1}y_{t-1}$, and $N^{-1}\widehat F_{6,t-1}y_{t-1}$ which are mildly significant and not for all estimates.
	
	\begin{table}[t]
		\caption{FNAR coefficient estimates and standard errors.}
		\centering
		\scriptsize{
			\scalebox{0.9}[0.8]{	\begin{tabular}{l cccc}
					\hline
					\hline
					\\[-8pt]
					& $\widehat\theta^{\text{\tiny OLS}}$& $\widehat\theta^{\text{\tiny GLS}}$& $\widehat\theta^\dag$& $\widehat\theta^*$\\
					\hline
					network effects & & \\
					\hline
					\\[-8pt]
					$\widehat{\beta}_1N^{-1}\widehat F_{1,t-1}y_{t-1}$&1.1641***&	1.0805***&1.0005***&1.0012***\\
					&(0.1220)&	(0.0877)&(0.1088)&(0.1088)\\
					$\widehat{\beta}_2N^{-1}\widehat F_{2,t-1}y_{t-1}$&-0.3529**&	0.0348&0.1156&0.1158\\
					&(0.1889)&	 (0.1163)&(0.1133)&(0.1133)\\
					$\widehat{\beta}_3N^{-1}\widehat F_{3,t-1}y_{t-1}$&-0.1217&	0.0385&0.0164&0.0162\\
					&(0.1919)&	(0.1369)&(0.1442)&(0.1442)\\
					$\widehat{\beta}_4N^{-1}\widehat F_{4,t-1}y_{t-1}$&-0.0597&	0.0977&0.1605*&0.1607*\\
					&(0.1911)	&(0.1093)&(0.1098)&(0.1098)\\
					$\widehat{\beta}_5N^{-1}\widehat F_{5,t-1}y_{t-1}$&0.8145***&		0.3700***&0.5506***&0.5506***\\
					&(0.1893)&	(0.1154)&(0.1128)&(0.1128)\\
					$\widehat{\beta}_6N^{-1}\widehat F_{6,t-1}y_{t-1}$&0.2669**&		0.0042&0.1050&0.1051\\
					&(0.1552)&	(0.0974)&(0.1118)&(0.1118)\\
					\hline
					momentum effects  & &\\
					\hline
					\\[-8pt]
					$\widehat{\rho}^{(1)}y_{t-1}^{(1)}$&0.0658*&	0.1504***&0.0802**&0.0802**\\
					&(0.0424)&	(0.0306)&(0.0350)&(0.0350)\\
					$\widehat{\rho}^{(2)}y_{t-1}^{(2)}$&0.2407***&	 0.3365***&0.3119***&0.3121***\\
					&(0.0348)&	(0.0328)&(0.0277)&(0.0277)\\
					\hline
					nodal effects & & \\
					\hline
					\\[-8pt]
					$\widehat{\alpha}^{(1)}$&0.0023***&	0.0875*&0.0021***&0.0021***\\
					&(0.0007)&	(0.0548)&(0.0004) &(0.0004)\\
					$\widehat{\alpha}^{(2)}$& 0.0054***&	0.2741*** & 0.0049*** & 0.0049***\\
					&(0.0010)&	(0.0729)&(0.0006)&(0.0006)\\
					\hline \hline
				\end{tabular}
			}
		}
		\label{tab:theta}
	\end{table}	
	
	Based on the interpretation of the first network factor, the coefficient on $N^{-1}\widehat F_{1,t-1}y_{t-1}$ captures a general network effect operating through aggregate economic linkages. 
	Given the loadings of factor 5 in Figure G.10, the coefficient on $N^{-1}\widehat F_{5,t-1}y_{t-1}$ indicates that trade in mineral fuels and in animal and vegetable oils (major inputs of chemical industry) has the main impact on GDP growth.
	Apart from this, trade linkages tend to generate larger spillovers in manufacturing sectors (layers 6-9) than in non-manufacturing sectors (layers 1-3), and financial linkages tend to generate larger spillovers when they take the form of M\&A or flows of banking assets (rather than portfolio holdings).

\begin{figure}[t!]
		\centering
		\caption{Network effects by original layer.}
		\includegraphics[scale=0.45]{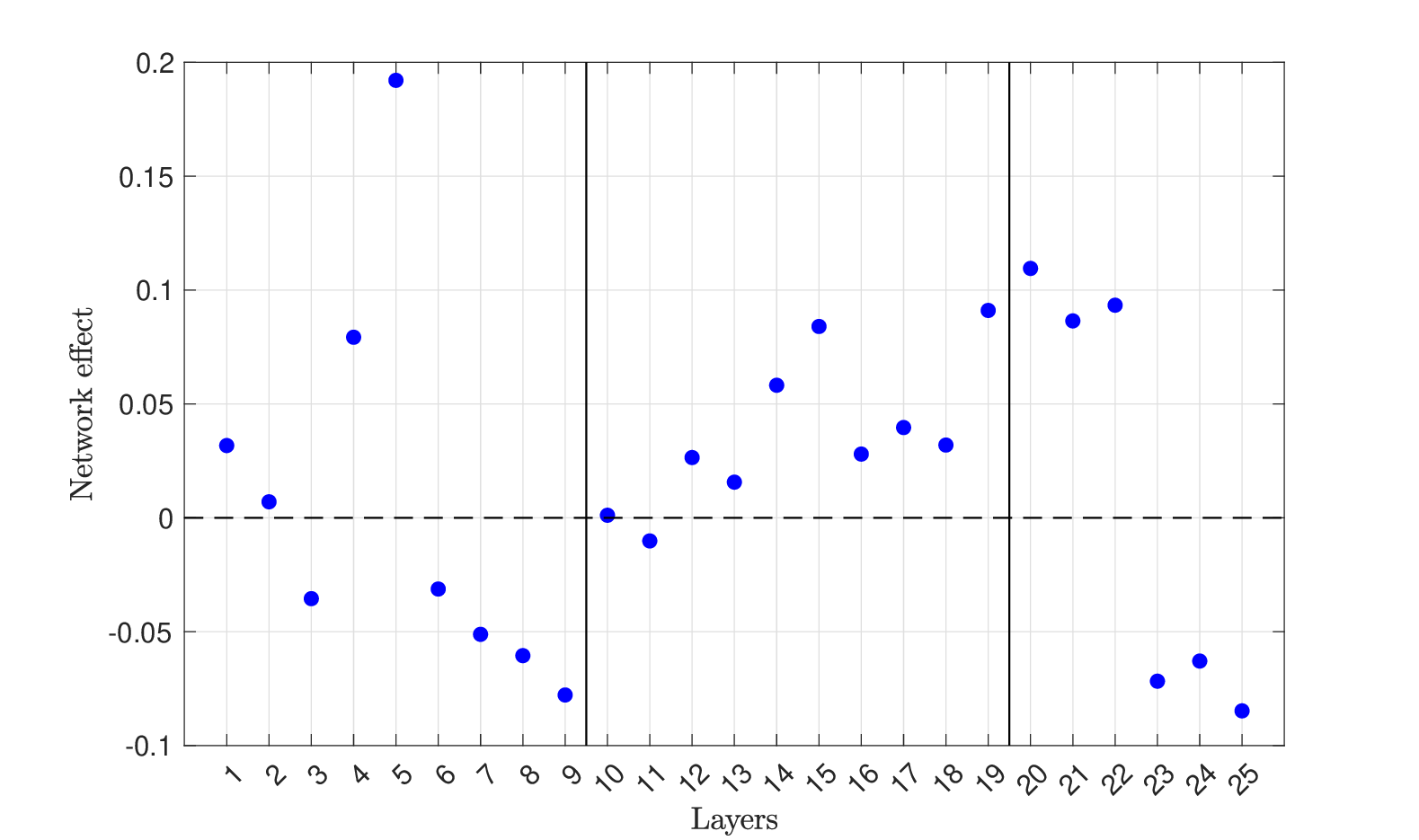} 
		\vspace{7pt}
		\subcaption*{\footnotesize The figure plots the estimated network effects associated with the original (and row-normalized) layers of the network computed as $N^2 \widehat{U} (\widehat{M}^{\mathcal{W}})^{-1} \widehat{\beta}$, when using the iterated estimator. The vertical lines separate layers related to trade in goods (1-9), layers related to trade in services (10-19) and financial layers (20-25). See Table G.9 for the complete list of layers. The estimation sample is 2001Q1-2019Q4.}
		\label{fig:weighted_beta}
	\end{figure}

	Finally, based on these estimates, we can approximate the network effects associated with the original layers of the network, by appropriately rescaling the estimated network effect coefficients $\beta$. Specifically, given the definition of estimated loadings in \eqref{eq:est_loadings} and the properties of tensor multiplication, and letting  $\widehat{\mathcal{W}}_{t} := \widehat{\mathcal{F}}_{t} \times_3 \widehat{U}$, for a given estimate $\widehat{\beta}$ we have that:
	\begin{align}
		\frac{\widehat{\mathcal{F}}_{t-1}}{N}  \times_2 y_{t-1}' \times_3 \widehat{\beta}' 
		&= \frac{\widehat{\mathcal{F}}_{t-1}}{N}  \times_2 y_{t-1}' \times_3 \left(\widehat{\beta}' (\widehat{M}^{\mathcal{W}})^{-1} \widehat{U}' \widehat{U} N^{2}\right)
		= \frac{\widehat{\mathcal{W}}_{t-1}}{N} \times_2 y_{t-1}' \times_3 \left(N^2 \widehat{U} (\widehat{M}^{\mathcal{W}})^{-1} \widehat{\beta} \right)'.\nonumber
	\end{align}
	Thus, $N^2 \widehat{U} (\widehat{M}^{\mathcal{W}})^{-1}\widehat{\beta}$ is the vector of network effects in terms of the row-normalized tensor $\widehat{\mathcal{W}}_{t-1} / N$ and its entries are shown in Figure \ref{fig:weighted_beta}, when computed using the iterated estimator $\widehat{\theta}^*$. 
	The figure confirms a substantial heterogeneity of effects across layers, reflecting their different loadings on the network factors. 	
	
	\noindent\textsc{Forecasting.}	 We conclude by studying the performance of our FNAR when producing 1-quarter-ahead forecasts of GDP growth rates based on a recursive  window exercise from 2002Q1 to 2019Q4. We consider the following competitors: the tensor-based estimators MLR and SHORR by \citet{wangetal21}; 
	a NAR estimated either using the $m$ layers of the common component (denoted as TUCKER COMMON and further regularized via LASSO or Ridge due to collinearity of the regressors) or just $r$ network factors (denoted as TUCKER FACTORS) both obtained from a full Tucker factor decomposition, estimated via TOPUP as in \citet{chenyangzhang22} (see Appendix H for more details); 
	a multilayer NAR estimated via LASSO or Ridge; and an ordinary VAR. Details on the adopted forecasting scheme and on the implementation of the alternative estimation methods are in Appendix G.2.4.
	
	Table \ref{tab:forecast_comparison} reports the root mean squared forecast errors (RMSFE) in terms of percentage points of GDP growth, for each country considered. In the last two rows of Table \ref{tab:forecast_comparison}, we report the average (across all countries) RMFSE and relative RMSFE (ReRMSFE) with respect to the FNAR, for all forecasting methods (values larger than one indicate a better performance of the FNAR). For most countries, and on average, our approach delivers the forecasts with smallest RMSFEs. In particular, we outperform the approaches based on a full Tucker decomposition which are our most natural competitors. Indeed, contrary to the case of factors extracted by means of a full Tucker decomposition, our network factors still contain terms which are idiosyncratic to the network nodes, i.e., countries, which are potentially relevant for predicting country specific GDP growth rates.
	
	\renewcommand{\arraystretch}{0.7}
	\begin{table}[H]
		\centering
		\caption{RMSFEs for GDP growth}
		{ \scriptsize
			\scalebox{0.9}[0.8]{\begin{tabular}{lp{1.25cm}p{1.25cm}p{1.25cm}p{1.25cm}p{1.25cm}p{1.25cm}p{1.25cm}p{1.25cm}p{1.25cm}}
					\hline \hline
					&      &   &  & TUCKER& TUCKER & &  &  \\   
					&      &   &  & COMMON& COMMON &TUCKER& &  &  \\   
					& FNAR     &  MLR & SHORR & +LASSO& +RIDGE & FACTORS & LASSO& RIDGE & VAR \\    
					\hline
					AUS	&	0.49\%	&	0.46\%	&	0.46\%	&	0.43\%	&	0.45\%	&0.62\%	&	0.44\%	&	0.45\%	&	0.56\%	\\
					BEL	&	0.45\%	&	0.48\%	&	0.50\%	&	0.62\%	&	0.66\%	&0.48\%	&	0.58\%	&	0.67\%	&	0.58\%	\\
					CAN	&	0.46\%	&	0.47\%	&	0.49\%	&	0.61\%	&	0.64\%	&0.60\%	&	0.58\%	&	0.64\%	&	0.48\%	\\
					FRA	&	0.46\%	&	0.48\%	&	0.48\%	&	0.64\%	&	0.69\%	&0.49\%	&	0.60\%	&	0.70\%	&	0.43\%	\\
					GER	&	0.80\%	&	0.83\%	&	0.84\%	&	0.93\%	&	0.96\%	&0.80\%	&	0.91\%	&	0.96\%	&	0.83\%	\\
					ITA	&	0.73\%	&	0.69\%	&	0.70\%	&	0.94\%	&	1.00\%	&0.72\%	&	0.90\%	&	1.01\%	&	0.57\%	\\
					JAP	&	1.05\%	&	1.06\%	&	1.06\%	&	1.16\%	&	1.19\%	&1.06\%	&	1.13\%	&	1.20\%	&	1.14\%	\\
					KOR	&	0.92\%	&	1.09\%	&	1.10\%	&	0.91\%	&	0.91\%	&1.02\%	&	0.92\%	&	0.91\%	&	1.08\%	\\
					NLD	&	0.54\%	&	0.56\%	&	0.58\%	&	0.76\%	&	0.80\%	&0.56\%	&	0.72\%	&	0.80\%	&	0.56\%	\\
					NOR	&	1.27\%	&	1.26\%	&	1.23\%	&	1.25\%	&	1.27\%	&1.28\%	&	1.25\%	&	1.27\%	&	1.21\%	\\
					ESP	&	0.46\%	&	0.56\%	&	0.56\%	&	0.66\%	&	0.71\%	&0.50\%	&	0.62\%	&	0.72\%	&	0.34\%	\\
					SWE	&	1.03\%	&	1.06\%	&	1.05\%	&	1.08\%	&	1.09\%	&1.02\%	&	1.07\%	&	1.09\%	&	1.16\%	\\
					CHE	&	0.78\%	&	0.83\%	&	0.83\%	&	0.85\%	&	0.88\%	&0.82\%	&	0.83\%	&	0.88\%	&	0.96\%	\\
					GBR	&	0.68\%	&	0.72\%	&	0.72\%	&	0.78\%	&	0.82\%	&0.70\%	&	0.76\%	&	0.82\%	&	0.71\%	\\
					USA	&	0.50\%	&	0.55\%	&	0.55\%	&	0.60\%	&	0.63\%	&0.54\%	&	0.58\%	&	0.63\%	&	0.57\%	\\
					\hline
					\textit{avg. advanced}    &	\it 0.71\%	&	\it 0.74\%	&	\it 0.74\%	&	\it 0.81\%	&	\it 0.85\%	&\it 0.75\%	&	\it 0.79\%	&	\it 0.85\%	&	\it 0.74\%	\\
					\hline
					BRA	&	1.27\%	&	1.35\%	&	1.40\%	&	1.33\%	&	1.28\%	&1.28\%	&	1.36\%	&	1.27\%	&	1.51\%	\\
					CHN	&	1.61\%	&	1.15\%	&	1.16\%	&	1.88\%	&	1.95\%	&1.80\%	&	1.82\%	&	1.96\%	&	1.22\%	\\
					HKG	&	1.43\%	&	1.74\%	&	1.82\%	&	1.41\%	&	1.37\%	&1.39\%	&	1.40\%	&	1.37\%	&	1.69\%	\\
					IND	&	1.36\%	&	1.19\%	&	1.18\%	&	1.42\%	&	1.49\%	&1.55\% 	&	1.39\%	&	1.50\%	&	1.51\%	\\
					IDN	&	0.48\%	&	0.65\%	&	0.67\%	&	0.54\%	&	0.64\%	&1.80\%	&	0.50\%	&	0.65\%	&	0.87\%	\\
					MEX	&	1.03\%	&	1.01\%	&	1.02\%	&	1.18\%	&	1.13\%	&0.99\% 	&	1.18\%	&	1.13\%	&	1.15\%	\\
					SAU	&	1.03\%	&	1.27\%	&	1.23\%	&	1.13\%	&	1.13\%	&1.07\% 	&	1.11\%	&	1.14\%	&	1.24\%	\\
					ZAF	&	0.60\%	&	0.58\%	&	0.58\%	&	0.77\%	&	0.67\%	&0.55\% 	&	0.77\%	&	0.66\%	&	0.59\%	\\
					TUR	&	2.35\%	&	2.39\%	&	2.41\%	&	2.32\%	&	2.33\%	&2.38\%	&	2.33\%	&	2.33\%	&	2.30\%	\\						
					\hline
					\textit{avg. emerging} &	\it 1.24\%	&	\it 1.26\%	&	\it 1.28\%	&	\it 1.33\%	&	\it 1.33\%	&\it 1.31\%	&	\it 1.32\%	&	\it 1.33\%	&	\it 1.34\%	\\
					\hline
					\textit{avg. all}    &	\it 0.91\%	&	\it 0.93\%	&	\it 0.94\%	&	\it 1.01\%	&	\it 1.03\%	&\it 0.96\%&	\it 0.99\%	&\it 	1.03\%	&\it 	0.97\%	\\ 
					\textbf{avg. ReRMSFE}& \bf 1.00&	\bf 1.05&\bf 	1.06&\bf  1.15&\bf 1.19&\bf 1.10&\bf  1.12&\bf 	1.19&\bf  1.09	 \\
					\hline \hline
				\end{tabular}
			}
			\vspace{7pt}
			\subcaption*{\footnotesize  The table reports the root mean squared forecast errors (RMSFE) of 1-quarter-ahead forecasts, based on a recursive forecasting scheme over the period 2002Q1-2019Q4. Advanced economies are in the upper panel; emerging economies are in the lower panel. See Appendix G.1 for the complete list of countries and their acronyms.  } 
			\label{tab:forecast_comparison}
		}
	\end{table}

	\section{Conclusions}\label{sec:conclusions}
	In this paper, we have introduced a factor network autoregression (FNAR) for time series characterized by multiple network effects. Estimation is based on two steps. First, we extract few network factors common across the layers of the underlying multilayer network. Second, we estimate a  factor-augmented NAR or FNAR where the network effects are determined by the latent network factors. The FNAR errors are allowed to have an underlying factor structure capturing common correlations across the network nodes.	 We prove consistency and asymptotic normality of the proposed estimators as the number of layers, nodes and time observations diverges to infinity.		 
	
	The results of an empirical application show that, by accounting for cross-country economic and financial linkages, the model provides a rich description of the dynamics of GDP growth rates and produces accurate forecasts.  
	
	We outline three possible extensions of this work, which we leave for further research. First, by adapting the works by \cite{chenetal2020cnar} and  \citet{zhu2022simultaneous} to the FNAR framework, we could consider a FNAR with momentum and nodal coefficients which are group specific, where the group structure is unknown and the number of groups $K$ can grow with the number of nodes $N$.
	Second, by generalizing to the tensor setting the three-pass regression filter by \citet{kelly2015three},
	we could improve the performance of our estimator by accounting also for the information contained in the vector of dependent variables $y_t$ when extracting the network factors.	
	Third, by extending to the tensor case an approach similar to the one proposed by \citet{wu2020adaptive}, we could allow for time-varying network factor loadings under the standard assumption of local stationarity.

	
	\newpage
	
	\begin{center}
	{\Large \textsc{Factor Network Autoregressions\\
			Supplementary material\\}}
	
		\author{\large Matteo Barigozzi \hskip 1cm
		\and Giuseppe Cavaliere \hskip 1cm
		\and Graziano Moramarco}
	\maketitle
	\end{center}

		\newcounter{lettersection}
		\setcounter{lettersection}{0}
		
		\renewcommand{\thesection}{\Alph{lettersection}}
		\let\oldsection\section
		\renewcommand{\section}[1]{%
			\refstepcounter{lettersection}%
			\oldsection{#1}%
		}

		\renewcommand{\thefigure}{\thesection.\arabic{figure}}
		\renewcommand{\thetable}{\thesection.\arabic{table}}

		\setcounter{figure}{0}%
		\setcounter{table}{0}%

		\noindent This supplemental material contains several appendices to our paper.
		In Appendix  \ref{app:node factors} we describe how to estimate the node factors by Principal Components Analysis (PCA) on the FNAR residuals.
		In Appendices \ref{app:OLS} and \ref{app:GLS}, we introduce the OLS estimator and a GLS-type estimator of the FNAR coefficients. In Appendix \ref{app:mainproof}, we provide the proofs for Propositions and Theorems. In Appendix \ref{app:lemma}, we provide and prove all auxiliary results. 
		In Appendix \ref{app:MC_appendix} we provide additional simulation results. 
		In Appendix \ref{app:data} we provide additional information on the data used in our empirical application, as well as additional empirical results.
		Finally, in Appendix \ref{app:Tucker} we consider alternative approaches for the estimation of a multilayer NAR, based on full Tucker decompositions.

		\section{Estimation of node factors}\label{app:node factors}

		Once we compute the OLS estimator, let $\widehat\nu:=(\widehat\nu_1\cdots \widehat\nu_T)'$ be the $T\times N$ matrix of residuals of the FNAR such that
		$\widehat \nu_t:=y_t -\widehat X_t \widehat{\theta}^{\text{\upshape\tiny OLS}}$, $t=1,\ldots, T$. Then, the node factors $G_t$ and their loadings $\Lambda$ can be estimated by PCA in two equivalent ways.
		Specifically, consider either the $N\times N$ or $T\times T$ sample covariance matrices 
		\begin{equation}\label{eq:Gammanu}
			\widehat {\Gamma}^{\widehat \nu} := \frac {\widehat \nu'\widehat \nu}T, \qquad \widetilde {\Gamma}^{\widehat \nu} := \frac {\widehat \nu\widehat \nu'}N.
		\end{equation} 
		Then, letting $\widehat G:=(\widehat G_1,\cdots, \widehat G_T)'$ and $\widetilde G:=(\widetilde G_1,\cdots, \widetilde G_T)'$ be the estimated $T\times q$ matrices of factors, the PC estimators are given by: 
		\begin{align}\label{eq:PCAerrors}
			\left\{
			\begin{array}{l}
				\widehat \Lambda:= \widehat V^{\widehat\nu}(\widehat M^{\widehat\nu})^{1/2},\\ 	
				\widehat G :=\widehat \nu \widehat \Lambda(\widehat \Lambda'\widehat \Lambda)^{-1},
			\end{array}
			\right.
			\quad
			\left\{
			\begin{array}{l}
				\widetilde G := \widetilde V^{\widehat\nu}\sqrt T, \\	
				\widetilde\Lambda:= \widehat \nu' \widetilde G (\widetilde G'\widetilde G)^{-1},
			\end{array}
			\right.
		\end{align}
		where $\widehat M^{\widehat\nu}$ is the $q\times q$ diagonal matrix of eigenvalues of $\widehat {\Gamma}^{\widehat \nu}$ with corresponding normalized eigenvectors being the columns of the $N\times q$ matrix $\widehat V^{\widehat\nu}$, and $\widetilde V^{\widehat\nu}$ is the $T\times q$ matrix of normalized eigenvectors of $\widetilde {\Gamma}^{\widehat \nu}$. It is easy to verify that, regardless of the choice made in \eqref{eq:PCAerrors}, $\widehat G\widehat \Lambda'=\widetilde G\widetilde \Lambda'$.
		
		Notice that the estimated loadings and factors are such that  $\widehat \Lambda'\widehat \Lambda /N$ and $\widetilde \Lambda'\widetilde \Lambda /N$ are diagonal and
		\[
		\frac 1T\sum_{t=1}^T \widehat G_t\widehat G_t'= I_q,\quad \frac 1T\sum_{t=1}^T \widetilde G_t\widetilde G_t'= I_q
		\]
		so that the estimated factors are orthonormal. 
		\section{OLS estimator and its asymptotic properties}\label{app:OLS}	
		
		The OLS estimator of $\theta$ is given by:
		\begin{equation}
			\widehat{\theta}^{\text{\upshape\tiny OLS}} := \left( \sideset{}{_{t=1}^{T}}\sum \widehat{X}_{t}' \widehat{X}_t \right)^{-1} 
			\left( \sideset{}{_{t=1}^{T}}\sum \widehat{X}_t' y_t \right).\label{eq:thetaOLS}
		\end{equation}
		This is the estimator proposed by \citet{zhuetal17} and \citet{chenetal2020cnar} for the case in which the network is observed and, respectively, when the FNAR errors are either uncorrelated or have a factor structure with serially uncorrelated node factors.
		
		We start by making the following standard assumption.
		
		\begin{assumption}[CLTs for FNAR - OLS]
			\label{as:X_nu}\
			\vspace{-10pt}
			\begin{enumerate}[label=(\roman*)]
				\item For all $t,s\in\mathbb Z$ with $t\ne s$, $\mathbb E[G_tG_s']=0$.
				\label{as:errors_nu_common_acf_OLS}
				\vspace{-10pt}
				\item As $N,T\to\infty$,
				$
				\frac{1}{N\sqrt{T}} \sum_{t=1}^{T} X_t' \nu_{t} \overset{d}{\to} N(0_{r+2},\Omega_0)
				$, 
				where 
				$\Omega_0 := \lim_{N \to \infty}  \frac{1}{N^2} \mathbb{E} \left[ X_t' V X_t \right]$  is an $(r+2)\times (r+2)$ positive definite matrix.
				\label{as:X_nu_i}
				\vspace{-10pt}
				\item As $N,T\to\infty$,
				$
				\frac{1}{NT} \sum_{t=1}^{T} X_t' X_{t} \overset{p}{\to}\Sigma_{XX}
				$,
				where 
				$\Sigma_{XX}:=\lim_{N\to\infty}\frac 1N \mathbb E[X_t' X_{t}]$ is an $(r+2)\times (r+2)$ positive definite matrix.
				\label{as:X_nu_ii}
			\end{enumerate}
		\end{assumption}

		Because of Assumptions \ref{as:X_nu}\ref{as:errors_nu_common_acf_OLS}, \ref{as:errors_nu}\ref{as:errors_nu_idio_ii}, and \ref{as:MB_Assumption_9_nu_common_idio}, we have that the FNAR errors $\{\nu_t\}$ are not autocorrelated. This is necessary for the CLT in the next part of this assumption to hold. Assumption \ref{as:X_nu}\ref{as:X_nu_i} is also found in \citet{baiandng2006}. In fact, Assumptions \ref{as:X_nu}\ref{as:X_nu_i} and \ref{as:X_nu}\ref{as:X_nu_ii} are made for simplicity and could be proved in a similar way as in \citet[Theorem 4.1 in the degenerate case of just one community, i.e., $K=1$ therein]{chenetal2020cnar}. 
		
		The OLS estimator in \eqref{eq:thetaOLS} satisfies:
		\begin{align}
			\widehat{\theta}^{\text{\upshape\tiny OLS}}-\bar J\theta =&\, \left(\frac 1{NT} \sideset{}{_{t=1}^{T}}\sum \widehat{X}_{t}'\widehat X_{t}\right)^{-1} \left\{\left(\frac1{NT}\sideset{}{_{t=1}^{T}}\sum \widehat X_{t}' \nu_t\right)+ \left(\frac1{NT}\sideset{}{_{t=1}^{T}}\sum \widehat X_{t}' u_t\right)\right\}. 
			\label{eq:theta_hat_terms}
		\end{align}
		The asymptotic properties of the terms of \eqref{eq:theta_hat_terms} are given in the following Proposition.
		
		\begin{proposition} \label{lemma:part2_lemma_1}
			Under Assumptions \ref{as:common_component}-\ref{as:X_nu_v} and \ref{as:X_nu}, as $m,N,T\to\infty$,
			\begin{enumerate}[label=(\roman*)]
				\item $\frac1{NT}\sum_{t=1}^T\widehat X_{t}' u_t = O_p \left( \max\left(\frac 1{N^{2}T},\frac 1{m},\frac 1{\sqrt {mT}}\right)\right) $.
				\label{lemma:part2_lemma_1_i}		
				\vspace{-10pt}
				\item $\frac1{NT}\sum_{t=1}^T\widehat X_{t}' \nu_t = O_p\left(\max\left(\frac 1{\sqrt {T}},\frac 1{N^{2}T},\frac 1{m},\frac 1{\sqrt {mT}}\right)\right)$. 
				\label{lemma:part2_lemma_1_ii}
				\vspace{-10pt}
				\item $ \norm {\frac 1{NT}\sum_{t=1}^T \widehat{X}_{t}'\widehat X_{t}-\frac 1{NT}\sum_{t=1}^T\bar J {X}_{t}' X_{t}\bar J} = O_p \left( \max\left(\frac 1{N^{2}T},\frac 1{m},\frac 1{\sqrt {mT}}\right)\right)$.  					
				\label{lemma:part2_lemma_1_iii}		
			\end{enumerate}		
		\end{proposition}
		The next theorem follows.
		
		\begin{theorem}[CLT for FNAR coefficients estimated by OLS]\label{th:CLT_FNAR}
			Under Assumptions \ref{as:common_component}-\ref{as:X_nu_v} and \ref{as:X_nu}, 
			if ${ \sqrt T}/ m \to 0$, as $m,N,T\to\infty$,
			\begin{equation}
				\sqrt T(\widehat{\theta}^{\text{\upshape\tiny OLS}}-\bar J\theta)
				\overset{d}{\to}
				\mathcal{N} ( 0_{r+2},
				\Sigma_{XX}^{-1}
				\Omega_0
				\Sigma_{XX}^{-1}
				),\nonumber
			\end{equation}
			where $\Omega_0$  and $\Sigma_{XX}$ are defined in Assumptions \ref{as:X_nu}\ref{as:X_nu_i} and \ref{as:X_nu}\ref{as:X_nu_ii}, respectively, and
			$\bar J$  is defined in \eqref{eq:barJ}.
		\end{theorem}

		From Proposition \ref{lemma:part2_lemma_1}, we see that if $\sqrt{ T}/ m \to 0$, as $m,N,T\to\infty$, then the network factors can be treated as observed and Theorem \ref{th:CLT_FNAR} follows. In particular, by virtue of Assumptions \ref{as:X_nu}\ref{as:X_nu_i} and  \ref{as:X_nu}\ref{as:X_nu_ii}, the OLS estimator is $\sqrt T$-consistent and asymptotically normal. Notice that 
		the requirement $\sqrt T/m\to 0$ for Theorem \ref{th:CLT_FNAR} to hold is analogous to the one assumed in the vector case by \citet{baiandng2006}. 
		
		An estimator of the asymptotic variance-covariance matrix of the OLS estimator is then given by (see \citealp[Theorem 1]{baiandng2006}):
		\[
		\widehat{\Avar}\left[ \sqrt{T} (\widehat{\theta}^{\text{\upshape\tiny OLS}} -\bar J\theta) \right]
		=
		\left(\frac{1}{NT} \sum_{t=1}^{T} \widehat{X}_{t}^' \widehat{X}_t \right)^{-1} 
		\left(\frac{1}{N^2T} \sum_{t=1}^{T} \widehat{X}_t^' 
		\widehat \nu_t   \widehat \nu_t' 
		\widehat{X}_t  \right) 
		\left( \frac{1}{NT} \sum_{t=1}^{T} \widehat{X}_{t}^' \widehat{X}_t \right)^{-1}.
		\]

		\section{GLS estimator and its asymptotic properties}\label{app:GLS}
		
		A GLS extension of the OLS estimator of the FNAR considered in Appendix \ref{app:OLS} can be computed by means of the following procedure, initially proposed by \citet{chenetal2020cnar} for the special case where the network is observed. 
		
		Once we estimate the node factors and their loadings as described in Appendix \ref{app:node factors}, let $\widehat{\epsilon}:=\widehat \nu-\widehat G\widehat \Lambda' = \widehat \nu-\widetilde G\widetilde \Lambda'$ and $\widehat S$ be the diagonal matrix with entries the diagonal entries of $T^{-1}\widehat{\epsilon}'\widehat{\epsilon}$. We can estimate the covariance matrix $V$ of $\nu_t$ as $\widehat V :=\widehat \Lambda\widehat \Lambda'+\widehat S$ and, by applying the Sherman-Morrison-Woodbury formula, its inverse as:
		\begin{equation}\label{eq:invVsherman}
			\widehat V^{-1}:=\widehat S^{-1}-\widehat S^{-1}\widehat \Lambda(I_q+\widehat \Lambda'\widehat S^{-1}\widehat \Lambda)^{-1}\widehat \Lambda' \widehat S^{-1},
		\end{equation}
		where $\widehat S^{-1}$ is a diagonal matrix and hence easy to compute.

		The GLS estimator of $\theta$ is then given by:
		\begin{equation}
			\widehat{\theta}^{\text{\upshape\tiny GLS}} := \left(  \sum_{t=1}^{T} \widehat{X}_{t}' \widehat V^{-1} \widehat{X}_t \right)^{-1} 
			\left( \sum_{t=1}^{T} \widehat{X}_t' \widehat V^{-1} y_t \right).\label{eq:thetaGLS}
		\end{equation}
		
		To study the asymptotic properties of the GLS estimator, we make two more assumptions. First, we extend Assumptions \ref{as:X_nu}\ref{as:X_nu_i} and \ref{as:X_nu}\ref{as:X_nu_ii}  to the following.
		
		\begin{assumption}[CLTs for FNAR - GLS]
			\label{as:X_nu_app}\
			\vspace{-10pt}
			\begin{enumerate}[label=(\roman*)]
				\item For all $t,s\in\mathbb Z$ with $t\ne s$, $\mathbb E[G_tG_s']=0$.
				\label{as:errors_nu_common_acf}
				\vspace{-10pt}
				\item As $N,T\to\infty$,
				$
				\frac{1}{\sqrt{NT}} \sum_{t=1}^{T} X_t' V^{-1}\nu_{t} \overset{d}{\to} N(0_{r+2},\Omega_1)
				$, where
				$\Omega_1 := \lim_{N \to \infty}  \frac{1}{N} \mathbb{E} \left[ X_t' V^{-1} X_t \right]$  is an $(r+2)\times (r+2)$ positive definite matrix.
				\label{as:X_nu_iii}
				\vspace{-10pt}
				\item As $N,T\to\infty$,
				$
				\frac{1}{NT} \sum_{t=1}^{T} X_t' V^{-1} X_{t}\overset{p}{\to}\Omega_1
				$,
				where 
				$\Omega_1$ is defined in part \ref{as:X_nu_iii}.
				\label{as:X_nu_iv}
			\end{enumerate}
		\end{assumption}
		
		Because of Assumptions \ref{as:X_nu_app}\ref{as:errors_nu_common_acf}, \ref{as:errors_nu}\ref{as:errors_nu_idio_ii}, and \ref{as:MB_Assumption_9_nu_common_idio}, we have that the FNAR errors $\{\nu_t\}$ are not autocorrelated. This is necessary for the CLT in the next part of this assumption to hold.
		Assumptions \ref{as:X_nu_app}\ref{as:X_nu_iii} and \ref{as:X_nu_app}\ref{as:X_nu_iv}   follow directly from \ref{as:X_nu}\ref{as:X_nu_i} and \ref{as:X_nu}\ref{as:X_nu_ii} since we know that $V^{-1}$ is finite for all $N\in\mathbb N$. Notice, however, the different role of $N$ in the definition of $\Omega_0$ and $\Omega_1$, indeed, as $N\to\infty$ we have $X_t'VX_t=O_p(N^2)$, but $X_t' V^{-1} X_t=O_p(N)$, since $X_t=O_p(\sqrt N)$, $V=O(N)$ and $V^{-1}=O(1)$.

		Second, to study the properties of the GLS estimator \eqref{eq:thetaGLS} we need to prove consistency of the estimated inverse of the FNAR errors covariance $\widehat V^{-1}$ defined in \eqref{eq:invVsherman}. This is not an easy task, for at least three reasons: first, the FNAR errors are estimated and not observed; second, the matrix $V$ is $N\times N$ so it is a high-dimensional one; third, to study $\widehat V^{-1}$, we need uniform consistency over all $N^2$ entries of the estimated covariance $\widehat V$. These difficulties are reduced if we assume that all considered random variables are sub-Gaussian, which is a classical assumption in high-dimensional statistics \citep[see, e.g.,][Chapter 2]{vershynin2018high}. 
		
		\begin{assumption}[Sub-Gaussianity]\label{as:subgauss}\
			\vspace{-10pt}
			\begin{enumerate}[label=(\roman*)]
				\item For all $i\in\mathbb N$, all $j=1,\ldots, r+2$, and all $t\in\mathbb Z$, $\text{\upshape P}(\abs{X_{ijt}-\mathbb E[X_{ijt}]}>s)\le 2\exp(-s^2/c_1^2)$ 
				for some finite $c_1$ independent of $i,j$, and $t$.
				\label{as:subgauss_i}
				\vspace{-10pt}
				\item For all $i,j,k\in\mathbb N$ and all $t\in\mathbb Z$, $\text{\upshape P}(\abs{\mathcal E_{tijk}}>s)\le 2\exp(-s^2/c_2^2)$ for some finite $c_2$ independent of $i,j,k$, and $t$.
				\label{as:subgauss_ii}
				\vspace{-10pt}
				\item For all $i\in\mathbb N$ and all $t\in\mathbb Z$, $\text{\upshape P}(\abs{\epsilon_{it}}>s)\le 2\exp(-s^2/c_3^2)$ for some finite $c_3$ independent of $i$ and $t$.
				\label{as:subgauss_iii}
			\end{enumerate}		
		\end{assumption}
		
		This approach is similar to the one adopted in a vector factor model context by \citet{fan2013large}. Instead, \citet{chenetal2020cnar} assume a set of moment conditions on the regressors matrix $X_t$, which amount to bounding up the 8th order cross-cumulants (in addition, they also make use of Hanson-Wright concentration inequalities which are based
		on the assumption of sub-Gaussianity).

		The GLS estimator in \eqref{eq:thetaGLS} satisfies:
		\begin{align}
			\widehat{\theta}^{\text{\upshape\tiny GLS}}-\bar J\theta =&\, \left(\frac 1{NT} \sum_{t=1}^T \widehat{X}_{t}'\widehat V^{-1}\widehat X_{t}\right)^{-1} \left\{\left(\frac1{NT}\sum_{t=1}^T\widehat X_{t}' \widehat V^{-1}\nu_t\right)+ \left(\frac1{NT}\sum_{t=1}^T\widehat X_{t}' \widehat V^{-1}u_t\right)\right\}. 
			\label{eq:thetaGLS_hat_terms}
		\end{align}	
		The asymptotic properties of the terms of \eqref{eq:thetaGLS_hat_terms} are then given in the following Proposition.
		
		\begin{proposition} \label{lemma:GLS}
			Under Assumptions \ref{as:common_component}-\ref{as:X_nu_v}, \ref{as:X_nu_app}, and \ref{as:subgauss}, if $\sqrt T/m\to 0$, as $m,N,T\to\infty$,
			\begin{enumerate}[label=(\roman*)]
				\item $\frac1{NT}\sum_{t=1}^T\widehat X_{t}'\widehat V^{-1} u_t = O_p \left( \max\left(\frac 1{N^{2}T},\frac 1{m},\frac 1{\sqrt {mT}}\right)\right)$.
				\label{lemma:GLS_i}		
				\vspace{-10pt}
				\item $\frac1{NT}\sum_{t=1}^T\widehat X_{t}' \widehat V^{-1}\nu_t = O_p \left( \max\left(\frac 1{\sqrt{NT}},\frac 1{N^{2}T},\frac 1{m},\frac 1{\sqrt {mT}}\right)\right)$. 
				\label{lemma:GLS_ii}
				\vspace{-10pt}
				\item $ \norm {\frac 1{NT}\sum_{t=1}^T \widehat{X}_{t}'\widehat V^{-1}\widehat X_{t}-\frac 1{NT}\sum_{t=1}^T\bar J {X}_{t}' V^{-1} X_{t}\bar J} = O_p\left(\max\left(\frac 1{\sqrt N},\sqrt{\frac{\log N}{T}},\frac 1{m},\frac 1{\sqrt {mT}}\right)\right)$. 
				\label{lemma:GLS_iii}		
			\end{enumerate}		
		\end{proposition}
		The next theorem follows.
		
		\begin{theorem}[CLT for FNAR coefficients estimated by GLS]\label{th:CLT_FNAR_GLS}
			Under Assumptions  \ref{as:common_component}-\ref{as:X_nu_v}, \ref{as:X_nu_app}, and \ref{as:subgauss}, if ${ \sqrt {NT}}/ m \to 0$ and $N/m \to 0$, as $m,N,T\to\infty$,
			\begin{equation}
				\sqrt {NT}(\widehat{\theta}^{\text{\upshape\tiny GLS}}-\bar J\theta)
				\overset{d}{\to}
				\mathcal{N} ( 0_{r+2},
				\Omega_1^{-1}
				),\nonumber
			\end{equation}
			where $\Omega_1$  is defined in Assumptions \ref{as:X_nu_app}\ref{as:X_nu_iii} and
			$\bar J$  is defined in \eqref{eq:barJ}.
		\end{theorem}
		
		For observed network factors, the GLS estimator has a faster rate of convergence than the OLS estimator and it is more efficient; see also \citet[Theorem 4.3]{chenetal2020cnar}. The different rates depend on the different scaling needed for Assumptions \ref{as:X_nu}\ref{as:X_nu_i} and \ref{as:X_nu_app}\ref{as:X_nu_iii} to hold. Indeed, on the one hand $\mathbb{E} \left[ X_t' V X_t \right]=O(N^2)$, while, on the other hand $\mathbb{E} \left[ X_t' V^{-1} X_t \right]=O(N)$. This is because, by Assumption \ref{as:errors_nu}, $\norm{V}=O(N)$  but $\norm{V^{-1}}=O(1)$.
		
		Now, from Proposition \ref{lemma:GLS}, we see that the conditions ${ \sqrt {NT}}/ m \to 0$ and $N/m\to 0$ allow us to treat the factors as observed. These conditions are stronger than in the OLS case. 
		
		An estimator of the asymptotic variance-covariance matrix of the GLS estimator is then given by:
		\[
		\widehat{\Avar}\left[ \sqrt{NT} (\widehat{\theta}^{\text{\upshape\tiny GLS}} -\bar J\theta) \right]
		=
		\left(\frac{1}{NT} \sum_{t=1}^{T} \widehat{X}_{t}^' \widehat{V}^{-1} \widehat{X}_t \right)^{-1}, 
		\]	
		where $ \widehat{V}^{-1}$ is defined in \eqref{eq:invVsherman}. Alternatively, to address possible residual cross-correlation of the node idiosyncratic components, we could use:
		\[
		\widehat{\Avar}\left[ \sqrt{NT} (\widehat{\theta}^{\text{\upshape\tiny GLS}} -\bar J\theta) \right]
		=
		\left(\frac{1}{NT} \sum_{t=1}^{T} \widehat{X}_{t}^' \widehat{V}^{-1} \widehat{X}_t \right)^{-1}
		\left(\frac{1}{NT} \sum_{t=1}^{T} \widehat{X}_{t}^' \widehat{V}^{-1}  \widehat\nu_t\widehat\nu_t' \widehat{V}^{-1} \widehat{X}_t \right)
		\left(\frac{1}{NT} \sum_{t=1}^{T} \widehat{X}_{t}^' \widehat{V}^{-1} \widehat{X}_t \right)^{-1}.
		\]	
		
		

		\newpage
		\section{Proofs of the main results}\label{app:mainproof}
		
		\subsection{General statements of Assumptions \ref{as:idiosyncratic_component}, 
			\ref{as:MB_Indep}, 
			\ref{as:CLT},
			and \ref{as:X_nu_v} }
		All the following results in Appendices \ref{app:mainproof} and \ref{app:lemma} are proved under a more general version of the assumptions in the main text. Namely, we replace Assumptions 
		\ref{as:idiosyncratic_component}\ref{as:idiosyncratic_component_ii},
		\ref{as:idiosyncratic_component}\ref{as:idiosyncratic_component_iv},
		\ref{as:idiosyncratic_component}\ref{as:idiosyncratic_component_v}, 
		\ref{as:MB_Indep}, 
		\ref{as:CLT}\ref{as:CLT_i},
		and \ref{as:X_nu_v} with:

		\textsc{Assumption 2} (General statement). 
		{\it 
			\begin{enumerate}
				\item [(ii)] \vskip -.3cm
				For all $N \in \mathbb{N}$, all $t,s \in \mathbb{Z}$, and all $i,j=1,\ldots N^2$, 
				$
				{N^{-\gamma}}
				\sum_{h=1}^{N^2} \sum_{k=1}^{N^2}
				\abs{
					\mathbb{E} \left[ \mathcal{E}_{(3) t i h} \mathcal{E}_{(3) s j k} \right]	 
				}
				\leq \rho_{\mathcal{E}}^{\abs{t-s}} M_{ij}  
				$
				and, for all $N \in \mathbb{N}$, all $t,s \in \mathbb{Z}$, and all $i,j,k=1, \dots, N^2$,
				$
				{N^{-\gamma}}
				\sum_{h=1}^{N^2}
				\abs{
					\mathbb{E} \left[ \mathcal{E}_{(3) t i h} \mathcal{E}_{(3) s j k} \right]	 
				}
				\leq \rho_{\mathcal{E}}^{\abs{t-s}} M_{ij}  
				$
				for some $\gamma \in [0,2]$ and some finite $\rho_{\mathcal{E}}$ and $M_{ij}$ independent of $t,s,k$ and $N$ such that
				$0 \leq \rho_{\mathcal{E}} <1 $, $\sum_{i=1, i\ne j}^m M_{ij}  \leq  M_{\mathcal{E}}$ and  $\sum_{j=1,j\ne i}^m M_{ij}  \leq  M_{\mathcal{E}}$, for some finite $ M_{\mathcal{E}}$ independent of $i,j$ and $m$.
				\vspace{-10pt}
				\item [(iv)]
				For all $m,T, N \in \mathbb{N}$ and all $j=1,\ldots, N^2$ and all $s=1,\ldots, T$,
				\begin{equation*}
					\mathbb{E} \left[  \abs{ \frac{1}{\sqrt{mT} N^{\gamma} } \sum_{i=1}^{m} \sum_{t=1}^{T} \sum_{h=1}^{N^2} \sum_{k=1}^{N^2} 
						\left\{ 
						\mathcal{E}_{(3)t i h} \mathcal{E}_{(3)t j k}
						-
						\mathbb{E} \left[  \mathcal{E}_{(3)t i h} \mathcal{E}_{(3)t j k} \right]
						\right\} }^2 \right]
					\leq C_{\mathcal{E}}
				\end{equation*}		
				and
				\begin{equation*}
					\mathbb{E} \left[  \abs{ \frac{1}{\sqrt{mT} N^{\gamma} } \sum_{i=1}^{m} \sum_{t=1}^{T} \sum_{h=1}^{N^2} \sum_{k=1}^{N^2} 
						\left\{ 
						\mathcal{E}_{(3)t i h} \mathcal{E}_{(3)s i k} 
						-
						\mathbb{E} \left[  \mathcal{E}_{(3)t i h} \mathcal{E}_{(3)s i k} \right]
						\right\} }^2 \right]
					\leq C_{\mathcal{E}}
				\end{equation*}		
				for some finite $C_{\mathcal{E}}$ independent of $j, s, m, T, N$ and some $\gamma \in [0,2]$.
				\vspace{-10pt}
				\item [(v)]
				For all $m,T, N \in \mathbb{N}$ and all $j=1,\ldots, N^2$ and all $s=1,\ldots, T$,
				\begin{equation*}
					\mathbb{E} \left[  \abs{ \frac{1}{\sqrt{mT} N^{\gamma/2} } \sum_{i=1}^{m} \sum_{t=1}^{T} \sum_{h=1}^{N^2}
						\left\{ 
						\mathcal{E}_{(3)t i h} \mathcal{E}_{(3)t j h}
						-
						\mathbb{E} \left[  \mathcal{E}_{(3)t i h} \mathcal{E}_{(3)t j h} \right]
						\right\} }^2 \right]
					\leq C_{\mathcal{E}}
				\end{equation*}		
				and
				\begin{equation*}
					\mathbb{E} \left[  \abs{ \frac{1}{\sqrt{mT} N^{\gamma/2} } \sum_{i=1}^{m} \sum_{t=1}^{T} \sum_{h=1}^{N^2}
						\left\{ 
						\mathcal{E}_{(3)t i h} \mathcal{E}_{(3)s i h} 
						-
						\mathbb{E} \left[  \mathcal{E}_{(3)t i h } \mathcal{E}_{(3)s i h} \right]
						\right\} }^2 \right]
					\leq C_{\mathcal{E}}
				\end{equation*}		
				for some finite $C_{\mathcal{E}}$ independent of $j, s, m, T, N$ and some $\gamma \in [0,2]$.
			\end{enumerate}
		}
		
		\noindent
		\textsc{Assumption 3} (General statement).  
		{\it 
			For all $i,j\in\mathbb N$, all $k=1,\ldots,r$, and all $t\in\mathbb Z$,  $\mathbb E[\mathcal{F}_{(3)tkj}\mathcal{E}_{(3)tij}]=0$,
			and, for all $m,N,T\in\mathbb N$ and all $t=1,\ldots, T$,
			\begin{align}
				&\mathbb{E} \left[\frac 1{m N^{\gamma}} \sum_{i=1}^{m} \norm{ \frac 1{\sqrt T} \sum_{t=1}^{T}  \mathcal{F}_{(3)t} \mathcal{E}_{(3)t i\cdot}^\prime  }_F^2 \right] \leq C_{\mathcal{F}\mathcal{E}},\nonumber\\
				&\mathbb{E}\left[\frac 1{ N^{2\gamma} }\left\Vert\frac 1{\sqrt{mT}}\sum_{i=1}^{m}  \sum_{s=1}^{T}\mathcal{F}_{(3)s}\left\{\mathcal{E}_{(3)s i \cdot}^\prime \mathcal{E}_{(3)t i \cdot}-\mathbb E[\mathcal{E}_{(3)s i \cdot}^\prime \mathcal{E}_{(3)t i \cdot}]\right\}  \right\Vert_F^2\right]\le C_{\mathcal{F}\mathcal{E}}^\prime,\nonumber
			\end{align}
			for some finite $C_{\mathcal{F}\mathcal{E}}$ and  $C_{\mathcal{F}\mathcal{E}}^\prime$ independent of $t$, $m, N$, and $T$ and some $\gamma \in [0,2]$.}\medskip

		\noindent
		\textsc{Assumption 5} (General statement). 
		{\it
			\begin{enumerate}
				\item [(i)] \vskip -.3cm For any given $i=1,\ldots, m$, 
				as $N,T\to\infty$, 
				$
				\frac{1}{\sqrt {N^\gamma T}}  \sum_{t=1}^{T} \mathcal{F}_{(3)t} \mathcal{E}_{(3)ti\cdot}' \overset{d}{\to}\mathcal N(0_r,\Phi_i),
				$
				where
				$
				\Phi_i:=\lim_{N,T\to\infty}  \mathbb E\left[\left (\frac 1{\sqrt {N^{\gamma} T }}\sum_{t=1}^{T} \mathcal{F}_{(3)t} \mathcal{E}_{(3)t i \cdot}' \right)
				\left(\frac 1{\sqrt {N^{\gamma} T }}\sum_{t=1}^{T} \mathcal{F}_{(3)t} \mathcal{E}_{(3)t i \cdot}' \right)'
				\right],
				$
				for some $\gamma \in [0,2]$.
			\end{enumerate}	
		}

		\noindent
		\textsc{Assumption 10} (General statement). 
		{\it For all $m,N,T\in\mathbb N$,
			\begin{align}
				&\mathbb E\left[\norm{\frac 1{\sqrt {mT}N^{1+\gamma/2}}\sum_{t=1}^T\sum_{i=1}^m u_i \mathcal E_{(3)ti\cdot} (y_{t-1}\otimes X_t)}^2 \right]\le\mathfrak K_1,\nonumber\\
				&\mathbb E\left[\norm{\frac 1{\sqrt {mT}N^{1+\gamma/2}}\sum_{t=1}^T\sum_{i=1}^m u_i \mathcal E_{(3)ti\cdot} (y_{t-1}\otimes \nu_t)}^2 \right]\le\mathfrak K_2,\nonumber
			\end{align} 
			for some finite $\mathfrak K_1$ and $\mathfrak K_2$ independent of $m,N$, and $T$ and some $\gamma\in[0,2]$.}

		These assumptions all depend on a generic $\gamma\in[0,2]$, thus are more general than those stated in the main text, which correspond to the case $\gamma=2$. All Theorems and Propositions in the main text and in Appendices \ref{app:OLS} and \ref{app:GLS} are stated in the case $\gamma=2$, which is the least favorable one, meaning it is the case giving the slowest possible convergence rates.			
		
		\subsection{Proof of Proposition \ref{prop:station}}
		
		\begin{proof}
			First, let $A_{t}:= N^{-1}\sum_{j=1}^r\beta_j F_{j,t-1}+\rho$ and
			notice that \eqref{eq:famnvar}  can be rewritten as 
			\begin{equation}\label{eq:FNAR_as_VAR}
				y_t = \alpha + A_{t} y_{t-1}+\nu_t.
			\end{equation}
			Thus, for any given $N\in\mathbb N$,
			letting  $y_{-\infty}=0$, if there exists a causal solution it is given by:
			\begin{align}
				y_t 
				&=\left\{\prod_{k=0}^\ell(\alpha+A_{t-k})\right\}y_{t-\ell-1}+\sum_{j=1}^\ell\left\{\prod_{k=0}^{j-1}(\alpha+A_{t-k}) \right\}\nu_{t-j}+\nu_t
				\nonumber\\
				&=\sum_{j=1}^\infty\left\{\prod_{k=0}^{j-1}(\alpha+A_{t-k}) \right\}\nu_{t-j}+\nu_t.\label{eq:FNAR_as_MA}
			\end{align}
			To this end first notice that since by Assumption \ref{as:MB_Assumption_9_nu_common_idio}, $\{\mathcal F_t\}$ is independent of $\{\nu_t\}$, for any given $N\in\mathbb N$,
			\[
			\mathbb E[y_t]= \alpha+\rho \mathbb E[ y_{t-1}]+\frac 1N \sum_{j=1}^r\beta_j\mathbb E[F_{j,t-1}]  \mathbb E[ y_{t-1}]
			\]
			hence, a stationary solution must have mean
			\[
			\mathbb E[y_t]=\left(I_N-\rho I_N- \frac 1N\sum_{j=1}^r\beta_j\mathbb E[F_{j,t}]\right)^{-1}\alpha,
			\]
			which is finite and independent of $t$ because of Assumptions \ref{as:common_component}\ref{as:common_component_ii} and \ref{as:stability}\ref{as:stability_i}. Clearly if $\alpha=0$ then $	\mathbb E[y_t]=0$ and vice versa.
			
			Let then $\alpha=0$ for simplicity and define $\Sigma_{t,s} := \mathbb E[y_ty_s']$ and recall that $V:=\mathbb E[\nu_t\nu_t']$, then 
			\begin{align}
				\text{vec}(\Sigma_{t,t})&= \rho^2 \text{vec}(\Sigma_{t-1,t-1}) +\frac 1{N^{2}}\sum_{j=1}^r\beta_j^2 \mathbb E [ F_{j,t-1}\otimes F_{j,t-1}] \text{vec}(\Sigma_{t-1,t-1})+ \text{vec}(V)\nonumber\\
				&=  \left\{\rho^2 I_{N^2}+ \frac1 {N^2}\sum_{j=1}^r \beta_j^2 \mathbb E [ F_{j,t-1}\otimes F_{j,t-1}]\right\} \text{vec}(\Sigma_{t-1,t-1})+ \text{vec}(V)\nonumber\\
				& = \sum_{k=0}^\ell \left\{\rho^2 I_{N^2}+ \frac1 {N^2}\sum_{j=1}^r\beta_j^2 \mathbb E [ F_{j,t-1}\otimes F_{j,t-1}]\right\}^k  \text{vec}(V)\nonumber\\
				&+ \left\{\rho^2 I_{N^2}+ \frac 1{N^2}\sum_{j=1}^r\beta_j^2 \mathbb E [ F_{j,t-1}\otimes F_{j,t-1}]\right\}^{\ell+1}\text{vec}(\Sigma_{t-\ell-1,t-\ell-1}).
			\end{align}
			Notice also that  $V$ is positive definite, indeed its smallest eigenvalue is such that, by Weyl's inequality,
			\[
			\mu_N(V) \ge \mu_N(\Lambda\Lambda')+\mu_N(S) = \min_{i=1,\ldots, N}\mathbb E[\epsilon_{it}^2]\ge  \underline M_{\epsilon},
			\]
			because of Assumption \ref{as:errors_nu}\ref{as:errors_nu_idio_i} and where $\mu_N(\Lambda\Lambda')$, $\mu_N(V)$, and $\mu_N(S)$ are the smallest eigenvalues of $\Lambda\Lambda'$, $V$, and $S$, respectively, and $\mu_N(\Lambda\Lambda')=0$.
			
			Hence, letting $\ell\to\infty$ we see that $\text{vec}(\Sigma_{t,t})$ is finite and independent of $t$, because of Assumptions \ref{as:common_component}\ref{as:common_component_ii} and \ref{as:stability}\ref{as:stability_ii}, and since  $V$ is 		positive definite, see also Theorem 2.1 in \citet{nicholls1981multiple}. 		
			
			To show that the above arguments imply that also $y_t^\omega=\lim_{N\to\infty} \omega'y_t$ exists almost surely and is stationary and causal it is enough to follow the same steps as in Theorem 2 by \citet{zhuetal17}. 
		\end{proof}
		\subsection{Proof of Theorem \ref{theorem:CLT_loadings}}
		
		\begin{proof}
			The proof of part \ref{theorem:CLT_loadings_i} follows directly from Lemmas \ref{lemma:MB_proposition_2}\ref{lemma:MB_proposition_2_ii} and \ref{lemma:H_J}.
			
			For part \ref{theorem:CLT_loadings_ii}, from \eqref{eq:MB_eq_29} in the proof of Lemma \ref{lemma:MB_proposition_2}\ref{lemma:MB_proposition_2_i} and Lemma \ref{lemma:H_J}\ref{lemma:H_J_i}, 	if $N^{\gamma/2}\sqrt T/m \to 0$,	
			as $m,N,T\to\infty$, we have
			\begin{align}
				N^{2-\gamma/2}\sqrt{T}\left(  \widehat{u}_i' - u_i' J \right) &=
				\frac{N^{2-\gamma/2}\sqrt{T}}{N^2 T}  \sum_{t=1}^{T} \mathcal{E}_{(3)ti\cdot} \mathcal{F}_{(3)t}' \left(\frac 1m \sum_{j=1}^{m}  u_j u_j'
				\right) 
				\left( \frac{\widehat{M}^{\mathcal{W}}}{m N^2} \right)^{-1}J  +o_p(1)\nonumber\\
				&=\frac 1{\sqrt{N^{\gamma}T}}\sum_{t=1}^{T} \mathcal{E}_{(3)ti\cdot} \mathcal{F}_{(3)t}'  \Sigma_U
				\left( \frac{{M}^{\chi}}{m N^2} \right)^{-1}J  +o_p(1)\nonumber\\
				&=\frac 1{\sqrt{N^{\gamma}T}}\sum_{t=1}^{T} \mathcal{E}_{(3)ti\cdot} \mathcal{F}_{(3)t}' J  +o_p(1)\nonumber\\
				&\overset{d}{\to}\mathcal N(0_r, J_0\Phi_i J_0),\nonumber
			\end{align} 
			because of Assumption \ref{as:MB_Assumption_9}\ref{as:MB_Assumption_9_i} and Lemma \ref{lemma:MB_Lemma_5}\ref{lemma:MB_Lemma_5_iii}, Assumption \ref{as:CLT}\ref{as:CLT_i}, and Slutsky's Theorem, and where $J_0=\plim_{m,N,T\to\infty} J$. Notice that $J\Phi_i J=\Phi_i$. This proves part \ref{theorem:CLT_loadings_ii}.
			The final statement follows by setting $\gamma=2$.
		\end{proof}
		\subsection{Proof of Theorem \ref{theorem:CLT_factors}}	
		
		\begin{proof}
			For part \ref{theorem:CLT_factors_i}, consider the estimated factors $\widehat{\mathcal{F}}_{(3)}$. We have that:
			\begin{align}
				\widehat{\mathcal{F}}_{t} &=  \mathcal{W}_{t} \times_3  \left( \widehat{U}'\widehat{U} \right)^{-1}  \widehat{U}' \nonumber\\
				&= \left( \mathcal{F}_{t}  \times_3 U + \mathcal{E}_t  \right) \times_3 \left( \widehat{U}'\widehat{U} \right)^{-1}  \widehat{U}' \nonumber \\
				&=    \mathcal{F}_{t}  \times_3  \left( \widehat{U}'\widehat{U} \right)^{-1}  \widehat{U}' \left( U - \widehat{U}J + \widehat{U} J \right)  \nonumber
				+ \mathcal{E}_t \times_3  \left( \widehat{U}'\widehat{U} \right)^{-1}   \left( \widehat{U} - \widehat{U}J + \widehat{U} J \right)',\nonumber
			\end{align}
			and
			\begin{align}
				\widehat{\mathcal{F}}_{(3)t} &=  
				\left( \widehat{U}'\widehat{U} \right)^{-1} 
				\left[  
				\widehat{U}' \left(U -  \widehat{U} J +  \widehat{U} J \right) \mathcal{F}_{(3)t} 
				\right]  
				+ \left(\widehat{U} - UJ +UJ \right)'  \mathcal{E}_{(3)t}\nonumber \\
				&=  \left( \frac{\widehat{M}^{\mathcal{W}}}{m N^2} \right)^{-1}
				\left[
				\frac{   \widehat{U}' \left(U - \widehat{U} J \right) }{m} \mathcal{F}_{(3)t}
				+ \frac{  \widehat{U}'\widehat{U}}{m} J  \mathcal{F}_{(3)t} 
				+ \frac{  \left(\widehat{U} - U J \right)' }{m} \mathcal{E}_{(3)t} 
				+  \frac{ J U' \mathcal{E}_{(3)t}}{m} 
				\right].\nonumber
			\end{align}
			Thus, because of Lemma \ref{lemma:MB_Lemma_7}\ref{lemma:MB_Lemma_7_iv},
			\begin{align}
				\frac{\widehat{\mathcal{F}}_{(3)t} - J  \mathcal{F}_{(3)t}}{N} 
				&=    \left( \frac{\widehat{M}^{\mathcal{W}}}{m N^2} \right)^{-1}
				\left[
				\frac{   \widehat{U}' \left(U - \widehat{U} J \right) }{mN} \mathcal{F}_{(3)t}
				+ \frac{    \left(\widehat{U} - U J \right)' }{mN} \mathcal{E}_{(3)t} 
				+ \frac{   J U' \mathcal{E}_{(3)t}}{mN} 
				\right] \nonumber \\
				&= O_p(1) \left[ A + B + C \right].\label{eq:F_HF}
			\end{align}
			
			For term (A), given \eqref{eq:MB_eq_125} in the proof of Lemma \ref{lemma:H_J} when $\widehat{H}=J$ and Lemma \ref{lemma:MB_Lemma_2}\ref{lemma:MB_Lemma_2_ii}
			\begin{equation}
				\norm{\frac{   \widehat{U}' \left(U - \widehat{U} J \right) }{mN} \mathcal{F}_{(3)t}}	
				\leq
				\norm{ \frac{   \widehat{U}' \left(U - \widehat{U} J \right) }{m}  }\
				\norm{\frac{\mathcal{F}_{(3)t}}{N}}
				= O_p \left(\frac{1}{\xi}\right),\label{eq:th2_termA}
			\end{equation}
			where $\xi = 
			\min \left(
			\sqrt{mT} N^{2-\gamma/2},
			N^{3-\gamma/2} T, 
			m \sqrt{T} N^{2-\gamma}, 
			m N^{2-\gamma}
			\right)
			$.
			For term (B), from \eqref{eq:Uhat_UHhat} in the proof of Lemma \ref{lemma:MB_proposition_2}\ref{lemma:MB_proposition_2_i} with $\widehat{H}=J$, we have 
			\begin{align}
				\frac{    \left(\widehat{U} - U J \right)' \mathcal{E}_{(3)t} }{mN}  
				=& \left(\frac{\widehat{M}^{\mathcal{W}}}{mN^2}\right)^{-1} J 
				\frac{U'\mathcal{E}_{(3)} \mathcal{F}_{(3)}' U' \mathcal{E}_{(3)t} }{m^2N^3T}
				+
				\left(\frac{\widehat{M}^{\mathcal{W}}}{mN^2}\right)^{-1} J 
				\frac{U'U \mathcal{F}_{(3)} \mathcal{E}_{(3)}' \mathcal{E}_{(3)t} }{m^2N^3T}\nonumber\\
				&+	\left(\frac{\widehat{M}^{\mathcal{W}}}{mN^2}\right)^{-1} J 
				\frac{U' \mathcal{E}_{(3)} \mathcal{E}_{(3)}' \mathcal{E}_{(3)t} }{m^2N^3T}\nonumber\\
				&+
				\left(\frac{\widehat{M}^{\mathcal{W}}}{mN^2}\right)^{-1} (\widehat{U} - UJ)'
				\frac{ \mathcal{E}_{(3)} \mathcal{F}_{(3)}' U' \mathcal{E}_{(3)t} }{m^2N^3T}
				+	
				\left(\frac{\widehat{M}^{\mathcal{W}}}{mN^2}\right)^{-1} (\widehat{U} - UJ)'
				\frac{ U \mathcal{F}_{(3)} \mathcal{E}_{(3)}' \mathcal{E}_{(3)t} }{m^2N^3T}\nonumber\\
				&+
				\left(\frac{\widehat{M}^{\mathcal{W}}}{mN^2}\right)^{-1} (\widehat{U} - UJ)'
				\frac{ \mathcal{E}_{(3)} \mathcal{E}_{(3)}' \mathcal{E}_{(3)t} }{m^2N^3T}\nonumber\\
				=& III_a + III_b + III_c + III_d + III_e + III_f.\nonumber
			\end{align}
			Then, because of Lemma \ref{lemma:MB_Lemma_2}\ref{lemma:MB_Lemma_2_i}, \ref{lemma:MB_Lemma_2}\ref{lemma:MB_Lemma_2_iii}, and \ref{lemma:MB_Lemma_7}\ref{lemma:MB_Lemma_7_iv}, and using \eqref{eq:MB_eq_126} and $\norm{J}=O(1)$,
			\begin{align}
				\norm{III_a} 
				&\leq 
				\norm{\left(  \frac{\widehat{M}^{\mathcal{W}}}{mN^2}  \right)^{-1}} \ 
				\norm{J} \ 
				\norm{  \frac{U'  \mathcal{E}_{(3)} \mathcal{F}_{(3)}'}{m N^2 T} } \
				\norm{\frac{U}{\sqrt{m}}} \
				\norm{\frac{\mathcal{E}_{(3)t}}{\sqrt{m}N}}\nonumber \\
				&= O_p(1) O(1) O_p\left(\frac{1}{\sqrt{mT}N^{2-\gamma/2}}\right)
				O_p(1) O\left(\frac{1}{N^{1-\gamma/2}}\right)\nonumber
				\\
				&= O \left(  \frac{1}{\sqrt{mT}N^{3-\gamma}}  \right).\nonumber
			\end{align}
			Turning to $III_b$ we have
			\begin{equation}
				\norm{\frac{\mathcal{F}_{(3)}\mathcal{E}_{(3)}' \mathcal{E}_{(3)t} }{m N^3 T}}\le \norm{\frac{\mathcal{F}_{(3)}\left\{\mathcal{E}_{(3)}' \mathcal{E}_{(3)t}-\mathbb {E}\left[\mathcal{E}_{(3)}' \mathcal{E}_{(3)t}\right]\right\}}{mN^3 T}}+\norm{\frac{\mathcal{F}_{(3)}\mathbb{E}\left[\mathcal{E}_{(3)}' \mathcal{E}_{(3)t}\right]  }{m N^3 T}}.\label{eq:DCPD0}
			\end{equation}
			By Assumption 	\ref{as:MB_Indep}, we have
			\begin{align}
				\mathbb{E}&\left[
				\norm{\frac{\mathcal{F}_{(3)}\left\{\mathcal{E}_{(3)}' \mathcal{E}_{(3)t}-\mathbb {E}\left[\mathcal{E}_{(3)}' \mathcal{E}_{(3)t}\right]\right\}}{mN^3 T}}^2
				\right]\le 			\mathbb{E}\left[
				\norm{\frac{\mathcal{F}_{(3)}\left\{\mathcal{E}_{(3)}' \mathcal{E}_{(3)t}-\mathbb {E}\left[\mathcal{E}_{(3)}' \mathcal{E}_{(3)t}\right]\right\}}{mN^3 T}}^2_F\right]\nonumber\\
				&\le \frac{1}{m^2 N^6 T^2}\mathbb E\left[\left\Vert\sum_{i=1}^{m}  \sum_{s=1}^{T}\mathcal{F}_{(3)s}\left\{\mathcal{E}_{(3)s i \cdot}^\prime \mathcal{E}_{(3)t i \cdot}-\mathbb E[\mathcal{E}_{(3)s i \cdot}^\prime \mathcal{E}_{(3)t i \cdot}]\right\}  \right\Vert^2\right]\le  \frac{C_{\mathcal{F}\mathcal{E}}^\prime }{m N^{6-2\gamma}T}.\label{eq:DCPD1}
			\end{align}
			Moreover, by Assumption  \ref{as:common_component}\ref{as:common_component_v} and Lemma \ref{lemma:MB_Lemma_1}\ref{lemma:Lemma_1_ii}, 
			\begin{align}
				\mathbb E&\left[\norm{\frac{\mathcal{F}_{(3)}\mathbb{E}\left[\mathcal{E}_{(3)}' \mathcal{E}_{(3)t}\right]  }{m N^3 T}}\right]\le \mathbb E\left[\norm{\frac{\mathcal{F}_{(3)}\mathbb{E}\left[\mathcal{E}_{(3)}' \mathcal{E}_{(3)t}\right]  }{m N^3 T}}_F\right]\nonumber\\
				&=\frac 1{mN^3 T} \mathbb E\left[\sqrt{ \sum_{l=1}^r \sum_{k=1}^{N^2} \left(\sum_{s=1}^T\sum_{h=1}^{N^2}\sum_{i=1}^m 
					\mathcal{F}_{(3)slh} \mathbb E\left[\mathcal{E}_{(3)sih}\mathcal{E}_{(3)tik}\right]
					\right)^2 }\right]\nonumber\\
				&\le\frac 1{mN^3 T} \mathbb E\left[ \sum_{l=1}^r \sum_{k=1}^{N^2} \left\vert \sum_{s=1}^T\sum_{h=1}^{N^2}\sum_{i=1}^m 
				\mathcal{F}_{(3)slh} \mathbb E\left[\mathcal{E}_{(3)sih}\mathcal{E}_{(3)tik}\right]
				\right\vert \right]\nonumber\\
				&\le\frac 1{mN^3 T}  \sum_{l=1}^r \sum_{k=1}^{N^2} \sum_{s=1}^T\sum_{h=1}^{N^2}\sum_{i=1}^m \mathbb E\left[\left\vert 
				\mathcal{F}_{(3)slh} \mathbb E\left[\mathcal{E}_{(3)sih}\mathcal{E}_{(3)tik}\right]
				\right\vert \right]\nonumber\\
				&\le\frac r{mN^3 T} \max_{l=1,\ldots,r} \max_{s=1,\ldots, T}\max_{h=1,\ldots, N^2} \mathbb E\left[\left\vert 
				\mathcal{F}_{(3)slh}\right\vert\right] \sum_{k=1}^{N^2} \sum_{s=1}^T\sum_{h=1}^{N^2}\sum_{i=1}^m 
				\left\vert\mathbb E\left[\mathcal{E}_{(3)sih}\mathcal{E}_{(3)tik}\right]
				\right\vert \nonumber\\
				&\le \frac {r C_{\mathcal F}^\prime N^\gamma m }{mN^3 T}.\label{eq:DCPD2}
			\end{align}
			From \eqref{eq:DCPD0}, \eqref{eq:DCPD1}, and \eqref{eq:DCPD2},
			\begin{align}
				\norm{III_b} 
				&\leq 
				\norm{\left(\frac{\widehat{M}^{\mathcal{W}}}{mN^2}\right)^{-1}} \ 
				\norm{J} \
				\norm{\frac{U'U }{m}} \
				\norm{\frac{\mathcal{F}_{(3)}\mathcal{E}_{(3)}' \mathcal{E}_{(3)t}  }{m N^3 T}}
				= O_p\left( \frac{1}{ \sqrt{mT} N^{3-\gamma} }\right)+O_p\left( \frac{1}{ T N^{3-\gamma} }\right).\nonumber 
			\end{align}
			Because of Lemma \ref{lemma:MB_Lemma_2}\ref{lemma:MB_Lemma_2_i}, \ref{lemma:MB_Lemma_2}\ref{lemma:MB_Lemma_2_iii}, and Lemma \ref{lemma:MB_Lemma_3}\ref{lemma:MB_Lemma_3_ii}, \ref{lemma:MB_Lemma_7}(iv), and since $\norm{J}=O(1)$,
			\begin{align}
				\norm{III_c} 
				&\leq 
				\norm{\left(\frac{\widehat{M}^{\mathcal{W}}}{mN^2}\right)^{-1}} \ 
				\norm{J} \
				\norm{\frac{U \mathcal{E}_{(3)} \mathcal{E}_{(3)}' }{m^{3/2} N^2T}}
				\norm{\frac{\mathcal{E}_{(3)t} }{\sqrt{m} N}}
				= O_p \left( \max \left(  \frac{1}{\sqrt{mT}N^{3-\gamma}}, \frac{1}{mN^{3(1-\gamma/2)}} \right)  \right).\nonumber
			\end{align}
			Next, notice that, by Assumption \ref{as:common_component}\ref{as:common_component_i}
			and Lemma \ref{lemma:MB_Lemma_1}\ref{lemma:Lemma_1_iii}
			\begin{align}
				\mathbb{E} \left[
				\norm{\frac{U' \mathcal{E}_{(3)t}}{mN}}^2 
				\right]
				&\leq
				\mathbb{E} \left[
				\norm{\frac{U' \mathcal{E}_{(3)t}}{mN}}^2_F 
				\right]
				=
				\frac{1}{m^2 N^2} \sum_{k=1}^{r}
				\mathbb{E} \left[
				\norm{u_k' \mathcal{E}_{(3)t}}^2 
				\right] \nonumber\\
				&= \frac{1}{m^2 N^2} \sum_{k=1}^{r}
				\mathbb{E} \left[
				\left(\sum_{i=1}^{m} \sum_{h=1}^{N^2} U_{ik} \mathcal{E}_{(3)t i h}
				\right)^2 
				\right]\nonumber \\
				&\leq 
				\frac{r}{m^2 N^2}
				\max_{k=1, \dots, r}
				\sum_{i=1}^{m} \sum_{j=1}^{m}
				\sum_{h=1}^{N^2} \sum_{l=1}^{N^2}
				\abs{ U_{ik} } \abs{ U_{jk} } 
				\abs{\mathbb{E}\left[ \mathcal{E}_{(3)t i h} \mathcal{E}_{(3)t j l} \right]}\nonumber \\
				&\leq
				\frac{r M_U^2}{m^2 N^2}
				\sum_{i=1}^{m} \sum_{j=1}^{m}
				\sum_{h=1}^{N^2} \sum_{l=1}^{N^2}
				\abs{\mathbb{E}\left[ \mathcal{E}_{(3)t i h}
					\mathcal{E}_{(3)t j l} \right]}
				\leq
				\frac{r M_U^2 M_{\mathcal{E}}}{m N^{2-\gamma}} 
				\label{eq:MB_eq_141}
			\end{align}
			since $M_{\mathcal{E}}$ does not depend on $t$.
			Therefore, by Theorem \ref{theorem:CLT_loadings}\ref{theorem:CLT_loadings_i}, Lemma \ref{lemma:MB_Lemma_3}\ref{lemma:MB_Lemma_3_i}, \ref{lemma:MB_Lemma_7}\ref{lemma:MB_Lemma_7_iv}, and
			using \eqref{eq:MB_eq_141}
			and  $\norm{J}=O(1)$
			\begin{align}
				\norm{III_d} 
				&\leq 
				\norm{\left(\frac{\widehat{M}^{\mathcal{W}}}{mN^2}\right)^{-1}}  \
				\norm{\frac{\widehat{U} - UJ}{\sqrt{m}}} \
				\norm{\frac{ \mathcal{E}_{(3)} \mathcal{F}_{(3)}'}{\sqrt{m}N^2T}}
				\norm{\frac{  U' \mathcal{E}_{(3)t} }{m N }}\nonumber\\
				&= O_p(1)  
				O_p \left(\max \left( \frac{1}{N^{2-\gamma/2} \sqrt{T}}, \frac{1}{ N^{2-\gamma} m} \right) \right) 
				O_p \left(\frac{1}{\sqrt{T}N^{2-\gamma/2}}\right) 
				O_p \left(\frac{1}{\sqrt{m}N^{1-\gamma/2}}\right) \nonumber\\
				&= 	
				O_p \left(\max \left( \frac{1}{N^{5-3\gamma/2} T \sqrt{m}}, \frac{1}{ N^{5-2\gamma} m^{3/2} \sqrt{T}} \right) \right) .\nonumber
			\end{align}
			Note that the term $III_d$ is clearly dominated by $III_a$. 	Analogously, $III_e$ and $III_f$ are dominated by $III_b$ and $III_c$, respectively. Thus, (B) is $O_p(1/(\sqrt{mT}N^{3-\gamma}, mN^{3-3\gamma/2}, TN^{3-\gamma}))$, hence it is dominated by term (A).
			
			For term (C), using \eqref{eq:MB_eq_141}
			\begin{equation}\label{eq:F_consistency_C}
				\norm{\frac{ J U' \mathcal{E}_{(3)t}  }{mN}}
				\leq 
				\norm{ J}	\norm{\frac{U' \mathcal{E}_{(3)t}}{mN}}
				=O_p \left(\frac{1}{\sqrt{m}N^{1-\gamma/2}}\right).
			\end{equation}
			By noticing that
			\[
			\max\left(\frac 1\xi, \frac{1}{\sqrt{m}N^{1-\gamma/2}}\right)=\max\left(\frac 1{N^{3-\gamma/2} T},\frac 1{\sqrt{m}N^{1-\gamma/2}}\right),
			\]
			we prove part \ref{theorem:CLT_factors_i}.

			For part \ref{theorem:CLT_factors_iii}, for any given $j=1,\ldots, N^2$, following the same reasoning as in part (i), we have
			\[
			{\widehat{\mathcal{F}}_{(3)t\cdot j} - J  \mathcal{F}_{(3)t\cdot j}} =N \left(\frac{\widehat{M}^{\mathcal{W}}}{m N^2} \right)^{-1}J \left[\frac{U' \mathcal{E}_{(3)t\cdot j}}{mN}+ O_p\left(\frac 1{N^3T}\right)\right].
			\]
			Moreover, 
			\[
			\norm{\frac{U' \mathcal{E}_{(3)t\cdot j}}{m}}=O_p\left(\frac 1{\sqrt m}\right).
			\]
			Therefore, if $\sqrt m/(N^2T)\to 0$ as $m,N,T\to\infty$, 
			\begin{align}
				\sqrt{m} \left({\widehat{\mathcal{F}}_{(3)t\cdot j} - J  \mathcal{F}_{(3)t\cdot j}} \right)
				&=\left(\frac{\widehat{M}^{\mathcal{W}}}{m N^2} \right)^{-1} J
				\left[\frac{U' \mathcal{E}_{(3)t\cdot j}}{\sqrt m}\right]			+o_p(1)\nonumber\\
				&=
				\Sigma_U^{-1} J
				\left[ \frac{   1}{\sqrt{m}}\sum_{i=1}^m u_i \mathcal E_{(3)tij} \right] + o_p(1)\nonumber\\
				&\overset{d}{\to}\mathcal N\left (0_{r},  \Sigma_U^{-1}J_0\Pi_{tj} J_0\Sigma_U^{-1}\right),\nonumber
			\end{align}
			because of Assumption \ref{as:MB_Assumption_9}\ref{as:MB_Assumption_9_i}, Lemma \ref{lemma:MB_Lemma_5}\ref{lemma:MB_Lemma_5_iii},
			Assumption \ref{as:CLT}\ref{as:CLT_iii}, and Slutsky's theorem, and $J_0=\plim_{m,N,T\to\infty} J$. Notice that $ \Sigma_U^{-1}J_0\Pi_{tj} J_0\Sigma_U^{-1}= \Sigma_U^{-1}\Pi_{tj}\Sigma_U^{-1}$. This proves part \ref{theorem:CLT_factors_iii}. 
		\end{proof}

		\subsection{Proof of Theorem \ref{th:CLT_FNAR_BAI}}
		
		\begin{proof} First, notice that 
			\begin{equation}\label{eq:MLstar}
				\norm{M_{\widehat G^\dag}} = \norm{I_T-\widehat G^\dag\widehat G^{\dag'}/T}= O(1),\qquad \norm{M_{\widehat \Lambda^*}} = \norm{I_N-\widehat V^{\widehat\nu^*}\widehat V^{\widehat{\nu}^{*'}} }= O(1)
			\end{equation}
			since $\norm{\widehat G^\dag}=O_p(\sqrt T)$ and eigenvectors are normalized.
			Then, because of \eqref{eq:MLstar} and by the same arguments used in the proof of Proposition \ref{lemma:GLS}\ref{lemma:GLS_i}, we have
			\begin{align}
				\norm{\frac 1{NT}\sum_{i=1}^N\widehat{X}_i'M_{\widehat G^\dag} u_i }=&\, O_p \left( \max\left(\frac 1{N^{3-\gamma/2}T},\frac 1{mN^{2-\gamma}},\frac 1{\sqrt {mT}N^{1-\gamma/2}}\right)\right),\label{eq:pezzo1}\\
				\norm{\frac 1{NT}\sum_{t=1}^T\widehat{X}_t'M_{\widehat \Lambda^*} u_t }=&\, O_p \left( \max\left(\frac 1{N^{3-\gamma/2}T},\frac 1{mN^{2-\gamma}},\frac 1{\sqrt {mT}N^{1-\gamma/2}}\right)\right).\label{eq:pezzo1_t}
			\end{align}
			Moreover,
			\begin{align}		
				&\norm{\frac 1{NT}\sum_{i=1}^N  \widehat{X}_{i}' M_{\widehat G^\dag} \widehat X_{i}-\frac 1{NT} \sum_{i=1}^N \bar J \widehat{X}_{i}' M_{\widehat G^\dag} \widehat X_{i}\bar J}= O_p \left( \max\left(\frac 1{N^{3-\gamma/2}T},\frac 1{mN^{2-\gamma}},\frac 1{\sqrt {mT}N^{1-\gamma/2}}\right)\right),\label{eq:pezzo2}\\
				&\norm{\frac 1{NT}\sum_{t=1}^T  \widehat{X}_{t}' M_{\widehat \Lambda^*} \widehat X_{t}-\frac 1{NT} \sum_{t=1}^T \bar J \widehat{X}_{t}' M_{\widehat \Lambda^*} \widehat X_{t}\bar J}= O_p \left( \max\left(\frac 1{N^{3-\gamma/2}T},\frac 1{mN^{2-\gamma}},\frac 1{\sqrt {mT}N^{1-\gamma/2}}\right)\right),\label{eq:pezzo2_t}
			\end{align}	
			because of \eqref{eq:MLstar}  and following Proposition \ref{lemma:GLS}\ref{lemma:GLS_iii}.
			And also, 
			\begin{align}		
				{\frac 1{NT}\sum_{i=1}^N\widehat{X}_{i}'M_{\widehat G^\dag}\, \left[G\Lambda'+\epsilon\right]} =&\, {\frac 1{NT}\sum_{i=1}^N\bar J X_i' M_{\widehat G^\dag} \left\{G\Lambda_i+\epsilon_i\right\}}\nonumber\\
				&+{\frac 1{NT}\sum_{i=1}^N\left(\widehat X_i'-\bar J X_i'\right) M_{\widehat G^\dag} \left\{G\Lambda_i+\epsilon_i\right\}} =:	A +B,\label{eq:pezzo3}\\
				{\frac 1{NT}\sum_{t=1}^T\widehat{X}_{t}'M_{\widehat \Lambda^*}\, \left[\Lambda G_t+\epsilon_t\right]} =&\, {\frac 1{NT}\sum_{t=1}^T\bar J X_t' M_{\widehat \Lambda^*} \left\{\Lambda G_t+\epsilon_t\right\}}\nonumber\\
				&+{\frac 1{NT}\sum_{t=1}^T\left(\widehat X_t'-\bar J X_t'\right) M_{\widehat \Lambda^*} \left\{\Lambda G_t+\epsilon_t\right\}} =:	C +D,\label{eq:pezzo3_t}.
			\end{align}
			Then, because of \eqref{eq:pezzo1}, \eqref{eq:pezzo1_t}, \eqref{eq:pezzo2}, and \eqref{eq:pezzo2_t}
			\begin{align}	
				\norm {B} &=O_p \left( \max\left(\frac 1{N^{3-\gamma/2}T},\frac 1{mN^{2-\gamma}},\frac 1{\sqrt {mT}N^{1-\gamma/2}}\right)\right),\label{eq:BBai09}\\
				\norm {D} &=O_p \left( \max\left(\frac 1{N^{3-\gamma/2}T},\frac 1{mN^{2-\gamma}},\frac 1{\sqrt {mT}N^{1-\gamma/2}}\right)\right).\label{eq:DBai09}
			\end{align}
			
			Consider first part \ref{th:CLT_FNAR_BAI_i}. If ${ \sqrt {NT}}/ (mN^{2-\gamma}) \to 0$ and ${ \sqrt {N}}/ (\sqrt{m}N^{1-\gamma/2}) \to 0$, as $m,N,T\to\infty$,
			by substituting \eqref{eq:BBai09} into \eqref{eq:pezzo3}, from \eqref{eq:thetastarAE} we get:
			\begin{align}
				\sqrt{NT}\left(\widehat\theta^\dag-\bar J\theta\right) =&\, \left(\frac 1{NT} \sum_{i=1}^N \bar J{X}_{i}' M_{\widehat G^\dag}  X_{i}\bar J\right)^{-1}  \left(\sqrt{NT}
				\, A\right) + o_p(1) =: I+o_p(1).	\label{eq:uffa}
			\end{align}
			By similar arguments to those used in \eqref{eq:pezzo2}, 
			\begin{align}
				&\norm{\frac 1{NT} \sum_{i=1}^N \left(y_i-\widehat X_i\widehat\theta^{\dag}\right) \left(y_i-\widehat X_i\widehat\theta^{\dag}\right)'- 
					\frac 1{NT} \sum_{i=1}^N \left(y_i- X_i\bar J\widehat\theta^{\dag}\right) \left(y_i- X_i\bar J\widehat\theta^{\dag}\right)'}\nonumber\\
				&\phantom{cippirimerlo}= O_p \left( \max\left(\frac 1{N^{3-\gamma/2}T},\frac 1{mN^{2-\gamma}},\frac 1{\sqrt {mT}N^{1-\gamma/2}}\right)\right).\label{eq:pezzo4}
			\end{align}
			Thus, from \eqref{eq:pezzo4} and the definition of $\widehat{G}^\dag$ in \eqref{eq:lambdastar} it is clear that, if ${ \sqrt {NT}}/ (mN^{2-\gamma}) \to 0$ and ${ \sqrt {N}}/ (\sqrt{m}N^{1-\gamma/2}) \to 0$, as $m,N,T\to\infty$, it solves
			\begin{align}
				&\left\{\frac 1{NT} \sum_{i=1}^N \left(y_i- X_i\bar J\widehat\theta^{\dag}\right) \left(y_i- X_i\bar J\widehat\theta^{\dag}\right)' + o_p\left (\frac 1{\sqrt {NT}}\right) \right\}\widehat G^\dag = \widehat G^\dag\frac{\widehat M^{\widehat\nu^\dag}}{T},\label{eq:lambdastarevec}
			\end{align}
			where $\widehat M^{\widehat\nu^\dag}$ is the $q\times q$ diagonal matrix of eigenvalues of $N^{-1}\widehat \nu^{\dag}\widehat \nu^{\dag'}$.
			
			Now, define the $T\times (r+2)$ matrices:
			\begin{align}
				\widehat Z_i^\dag &:= \left\{
				M_{\widehat G^\dag}X_i-\frac 1N\sum_{k=1}^N\left(\Lambda_i'\left(\frac{\Lambda'\Lambda} N\right)^{-1}\Lambda_k \right)M_{\widehat G^\dag}X_k
				\right\},\nonumber\\
				Z_i &:= \left\{
				M_{ G}X_i-\frac 1N\sum_{k=1}^N\left(\Lambda_i'\left(\frac{\Lambda'\Lambda} N\right)^{-1}\Lambda_k \right)M_{ G}X_k
				\right\}.\nonumber
			\end{align}	
			Then, following the same steps as in \citet[Proposition A.2, Lemma A.8, Corollary A.1, and Lemma A.9]{bai2009panel},  if ${ \sqrt {NT}}/ (mN^{2-\gamma}) \to 0$ and ${ \sqrt {N}}/ (\sqrt{m}N^{1-\gamma/2}) \to 0$, as $m,N,T\to\infty$, we have that $I$ in \eqref{eq:uffa} is such that
			\begin{align}
				I =&\,\left(\frac 1{NT}\sum_{i=1}^N\bar J\widehat Z_i^{\dag'}\widehat Z_i^\dag\bar J\right)^{-1}
				\left\{ \frac 1{\sqrt {NT}}\sum_{i=1}^N
				\bar J \widehat Z_i^\dag \epsilon_i\right.\nonumber\\
				&\left.-\sqrt{\frac NT}\left[\frac 1{NT}\sum_{i=1}^N \bar J X_i'M_{\widehat G^\dag}\left(\frac 1N \sum_{k=1}^N \mathbb E[\epsilon_k\epsilon_k']\right)\widehat G^\dag \left(\frac{G'\widehat G^\dag}{T}\right)^{-1} \left(\frac{\Lambda'\Lambda}{N}\right)^{-1} \Lambda_i\right]\right\}+  o_p(1)\nonumber\\
				=&\,\left(\frac 1{NT}\sum_{i=1}^N\bar J \widehat{Z}_i^{\dag'} \widehat{Z}_i^{\dag}\bar J\right)^{-1}
				\left\{ \frac 1{\sqrt {NT}}\sum_{i=1}^N
				\bar J  Z_i \epsilon_i\right.\nonumber\\
				&\left.-\sqrt{\frac NT}\left[\frac 1{NT}\sum_{i=1}^N \bar J X_i'M_{\widehat G^\dag}\left(\frac 1N \sum_{k=1}^N \mathbb E[\epsilon_k\epsilon_k']\right)\widehat G^\dag \left(\frac{G'\widehat G^\dag}{T}\right)^{-1} \left(\frac{\Lambda'\Lambda}{N}\right)^{-1}\Lambda_i\right]\right.\nonumber\\
				&\left.- \sqrt {\frac TN} \left[\frac 1{NT}\sum_{i=1}^N\bar J\left(X_i-\frac 1N\sum_{k=1}^N\left(\Lambda_i'\left(\frac{\Lambda'\Lambda}{N}\right)^{-1}\Lambda_k \right) X_k\right)' G \left(\frac{\Lambda'\Lambda}{N}\right)^{-1}\sum_{j=1}^N\Lambda_j \left(\frac 1T\sum_{t=1}^T\epsilon_{jt}\epsilon_{it}\right)\right]\right\}+o_p(1)\nonumber\\
				=&\,\left(\frac 1{NT}\sum_{i=1}^N\bar J Z_i' Z_i\bar J\right)^{-1}
				\left\{ \frac 1{\sqrt {NT}}\sum_{i=1}^N
				\bar J  Z_i \epsilon_i\right.\nonumber\\
				&\left.-\sqrt{\frac NT}\left[\frac 1{NT}\sum_{i=1}^N \bar J X_i'M_{G}\left(\frac 1N \sum_{k=1}^N \mathbb E[\epsilon_k\epsilon_k']\right)G \left(\frac{\Lambda'\Lambda}{N}\right)^{-1}\Lambda_i\right]\right.\nonumber\\
				&\left.- \sqrt {\frac TN} \left[\frac 1{NT}\sum_{i=1}^N\bar J\left(X_i-\frac 1N\sum_{k=1}^N\left(\Lambda_i'\left(\frac{\Lambda'\Lambda}{N}\right)^{-1}\Lambda_k \right) X_k\right)' G\left(\frac{\Lambda'\Lambda}{N}\right)^{-1}\sum_{j=1}^N\Lambda_j \left(\frac 1T\sum_{t=1}^T\mathbb E[\epsilon_{jt}\epsilon_{it}]\right)\right]\right\}\nonumber\\
				&+ O_p\left(\frac {\sqrt T}{N}\right)+O_p\left(\frac {\sqrt N}{T}\right)+o_p(1)\nonumber\\
				=:&\,\left(\frac 1{NT}\sum_{i=1}^N\bar J Z_i' Z_i\bar J\right)^{-1}
				\left\{ \frac 1{\sqrt {NT}}\sum_{i=1}^N
				\bar J  Z_i \epsilon_i +\sqrt{\frac NT} I_a+\sqrt{\frac TN}I_b \right\}\nonumber\\
				&+  O_p\left(\frac {\sqrt T}{N}\right)+O_p\left(\frac {\sqrt N}{T}\right)+o_p(1).\label{eq:A1A2_th5}
			\end{align}
			Since by Assumptions \ref{as:errors_nu}\ref{as:errors_nu_idio_i} and \ref{as:errors_nu}\ref{as:errors_nu_idio_ii}, $\mathbb E[\epsilon_k\epsilon_k'] = \mathbb E[\epsilon_{kt}^2] I_T=\sigma^2_k I_T$ and since $M_{G} G=0_{T\times q}$, we have $I_a=0_{r+2}$. Moreover, since we assumed $T/N\to 0$ and $\sqrt N/T\to 0$, as $N,T\to\infty$, by using \eqref{eq:A1A2_th5} into \eqref{eq:uffa}
			we have
			\begin{align}
				\sqrt{NT}\left(\widehat\theta^\dag-\bar J\theta\right) &= \left(\frac 1{NT}\sum_{i=1}^N\bar J Z_i' Z_i\bar J\right)^{-1}\left(\frac 1{\sqrt {NT}}\sum_{i=1}^N
				\bar JZ_i \epsilon_i\right) + o_p(1)\nonumber\\
				&\overset{p}{\to}\mathcal N\left(0_{r+2}, \bar J_0 \Sigma_{ZZ}^{-1} \bar J_0 \bar J_0D_1 \bar J_0\bar J_0\Sigma_{ZZ}^{-1}\bar J_0\right),
			\end{align}
			because of Assumptions \ref{as:Z}\ref{as:Z_i} and \ref{as:Z}\ref{as:Z_ii}, and Slutsky's theorem and where $\bar J_0:=\plim_{m,N,T\to\infty} \bar J$. We complete the proof by noticing that $\bar J_0 \Sigma_{ZZ}^{-1} \bar J_0 \bar J_0D_1 \bar J_0\bar J_0\Sigma_{ZZ}^{-1}\bar J_0:= \Sigma_{ZZ}^{-1}D_1 \Sigma_{ZZ}^{-1}$.
			
			For part \ref{th:CLT_FNAR_BAI_ii}, notice that if $\sigma_i^2=\sigma^2$ for all $i=1,\ldots, N$, then in \eqref{eq:A1A2_th5} we have
			\begin{align}
				I_b=&\, \frac 1{NT}\sum_{i=1}^N\bar JX_i' G\left(\frac{\Lambda'\Lambda}{N}\right)^{-1}\Lambda_i \sigma^2
				-\frac 1{NT}\sum_{i=1}^N\bar J\frac 1N\sum_{k=1}^N\left(\Lambda_i'\left(\frac{\Lambda'\Lambda}{N}\right)^{-1}\Lambda_k \right) X_k' G\left(\frac{\Lambda'\Lambda}{N}\right)^{-1}\Lambda_i\sigma^2\nonumber\\
				=&\,\frac 1{NT}\sum_{i=1}^N\bar JX_i' G\left(\frac{\Lambda'\Lambda}{N}\right)^{-1}\Lambda_i \sigma^2
				-\frac 1{NT}\sum_{k=1}^N\bar J X_k'G\left(\frac{\Lambda'\Lambda}{N}\right)^{-1}\frac 1N\sum_{i=1}^N \Lambda_i\Lambda_i'\left(\frac{\Lambda'\Lambda}{N}\right)^{-1}\Lambda_k \sigma^2\nonumber\\
				=&\,\frac 1{NT}\sum_{i=1}^N\bar JX_i' G\left(\frac{\Lambda'\Lambda}{N}\right)^{-1}\Lambda_i \sigma^2
				-\frac 1{NT}\sum_{k=1}^N\bar J X_k'G\left(\frac{\Lambda'\Lambda}{N}\right)^{-1}\left(\frac{\Lambda'\Lambda}{N}\right)\left(\frac{\Lambda'\Lambda}{N}\right)^{-1}\Lambda_k \sigma^2\nonumber\\
				=&\, 0_{r+2}.\nonumber
			\end{align}
			The proof then follows as in part \ref{th:CLT_FNAR_BAI_i} and by noticing that in this case $ \bar J_0 \Sigma_{ZZ}^{-1} \bar J_0 \bar J_0D_1 \bar J_0\bar J_0\Sigma_{ZZ}^{-1}\bar J_0= \sigma^2\Sigma_{ZZ}^{-1}$.
			
			For part \ref{th:CLT_FNAR_BAI_iii}. By similar arguments to those used in \eqref{eq:pezzo2}, 
			\begin{align}
				&\norm{\frac 1{NT} \sum_{t=1}^T \left(y_t-\widehat X_t\widehat\theta^{*}\right) \left(y_t-\widehat X_t\widehat\theta^{*}\right)'- 
					\frac 1{NT} \sum_{t=1}^T \left(y_t- X_t\bar J\widehat\theta^{*}\right) \left(y_t- X_t\bar J\widehat\theta^{*}\right)'}\nonumber\\
				&\phantom{cippirimerlo}= O_p \left( \max\left(\frac 1{N^{3-\gamma/2}T},\frac 1{mN^{2-\gamma}},\frac 1{\sqrt {mT}N^{1-\gamma/2}}\right)\right).\label{eq:pezzo4iii}
			\end{align}
			Thus, from \eqref{eq:pezzo4iii} and the definition of $\widehat{\Lambda}^*$ in \eqref{eq:lambdastar} it is clear that,  if ${ \sqrt {NT}}/ (mN^{2-\gamma}) \to 0$ and ${ \sqrt {N}}/ (\sqrt{m}N^{1-\gamma/2}) \to 0$, as $m,N,T\to\infty$, it solves
			\begin{align}
				&\left\{\frac 1{NT} \sum_{t=1}^T \left(y_t- X_t\bar J\widehat\theta^{*}\right) \left(y_t- X_t\bar J\widehat\theta^{*}\right)' + o_p\left (\frac 1{\sqrt {NT}}\right) \right\}\widehat \Lambda^* = \widehat \Lambda^*\frac{\widehat M^{\widehat\nu^*}}{N},\label{eq:lambdastareveciii}
			\end{align}
			where $\widehat M^{\widehat\nu^*}$ is the $q\times q$ diagonal matrix of eigenvalues of $T^{-1}\widehat \nu^{*'}\widehat \nu^{*}$.
			Moreover, since $y_t- X_t\bar J\widehat\theta^{*}=X_t\bar J(\bar J\theta-\widehat{\theta}^*)+\Lambda G_t+\epsilon_t$, from \eqref{eq:lambdastareveciii} we get
			\begin{align}
				\widehat \Lambda^*\frac{\widehat M^{\widehat\nu^*}}{N}=&\,\frac 1{NT}\sum_{t=1}^TX_t\bar J(\bar J\theta-\widehat{\theta}^*)(\bar J\theta-\widehat{\theta}^*)'\bar JX_t'\widehat{\Lambda}^*+\frac 1{NT}\sum_{t=1}^TX_t\bar J(\bar J\theta-\widehat{\theta}^*)G_t'\Lambda'\widehat{\Lambda}^*\nonumber\\
				&+\frac 1{NT}\sum_{t=1}^TX_t\bar J(\bar J\theta-\widehat{\theta}^*)\epsilon_t'\widehat{\Lambda}^*
				+\frac 1{NT}\sum_{t=1}^T\Lambda G_t(\bar J\theta-\widehat{\theta}^*)' \bar JX_t'\widehat{\Lambda}^*
				\nonumber\\
				&+\frac 1{NT}\sum_{t=1}^T\epsilon_t(\bar J\theta-\widehat{\theta}^*)' \bar JX_t'\widehat{\Lambda}^*+\frac 1{NT}\sum_{t=1}^T\Lambda G_t\epsilon_t'\widehat{\Lambda}^*+\frac 1{NT}\sum_{t=1}^T\epsilon_tG_t'\Lambda'\widehat{\Lambda}^*\nonumber\\
				&+\frac 1{NT}\sum_{t=1}^T\epsilon_t\epsilon_t'\widehat{\Lambda}^*+\frac 1{NT}\sum_{t=1}^T \Lambda G_tG_t'\Lambda'\widehat{\Lambda}^*\nonumber\\
				=:&\, I_1+I_2+I_3+I_4+I_5+I_6+I_7+I_8+ \Lambda\frac{\Lambda'\widehat{\Lambda}^*}{N} + o_p\left (\frac 1{\sqrt {NT}}\right).\nonumber
			\end{align}
			since $T^{-1}G'G=I_q$ by Assumption \ref{as:errors_nu_id}\ref{as:errors_nu_id_ii}.	So,
			\begin{align}
				\widehat \Lambda^*\frac{\widehat M^{\widehat\nu^*}}{N}\left(\frac{\Lambda'\widehat{\Lambda}^*}{N}\right)^{-1}-\Lambda=
				\sum_{j=1}^8 I_j
				\left(\frac{\Lambda'\widehat{\Lambda}^*}{N}\right)^{-1}+o_p\left (\frac 1{\sqrt {NT}}\right),\label{8indiani}
			\end{align}
			and notice that 
			\begin{align}
				\left\Vert\left (\frac{\Lambda'\widehat{\Lambda}^*}N\right)^{-1}\right\Vert=O_p(1), \label{jsm}
			\end{align}
			because of \eqref{eq:Lhat_MB_V} in Lemma \ref{lemma:VhatCNAR} (where $\widehat{\Lambda}^*$ is simply denoted as $\widehat{\Lambda}$) and Assumption \ref{as:errors_nu}\ref{as:errors_nu_common_i}. 
			It follows that, 
			\begin{align}
				\frac 1{\sqrt N}
				\left\Vert
				\widehat \Lambda^*\frac{\widehat M^{\widehat\nu^*}}{N}\left(\frac{\Lambda'\widehat{\Lambda}^*}{N}\right)^{-1}-\Lambda
				\right\Vert \le \frac 1{\sqrt N}\sum_{j=1}^8\Vert I_j\Vert\,\left\Vert\left(\frac{\Lambda'\widehat{\Lambda}^*}{N}\right)^{-1}\right\Vert +
				o_p\left (\frac 1{\sqrt {NT}}\right).\label{melone}
			\end{align}
			Now, 
			\begin{align}
				\frac 1{\sqrt N}
				\left\Vert I_1\right\Vert \le\left\Vert \frac {\widehat \Lambda^*}{\sqrt N}\right\Vert \frac 1T\sum_{t=1}^T\frac{\Vert X_t\Vert ^2}{N}\left\Vert\widehat{\theta}^*-\bar J\theta\right\Vert^2= o_p\left(\Vert\widehat{\theta}^*-\bar J\theta\Vert\right),\label{indiano1}
			\end{align}
			because of \eqref{eq:Lhat_MB_V} in Lemma \ref{lemma:VhatCNAR} and Assumption \ref{as:errors_nu}\ref{as:errors_nu_common_i}, and because 
			\begin{align}
				\mathbb E\left[\left( \frac 1T\sum_{t=1}^T\frac{\Vert X_t\Vert ^2}{N}\right)^2\right]&=\frac 1{T^2N^2}\sum_{t=1}^T\sum_{s=1}^T\mathbb E\left[\Vert X_t\Vert ^2\Vert X_s\Vert ^2\right] \le \frac 1{N^2} \mathbb E\left[\Vert X_t\Vert ^4\right] =  \frac 1{N^2}\mathbb E\left[\left( \sum_{i=1}^N X_{it}^2\right)^2\right]\nonumber\\
				& = \frac 1{N^2}\sum_{i=1}^N\sum_{j=1}^N \mathbb E[X_{it}^2X_{jt}^2] \le \mathcal K_X,\label{X44}
			\end{align}
			for some finite $\mathcal K_X$ independent of $N$, $i$, and $t$, due to Assumptions \ref{as:common_component}\ref{as:common_component_4}, \ref{as:errors_nu}\ref{as:errors_nu_common_iii}, and \ref{as:errors_nu}\ref{as:errors_nu_idio_iii}.
			
			Similarly to \eqref{indiano1} we have
			\begin{align}
				\frac 1{\sqrt N}\left\Vert I_2\right\Vert
				=			\frac 1{\sqrt N}\left\Vert I_3\right\Vert
				=			\frac 1{\sqrt N}\left\Vert I_4\right\Vert
				=			\frac 1{\sqrt N}\left\Vert I_5\right\Vert
				= 
				O_p\left(\Vert\widehat{\theta}^*-\bar J\theta\Vert\right),\label{indiano2-5}
			\end{align}
			Furthermore, by \citet[Proposition 3]{barigozzi2022}
			\begin{align}
				\frac 1{\sqrt N}\left\Vert I_6\right\Vert
				=			\frac 1{\sqrt N}\left\Vert I_7\right\Vert
				=			\frac 1{\sqrt N}\left\Vert I_8\right\Vert
				= 
				O_p\left(\max\left(\frac 1{\sqrt N},\frac 1{\sqrt T}\right)\right),\label{indiano6-8}
			\end{align}
			and, by \citet[Proposition 8]{barigozzi2022}
			\begin{align}
				\left\Vert\frac{\widehat M^{\widehat\nu^*}}{N}\left(\frac{\Lambda'\widehat{\Lambda}^*}{N}\right)^{-1}-J\right\Vert = O_p\left(\max\left(\frac 1{\sqrt {NT}},\frac 1{ T}\right)\right).\label{indianata}
			\end{align}
			By substituting \eqref{jsm}, \eqref{indiano1}, \eqref{indiano2-5}, \eqref{indiano6-8}, and \eqref{indianata} into \eqref{melone} and since we assumed $\sqrt N/T\to 0$ as $N,T\to\infty$,
			\begin{align}
				\frac 1{\sqrt N}
				\left\Vert
				\widehat \Lambda^*-\Lambda J
				\right\Vert = O_p\left(\Vert\widehat{\theta}^*-\bar J\theta\Vert\right)+ O_p\left(\max\left(\frac 1{\sqrt N},\frac 1{\sqrt T}\right)\right).\label{A1Bai09}
			\end{align}
			From \eqref{X44} and \eqref{A1Bai09}, it follows also that
			\begin{align}
				\frac 1{NT}\sum_{t=1}^T\bar JX_t' M_{\widehat{\Lambda}^*} (\Lambda-\widehat \Lambda^* J) = O_p\left(\widehat{\theta}^*-\bar J\theta\right)+ O_p\left(\max\left(\frac 1{\sqrt N},\frac 1{\sqrt T}\right)\right).\label{A1Bai09T}
			\end{align}
			
			Now, from \eqref{8indiani} and \eqref{indianata}, and noticing that $M_{\widehat{\Lambda}^*}\widehat{\Lambda}^*=0_{r\times r}$, it follows that:
			\begin{align}
				\frac 1{NT}\sum_{t=1}^T \bar JX_t'M_{\widehat{\Lambda}^*}\Lambda G_t =&\,\frac 1{NT}\sum_{t=1}^T \bar JX_t'M_{\widehat{\Lambda}^*}(\Lambda-\widehat{\Lambda}^*J) G_t +O_p\left(\max\left(\frac 1{\sqrt {NT}},\frac 1{ T}\right)\right) \nonumber\\
				=&\,- \frac 1{NT}\sum_{t=1}^T \bar JX_t'M_{\widehat{\Lambda}^*}
				\left\{I_1+I_2+I_3+I_4+I_5+I_6+I_7+I_8\right\}\left(\frac{\Lambda'\widehat{\Lambda}^*}{N}\right)^{-1}G_t\nonumber\\
				&+o_p\left (\frac 1{\sqrt {NT}}\right)
				+o_p\left(\max\left(\frac 1{\sqrt {NT}},\frac 1{ T}\right)\right)\nonumber\\
				=:&\, J_1+J_2+J_3+J_4+J_5+J_6+J_7+J_8+O_p\left(\max\left(\frac 1{\sqrt {NT}},\frac 1{T}\right)\right).\label{8nani}
			\end{align}
			Then, for $J_1$ by \eqref{indiano1} we have
			\begin{align}
				\Vert J_1\Vert &=\left\Vert- \frac 1{NT}\sum_{t=1}^T \bar JX_t'M_{\widehat{\Lambda}^*}I_1 \left(\frac{\Lambda'\widehat{\Lambda}^*}{N}\right)^{-1}G_t\right\Vert\le \frac 1{T}\sum_{t=1}^T \frac{\Vert \bar JX_t'M_{\widehat{\Lambda}^*}\Vert}{\sqrt N} \,\frac{\Vert I_1\Vert}{\sqrt N}\left\Vert\left(\frac{\Lambda'\widehat{\Lambda}^*}{N}\right)^{-1}\right\Vert\, \Vert G_t\Vert\nonumber\\
				&= O_p(1) \left\Vert\widehat{\theta}^*-\bar J\theta\right\Vert^2= o_p(1) \left\Vert\widehat{\theta}^*-\bar J\theta\right\Vert,\label{nano1}
			\end{align}
			since $\Vert J\Vert=1$, $\Vert M_{\widehat{\Lambda}^*}\Vert=1$ (it is a projector), $\Vert G_t\Vert=O_p(1)$ due to Assumption \ref{as:errors_nu}\ref{as:errors_nu_common_ii}, and because of \eqref{jsm} and \eqref{X44}.
			For $J_2$ we have
			\begin{align}
				J_2&=-\frac 1{NT}\sum_{t=1}^T \bar JX_t'M_{\widehat{\Lambda}^*}I_2 \left(\frac{\Lambda'\widehat{\Lambda}^*}{N}\right)^{-1}G_t\nonumber\\
				&= -\frac 1{NT}\sum_{t=1}^T \bar JX_t'M_{\widehat{\Lambda}^*}\left[
				\frac 1{T}\sum_{s=1}^TX_s\bar J(\bar J\theta-\widehat{\theta}^*)G_s'\frac{\Lambda'\widehat{\Lambda}^*}N
				\right] \left(\frac{\Lambda'\widehat{\Lambda}^*}{N}\right)^{-1}G_t\nonumber\\
				&= \frac 1{NT^2}\sum_{t=1}^T\sum_{s=1}^T\bar JX_t'M_{\widehat{\Lambda}^*}X_s\bar J G_s'G_t(\widehat{\theta}^*-\bar J\theta).\label{nano2}
			\end{align}
			For $J_3$ we have
			\begin{align}
				J_3&=-\frac 1{NT}\sum_{t=1}^T \bar JX_t'M_{\widehat{\Lambda}^*}I_3 \left(\frac{\Lambda'\widehat{\Lambda}^*}{N}\right)^{-1}G_t\nonumber\\
				&=  -\frac 1{NT}\sum_{t=1}^T \bar JX_t'M_{\widehat{\Lambda}^*}\left[
				\frac 1{NT}\sum_{s=1}^TX_s\bar J(\bar J\theta-\widehat{\theta}^*)\epsilon_s'\widehat{\Lambda}^*
				\right] \left(\frac{\Lambda'\widehat{\Lambda}^*}{N}\right)^{-1}G_t\nonumber\\
				&=\frac 1{NT^2}\sum_{t=1}^T\sum_{s=1}^T\bar JX_t'M_{\widehat{\Lambda}^*}X_s\bar J\left[\frac{\epsilon_s'\widehat{\Lambda}^*}{N}\left(\frac{\Lambda'\widehat{\Lambda}^*}{N}\right)^{-1}\right](\widehat{\theta}^*-\bar J\theta)\nonumber\\
				&= o_p(1)(\widehat{\theta}^*-\bar J\theta),\label{nano3}
			\end{align}
			since, by Assumption \ref{as:errors_nu}\ref{as:errors_nu_idio_ii} and \eqref{A1Bai09},
			\begin{align}
				\frac{\epsilon_s'\widehat{\Lambda}^*}{N}=\frac{\epsilon_s'{\Lambda}J}{N}+\frac{\epsilon_s'(\widehat{\Lambda}^*-\Lambda J)}{N} = O_p\left(\frac 1{\sqrt N}\right)+
				O_p\left(\Vert\widehat{\theta}^*-\bar J\theta\Vert\right)+ O_p\left(\max\left(\frac 1{\sqrt N},\frac 1{\sqrt T}\right)\right).\nonumber
			\end{align}
			For $J_4$, because of \eqref{A1Bai09T}, we have
			\begin{align}
				J_4&=-\frac 1{NT}\sum_{t=1}^T \bar JX_t'M_{\widehat{\Lambda}^*}I_4 \left(\frac{\Lambda'\widehat{\Lambda}^*}{N}\right)^{-1}G_t\nonumber\\
				&= - \frac 1{NT}\sum_{t=1}^T \bar JX_t'M_{\widehat{\Lambda}^*}\left[
				\frac 1{NT}\sum_{s=1}^T\Lambda G_s(\bar J\theta-\widehat{\theta}^*)'\bar JX_s'\widehat{\Lambda}^*
				\right] \left(\frac{\Lambda'\widehat{\Lambda}^*}{N}\right)^{-1}G_t\nonumber\\
				&=-\frac 1{\sqrt NT^2}\sum_{t=1}^T\sum_{s=1}^T\bar JX_t'M_{\widehat{\Lambda}^*}\frac{(\Lambda-\widehat{\Lambda}^*J)}{\sqrt N}G_s(\bar J\theta-\widehat{\theta}^*)'\bar J\frac{X_s'\widehat{\Lambda}^*}{N}
				\left(\frac{\Lambda'\widehat{\Lambda}^*}{N}\right)^{-1}G_t\nonumber\\
				&= o_p(1)(\widehat{\theta}^*-\bar J\theta).\label{nano4}
			\end{align}
			Likewise, because of \eqref{indiano2-5}, for $J_5$ we have
			\begin{align}
				J_5&=-\frac 1{NT}\sum_{t=1}^T \bar JX_t'M_{\widehat{\Lambda}^*}I_5 \left(\frac{\Lambda'\widehat{\Lambda}^*}{N}\right)^{-1}G_t
				= o_p(1)(\widehat{\theta}^*-\bar J\theta).\label{nano5}
			\end{align}
			Then, for $J_6$ we have
			\begin{align}
				J_6&= -\frac 1{NT}\sum_{t=1}^T \bar JX_t'M_{\widehat{\Lambda}^*}I_6 \left(\frac{\Lambda'\widehat{\Lambda}^*}{N}\right)^{-1}G_t\nonumber\\
				&= -\frac 1{NT}\sum_{t=1}^T \bar JX_t'M_{\widehat{\Lambda}^*}\left[\frac 1{NT}\sum_{s=1}^T\Lambda G_s\epsilon_s'\widehat{\Lambda}^*\right] \left(\frac{\Lambda'\widehat{\Lambda}^*}{N}\right)^{-1}G_t\nonumber\\
				&= -\frac 1{NT}\sum_{t=1}^T\bar JX_t'M_{\widehat{\Lambda}^*}(\Lambda-\widehat{\Lambda}^*J)\frac 1T \sum_{s=1}^T G_s\frac{\epsilon_s'\widehat{\Lambda}^*}{N}
				\left(\frac{\Lambda'\widehat{\Lambda}^*}{N}\right)^{-1}G_t\nonumber\\
				&= o_p(\widehat{\theta}^*-\bar J\theta)+o_p\left(\frac 1{\sqrt{NT}}\right),\label{nano6}
			\end{align}
			because of \eqref{A1Bai09T} and since from  \citet[Lemma 3]{barigozzi2022} and \eqref{A1Bai09} we have
			\begin{align}
				\frac 1T \sum_{s=1}^T G_s\frac{\epsilon_s'\widehat{\Lambda}^*}{N}&=\frac 1T \sum_{s=1}^T G_s\frac{\epsilon_s'{\Lambda}J}{N}+
				\frac 1T \sum_{s=1}^T G_s\frac{\epsilon_s'(\widehat{\Lambda}^*-\Lambda J)}{N}\nonumber\\ 
				&= O_p\left(\frac 1{\sqrt{NT}}\right)+O_p\left(\frac 1{\sqrt{NT}}\right)\left\{O_p(\widehat{\theta}^*-\bar J\theta)+O_p\left(\max\left(\frac 1{\sqrt N},\frac 1{\sqrt T}\right)\right)\right\}.\nonumber
			\end{align}
			For $J_7$ we have
			\begin{align}
				J_7&=  -\frac 1{NT}\sum_{t=1}^T \bar JX_t'M_{\widehat{\Lambda}^*}I_7 \left(\frac{\Lambda'\widehat{\Lambda}^*}{N}\right)^{-1}G_t\nonumber\\
				&=  -\frac 1{NT}\sum_{t=1}^T \bar JX_t'M_{\widehat{\Lambda}^*}\left[\frac 1T\sum_{s=1}^T\epsilon_s G_s'\frac{\Lambda'\widehat{\Lambda}^*}{N}\right] \left(\frac{\Lambda'\widehat{\Lambda}^*}{N}\right)^{-1}G_t\nonumber\\
				&= -\frac 1{NT^2}\sum_{t=1}^T\sum_{s=1}^T G_s'G_t\bar JX_t'M_{\widehat{\Lambda}^*}\epsilon_s.\label{nano7}
			\end{align}
			Finally, for $J_8$ we have
			\begin{align}
				J_8=&\,  -\frac 1{NT}\sum_{t=1}^T \bar JX_t'M_{\widehat{\Lambda}^*}I_8 \left(\frac{\Lambda'\widehat{\Lambda}^*}{N}\right)^{-1}G_t\nonumber\\
				=&\,  -\frac 1{NT}\sum_{t=1}^T \bar JX_t'M_{\widehat{\Lambda}^*}\left[
				\frac 1{NT}\sum_{s=1}^T\left\{\epsilon_s\epsilon_s'-\mathbb E[\epsilon_s\epsilon_s']+\mathbb E[\epsilon_s\epsilon_s']\right\}\widehat{\Lambda}^*
				\right]\left(\frac{\Lambda'\widehat{\Lambda}^*}{N}\right)^{-1}G_t\nonumber\\
				=&\,  -\frac 1{N^2T^2}\sum_{t=1}^T\sum_{s=1}^T \bar JX_t'M_{\widehat{\Lambda}^*}
				\mathbb E[\epsilon_s\epsilon_s']\widehat{\Lambda}^*\left(\frac{\Lambda'\widehat{\Lambda}^*}{N}\right)^{-1}G_t\nonumber\\
				&  -\frac 1{NT}\sum_{t=1}^T \bar JX_t'M_{\widehat{\Lambda}^*}\left[
				\frac 1{NT}\sum_{s=1}^T\left\{\epsilon_s\epsilon_s'-\mathbb E[\epsilon_s\epsilon_s']\right\}\widehat{\Lambda}^*
				\right]\left(\frac{\Lambda'\widehat{\Lambda}^*}{N}\right)^{-1}G_t\nonumber\\
				=&\, A_{NT}-\frac 1{NT}\sum_{t=1}^T \bar JX_t'M_{\widehat{\Lambda}^*}\left[
				\frac 1{NT}\sum_{s=1}^T\left\{\epsilon_s\epsilon_s'-\mathbb E[\epsilon_s\epsilon_s']\right\}\widehat{\Lambda}^*
				\right]\left(\frac{\Lambda'\widehat{\Lambda}^*}{N}\right)^{-1}G_t\nonumber\\
				=&\, A_{NT}+o_p(\widehat{\theta}^*-\bar J\theta)+o_p\left(\frac 1{\sqrt{NT}}\right),\label{nano8}
			\end{align}
			since, by \citet[Lemmas 3 and 4]{barigozzi2022} and \eqref{A1Bai09},
			\begin{align}
				\frac 1{NT}\sum_{s=1}^T\left\{\epsilon_s\epsilon_s'-\mathbb E[\epsilon_s\epsilon_s']\right\}\widehat{\Lambda}^*&=
				\frac 1{NT}\sum_{s=1}^T\left\{\epsilon_s\epsilon_s'-\mathbb E[\epsilon_s\epsilon_s']\right\}{\Lambda}J +
				\frac 1{NT}\sum_{s=1}^T\left\{\epsilon_s\epsilon_s'-\mathbb E[\epsilon_s\epsilon_s']\right\}(\widehat{\Lambda}^*-\Lambda J)\nonumber\\
				&=O_p\left( \frac 1{\sqrt T}\frac 1{\sqrt{NT}}\right) + O_p\left(\frac 1{\sqrt{NT}}\right)O_p(\widehat{\theta}^*-\bar J\theta).\nonumber
			\end{align}
			Therefore, by substituting \eqref{nano1},  \eqref{nano2}, \eqref{nano3}, \eqref{nano4}, \eqref{nano5}, \eqref{nano6}, \eqref{nano7}, and \eqref{nano8}
			into \eqref{8nani} we get
			\begin{align}
				\frac 1{NT}\sum_{t=1}^T \bar JX_t'M_{\widehat{\Lambda}^*}\Lambda G_t = J_2+J_7+A_{NT}+o_p(\widehat{\theta}^*-\bar J\theta)+
				O_p\left(\max\left(\frac 1{\sqrt {NT}},\frac 1{ T}\right)\right).\label{eq:uffa3}
			\end{align}
			
			Now, if ${ \sqrt {NT}}/ (mN^{2-\gamma}) \to 0$ and ${ \sqrt {N}}/ (\sqrt{m}N^{1-\gamma/2}) \to 0$, as $m,N,T\to\infty$,
			by substituting \eqref{eq:DBai09} into \eqref{eq:pezzo3_t}, from \eqref{eq:thetastarAE} we get:
			\begin{align}
				\left(\widehat\theta^*-\bar J\theta\right) =&\, \left(\frac 1{NT} \sum_{t=1}^T \bar J{X}_{t}' M_{\widehat \Lambda^*}  X_{t}\bar J\right)^{-1}  
				\, C + o_p\left(\frac 1{\sqrt{NT}}\right).\label{eq:uffadoppia}
			\end{align}
			Thus, from \eqref{eq:uffa3} and \eqref{eq:uffadoppia} and given the definition of $C$ in \eqref{eq:pezzo3_t}, 
			\begin{align}
				\left(\frac 1{NT} \sum_{t=1}^T \bar J{X}_{t}' M_{\widehat \Lambda^*}  X_{t}\bar J+o_p(1)\right)&\left(\widehat\theta^*-\bar J\theta\right) -J_2\nonumber\\
				&= 
				\frac 1{NT}\sum_{t=1}^T \bar JX_t'M_{\widehat{\Lambda}^*}\epsilon_t +J_7 +A_{NT}+O_p\left(\max\left(\frac 1{\sqrt {NT}},\frac 1{ T}\right)\right).\label{cinesecheurla}
			\end{align}
			Then, define the $T\times (r+2)$ matrices:
			\begin{align}
				\widehat W_t^* &:= \left\{
				M_{\widehat \Lambda^*}X_t-\frac 1T\sum_{s=1}^T \left(G_t'G_s \right)M_{\widehat \Lambda^*}X_s
				\right\},\nonumber\\
				W_t &:= \left\{
				M_{ \Lambda}X_t-\frac 1T\sum_{s=1}^T\left(G_t'G_s \right)M_{ \Lambda}X_s.
				\right\}.\nonumber
			\end{align}
			Using these definitions, by multiplying \eqref{cinesecheurla} by $\sqrt{NT}$, from \eqref{nano2}, \eqref{nano7} we have
			\begin{align}
				\left(\frac 1{NT} \sum_{t=1}^T \bar J\widehat{W}_{t}^{*'} \widehat{W}_{t}^{*}\bar J+o_p(1)\right)\sqrt{NT}\left(\widehat\theta^*-\bar J\theta\right)=
				\frac 1{\sqrt{NT}}\sum_{t=1}^T \bar J\widehat{W}_t^{*'}\epsilon_t+
				\sqrt{NT} A_{NT}+ o_p(1),	
				\nonumber
			\end{align}
			which implies
			\begin{align}
				\sqrt{NT}\left(\widehat\theta^*-\bar J\theta\right)&=: II+o_P(1).\label{uffissima}
			\end{align}
			Now, by the definition of $A_{NT}$ in \eqref{nano8}  term $II$ is such that:
			\begin{align}
				II=&\, \left(\frac 1{NT} \sum_{t=1}^T \bar J\widehat{W}_{t}^{*'} \widehat{W}_{t}^{*}\bar J\right)^{-1} \left\{\frac 1{\sqrt{NT}}\sum_{t=1}^T \bar J\widehat{W}_t^{*'}\epsilon_t\right.\nonumber\\
				&\left.-\sqrt{\frac TN}\left[\frac {1}{NT}\sum_{t=1}^T \bar JX_t'M_{\widehat{\Lambda}^*}\left(\frac 1T\sum_{s=1}^T
				\mathbb E[\epsilon_s\epsilon_s']\right)\widehat{\Lambda}^*\left(\frac{\Lambda'\widehat{\Lambda}^*}{N}\right)^{-1}G_t\right]
				\right\}+o_p(1)\nonumber\\
				=&\, \left(\frac 1{NT} \sum_{t=1}^T \bar J\widehat{W}_{t}^{*'} \widehat{W}_{t}^{*}\bar J\right)^{-1} \left\{\frac 1{\sqrt{NT}}\sum_{t=1}^T \bar J{W}_t^{'}\epsilon_t\right.\nonumber\\
				&-\sqrt{\frac TN}\left[\frac {1}{NT}\sum_{t=1}^T \bar JX_t'M_{\widehat{\Lambda}^*}\left(\frac 1T\sum_{s=1}^T
				\mathbb E[\epsilon_s\epsilon_s']\right)\widehat{\Lambda}^*\left(\frac{\Lambda'\widehat{\Lambda}^*}{N}\right)^{-1}G_t\right]\nonumber\\
				&\left.-\sqrt{\frac NT} \left[\frac 1{NT}\sum_{t=1}^T\bar J\left(X_t-\frac 1T\sum_{s=1}^T (G_t'G_s)X_s\right)'\Lambda\left(\frac{\Lambda'\Lambda}{N}\right)^{-1}\sum_{u=1}^T G_u\left(\frac 1N\sum_{i=1}^N\epsilon_{iu}\epsilon_{it} \right)
				\right]
				\right\}+o_p(1)\nonumber\\
				=&\, \left(\frac 1{NT} \sum_{t=1}^T \bar J{W}_{t}^{'} {W}_{t}\bar J\right)^{-1} \left\{\frac 1{\sqrt{NT}}\sum_{t=1}^T \bar J{W}_t^{'}\epsilon_t\right.\nonumber\\
				&-\sqrt{\frac TN}\left[\frac {1}{NT}\sum_{t=1}^T \bar JX_t'M_{{\Lambda}}\left(\frac 1T\sum_{s=1}^T
				\mathbb E[\epsilon_s\epsilon_s']\right){\Lambda}\left(\frac{\Lambda'{\Lambda}}{N}\right)^{-1}G_t\right]\nonumber\\
				&\left.-\sqrt{\frac NT} \left[\frac 1{NT}\sum_{t=1}^T\bar J\left(X_t-\frac 1T\sum_{s=1}^T (G_t'G_s)X_s\right)'\Lambda\left(\frac{\Lambda'\Lambda}{N}\right)^{-1}\sum_{u=1}^T G_u\left(\frac 1N\sum_{i=1}^N\mathbb E[\epsilon_{iu}\epsilon_{it}] \right)
				\right]
				\right\}\nonumber\\
				&+O_p\left(\frac{\sqrt N}{T}\right)+O_p\left(\frac{\sqrt T}{N}\right)+o_p(1)\nonumber\\
				=:&\, \left(\frac 1{NT} \sum_{t=1}^T \bar J{W}_{t}^{'} {W}_{t}\bar J\right)^{-1} \left\{\frac 1{\sqrt{NT}}\sum_{t=1}^T \bar J{W}_t^{'}\epsilon_t
				+\sqrt{\frac TN} II_a+\sqrt{\frac NT} II_b\right\}
				\nonumber\\
				&+O_p\left(\frac{\sqrt N}{T}\right)+O_p\left(\frac{\sqrt T}{N}\right)+o_p(1).\label{toronto}
			\end{align}
			To derive \eqref{toronto} in the first step we used Lemma \ref{lem:mancavasololui} and the fact that $\sqrt NO_p(\Vert\widehat{\theta}^*-\bar J\theta \Vert^2)$ and $O_p(\Vert\widehat{\theta}^*-\bar J\theta \Vert)$ are dominated by $\sqrt {NT} O_p(\Vert\widehat{\theta}^*-\bar J\theta \Vert)$, and $\sqrt NO_p\left(\max\left(\frac 1N, \frac 1T\right)\right)=o_p(1)$ since $\sqrt N/T\to 0$, as $N,T\to\infty$, by assumption. Furthermore, in the second step we used Lemma \ref{118}\ref{118_i} for the denominator, Lemma \ref{rimini}\ref{rimini_i} for the second term at the numerator and Lemma \ref{rimini}\ref{rimini_ii} for the third term at the numerator.
			
			Since by Assumptions \ref{as:errors_nu}\ref{as:errors_nu_idio_i} and \ref{as:errors_nu}\ref{as:errors_nu_idio_ii}, $\mathbb E[\epsilon_{iu}\epsilon_{it}] = \sigma_i^2 \mathbb I(t=u)$ letting $\bar {\sigma}^2=N^{-1}\sum_{i=1}^N \sigma_i^2$, we get
			\begin{align}
				II_b& = - \frac 1{NT}
				\sum_{t=1}^T
				\bar J\left(X_t-\frac 1T\sum_{s=1}^T (G_t'G_s)X_s\right)'
				\Lambda\left(\frac{\Lambda'\Lambda}{N}\right)^{-1}G_t\bar{\sigma}^2\nonumber\\
				&=- \frac 1{NT}
				\sum_{t=1}^T
				\bar JX_t'
				\Lambda\left(\frac{\Lambda'\Lambda}{N}\right)^{-1}G_t\bar{\sigma}^2+
				- \frac 1{NT}
				\sum_{t=1}^T
				\bar J\left(\frac 1T\sum_{s=1}^T X_s\right)'
				\Lambda\left(\frac{\Lambda'\Lambda}{N}\right)^{-1}G_t'G_sG_t\bar{\sigma}^2\nonumber\\
				&=- \frac 1{NT}
				\sum_{t=1}^T
				\bar JX_t'
				\Lambda\left(\frac{\Lambda'\Lambda}{N}\right)^{-1}G_t\bar{\sigma}^2+
				- \frac 1{NT}
				\sum_{s=1}^T
				\bar J X_s' \Lambda\left(\frac{\Lambda'\Lambda}{N}\right)^{-1}
				\frac 1T\sum_{t=1}^T 
				G_tG_t'G_s\bar{\sigma}^2\nonumber\\
				&=- \frac 1{NT}
				\sum_{t=1}^T
				\bar JX_t'
				\Lambda\left(\frac{\Lambda'\Lambda}{N}\right)^{-1}G_t\bar{\sigma}^2+
				- \frac 1{NT}
				\sum_{s=1}^T
				\bar J X_s' \Lambda\left(\frac{\Lambda'\Lambda}{N}\right)^{-1}
				G_s\bar{\sigma}^2 = 0_{r+2}.\nonumber
			\end{align}
			Moreover, since we assumed $T/N\to 0$ and $\sqrt N/T\to 0$, as $N,T\to\infty$, by using \eqref{toronto} into \eqref{uffissima} we have
			\begin{align}
				\sqrt{NT}\left(\widehat\theta^*-\bar J\theta\right) &= \left(\frac 1{NT}\sum_{t=1}^T\bar J W_t' W_t\bar J\right)^{-1}\left(\frac 1{\sqrt {NT}}\sum_{t=1}^T
				\bar JW_t \epsilon_t\right) + o_p(1)\nonumber\\
				&\overset{p}{\to}\mathcal N\left(0_{r+2}, \bar J_0 \Sigma_{WW}^{-1} \bar J_0 \bar J_0D_2 \bar J_0\bar J_0\Sigma_{WW}^{-1}\bar J_0\right),
			\end{align}
			because of Assumptions \ref{as:Z}\ref{as:Z_iii} and \ref{as:Z}\ref{as:Z_iv}, and Slutsky's theorem and where $\bar J_0:=\plim_{m,N,T\to\infty} \bar J$. We complete the proof by noticing that $\bar J_0 \Sigma_{WW}^{-1} \bar J_0 \bar J_0D_2 \bar J_0\bar J_0\Sigma_{WW}^{-1}\bar J_0= \Sigma_{WW}^{-1}D_2 \Sigma_{WW}^{-1}$.
			
			For part \ref{th:CLT_FNAR_BAI_iv}, if $\sigma_i^2=\sigma^2$ for all $i=1,\ldots, N$, then in \eqref{toronto} we have $\mathbb E[\epsilon_s\epsilon_s'] = \sigma^2 I_N$ and since $M_{\Lambda} \Lambda=0_{N\times q}$, we have $II_a=0_{r+2}$.  The proof then follows as in part \ref{th:CLT_FNAR_BAI_iii} and by noticing that in this case $ \bar J_0 \Sigma_{WW}^{-1} \bar J_0 \bar J_0D_2 \bar J_0\bar J_0\Sigma_{WW}^{-1}\bar J_0= \sigma^2\Sigma_{WW}^{-1}$. This completes the proof. 	
		\end{proof}

		\subsection{Proof of Proposition \ref{lemma:part2_lemma_1}}
		
		\begin{proof}
			First, consider part \ref{lemma:part2_lemma_1_i}.
			We have
			\begin{align}
				\frac1{NT}\sum_{t=1}^T\widehat X_t' u_t
				=& 
				\frac1{NT}\sum_{t=1}^T  \bar J X_t' u_t 
				+\frac1{NT}\sum_{t=1}^T \left(\widehat X_t' - \bar J X_t' \right) u_t  = a + b.\label{eq:ab_prop1}
			\end{align}
			Then, notice that we can write
			\[
			u_t =(X_t\bar J-\widehat X_t)\bar J\theta= \text{mat}_1 \left( N^{-1}(\mathcal F_{t-1}\times_3 J -\widehat {\mathcal F}_{t-1})\times_2 y'_{t-1}\times_3 \beta' J \right).
			\]
			For term $a$, we have that 
			\begin{align}
				\norm{a} 
				&= 
				\norm{  \bar J X_t' \text{mat}_1 \left(  \frac{1}{N^2T} \sum_{t=1}^T \left(\widehat{\mathcal{F}}_{t-1} - \mathcal{F}_{t-1} \times_3 J \right)  \times_2 y_{t-1}' \times_3  \beta' J \right)}\nonumber  \\
				&= 
				\norm{ \text{mat}_1 \left(  \frac{1}{N^2T} \sum_{t=1}^T \left(\widehat{\mathcal{F}}_{t-1} - \mathcal{F}_{t-1} \times_3 J \right)  \times_1  \bar J X_t' \times_2 y_{t-1}' \times_3  \beta' J \right)}\nonumber  \\
				&= 
				\norm{ \text{mat}_3 \left(  \frac{1}{N^2T} \sum_{t=1}^T \left(\widehat{\mathcal{F}}_{t-1} - \mathcal{F}_{t-1} \times_3 J \right)  \times_1  \bar J X_t' \times_2 y_{t-1}' \times_3  \beta' J \right)^\prime}\nonumber  \\
				&= 
				\norm{ \beta' J  \text{mat}_3 \left(  \frac{1}{N^2T} \sum_{t=1}^T \left(\widehat{\mathcal{F}}_{t-1} - \mathcal{F}_{t-1} \times_3 J \right)  \times_1  \bar J X_t' \times_2 y_{t-1}'  \right)}\nonumber  \\
				&\leq \norm{\beta J} \cdot \norm{\text{mat}_3 \left(  \frac{1}{N^2T} \sum_{t=1}^T \left(\widehat{\mathcal{F}}_{t-1} - \mathcal{F}_{t-1} \times_3 J \right) \times_1  \bar J X_t'  \times_2 y_{t-1}' \right)}\nonumber  \\
				&= \norm{a_1} \cdot \norm{a_2}.\label{eq:a1a2_prop1}
			\end{align}
			Clearly, $\norm{a_1} = O(1)$.
			As for $\norm{a_2}$, let us first define $z_t :=  y_{t-1} \otimes  X_t \bar J$.
			Then, recall that for a generic tensor  $\mathcal{Z}$ and matrices $A,B$, and $C$ such that $\mathcal{Z} = \mathcal{X} \times_1 A \times_2 B \times_3 C$, we have
			$\text{mat}_3(\mathcal{Z}) = C \text{mat}_3(\mathcal{X}) \left( B \otimes A \right)'$. Therefore, by using also \eqref{eq:F_HF}, we have that: 
			\begin{align}
				a_2 = &  \text{mat}_3 \left(  \frac{1}{N^2T} \sum_{t=1}^T \left(\widehat{\mathcal{F}}_{t-1} - \mathcal{F}_{t-1} \times_3 J \right) \times_1 \bar J X_t'  \times_2 y_{t-1}' \right)\nonumber\\
				&= \frac{1}{N^2T} \sum_{t=1}^T \text{mat}_3 \left(\widehat{\mathcal{F}}_{t-1} - \mathcal{F}_{t-1} \times_3 J  \right) \left(  y_{t-1}' \otimes \bar J X_t' \right)' 
				\nonumber\\
				&=  
				\frac{1}{N^2T} \sum_{t=1}^T \left( \widehat{\mathcal{F}}_{(3)t-1} - J  \mathcal{F}_{(3)t-1} \right)  \left( y_{t-1} \otimes  X_t\bar J \right)\nonumber \\
				&=  
				\frac{1}{N^2T} \sum_{t=1}^T \left( \widehat{\mathcal{F}}_{(3)t-1} - J  \mathcal{F}_{(3)t-1} \right) z_t\nonumber\nonumber \\
				&= \left( \frac{\widehat{M}^{\mathcal{W}}}{mN^2} \right)^{-1} 
				\Bigg\{ 
				\frac{1}{N^2T} \sum_{t=1}^T \frac{ \widehat{U}' \left(U - \widehat{U}J \right) }{m} \mathcal{F}_{(3)t} z_t \nonumber\\ 
				&+ \frac{1}{N^2T} \sum_{t=1}^T \frac{ \left(\widehat{U} - U J\right)' }{m} \mathcal{E}_{(3)t} z_t \nonumber\\
				&+ \frac{1}{N^2T} \sum_{t=1}^T \frac{ J' U'}{m}   \mathcal{E}_{(3)t} z_t \Bigg\} \nonumber \\
				&= \left( \frac{\widehat{M}^{\mathcal{W}}}{mN^2} \right)^{-1} \left\{ a_I + a_{II} + a_{III} \right\}.\label{eq:a2_prop1}
			\end{align}
			Now, because of \eqref{eq:MB_eq_125} in the proof of Lemma \ref{lemma:H_J} when $\widehat{H}=J$, using Cauchy-Schwarz inequality
			\begin{align}
				\norm{a_I} &= \norm{\frac{1}{N^2T} \sum_{t=1}^T \frac{ \widehat{U}' \left(U - \widehat{U}J \right) }{m} \mathcal{F}_{(3)t} z_t}_F\nonumber\\
				&\le \left(\frac{1}{N^2T} \sum_{t=1}^T \norm{\frac{ \widehat{U}' \left(U - \widehat{U}J \right) }{m} \mathcal{F}_{(3)t}}_F^2 \right)^{1/2}
				\left(\frac{1}{N^2T} \sum_{t=1}^T \norm{z_t}_F^2\right)^{1/2}\nonumber\\
				&\le \sqrt r \norm{\frac{ \widehat{U}' \left(U - \widehat{U}J \right) }{m} }\left(\frac{1}{N^2T} \sum_{t=1}^T \norm{\mathcal{F}_{(3)t}}_F^2 \right)^{1/2}
				\left(\frac{1}{N^2T} \sum_{t=1}^T \norm{z_t}_F^2\right)^{1/2}
				&= O_p\left(\frac 1{\xi}\right),\label{eq:a2_I_prop1}
			\end{align}
			where $\xi = 
			\min \left(
			\sqrt{mT} N^{2-\gamma/2},
			N^{3-\gamma/2} T, 
			m \sqrt{T} N^{2-\gamma}, 
			m N^{2-\gamma}
			\right)
			$,
			and since
			\begin{align} 
				\frac{1}{N^2T} \sum_{t=1}^T 
				\mathbb{E} \left[ \norm{\mathcal{F}_{(3)t}}_F^2 \right] 
				&= 
				\frac{1}{N^2T} \sum_{t=1}^T  \sum_{i=1}^{N^2} \sum_{j=1}^{r}
				\mathbb{E} \left[ \mathcal{F}_{(3)tji}^2 \right] \leq 
				r
				\max_{t=1,\dots,T} \max_{i=1, \dots, N^2}
				\max_{j=1, \dots, r}
				\mathbb{E} \left[ \mathcal{F}_{(3)tji}^2 \right]=r O(1),\nonumber
			\end{align}	
			because of Assumption \ref{as:common_component}\ref{as:common_component_ii}, and
			\begin{align} 
				\frac{1}{N^2T} \sum_{t=1}^T 
				\mathbb{E} \left[ \norm{z_t }_F^2 \right] 
				&= 
				\frac{1}{N^2T} \sum_{t=1}^T  \sum_{i=1}^{N^2} \sum_{j=1}^{r+2}
				\mathbb{E} \left[ z_{ijt}^2 \right] \leq 
				\frac{1}{N^2T} \sum_{t=1}^T 
				\sum_{i,j=1}^{N} 
				\sum_{k=1}^{r+2}
				\sqrt{\mathbb{E} \left[ y_{it}^4 \right]} \sqrt{\mathbb{E} \left[ X_{jkt}^4 \right]}\nonumber \\
				&\leq 
				(r+2)
				\max_{t=1,\dots,T} \max_{i,j=1, \dots, N}
				\max_{k=1, \dots, r+2}
				\sqrt{\mathbb{E} \left[ y_{it}^4 \right]} \sqrt{\mathbb{E} [ X_{jkt}^4 ]} \nonumber\\
				&=
				(r+2) O(1)O(1),\nonumber
			\end{align}	
			following from Assumptions \ref{as:common_component}\ref{as:common_component_4}, the MA representation  \eqref{eq:FNAR_as_MA} of the FNAR, and Assumptions \ref{as:errors_nu}\ref{as:errors_nu_common_iii} and \ref{as:errors_nu}\ref{as:errors_nu_idio_iii}.
			
			Moreover,
			\begin{align}
				\norm {a_{III}}&\le \norm{J} \norm{\frac 1{mN^2T}\sum_{t=1}^T\sum_{i=1}^m u_i \mathcal E_{(3)ti\cdot} z_t} = O_p\left(\frac 1{\sqrt {mT}N^{1-\gamma/2}}\right),\label{eq:a2_III_prop1}
			\end{align}
			since
			\[
			\mathbb E\left[\norm{\frac 1{mN^2T}\sum_{t=1}^T\sum_{i=1}^m u_i \mathcal E_{(3)ti\cdot} z_t}^2\right]\le \frac{\mathfrak K_1N^{2+\gamma}}{mTN^4} = \frac{\mathfrak K_1}{mTN^{2-\gamma}},
			\]
			because of Assumption \ref{as:X_nu_v}. Finally, term $a_{II}$ is dominated by term $a_{III}$ because of Theorem \ref{theorem:CLT_loadings}\ref{theorem:CLT_loadings_i}. 
			Therefore, by using \eqref{eq:a2_I_prop1} and \eqref{eq:a2_III_prop1} into \eqref{eq:a2_prop1},  and since by Lemma \ref{lemma:MB_Lemma_7}\ref{lemma:MB_Lemma_7_iv}, $\norm { \left( \frac{\widehat{M}^{\mathcal{W}}}{mN^2} \right)^{-1}}=O_p(1)$, we have
			\[
			\norm{a}=O_p \left( \max\left(\frac 1{\xi},\frac 1{\sqrt {mT}N^{1-\gamma/2}}\right)\right) = O_p \left( \max\left(\frac 1{N^{3-\gamma/2}T},\frac 1{mN^{2-\gamma}},\frac 1{\sqrt {mT}N^{1-\gamma/2}}\right)\right) .
			\]
			As for term $b$, by analogy with term $a$ above, we have
			\begin{align}
				\norm{b} \leq& \norm{\beta J} \cdot \norm{\text{mat}_3 \left(  \frac{1}{N^2T} \sum_{t=1}^T \left(\widehat{\mathcal{F}}_{t-1} - \mathcal{F}_{t-1} \times_3 J \right) \times_1  \left(\widehat X_{t}' - \bar J X'_t \right)  \times_2 y_{t-1}' \right)}\nonumber \\
				=& 
				\norm{\beta J} \cdot \norm{\frac{1}{N^2T} \sum_{t=1}^T \left( \widehat{\mathcal{F}}_{(3)t-1} - J  \mathcal{F}_{(3)t-1} \right)  \left( y_{t-1} \otimes  \left(\widehat X_{t}' - \bar J X'_t \right)  \right)}\nonumber \\ 
				=& 
				\norm{\beta J} \cdot \norm{\frac{1}{N^2T} \sum_{t=1}^T \left( \widehat{\mathcal{F}}_{(3)t-1} - J  \mathcal{F}_{(3)t-1} \right)  \left(\widehat z_{t} - z_t  \right)} \nonumber\\ 
				=& \norm{b_1}  \cdot \norm{b_2}.\label{eq:b1b2_prop1}
			\end{align}
			Now, $b_1$ is the same as $a_1$ defined before, while $b_2$ is like $a_2$ but with 
			$\widehat z_{t} - z_t = y_{t-1} \otimes  (\widehat X_{t}-  X_t\bar J )$ replacing $z_t$.
			Also,
			\begin{align}
				\widehat X_{t} -  X_t \bar J 
				=& 
				\left( \text{mat}_1 \left( \frac{1}{N} (\widehat{\mathcal{F}}_{t-1}  -\mathcal{F}_{t-1} \times_3 J )  \times_2 y_{t-1}' \right), 0_{N}, 0_{N} \right) \nonumber \\
				=&
				\left( \text{mat}_3 \left(\frac{1}{N} (\widehat{\mathcal{F}}_{t-1}  -\mathcal{F}_{t-1} \times_3 J)  \times_2 y_{t-1}' \right)', 0_{N \times 2} \right)\nonumber\\
				=& 
				\left( \frac{1}{N} \left( y_{t-1} \otimes  I_N  \right)' \left( \widehat{\mathcal{F}}_{(3)t-1} - J  \mathcal{F}_{(3)t-1} \right)', 0_{N \times 2}   \right).\label{eq:b2_prop1}
			\end{align}
			Thus, from term $a_{III}$ in \eqref{eq:a2_prop1} we see that the leading term in $b_2$ is given by
			\begin{equation}
				\frac 1{mN^2T}\sum_{t=1}^T\sum_{i=1}^m u_i \mathcal E_{(3)ti\cdot} \left(y_{t-1}\otimes\left\{y_{t-1}\otimes I_N\right\}' \frac{\mathcal E_{(3)t-1}'U}{mN}\right),\label{eq:b2lead_prop1}
			\end{equation}
			which is dominated by term $a_{III}$ because of \eqref{eq:F_consistency_C} in the proof of Theorem \ref{theorem:CLT_factors}\ref{theorem:CLT_factors_i}. Therefore, $b_2$ is dominated by $a_2$.
			By combining \eqref{eq:ab_prop1}, \eqref{eq:a1a2_prop1}, and \eqref{eq:b1b2_prop1}, we complete the proof of part \ref{lemma:part2_lemma_1_i}.

			Next, consider part \ref{lemma:part2_lemma_1_ii}.
			We have
			\begin{align}
				\frac1{TN}\sum_{t=1}^T\widehat X_{t}' \nu_t
				=& 
				\frac1{TN}\sum_{t=1}^T \bar JX_t' \nu_t 
				+\frac1{TN}\sum_{t=1}^T \left(\widehat X_{t}' - \bar J X'_t \right) \nu_t  = A + B.\label{eq:ab_prop1_ii}
			\end{align}
			By Assumption \ref{as:X_nu}\ref{as:X_nu_i},
			\[
			\norm{A}=O_p\left(\frac 1{\sqrt T}\right).
			\]
			As for term $B$, we have
			\begin{align}
				\norm{B}=\norm{\frac{1}{NT} \sum_{t=1}^{T}   \left(\widehat X_{t}' - \bar J X'_t \right) \nu_t }
				&=\norm{ \frac{1}{N^2T}  \sum_{t=1}^{T}  \left( \text{mat}_1 ( (\widehat{\mathcal{F}}_t  -\mathcal{F}_t \times_3 J )  \times_2 y_{t-1}' ), 0_{N \times N}, 0_{N \times N} \right)' \nu_t}\nonumber \\
				&=\norm{ \frac{1}{N^2T}  \sum_{t=1}^{T}  \left( \text{mat}_1 \left( (\widehat{\mathcal{F}}_t  -\mathcal{F}_t \times_3 J) \times_1 \nu'_t \times_2 y_{t-1}'  \right), 0_{N \times 2N} \right)' }
				\nonumber\\
				&= \norm{\frac{1}{N^2T}  \sum_{t=1}^{T}  \left( \text{mat}_3 \left( (\widehat{\mathcal{F}}_t  -\mathcal{F}_t \times_3 J ) \times_1 \nu'_t \times_2 y_{t-1}'  \right)' , 0_{N \times 2N} \right)' }
				\nonumber\\
				& = O_p \left( \max\left(\frac 1{N^{3-\gamma/2}T},\frac{1}{mN^{2-\gamma}},\frac{1}{\sqrt{mT} N^{1-\gamma/2}}  \right) \right).\nonumber
			\end{align}
			The proof is similar to that of term $a_2$ defined in part \ref{lemma:part2_lemma_1_i}, with $\nu_t$ replacing $X_t$.

			Finally, consider part \ref{lemma:part2_lemma_1_iii}.
			We have that 
			\begin{align}
				\norm{\frac{1}{TN}  \sum_{t=1}^T \widehat{X}_{t}'\widehat X_{t} - \frac{1}{TN} \sum_{t=1}^T \bar J X_{t}'X_{t}\bar J } \nonumber
				&\leq
				2\norm{  \frac{1}{TN} \sum_{t=1}^T (\widehat{X}_{t}' - \bar J X_t') X_{t}\bar J }  
				+
				\norm{  \frac{1}{TN} \sum_{t=1}^T (\widehat{X}_{t}' - \bar J X_t') (\widehat{X}_{t}' - \bar J X_{t}) }  
				\nonumber\\
				&= I + II.\label{eq:romani_prop1}
			\end{align}
			Now, 
			\begin{align}
				\norm{I} & =2 \norm{
					\frac{1}{N^2T}  \sum_{t=1}^{T}  \left( \text{mat}_3 \left( (\widehat{\mathcal{F}}_t  -\mathcal{F}_t \times_3 J ) \times_1 \bar J X'_t \times_2 y_{t-1}'  \right)', 0_{N \times 2N} \right)'
				}\nonumber \\
				& =   O_p \left( \max\left(\frac 1{N^{3-\gamma/2}T},\frac{1}{mN^{2-\gamma}},\frac{1}{\sqrt{mT} N^{1-\gamma/2}}  \right) \right),\nonumber
			\end{align}
			following the same steps as in the proof for term $a_2$ in part \ref{lemma:part2_lemma_1_i}.
			Moreover, $II$ is clearly dominated by $I$. This completes the proof.
		\end{proof}
		\subsection{Proof of Theorem \ref{th:CLT_FNAR}}
		\begin{proof}
			From \eqref{eq:theta_hat_terms}, Proposition \ref{lemma:part2_lemma_1}, Assumptions \ref{as:X_nu}\ref{as:X_nu_i} and \ref{as:X_nu}\ref{as:X_nu_ii}, and Slutsky's theorem:
			\begin{align}
				\sqrt T \left(\widehat{\theta}^{\text{\upshape\tiny OLS}}-\bar J\theta\right) =&\, \left(\frac 1{NT} \sum_{t=1}^T \widehat{X}_{t}'\widehat X_{t}\right)^{-1} \left\{\left(\frac1{N\sqrt T}\sum_{t=1}^T\widehat X_{t}' \nu_t\right)+ \left(\frac1{N\sqrt T}\sum_{t=1}^T\widehat X_{t}' u_t\right)\right\}\nonumber\\
				=&\, \left(\frac 1{NT} \sum_{t=1}^T \widehat{X}_{t}'\widehat X_{t}\right)^{-1} \left\{\left(\frac1{N\sqrt T}\sum_{t=1}^T \bar J X_{t}' \nu_t\right)+\left(\frac1{N\sqrt T}\sum_{t=1}^T(\widehat X_{t}-\bar J X_t)' \nu_t\right)\right.\nonumber\\
				&\left.+ \left(\frac1{N\sqrt T}\sum_{t=1}^T\widehat X_{t}' u_t\right)\right\}\nonumber\\
				=&\, \left(\frac 1{NT} \sum_{t=1}^T\bar J {X}_{t}' X_{t}\bar J \right)^{-1} \left(\frac1{N\sqrt T}\sum_{t=1}^T \bar J X_{t}' \nu_t\right) + O_p\left(\frac {\sqrt T}{ mN^{2-\gamma}}\right)+o_p(1)\nonumber\\
				&\overset{d}{\to} \mathcal N\left(0_{r+2},\bar J_0 \Sigma_{XX}^{-1}\bar J_0\bar J_0\Omega_0\bar J_0\bar J_0\Sigma_{XX}^{-1}\bar J_0\right),\nonumber
			\end{align}
			since we assumed $\sqrt { T}/ (mN^{2-\gamma}) \to 0$ and where and $\bar J_0:=\plim_{m,N,T\to\infty} \bar J$. Notice, finally  that 
			$\bar J_0 \Sigma_{XX}^{-1}\bar J_0\bar J_0\Omega_0\bar J_0\bar J_0\Sigma_{XX}^{-1}\bar J_0= \Sigma_{XX}^{-1}\Omega_0\Sigma_{XX}^{-1}$. This completes the proof.
		\end{proof}

		\subsection{Proof of Proposition \ref{lemma:GLS}}
		\begin{proof}
			First of all, notice that from Assumption \ref{as:errors_nu}\ref{as:errors_nu_idio_i} and 
			by Weyl's inequality,
			\begin{align}\label{eq:BEOBEO}
				\norm{V^{-1}} = \frac 1{\mu_N(V)} \le \frac 1{\mu_N(\Lambda\Lambda')+\mu_N(S)}=\frac 1{ \min_{i=1,\ldots, N}\mathbb E[\epsilon_{it}^2]}\le \frac 1{\underline M_{\epsilon}},
			\end{align}
			where $\mu_N(\Lambda\Lambda')=0$, $\mu_N(V)$, and $\mu_N(S)$ are the smallest eigenvalues of $\Lambda\Lambda'$, $V$, and $S$, respectively, and $\mu_N(\Lambda\Lambda')=0$.
			
			Consider part \ref{lemma:GLS_i}. We have
			\begin{align}
				\frac1{NT}\sum_{t=1}^T\widehat X_t'\widehat V^{-1} u_t
				=&\, 
				\frac1{NT}\sum_{t=1}^T  \bar J X_t' V^{-1} u_t 
				+\frac1{NT}\sum_{t=1}^T \left(\widehat X_t' - \bar J X_t' \right) V^{-1} u_t  \nonumber\\
				&+\frac1{NT}\sum_{t=1}^T  \bar J X_t'(\widehat V^{-1}-V^{-1}) u_t 
				+\frac1{NT}\sum_{t=1}^T \left(\widehat X_t' - \bar J X_t' \right)(\widehat V^{-1}-V^{-1}) u_t\nonumber\\
				=&\, a + b +c+d.\nonumber
			\end{align}
			Because of \eqref{eq:BEOBEO}, term $a$ behaves like term $a$ in \eqref{eq:ab_prop1} in the proof of Proposition \ref{lemma:part2_lemma_1}\ref{lemma:part2_lemma_1_i}, thus:
			\[
			\norm{a}= O_p \left( \max\left(\frac 1{N^{3-\gamma/2}T},\frac 1{mN^{2-\gamma}},\frac 1{\sqrt {mT}N^{1-\gamma/2}}\right)\right) .
			\]
			Likewise, term $b$ behaves like term $b$ in the proof of Proposition \ref{lemma:part2_lemma_1}\ref{lemma:part2_lemma_1_i}, thus term $b$ is dominated by term $a$. Moreover, because of Lemma \ref{lemma:VhatCNAR}, terms $c$ and $d$ are dominated by terms $a$ and $b$, respectively. This proves part \ref{lemma:GLS_i}.
			
			For part \ref{lemma:GLS_ii}, we have:
			\begin{align}
				\frac1{NT}\sum_{t=1}^T\widehat X_{t}'\widehat V^{-1} \nu_t =&\, 
				\frac1{NT}\sum_{t=1}^T\widehat X_{t}' V^{-1} \nu_t
				+\frac1{NT}\sum_{t=1}^T\widehat X_{t}'(\widehat V^{-1}-V^{-1}) \nu_t\nonumber\\
				=&\, \frac1{NT}\sum_{t=1}^T \bar J X_{t}' V^{-1} \nu_t+\frac1{NT}\sum_{t=1}^T(\widehat X_{t}-X_t\bar J)' V^{-1} \nu_t\nonumber\\
				&+\frac1{NT}\sum_{t=1}^T \bar J X_{t}'(\widehat V^{-1}-V^{-1}) \nu_t
				+\frac1{NT}\sum_{t=1}^T(\widehat X_{t}-X_t\bar J)'(\widehat V^{-1}-V^{-1}) \nu_t.\nonumber\\
				=&\, A+B+C+D.\nonumber
			\end{align}
			By Assumption \ref{as:X_nu_app}\ref{as:X_nu_iii}:
			\[
			\norm{A}=O_p\left(\frac1{\sqrt{NT}}\right).
			\]
			Term $B$, because of \eqref{eq:BEOBEO}, behaves like term $B$ in \eqref{eq:ab_prop1_ii} in the proof of Proposition \ref{lemma:part2_lemma_1}\ref{lemma:part2_lemma_1_ii}, i.e., it behaves like term $A$ in part \ref{lemma:GLS_i}: 
			\[
			\norm{B}= O_p \left( \max\left(\frac 1{N^{3-\gamma/2}T},\frac 1{mN^{2-\gamma}},\frac 1{\sqrt {mT}N^{1-\gamma/2}}\right)\right) .
			\]
			Terms $C$ and $D$ are dominated by terms $A$ and $B$, respectively, because of Lemma \ref{lemma:VhatCNAR}. This proves part \ref{lemma:GLS_ii}.
			
			For part  \ref{lemma:GLS_iii}, we have
			\begin{align}
				\norm{\frac 1{NT}\sum_{t=1}^T \widehat X_t'\widehat V^{-1}\widehat X_t- \frac 1{NT}\sum_{t=1}^T \bar J X_t' V^{-1} X_t\bar J}\le&\, 
				2\norm{  \frac{1}{TN} \sum_{t=1}^T (\widehat{X}_{t}' - \bar J X_t') V^{-1}X_{t}\bar J }  \nonumber\\
				&+\norm{  \frac{1}{TN} \sum_{t=1}^T (\widehat{X}_{t}' - \bar J X_t')V^{-1} (\widehat{X}_{t}' - \bar J X_{t}) }  \nonumber\\
				&+\norm{\frac{1}{TN} \sum_{t=1}^T \bar J X_t' (\widehat V^{-1}-V^{-1})X_t\bar J } \nonumber\\
				&+2\norm{\frac{1}{TN} \sum_{t=1}^T (\widehat{X}_{t}' - \bar J X_t')(\widehat V^{-1}-V^{-1})X_t\bar J} \nonumber\\
				&+\norm{\frac{1}{TN} \sum_{t=1}^T (\widehat{X}_{t}' - \bar J X_t')(\widehat V^{-1}-V^{-1})(\widehat{X}_{t}' - \bar J X_t')} \nonumber\\
				=&\, I+II+III+IV+V.\nonumber
			\end{align}
			Because of \eqref{eq:BEOBEO}, terms $I$ and $II$ behave like terms $I$ and $II$ in \eqref{eq:romani_prop1}  in the proof of Proposition \ref{lemma:part2_lemma_1}\ref{lemma:part2_lemma_1_iii}, thus
			\[
			\norm{I}=O_p \left( \max\left(\frac 1{N^{3-\gamma/2}T},\frac{1}{mN^{2-\gamma}},\frac{1}{\sqrt{mT} N^{1-\gamma/2}}  \right) \right),
			\]
			while $II$ is dominated by $I$. Then, by Lemma \ref{lemma:VhatCNAR}, we have
			\[
			\norm{III}=O_p\left(\max\left(\frac 1{\sqrt N},\sqrt{\frac{\log N}{T}}\right)\right).
			\]
			Finally, terms $IV$ and $V$ are dominated by terms $I$ and $II$, respectively, by Lemma \ref{lemma:VhatCNAR}. This completes the proof.
		\end{proof}

		\subsection{Proof of Theorem \ref{th:CLT_FNAR_GLS}}
		
		\begin{proof} From \eqref{eq:thetaGLS_hat_terms}, Proposition \ref{lemma:GLS}, Assumptions \ref{as:X_nu_app}\ref{as:X_nu_iii} and \ref{as:X_nu_app}\ref{as:X_nu_iv}, and Slutsky's theorem:
			\begin{align}
				\sqrt {NT} \left(\widehat{\theta}^{\text{\upshape\tiny GLS}}-\bar J\theta\right) =&\, \left(\frac 1{NT} \sum_{t=1}^T \widehat{X}_{t}'\widehat{V}^{-1}\widehat X_{t}\right)^{-1} \left\{\left(\frac1{\sqrt{N T}}\sum_{t=1}^T\widehat X_{t}' \widehat{V}^{-1}\nu_t\right)+ \left(\frac1{\sqrt{N T}}\sum_{t=1}^T\widehat X_{t}'\widehat{V}^{-1} u_t\right)\right\}\nonumber\\
				=&\, \left(\frac 1{NT} \sum_{t=1}^T\bar J {X}_{t}' {V}^{-1} X_{t}\bar J \right)^{-1} \left(\frac1{\sqrt{N T}}\sum_{t=1}^T \bar J X_{t}'{V}^{-1} \nu_t\right)\nonumber\\
				& + O_p\left(\frac {\sqrt{N T}}{ mN^{2-\gamma}}\right)+O_p\left(\frac {\sqrt{N }}{ \sqrt mN^{1-\gamma/2}}\right)+o_p(1)\nonumber\\
				&\overset{d}{\to} \mathcal N\left(0_{r+2},\bar J_0 \Omega_1^{-1}\bar J_0\bar J_0\Omega_1\bar J_0\bar J_0\Omega_1^{-1}\bar J_0\right),\nonumber
			\end{align}
			since we assumed $\sqrt { {NT}}/ (mN^{2-\gamma}) \to 0$ and $ {\sqrt{N }}/{ \sqrt mN^{1-\gamma/2}}\to 0$ and where and $\bar J_0:=\plim_{m,N,T\to\infty} \bar J$. Notice that $\bar J_0 \Omega_1^{-1}\bar J_0\bar J_0\Omega_1\bar J_0\bar J_0\Omega_1^{-1}\bar J_0=\Omega_1^{-1}$. This completes the proof.
		\end{proof}
		
		\newpage
		\section{Auxiliary lemmata}\label{app:lemma}

		We start with some notation. Let $\chi_t:=U\mathcal{F}_{(3)t}$ and recall the definition of the  $m\times m$ matrices:
		\begin{align}
			\widehat{\Gamma}^{\mathcal{W}} &:= \frac{1}{T} \sum_{t=1}^T \mathcal{W}_{(3)t} \mathcal{W}_{(3)t}',  \nonumber\\
			\Gamma^{\mathcal{W}} &:=\mathbb E \left[\mathcal{W}_{(3)t} \mathcal{W}'_{(3)t}\right],  \nonumber\\
			\Gamma^{\chi} &:=U\mathbb E \left[\mathcal{F}_{(3)t} \mathcal{F}'_{(3)t}\right] U',  \nonumber\\
			\Gamma^{\mathcal{E}}  &:=\mathbb E \left[\mathcal{E}_{(3)t} \mathcal{E}'_{(3)t}\right],  \nonumber
		\end{align} 
		with $j$ largest eigenvalues $\widehat{\mu}_j^{\mathcal{W}}$, ${\mu}_j^{\mathcal{W}}$, ${\mu}_j^{\chi}$, and ${\mu}_j^{\mathcal{E}}$, respectively.	
		
		\begin{lemma}\label{lemma:MB_Lemma_1}
			Under Assumptions \ref{as:common_component} and \ref{as:idiosyncratic_component}:
			\vspace{-7pt}
			\begin{enumerate}[label=(\roman*)]
				\item for all $j=1, \dots, r$,
				$\underline{C}_j < \underset{m, N \to \infty }{\text{lim inf }} \frac{\mu_j^{\chi}}{m N^2} \leq
				\underset{m, N \to \infty }{\text{lim sup }} \frac{\mu_j^{\chi}}{m N^2} < \overline{C}_{j}$
				for some finite $\underline{C}_{j}$ and $\overline{C}_{j}$ independent of $m$ and $N$.
				\label{lemma:Lemma_1_i}
				\item
				For all $m \in \mathbb{N}$, $N \in \mathbb{N}$ and
				$T \in \mathbb{N}$,
				\[ \frac{1}{m N^\gamma T} \sum_{i,j=1}^{m} \sum_{h,k=1}^{N^2} \sum_{t,s=1}^{T} 
				\abs{
					\mathbb{E} \left[ \mathcal{E}_{(3) t i h} \mathcal{E}_{(3) s j k} \right] 
				} \leq M_{2 \mathcal{E}}
				\]
				for some finite $M_{2 \mathcal{E}}$ independent of $m, N, T$ and some $\gamma \in [0,2]$.
				\label{lemma:Lemma_1_ii}
				\item 
				For all $m \in \mathbb{N}$ and $N \in \mathbb{N}$,
				\[ \frac{1}{m N^\gamma} \sum_{i,j=1}^{m} \sum_{h,k=1}^{N^2} 
				\abs{
					\mathbb{E} \left[ \mathcal{E}_{(3) t i h} \mathcal{E}_{(3) s j k} \right] 
				} \leq M_{3 \mathcal{E}}
				\]
				for some finite $M_{3 \mathcal{E}}$ independent of $m, N, T$ and some $\gamma \in [0,2]$.
				\label{lemma:Lemma_1_iii}
				\item 
				Let $\mu_1^{\mathcal{E}} $ denote the largest eigenvalue of $\Gamma^{\mathcal{E}}$. 
				Then, $\frac{\mu_1^{\mathcal{E}} }{N^{\gamma}} \leq M_{\mathcal{E}} $ for some $\gamma \in [0,2]$ and some finite $ M_{\mathcal{E}}$.
				\label{lemma:Lemma_1_iv}
			\end{enumerate}
		\end{lemma}
		
		\begin{proof}
			For part \ref{lemma:Lemma_1_i}, by \citet{merikoskikumar2004} (Theorem 7), for all $j=1, \dots, r$, we have 
			\begin{equation}
				\frac{\mu_r (U'U)}{m} \frac{\mu_j(\Gamma^{\mathcal{F}})}{N^2} 
				\leq \frac{\mu_j^{\chi}}{m N^2} 
				\leq \frac{\mu_j (U'U)}{m} \frac{\mu_1(\Gamma^{\mathcal{F}})}{N^2}, \nonumber
			\end{equation}
			where $\Gamma^{\mathcal{F}} = \mathbb{E} [\mathcal{F}_{(3)t}' \mathcal{F}_{(3)t}]$ and $\mu_j (\cdot)$ denotes the $j$-th eigenvalues of the matrix in parenthesis.
			The proof follows from Assumption \ref{as:common_component}\ref{as:common_component_i} which, by continuity of the eigenvalues, implies that, for any $j=1, \dots, r$, as $m \to \infty$
			\begin{equation}
				\lim_{m \to \infty} \frac{\mu_j (U'U)}{m} = \mu_j (\Sigma_U),\nonumber
			\end{equation}
			with 
			\begin{equation}
				0 < m_U^2 \leq \mu_r(\Sigma_U) \leq \mu_1 (\Sigma_U) \leq M^2_U < \infty, \nonumber
			\end{equation}
			and by Assumption \ref{as:common_component}\ref{as:common_component_ii}, which implies that 
			$\frac{\mu_r(\Gamma^{\mathcal{F}})}{N^2}$ and $\frac{\mu_1(\Gamma^{\mathcal{F}})}{N^2}$ are both finite and bounded away from zero.
			
			For part \ref{lemma:Lemma_1_ii}, by Assumption \ref{as:idiosyncratic_component}\ref{as:idiosyncratic_component_ii} we have
			\begin{align}
				\frac{1}{m N^\gamma T} \sum_{i,j=1}^{m} \sum_{h,k=1}^{N^2} \sum_{t,s=1}^{T} 
				\abs{
					\mathbb{E} \left[ \mathcal{E}_{(3) t i h} \mathcal{E}_{(3) s j k} \right] 
				}
				=&
				\frac{1}{m} \sum_{i,j=1}^{m} \sum_{h,k=1}^{N^2} 
				\sum_{l=-(T-1)}^{T-1} \left(1 - \frac{\abs{l}}{T}\right) \frac{\abs{
						\mathbb{E} \left[ \mathcal{E}_{(3) t i h} \mathcal{E}_{(3) s j k} \right] }}{N^\gamma}\nonumber \\
				\leq&
				\max_{i=1, \dots, m} \sum_{j=1}^{m} \sum_{l=-\infty}^{\infty}
				\rho_{\mathcal{E}}^{\abs{l}} M_{ij}
				\leq \frac{M_{\mathcal{E}} (1+\rho_{\mathcal{E}})}{1-\rho_{\mathcal{E}}}.\nonumber
			\end{align}

			Similarly, for part \ref{lemma:Lemma_1_iii}, 
			\begin{align}
				\frac{1}{m N^\gamma } \sum_{i,j=1}^{m} \sum_{h,k=1}^{N^2}
				\abs{
					\mathbb{E} \left[ \mathcal{E}_{(3) t i h} \mathcal{E}_{(3) t j k} \right] 
				}
				\leq&
				\max_{i=1, \dots, m} \sum_{j=1}^{m} 
				M_{ij}
				\leq M_{\mathcal{E}}.\nonumber
			\end{align}
			
			For part \ref{lemma:Lemma_1_iv}, 
			\begin{align}
				\frac{1}{N^{\gamma}} \mu_1^{\mathcal{E}}  =&  \frac{1}{N^\gamma} \norm{\mathbb{E}\left[\mathcal{E}_{(3) t} \mathcal{E}_{(3) t}' \right]} \\
				\leq& 
				\frac{1}{N^\gamma}
				\max_{i=1, \dots, m} \sum_{j=1}^{m} 
				\abs{\mathbb{E}\left[\mathcal{E}_{(3) t i \cdot}' \mathcal{E}_{(3) t \cdot j}' \right]  }\nonumber \\
				\leq& 
				\frac{1}{N^\gamma}
				\max_{i=1, \dots, m} \sum_{j=1}^{m} 
				\abs{ \sum_{h=1}^{N^2} \mathbb{E}\left[\mathcal{E}_{(3) t i h} \mathcal{E}_{(3) t h j}' \right]  }\nonumber \\
				\leq& 
				\frac{1}{N^\gamma}
				\max_{i=1, \dots, m} \sum_{j=1}^{m}  \sum_{h=1}^{N^2}
				\abs{ \mathbb{E}\left[\mathcal{E}_{(3) t i h} \mathcal{E}_{(3) t h j}' \right]  }\nonumber\\
				\leq& M_{\mathcal{E}},\nonumber
			\end{align}
			following from Assumption \ref{as:idiosyncratic_component}\ref{as:idiosyncratic_component_ii}
		\end{proof}
		
		\begin{lemma}\label{lemma:FE}
			Under Assumption \ref{as:MB_Indep},
			\begin{equation}
				\frac{1}{\sqrt{mT}N^{\gamma/2}}
				\norm{ \sum_{t=1}^{T} \mathcal{F}_{(3)t} \mathcal{E}_{(3)t}' }
				=
				O_p (1).\nonumber
			\end{equation}    	
		\end{lemma}
		\begin{proof}
			From Assumption \ref{as:MB_Indep},
			\begin{align}
				\mathbb{E} \left[
				\frac{1}{mTN^{\gamma}}
				\norm{ \sum_{t=1}^{T} \mathcal{F}_{(3)t} \mathcal{E}_{(3)t }' }^2
				\right] 
				&\leq
				\mathbb{E} \left[
				\frac{1}{mTN^{\gamma}}
				\norm{ \sum_{t=1}^{T} \mathcal{F}_{(3)t} \mathcal{E}_{(3)t }' }^2_F
				\right]\nonumber \\
				&=
				\mathbb{E} \left[\frac 1{m N^{\gamma} T} \sum_{i=1}^{m} \norm{ \sum_{t=1}^{T}  \mathcal{F}_{(3)t} \mathcal{E}_{(3)t i\cdot}^\prime  }^2 \right]
				\leq
				C_{\mathcal{FE}}.\nonumber
			\end{align}
		\end{proof}
		\begin{lemma}\label{lemma:MB_Proposition_7}
			Under Assumptions \ref{as:common_component} and \ref{as:MB_Assumption_9},
			\begin{enumerate}[label=(\roman*)]
				\item $U  = V^{\chi} (M^{\chi})^{1/2} N^{-1}$.
				\label{lemma:MB_Proposition_7_i}
				\item $\mathcal{F}_{(3)t} = N (M^{\chi})^{-\frac{1}{2}} V^{\chi}{}' \chi_t$.
				\label{lemma:MB_Proposition_7_ii}
			\end{enumerate}		
			where $ V^{\chi}$ is the $m \times r$ matrix whose $j$-th column is the normalized eigenvector corresponding to the $j$-th largest eigenvalue  $ \mu_j^{\chi}$ of the matrix $\Gamma^{\chi}=E(\chi_t \chi_t')$, 
			and  $M^{\chi}$ is the $r \times r$ diagonal matrix with $ \mu_j^{\chi}$ as its entry $(j,j)$.
		\end{lemma}
		
		\begin{proof}
			For part \ref{lemma:MB_Proposition_7_i},
			Assumption \ref{as:common_component}\ref{as:common_component_i} implies  $\frac{\Gamma^{\chi}}{N^2} =  U U' $. Therefore, since the non-zero eigenvalues of $\frac{\Gamma^{\chi}}{m N^2}$ are the same as the $r$ eigenvalues of $\frac{U'U}{m}$, which is diagonal by Assumption \ref{as:MB_Assumption_9}\ref{as:MB_Assumption_9_i}.
			Then, we must have, for all $m \in \mathbb{N}$,
			\begin{equation}
				\frac{U'U}{m} = \frac{M^{\chi}}{mN^2}.\nonumber
			\end{equation}
			Since $\Gamma^{\chi} = V^{\chi} M^{\chi} V^{\chi}{}'$, it must be that
			\begin{equation}
				U K = \frac{V^{\chi} (M^{\chi})^{1/2}}{N},\nonumber
			\end{equation}
			for some $r \times r$ invertible $K$. By multiplying on the right both sides by their transposed:
			\begin{equation}
				K' U' U K = \frac{M^{\chi}}{N^2},\nonumber
			\end{equation} 
			since eigenvectors are normalized. Thus, we must have $K = I_r$.
			This proves part \ref{lemma:MB_Proposition_7_i}.
			
			For part \ref{lemma:MB_Proposition_7_ii}, since $\chi = U \mathcal{F}_{(3)t}$, then by linear projection of $U$ onto $\chi_t$, and using part (i), for $t=1, \dots, T$, 
			\begin{equation}
				\mathcal{F}_{(3)t} = \left( U' U \right)^{-1} U' \chi_t = 
				N  (M^{\chi})^{-\frac{1}{2}} V^{\chi}{}' \chi_t.\nonumber
			\end{equation}
			This proves part \ref{lemma:MB_Proposition_7_ii}.	
		\end{proof}
		\begin{lemma}\label{lemma:MB_Lemma_2}
			Under Assumptions \ref{as:common_component} and \ref{as:idiosyncratic_component}, for all $t=1, \dots, T$ and all $m,N,T \in \mathbb{N}$, 
			\begin{enumerate}[label=(\roman*)]
				\item 
				$\norm{ \frac{U}{\sqrt{m}} } = O(1)$.
				\label{lemma:MB_Lemma_2_i}
				\item
				$\norm{\frac{\mathcal{F}_{(3)t}}{N}} = O_p(1)$ and
				$\norm{\frac{\mathcal{F}_{(3)}}{\sqrt{T}N}} = O_p(1)$. 
				\label{lemma:MB_Lemma_2_ii}
				\item
				$\norm{\frac{\mathcal{E}_{(3)t}}{\sqrt{m}N}} =  O_p\left(\frac{1}{N^{1-\gamma/2}}\right)$ and
				$\norm{\frac{\mathcal{E}_{(3)}}{\sqrt{mT}N}} =  O_p\left(\frac{1}{N^{1-\gamma/2}}\right)$ .
				\label{lemma:MB_Lemma_2_iii}
			\end{enumerate}
		\end{lemma}
		
		\begin{proof}
			By Assumption \ref{as:common_component}\ref{as:common_component_i}, which holds for all $m \in \mathbb{N}$, 
			\begin{equation}
				\sup_{m \in \mathbb{N}} \norm{\frac{U}{\sqrt{m}}}^2
				\leq 
				\sup_{m \in \mathbb{N}} \norm{\frac{U}{\sqrt{m}}}^2_F
				=
				\sup_{m \in \mathbb{N}}  \frac{1}{m} \sum_{j=1}^{r} \sum_{i=1}^{m}
				u_{ij}^2 
				\leq \sup_{m \in \mathbb{N}} \max_{i=1, \dots, m} \norm{u_i}^2
				\leq M_U^2,\nonumber
			\end{equation}
			since $M_U$ is independent of $i$. This proves part \ref{lemma:MB_Lemma_2_i}.
			
			For part \ref{lemma:MB_Lemma_2_ii}, by Assumption \ref{as:common_component}\ref{as:common_component_ii}
			\begin{align}
				\sup_{N,T \in \mathbb{N}} \max_{t=1, \dots, T} 
				\mathbb{E} \left[ \norm{\frac{\mathcal{F}_{(3)t}}{N}}^2  \right]
				&\leq  \sup_{N,T \in \mathbb{N}} \max_{t=1, \dots, T} 		\mathbb{E} \left[ \norm{\frac{\mathcal{F}_{(3)t}}{N}}^2_F  \right]\nonumber	 \\
				&=
				\sup_{N,T \in \mathbb{N}} \max_{t=1, \dots, T} 	
				\mathbb{E} \left[\tr\left( \frac{\mathcal{F}_{(3)t}\mathcal{F}_{(3)t}'}{N^2}\right)  \right]\nonumber\\
				&= \sup_{N,T \in \mathbb{N}} \max_{t=1, \dots, T}
				\tr( \frac{1}{N^2} \mathbb{E} \left[ \mathcal{F}_{(3)t}\mathcal{F}_{(3)t}' \right])
				= r.\nonumber
			\end{align}
			Therefore,
			\begin{align}
				\sup_{N,T \in \mathbb{N}} 
				\mathbb{E} \left[ \norm{\frac{\mathcal{F}_{(3)}}{\sqrt{T}N}}^2  \right]
				&\leq  \sup_{N,T \in \mathbb{N}}  		\mathbb{E} \left[ \norm{\frac{\mathcal{F}_{(3)}}{\sqrt{T}N}}^2_F  \right]	\nonumber \\
				&= \sup_{N,T \in \mathbb{N}} 
				\frac{1}{T} \sum_{t=1}^{T} \mathbb{E} \left[ \norm{\frac{\mathcal{F}_{(3)t}}{N}}^2_F  \right] \nonumber \\
				&=\sup_{N,T \in \mathbb{N}} \max_{t=1, \dots, T} 
				\mathbb{E} \left[ \norm{\frac{\mathcal{F}_{(3)t}}{N}}^2_F  \right]
				= r.\nonumber
			\end{align}
			For part \ref{lemma:MB_Lemma_2_iii}, by Lemma \ref{lemma:MB_Lemma_1}\ref{lemma:Lemma_1_iv}
			\begin{align}
				\sup_{N,T \in \mathbb{N}} \max_{t=1, \dots, T} 
				\mathbb{E} \left[ \norm{\frac{\mathcal{E}_{(3)t}}{\sqrt{m}N}}^2  \right]
				&\leq  \sup_{N,T \in \mathbb{N}} \max_{t=1, \dots, T} 		\mathbb{E} \left[ \norm{\frac{\mathcal{E}_{(3)t}}{\sqrt{m}N}}^2_F  \right]	\nonumber \\
				&=
				\sup_{N,T \in \mathbb{N}} \max_{t=1, \dots, T} 	
				\mathbb{E} \left[\tr\left( \frac{\mathcal{E}_{(3)t}\mathcal{E}_{(3)t}'}{mN^2}\right)  \right]\nonumber\\
				&= \sup_{N,T \in \mathbb{N}} \max_{t=1, \dots, T}
				\frac{1}{mN^2} \tr(  \Gamma^{\mathcal{E}})\nonumber\\
				&\leq \frac{ M_{\mathcal{E}} N^{\gamma}}{N^{2-\gamma}} .\nonumber 
			\end{align}
			Thus,
			\begin{align}
				\sup_{N,T \in \mathbb{N}} 
				\mathbb{E} \left[ \norm{\frac{\mathcal{E}_{(3)}}{\sqrt{mT}N}}^2  \right]
				&\leq  \sup_{N,T \in \mathbb{N}}		\mathbb{E} \left[ \norm{\frac{\mathcal{E}_{(3)}}{\sqrt{mT}N}}^2_F  \right]\nonumber	 \\
				&=
				\sup_{N,T \in \mathbb{N}}	
				\frac{1}{T} \sum_{t=1}^T	\mathbb{E} \left[ \norm{\frac{\mathcal{E}_{(3)t}}{\sqrt{m}N}}^2_F  \right]\nonumber	\\
				&= \sup_{N,T \in \mathbb{N}} \max_{t=1, \dots, T}
				\mathbb{E} \left[ \norm{\frac{\mathcal{E}_{(3)t}}{\sqrt{m}N}}^2_F  \right]
				\leq \frac{ M_{\mathcal{E}} N^{\gamma}}{N^{2-\gamma}} ,\nonumber
			\end{align}
			which completes the proof.
		\end{proof}
		\begin{lemma}\label{lemma:MB_Lemma_4}
			Under Assumptions \ref{as:common_component}-\ref{as:idiosyncratic_component}, for all $m,N,T \in \mathbb{N}$,
			\begin{enumerate}[label=(\roman*)]
				\item 
				$
				\norm{ \frac{1}{N^2 T} \sum_{t=1}^{T}  \mathcal{F}_{(3)t}  \mathcal{F}_{(3)t}' - \frac{\Gamma^{\mathcal{F}}}{N^2}} 
				= O_p \left(\frac{1}{N \sqrt{T}}\right).
				$	
				\label{lemma:MB_Lemma_4_i}
				
				\item 
				$
				\norm{ \frac{1}{ m N^2 T} \sum_{t=1}^{T} \mathcal{E}_{(3)t}\mathcal{E}_{(3)t}' - \frac{\Gamma^{\mathcal{E}}}{m N^2} } = O_p \left(\frac{1}{ N^{2-\gamma/2} \sqrt{T}}\right).
				$
				\label{lemma:MB_Lemma_4_ii}
			\end{enumerate}
		\end{lemma}
		
		\begin{proof}
			For part \ref{lemma:MB_Lemma_4_i}, because of Assumption \ref{as:common_component}\ref{as:common_component_iii},
			\begin{align}
				\mathbb{E} \left[  \norm{ \frac{1}{N^2 T}  \sum_{t=1}^{T}  \mathcal{F}_{(3)t}  \mathcal{F}_{(3)t}' - \frac{\Gamma^{\mathcal{F}}}{N^2}}^2 \right]
				&\leq
				\mathbb{E} \left[  \norm{ \frac{1}{N^2 T} \sum_{t=1}^{T}  \mathcal{F}_{(3)t}  \mathcal{F}_{(3)t}' - \frac{\Gamma^{\mathcal{F}}}{N^2}}^2_F \right]\nonumber\\
				&= \frac{1}{N^2 T} \sum_{j=1}^{r} \sum_{i=1}^{r}
				\mathbb{E} \left[ \left( 
				\frac{1}{N \sqrt{T}}
				\sum_{t=1}^{T} 
				\{  
				\mathcal{F}_{(3)t j \cdot} \mathcal{F}_{(3)t i \cdot}' 
				- 
				\mathbb{E} [ \mathcal{F}_{(3)t j \cdot} \mathcal{F}_{(3)t i \cdot}' ]
				\}  
				\right)^2  \right]\nonumber \\
				&\leq \frac{r^2 C_{\mathcal{F}}}{N^2 T} .\nonumber
			\end{align}
			This proves part \ref{lemma:MB_Lemma_4_i}.
			
			For part \ref{lemma:MB_Lemma_4_ii}, 
			because of Assumption \ref{as:idiosyncratic_component}\ref{as:idiosyncratic_component_v},
			\begin{align}
				&\mathbb{E} \left[  \norm{ \frac{1}{m N^2 T}  \sum_{t=1}^{T}  \mathcal{E}_{(3)t}  \mathcal{E}_{(3)t}' - \frac{\Gamma^{\mathcal{E}}}{m N^2}}^2 \right]\nonumber
				\\
				&\leq \frac{1}{m N^{4- \gamma} T}
				\sum_{j=1}^{m} 
				\mathbb{E} \left[ \left( 
				\frac{1}{\sqrt{mT} N^{\gamma/2} } 
				\sum_{i=1}^{m} \sum_{t=1}^{T} 
				\left\{ 
				\mathcal{E}_{(3)t i \cdot} \mathcal{E}_{(3)t j \cdot}'
				-
				\mathbb{E} \left[  \mathcal{E}_{(3)t i \cdot} \mathcal{E}_{(3)t j \cdot}' \right]
				\right\} 
				\right)^2  \right] \nonumber\\
				&\leq \frac{C_{\mathcal{E}}}{N^{4 - \gamma} T}, \nonumber
			\end{align}
			which completes the proof.
		\end{proof}
		\begin{lemma}\label{lemma:MB_Lemma_5}
			Under Assumptions \ref{as:common_component}-\ref{as:MB_Assumption_9}, for all $m,T, N \in \mathbb{N}$
			\begin{enumerate}[label=(\roman*)]
				\item 
				$\frac{1}{m N^2}  \norm{ \widehat{\Gamma}^{\mathcal{W}} - \Gamma^{\mathcal{W}} } = O_p \left(\frac{1}{N \sqrt{T}}\right)$.
				\label{lemma:MB_Lemma_5_i}
				\item 	
				$\frac{1}{m N^2} \norm{ \widehat{\Gamma}^{\mathcal{W}}  -  \Gamma^{\chi} } = O_p \left( \max \left( \frac{1}{N \sqrt{T}}, \frac{1}{N^{2-\gamma} m}\right)  \right)$.  
				\label{lemma:MB_Lemma_5_ii}
				\item 
				$\frac{1}{m N^2} \left| \widehat{\mu}_j^{\mathcal{W}} -  \mu_j^{\chi} \right| =  O_p \left(\max \left(\frac{1}{N \sqrt{T}},\frac{1}{N^{2-\gamma} m} \right) \right)$.
				\label{lemma:MB_Lemma_5_iii}
				for all $j$  			
				\item 
				$\norm{\widehat{V}^{\mathcal{W}} - V^{\chi} J} =   O_p \left(\max \left(\frac{1}{N \sqrt{T}},\frac{1}{N^{2-\gamma} m} \right) \right)$,
				\label{lemma:MB_Lemma_5_iv}
				where $\widehat{V}^{\mathcal{W}} $ is the $m \times r$ matrix whose $j$-th column is the normalized (unit-modulus) eigenvector corresponding to the $j$-th largest eigenvalue of $  \widehat{\Gamma}^{\mathcal{W}}$, $V^{\chi}$ is the $m \times r$ matrix whose $j$-th column is the normalized eigenvector corresponding to the $j$-th largest eigenvalue of  $ \Gamma^{\chi} $, and $J$ is a $r \times r$ diagonal matrix whose diagonal entries are $\pm 1$.
			\end{enumerate}
		\end{lemma}
		
		\begin{proof}
			For part \ref{lemma:MB_Lemma_5_i}, we have that
			\begin{align}
				\norm{\frac{1}{m N^2} \left(   \widehat{\Gamma}^{\mathcal{W}} - \Gamma^{\mathcal{W}}  \right) }
				&\leq
				\norm{ 
					\frac{1}{mN^2} 
					\left\{ 
					U \left( \frac{1}{T}   \sum_{t=1}^{T}  \mathcal{F}_{(3)t}  \mathcal{F}_{(3)t}' - \Gamma^{\mathcal{F}} \right) U' 
					+ \frac{1}{T} \sum_{t=1}^{T} \mathcal{E}_{(3)t} \mathcal{E}_{(3)t}' - \Gamma^{\mathcal{E}}
					\right\} }\nonumber \\
				&+ \norm{\frac{2}{mN^2 T} \sum_{t=1}^{T}  \mathcal{F}_{(3)t}   \mathcal{E}_{(3)t}^\prime}\nonumber\\
				&\leq  
				\norm{	\frac{U}{\sqrt{m}} }^2  \norm{ \frac{1}{T N^2} \sum_{t=1}^{T}  \mathcal{F}_{(3)t}  \mathcal{F}_{(3)t}' - \frac{\Gamma^{\mathcal{F}}}{N^2}}
				+ \norm{ \left( \frac{1}{m T N^2} \sum_{t=1}^{T} \mathcal{E}_{(3)t} \mathcal{E}_{(3)t}' - \frac{\Gamma^{\mathcal{E}} }{m N^2}\right) } \nonumber\\
				&+ \norm{\frac{2}{mN^2 T} \sum_{t=1}^{T}  \mathcal{F}_{(3)t}   \mathcal{E}_{(3)t}^\prime}.\nonumber
			\end{align} 
			By Lemma \ref{lemma:MB_Lemma_2}\ref{lemma:MB_Lemma_2_i}, \ref{lemma:MB_Lemma_4}\ref{lemma:MB_Lemma_4_i} and \ref{lemma:MB_Lemma_4}\ref{lemma:MB_Lemma_4_ii}, the first two terms are 
			$O_p \left(  \frac{1}{N \sqrt{T}} \right)$.
			By Lemma \ref{lemma:FE}, 
			$$
			\mathbb{E} \left[\frac 1{m^2 N^4 T^2} \norm{ \sum_{t=1}^T   \mathcal{F}_{(3)t}   \mathcal{E}_{(3)t}^\prime }^2 \right] =
			O \left(  \frac{1}{ m T N^{4-\gamma}} \right),
			$$
			so
			$$
			\norm{  \frac 1{m N^2 T} \sum_{t=1}^T   \mathcal{F}_{(3)t}  \mathcal{E}_{(3)t}^\prime } =
			O_p\left( \frac{1}{\sqrt
				{mT} N^{2-\gamma/2}}  \right).
			$$
			This proves part \ref{lemma:MB_Lemma_5_i}.
			
			For part \ref{lemma:MB_Lemma_5_ii}, under Assumption \ref{as:MB_Indep}, we have that
			\begin{equation} 
				\widehat{\Gamma}^{\mathcal{W}}  -  \Gamma^{\chi}  =   \widehat{\Gamma}^{\mathcal{W}}  - \Gamma^{\mathcal{W}} +  \Gamma^{\mathcal{E}} .\nonumber
			\end{equation}
			Hence,
			\begin{align} 
				\frac{1}{m N^2} \norm{ \widehat{\Gamma}^{\mathcal{W}}  -  \Gamma^{\chi} } &= \frac{1}{m N^2} \norm{ \widehat{\Gamma}^{\mathcal{W}}  -  \Gamma^{\mathcal{W}} }  + \frac{1}{m N^2} \norm{  \Gamma^{\mathcal{E}} }\nonumber \\
				&= O_p \left(\frac{1}{N \sqrt{T}}\right) + O_p\left(\frac{1}{m N^{2-\gamma}}\right) \label{eq:gamma_op_2},\nonumber
			\end{align}
			following from part (i) and Lemma \ref{lemma:MB_Lemma_1}\ref{lemma:Lemma_1_iv}, since $\norm{\Gamma^{\mathcal{E}}}$ is bounded by $\mu_1^{\mathcal{E}}$.

			For part \ref{lemma:MB_Lemma_5_iii}, given Weyl's inequality,
			\begin{equation}
				\left| \widehat{\mu}_j^{\mathcal{W}} -  \mu_j^{\chi} \right| 
				\leq \mu_1 \left( \widehat{\Gamma}^{\mathcal{W}}  -  \Gamma^{\chi} \right)
				=  
				\norm{ \widehat{\Gamma}^{\mathcal{W}}  -  \Gamma^{\chi} },\nonumber 
			\end{equation}
			so, given part \ref{lemma:MB_Lemma_5_ii},
			\begin{equation}\label{eq:mu_op}
				\frac{1}{mN^2} \left| \widehat{\mu}_j^{\mathcal{W}} -  \mu_j^{\chi} \right| =  O_p \left(\max \left(\frac{1}{N \sqrt{T}},\frac{1}{m N^{2-\gamma}} \right) \right).\nonumber
			\end{equation}
			
			Last, for part \ref{lemma:MB_Lemma_5_iv}, given part \ref{lemma:MB_Lemma_5_i}, Lemma \ref{lemma:MB_Lemma_1}\ref{lemma:Lemma_1_i},  and  Theorem 2 in \citet{yu2015useful}, which is a special case of Davis-Kahn Theorem						
			\begin{equation*}
				\norm{\widehat{V}^{\mathcal{W}} - V^{\chi} J}
				\leq
				\frac{ \norm{ \widehat{\Gamma}^{\mathcal{W}} - \Gamma^{\chi} } }{\mu^{\chi}_r}
				=
				\frac{N^2 m O_p\left( \frac{1}{N\sqrt{T}}, \frac{1}{N^{2-\gamma}m} \right) }{O(N^2 m)} = O_p\left( \frac{1}{N\sqrt{T}}, \frac{1}{N^{2-\gamma}m} \right),\nonumber
			\end{equation*}			
			and this completes the proof.
		\end{proof}
		\begin{lemma}\label{lemma:MB_Lemma_7}
			Under Assumptions \ref{as:common_component}-\ref{as:MB_Assumption_9}, for all $m,T, N \in \mathbb{N}$
			\begin{enumerate}[label=(\roman*)]
				\item $\norm{  \frac{M^{\chi}}{m N^2}} = O(1)$.
				\label{lemma:MB_Lemma_7_i}
				\item $\norm{ \left(\frac{M^{\chi}}{m N^2} \right)^{-1}} = O(1)$.
				\label{lemma:MB_Lemma_7_ii}
				\item $\norm{  \frac{\widehat{M}^{\mathcal{W}}}{m N^2}} = O_p(1)$.
				\label{lemma:MB_Lemma_7_iii}
				\item $\norm{ \left( \frac{\widehat{M}^{\mathcal{W}}}{m N^2} \right)^{-1}} = O_p(1)$.
				\label{lemma:MB_Lemma_7_iv}
			\end{enumerate}
		\end{lemma}
		
		\begin{proof}
			Parts \ref{lemma:MB_Lemma_7_i} and \ref{lemma:MB_Lemma_7_ii} follow directly from Lemma \ref{lemma:MB_Lemma_1}, indeed, 
			\begin{equation}
				\norm{\frac{M^{\chi}}{m N^2}} = \frac{\mu_1^{\chi}}{m N^2} \leq \overline{C}_1,\nonumber
			\end{equation}
			and 
			\begin{equation}
				\norm{ \left( \frac{M^{\chi}}{m N^2} \right)} = \frac{mN^2}{\mu_r^{\chi}} \leq \underline{C}_r.\nonumber
			\end{equation}
			Both statements hold for all $m \in \mathbb{N}$ since the eigenvalues are an increasing sequence in $m$ and $N$.
			
			For part \ref{lemma:MB_Lemma_7_iii}, because of Lemma \ref{lemma:MB_Lemma_5}\ref{lemma:MB_Lemma_5_iii}, 
			\begin{equation}
				\norm{ \frac{\widehat{M}^{\mathcal{W}}}{m N^2}} 
				\leq 	
				\norm{  \frac{M^{\chi}}{m N^2} } 
				+
				\norm{  \frac{\widehat{M}^{\mathcal{W}}}{m N^2} -  \frac{M^{\chi}}{m N^2} }
				\leq
				\overline{C}_1 + O_p\left(  \max \left(  \frac{1}{N \sqrt{T}}, \frac{1}{N^{2-\gamma} m } \right) \right).\nonumber
			\end{equation}	
			For part \ref{lemma:MB_Lemma_7_iv}, just notice that, because of Lemma \ref{lemma:MB_Lemma_5}\ref{lemma:MB_Lemma_5_iii} and part \ref{lemma:MB_Lemma_7_ii}, then $\frac{\widehat{M}^{\mathcal{W}}}{m N^2} $ is positive definite with probability tending to one as $m, N, T \to \infty$. This completes the proof.
		\end{proof}
		
		\begin{lemma}\label{lemma:MB_Lemma_6} 
			Under Assumptions \ref{as:common_component}-\ref{as:MB_Assumption_9}, for any given $i=1,\ldots ,m$ and 
			$m,T \to \infty$
			\begin{enumerate}[label=(\roman*)]
				\item $\frac{1}{m N^2} \norm{\varepsilon_i' \left(\widehat{\Gamma}^{\mathcal{W}}  -  \Gamma^{\chi}\right) } = O_p \left( \max \left( \frac{1}{N \sqrt{T}}, \frac{1}{\sqrt{N^{2-\gamma} m}}\right)  \right)$ . 
				\label{lemma:MB_Lemma_6_0}
				\item $\norm{\sqrt {m} v_i^\chi}=O_p(1)$.
				\label{lemma:MB_Lemma_6_i}
				\item $\norm{\sqrt {m}\, \widehat v_i^{\mathcal W'}-\sqrt{m} v_i^{\chi'}J }=O_p\left( \frac{1}{N\sqrt{T}}, \frac{1}{N^{2-\gamma}m} \right)$.\label{lemma:MB_Lemma_6_ii}
			\end{enumerate}
		\end{lemma}
		
		\begin{proof} Part \ref{lemma:MB_Lemma_6_0} follows directly from Lemma \ref{lemma:MB_Lemma_5}\ref{lemma:MB_Lemma_5_ii}.
			For part \ref{lemma:MB_Lemma_6_i} notice that for all $i=1,\ldots, m$, since $N^{-2}\Gamma^{\mathcal F}=I_r$ because of Assumption \ref{as:MB_Assumption_9}\ref{as:MB_Assumption_9_ii}, we 	must have
			\begin{equation}\label{eq:eccheccazzo}
				\frac{\text{Var}(\chi_{it})}{N^2}= u_i'u_i \le M_U^2,
			\end{equation}
			which is finite for all $i$ and $t$. Then, by Lemma \ref{lemma:MB_Lemma_1}\ref{lemma:Lemma_1_i}
			\begin{align}
				\lim\inf_{m\to\infty}\max_{i=1,\ldots, m} \text{Var}(\chi_{it})&= \lim\inf_{m\to\infty}\max_{i=1,\ldots, m}\sum_{j=1}^r\mu_j^\chi \abs{V_{ij}^\chi}^2
				\ge  \lim\inf_{m\to\infty}\max_{i=1,\ldots, m}\mu_r^\chi \sum_{j=1}^r\abs{V_{ij}^\chi}^2\nonumber\\
				&\ge \underline C_r mN^2\max_{i=1,\ldots, m}\norm {v_i^\chi}^2,\nonumber
			\end{align}
			and by \eqref{eq:eccheccazzo} we must have
			\[
			\underline C_r mN^2\max_{i=1,\ldots, m}\norm {v_i^\chi}^2\le M_U^2 N^2,
			\]
			which implies
			\[
			m\max_{i=1,\ldots, m}\norm {v_i^\chi}^2\le \frac{M_U^2}{\underline C_r }.
			\]
			This proves part \ref{lemma:MB_Lemma_6_i}.
			
			For part \ref{lemma:MB_Lemma_6_ii} we follow the same approach as in the proof of Lemma \ref{lemma:MB_Lemma_5}\ref{lemma:MB_Lemma_5_iv} but using part \ref{lemma:MB_Lemma_6_0}.
			
		\end{proof}
		\begin{lemma}\label{lemma:MB_Proposition_3}
			Under Assumptions \ref{as:common_component}-\ref{as:MB_Assumption_9}, for $m,T \to \infty$
			\begin{enumerate}[label=(\roman*)]
				\item $\frac{1}{\sqrt{m}}\norm{ \widehat{U} -  U J } = O_p \left(\max \left(\frac{1}{N \sqrt{T}},\frac{1}{N^{2-\gamma} m} \right) \right)$.\label{lemma:MB_Proposition_3_i}
				\item for any given $i=1,\ldots ,m$, $\norm{ \widehat{u_i}' -  u_i' J } = O_p \left(\max \left(\frac{1}{N \sqrt{T}},\frac{1}{\sqrt{N^{2-\gamma} m}} \right) \right)$.\label{lemma:MB_Proposition_3_ii}
			\end{enumerate}
		\end{lemma}
		
		\begin{proof}
			For part \ref{lemma:MB_Proposition_3_i}, first notice that $\text{rk} \left(\frac{U}{\sqrt{m}} \right) = r$ for all $m$, since $\text{rk} \left( \frac{\Gamma^{\mathcal{F}}}{N^2} \right) = r$ by Assumption \ref{as:common_component}\ref{as:common_component_ii} and $\text{rk} \left(\frac{\Gamma^{\chi}}{m} \right) = r$ by Lemma \ref{lemma:MB_Lemma_1}\ref{lemma:Lemma_1_i}.
			Indeed, 
			$\text{rk} \left( \frac{\Gamma^{\mathcal{F}}}{m N^2}  \right) \leq 
			\min 
			\left( 
			\text{rk} \left( \frac{\Gamma^{\mathcal{F}}}{N^2} \right), 
			\text{rk} \left(\frac{U}{\sqrt{m}} \right)  
			\right)$. 
			This holds for all $m>\overline{m}$ and since eigenvalues	are an increasing sequence in $m$.
			Therefore, $\left( \frac{U'U}{m}\right)^{-1}$ is well defined for all $m$ and $U$ admits a left inverse.
			
			Now, because of Lemma \ref{lemma:MB_Lemma_5}\ref{lemma:MB_Lemma_5_iii}, \ref{lemma:MB_Lemma_5}\ref{lemma:MB_Lemma_5_iv}, \ref{lemma:MB_Lemma_7}(i), using \eqref{eq:est_loadings}
			\begin{align}
				\norm{ \frac{\widehat{U} -  U J}{\sqrt{m}}   } 
				&=
				\norm{\widehat{V}^{\mathcal{W}} \left( \frac{\widehat{M}^{\mathcal{W}}}{m N^2} \right)^{1/2}
					-
					V^{\chi} \left( \frac{M^{\chi}}{m N^2} \right)^{1/2} J } \nonumber\\
				&\leq  \norm{\widehat{V}^{\mathcal{W}} - 	V^{\chi} J}	 \norm{\frac{M^{\chi}}{m N^2}}
				+  \norm{  \frac{1}{\sqrt{m} N} \left\{  \left( \widehat{M}^{\mathcal{W}} \right)^{1/2}  -  \left( M^{\chi} \right)^{1/2} \right\} }	 \norm{V^{\chi}}	\nonumber	\\
				&+  \norm{\widehat{V}^{\mathcal{W}} - 	V^{\chi} J}	  \norm{  \frac{1}{\sqrt{m} N} \left\{  \left( \widehat{M}^{\mathcal{W}} \right)^{1/2}  -  \left( M^{\chi} \right)^{1/2} \right\}  }\nonumber  \\
				& = O_p \left(\max \left( \frac{1}{N \sqrt{T}}, \frac{1}{m N^{2-\gamma}} \right) \right).\nonumber
			\end{align}
			
			For part \ref{lemma:MB_Proposition_3_ii}, because of Lemmas
			\ref{lemma:MB_Lemma_5}\ref{lemma:MB_Lemma_5_iii},
			\ref{lemma:MB_Lemma_7}\ref{lemma:MB_Lemma_7_iii},
			\ref{lemma:MB_Lemma_6}\ref{lemma:MB_Lemma_6_i}, and
			\ref{lemma:MB_Lemma_6}\ref{lemma:MB_Lemma_6_ii}
			\begin{align}
				\norm{\widehat{u}_i'-u_i'J} &= 
				\norm{\sqrt {m}\, \widehat v_i^{\mathcal W'}\left(\frac{\widehat M^{\mathcal W}}{mN^2}\right)^{1/2}-\sqrt{m} v_i^{\chi'}J 
					\left(\frac{M^\chi}{mN^2}\right)^{1/2}}\nonumber\\
				&\le \norm{\sqrt {m}\, \widehat v_i^{\mathcal W'}-\sqrt{m} v_i^{\chi'}J }\,\norm{\frac{\widehat M^{\mathcal W}}{mN^2}}
				+\norm{\frac 1{\sqrt{m}} \left\{\left(\widehat M^{\mathcal W}\right)^{1/2}-\left( M^{\chi}\right)^{1/2}\right\} }\,\norm{\sqrt {m} v_i^\chi}\nonumber\\
				&+\norm{\sqrt {m}\, \widehat v_i^{\mathcal W'}-\sqrt{m} v_i^{\chi'}J }\,\norm{\frac 1{\sqrt{mN^2}} \left\{\left(\widehat M^{\mathcal W}\right)^{1/2}-\left( M^{\chi}\right)^{1/2}\right\} }\nonumber\\
				&=O_p \left(\max \left( \frac{1}{N \sqrt{T}}, \frac{1}{\sqrt {m N^{2-\gamma}}} \right) \right).\nonumber
			\end{align}
		\end{proof}
		\begin{lemma}\label{lemma:MB_Lemma_3}
			Under Assumptions \ref{as:common_component}-\ref{as:MB_Assumption_9}, for all $m, N, T \in \mathbb{N}$
			\begin{enumerate}[label=(\roman*)]
				\item 
				$ 
				\norm{    \frac{ \mathcal{F}_{(3)} \mathcal{E}_{(3)}' }{\sqrt{m} N^2T}   }    
				= O_p\left( \frac{1}{ \sqrt{T} N^{2-\gamma/2}}  \right) $.
				\label{lemma:MB_Lemma_3_i}
				\item 
				$ 
				\norm{    \frac{ \mathcal{E}_{(3)} \mathcal{E}_{(3)}' }{mN^2T}   }    
				= O_p \left( \max \left(  \frac{1}{N^{2-\gamma/2}\sqrt{T}}, \frac{1}{mN^{2-\gamma}} \right)  \right)$
				and 
				$\norm{\frac{U' \mathcal{E}_{(3)} \mathcal{E}_{(3)}' }{m^{3/2} N^2T}} = 
				O_p \left( \max \left(  \frac{1}{\sqrt{mT}N^{2-\gamma/2}}, \frac{1}{mN^{2-\gamma}} \right)  \right)$.
				\label{lemma:MB_Lemma_3_ii}
				\item 
				$ \norm{\frac{ U'U }{m}} = O\left( 1 \right) $.
				\label{lemma:MB_Lemma_3_iii}
			\end{enumerate}
		\end{lemma}
		
		\begin{proof}
			Part \ref{lemma:MB_Lemma_3_i} follows directly from Assumption \ref{as:MB_Indep}.
			For part \ref{lemma:MB_Lemma_3_ii}, by Lemma \ref{lemma:MB_Lemma_4}\ref{lemma:MB_Lemma_4_ii} and Lemma \ref{lemma:MB_Lemma_1}\ref{lemma:Lemma_1_iv}
			\begin{align}
				\norm{    \frac{ \mathcal{E}_{(3)} \mathcal{E}_{(3)}' }{mN^2T} }    
				\leq& \norm{    \frac{ \mathcal{E}_{(3)} \mathcal{E}_{(3)}' }{mN^2T} 
					- \frac{\Gamma^{\mathcal{E}}}{mN^2} }   
				+ \norm{ \frac{\Gamma^{\mathcal{E}}}{mN^2}  }\nonumber  \\
				&= O_p \left(  \frac{1}{N^{2-\gamma/2}\sqrt{T}}\right) + O\left(\frac{1}{mN^{2-\gamma}} \right).\nonumber
			\end{align}
			Similarly, by Lemma \ref{lemma:MB_Lemma_1}\ref{lemma:Lemma_1_iv} and Lemma \ref{lemma:MB_Lemma_2}\ref{lemma:MB_Lemma_2_i} 
			\begin{align}
				\norm{\frac{U' \mathcal{E}_{(3)} \mathcal{E}_{(3)}' }{m^{3/2} N^2T}} 
				&\leq 
				\norm{\frac{U' \mathcal{E}_{(3)} \mathcal{E}_{(3)}' }{m^{3/2} N^2T} - \frac{U \Gamma^{\mathcal{E}}}{m^{3/2}N^2}} 
				+
				\norm{ \frac{U' }{\sqrt{m}}}
				\norm{ \frac{\Gamma^{\mathcal{E}}}{m N^2}}	\nonumber\\
				&=\norm{\frac{U \mathcal{E}_{(3)} \mathcal{E}_{(3)}' }{m^{3/2} N^2T} - \frac{U \Gamma^{\mathcal{E}}}{m^{3/2}N^2}} 
				+
				O\left(\frac{1}{mN^{2-\gamma}} \right).\nonumber
			\end{align}
			Then, because of Assumption \ref{as:idiosyncratic_component}\ref{as:idiosyncratic_component_v}, 
			\begin{align}
				\mathbb{E} \left[\norm{\frac{U' \mathcal{E}_{(3)} \mathcal{E}_{(3)}' }{m^{3/2} N^2T} - \frac{U \Gamma^{\mathcal{E}}}{m^{3/2}N^2}}^2 \right] 
				\leq&
				\mathbb{E} \left[\norm{\frac{U' \mathcal{E}_{(3)} \mathcal{E}_{(3)}' }{m^{3/2} N^2T} - \frac{U \Gamma^{\mathcal{E}}}{m^{3/2}N^2}}^2_F \right]\nonumber \\
				=&
				\sum_{k=1}^{r} \sum_{j=1}^{m}
				\mathbb{E} \left[ 
				\abs{
					\frac{1}{m^{3/2} N^2T}
					\sum_{i=1}^{m}
					\sum_{t=1}^{T}
					U_{ik} \mathcal{E}_{(3)t i \cdot} \mathcal{E}_{(3)t j \cdot}'
					- U_{ik} \mathbb{E} \left[ \mathcal{E}_{(3)t i \cdot} \mathcal{E}_{(3)t j \cdot}' \right]
				}^2
				\right] \nonumber\\
				\leq& 
				\frac{r M_U^2 m}{m^2N^{4-\gamma} T}
				\max_{j=1, \dots, m}
				\mathbb{E} \left[ 
				\abs{
					\frac{1}{\sqrt{mT} N^{\gamma/2}}
					\sum_{i=1}^{m}
					\sum_{t=1}^{T}
					\mathcal{E}_{(3)t i \cdot} \mathcal{E}_{(3)t j \cdot}'
					-  \mathbb{E} \left[ \mathcal{E}_{(3)t i \cdot} \mathcal{E}_{(3)t j \cdot}' \right]
				}^2
				\right]\nonumber\\
				\leq& 
				\frac{r M_U^2 C_{\mathcal{E}}}{mN^{4-\gamma} T}.\nonumber
			\end{align}
			This proves part \ref{lemma:MB_Lemma_3_ii}.
			
			Part (\ref{lemma:MB_Lemma_3_iii} follows from Lemma \ref{lemma:MB_Lemma_2}\ref{lemma:MB_Lemma_2_i}, since 
			\begin{equation}
				\norm{\frac{ U'U }{m}} \leq \norm{\frac{ U}{\sqrt{m}}}^2 \leq M_U^2 .\nonumber
			\end{equation}
		\end{proof}
		\begin{lemma}[Consistency of loadings]\label{lemma:MB_proposition_2}
			Under Assumptions \ref{as:common_component}-\ref{as:MB_Assumption_9}
			\begin{enumerate}[label=(\roman*)]
				\item 
				$
				\norm{   \widehat{u}_i' - u_i' \widehat{H}   } 
				= O_p \left(
				\max \left( \frac{1}{N^{2-\gamma/2} \sqrt{T}}, \frac{1}{m N^{2-\gamma}} \right)  	\right),
				$
				where $\widehat H = \frac{\mathcal{F}_{(3)} \mathcal{F}_{(3)}'}{N^2 T} \frac{U'\widehat{U}}{m} \left( \frac{\widehat{M}^{\mathcal{W}}}{m N^2} \right)^{-1}$.
				\label{lemma:MB_proposition_2_i}
				\item 
				$
				\norm{ \frac{\widehat{U} - U\widehat{H}}{\sqrt{m}} } 
				= O_p \left(
				\max \left( \frac{1}{N^{2-\gamma/2} \sqrt{T}}, \frac{1}{m N^{2-\gamma}} \right)  	\right).
				$
				\label{lemma:MB_proposition_2_ii}
			\end{enumerate}
		\end{lemma}
		
		\begin{proof}
			Substituting
			\begin{equation}
				\mathcal{W}_{(3)}\mathcal{W}_{(3)}' = 
				U \mathcal{F}_{(3)} \mathcal{F}_{(3)}' U' + U\mathcal{F}_{(3)}  \mathcal{E}_{(3)}' + 
				\mathcal{E}_{(3)} \mathcal{F}_{(3)}'U' +\mathcal{E}_{(3)} \mathcal{E}_{(3)}'\nonumber
			\end{equation}
			into
			\begin{equation}
				\frac{\mathcal{W}_{(3)}\mathcal{W}_{(3)}'}{m N^2 T} \widehat{U} = \widehat{U}  \frac{\widehat{M}^{\mathcal{W}}}{m N^2},\nonumber
			\end{equation}
			where $\widehat{M}^{\mathcal{W}}$ are the eigenvalues of $\mathcal{W}_{(3)}\mathcal{W}_{(3)}'/T$, we get:
			\begin{equation}
				\frac{U \mathcal{F}_{(3)} \mathcal{F}_{(3)}' U'\widehat{U}}{m N^2 T} 
				+ \frac{U\mathcal{F}_{(3)}  \mathcal{E}_{(3)}'\widehat{U}}{m N^2 T}  
				+ 
				\frac{\mathcal{E}_{(3)} \mathcal{F}_{(3)}'U'\widehat{U}}{m N^2 T}  +\frac{\mathcal{E}_{(3)} \mathcal{E}_{(3)}'\widehat{U}}{m N^2 T}  = \widehat{U} \frac{\widehat{M}^{\mathcal{W}}}{m N^2}. \nonumber
			\end{equation}
			Define
			\begin{equation}
				\widehat{H} := \frac{\mathcal{F}_{(3)} \mathcal{F}_{(3)}'}{N^2 T} \frac{U'\widehat{U}}{m} \left( \frac{\widehat{M}^{\mathcal{W}}}{m N^2} \right)^{-1}.\label{eq:HDEF_PD}
			\end{equation}
			Then, 
			\begin{align}
				\widehat{U} - U \widehat{H} =& 
				\left( 
				\frac{U \mathcal{F}_{(3)} \mathcal{E}_{(3)}' \widehat{U}}{m N^2 T} +
				\frac{\mathcal{E}_{(3)} \mathcal{F}_{(3)}' U'\widehat{U}}{m N^2 T} +
				\frac{\mathcal{E}_{(3)} \mathcal{E}_{(3)}' \widehat{U}}{m N^2 T} 
				\right) 
				\left( \frac{\widehat{M}^{\mathcal{W}}}{m N^2} \right)^{-1}\nonumber\\
				=& \left( 
				\frac{U \mathcal{F}_{(3)} \mathcal{E}_{(3)}' U}{m N^2 T} +
				\frac{\mathcal{E}_{(3)} \mathcal{F}_{(3)}' U'U}{m N^2 T} +
				\frac{\mathcal{E}_{(3)} \mathcal{E}_{(3)}' U}{m N^2 T} 
				\right) 
				J
				\left( \frac{\widehat{M}^{\mathcal{W}}}{m N^2} \right)^{-1}\nonumber \\
				& + \left( 
				\frac{U \mathcal{F}_{(3)} \mathcal{E}_{(3)}' }{m N^2 T} +
				\frac{\mathcal{E}_{(3)} \mathcal{F}_{(3)}' U'}{m N^2 T} +
				\frac{\mathcal{E}_{(3)} \mathcal{E}_{(3)}' }{m N^2 T} 
				\right) 
				\left( \widehat{U} - U  J \right)
				\left( \frac{\widehat{M}^{\mathcal{W}}}{m N^2} \right)^{-1}  \label{eq:Uhat_UHhat}
			\end{align}
			Taking the $i$-th row of  \eqref{eq:Uhat_UHhat}, we have that:
			\begin{align}
				&\left(  \widehat{u}_i' - u_i' \widehat{H} \right) \nonumber
				\\
				&=
				\left(
				\frac{1}{m N^2 T} u_i' \sum_{t=1}^{T} \sum_{j=1}^{m}  \mathcal{F}_{(3)t} \mathcal{E}_{(3)tj\cdot}' u_j'  
				+ \frac{1}{m N^2 T}  \sum_{t=1}^{T} \mathcal{E}_{(3)ti\cdot} \mathcal{F}_{(3)t}' \sum_{j=1}^{m}  u_j u_j'
				+ \frac{1}{m N^2 T}  \sum_{t=1}^{T} \sum_{j=1}^{m} \mathcal{E}_{(3)ti\cdot} \mathcal{E}_{(3)tj\cdot}'  u_j'
				\right) \nonumber\\
				&\times
				J
				\left( \frac{\widehat{M}^{\mathcal{W}}}{m N^2} \right)^{-1}\nonumber \\
				&+ \Bigg(
				\frac{1}{m N^2 T} u_i' \sum_{t=1}^{T} \sum_{j=1}^{m}  \mathcal{F}_{(3)t} \mathcal{E}_{(3)tj\cdot}' (\widehat{u}_j' - u_j' J)
				+ \frac{1}{m N^2 T}  \sum_{t=1}^{T} \mathcal{E}_{(3)ti\cdot} \mathcal{F}_{(3)t}' \sum_{j=1}^{m}  u_j  (\widehat{u}_j' - u_j' J)\nonumber\\
				&+ \frac{1}{m N^2 T}  \sum_{t=1}^{T} \sum_{j=1}^{m} \mathcal{E}_{(3)ti\cdot} \mathcal{E}_{(3)tj\cdot}'  (\widehat{u}_j' - u_j' J)
				\Bigg) 
				\left( \frac{\widehat{M}^{\mathcal{W}}}{m N^2} \right)^{-1}\nonumber\\
				&= (a+b+c) J
				\left( \frac{\widehat{M}^{\mathcal{W}}}{m N^2} \right)^{-1}
				+ (d+e+f)
				\left( \frac{\widehat{M}^{\mathcal{W}}}{m N^2} \right)^{-1}.  \label{eq:MB_eq_29}
			\end{align}			
			First consider part \ref{lemma:MB_proposition_2_i}.
			For term $a$, by Assumption \ref{as:common_component}\ref{as:common_component_i},
			\begin{equation}
				\norm{
					\frac{1}{m N^2 T} u_i' \sum_{t=1}^{T} \sum_{j=1}^{m}  \mathcal{F}_{(3)t} \mathcal{E}_{(3)tj\cdot}' u_j' }
				\leq
				\norm{u_i}
				\norm{
					\frac{1}{m N^2 T} \sum_{t=1}^{T} \sum_{j=1}^{m}  \mathcal{F}_{(3)t} \mathcal{E}_{(3)tj\cdot}' u_j' }
				\leq
				M_U
				\norm{
					\frac{1}{m N^2 T} \sum_{t=1}^{T} \sum_{j=1}^{m}  \mathcal{F}_{(3)t} \mathcal{E}_{(3)tj\cdot}' u_j' },\nonumber
			\end{equation}
			for any $i=1, \dots, m$.
			Then, by Assumption \ref{as:common_component}\ref{as:common_component_i} and Lemma \ref{lemma:FE}
			\begin{align}
				\mathbb{E}
				\left[
				\norm{
					\frac{1}{m N^2 T}  \sum_{t=1}^{T} \sum_{j=1}^{m}  \mathcal{F}_{(3)t} \mathcal{E}_{(3)tj\cdot}' u_j' }^2
				\right]
				&\leq 
				\mathbb{E}
				\left[
				\norm{
					\frac{1}{m N^2 T}  \sum_{t=1}^{T} \sum_{j=1}^{m}  \mathcal{F}_{(3)t} \mathcal{E}_{(3)tj\cdot}' u_j' }^2_F
				\right]\nonumber\\
				&= 
				\frac{1}{m^2 N^4 T^2}
				\sum_{h=1}^{r}
				\mathbb{E}
				\left[
				\norm{
					\sum_{t=1}^{T} \sum_{j=1}^{m}  \mathcal{F}_{(3)t} \mathcal{E}_{(3)tj\cdot}' [U_{jh}] }^2_F
				\right]\nonumber\\
				&=
				\max_{h=1, \dots, r}
				\frac{r}{m^2 N^4 T^2}
				\mathbb{E}
				\left[
				\norm{
					\sum_{t=1}^{T} \sum_{j=1}^{m}  \mathcal{F}_{(3)t} \mathcal{E}_{(3)tj\cdot}'  [U_{jh}] }^2_F
				\right]\nonumber\\
				&=
				\frac{r M_U^2}{m^2 N^4 T^2}
				\mathbb{E}
				\left[
				\norm{
					\sum_{t=1}^{T} \sum_{j=1}^{m}  \mathcal{F}_{(3)t} \mathcal{E}_{(3)tj\cdot}' }^2_F
				\right]\nonumber\\
				&=
				\frac{r M_U^2}{m N^{4-\gamma} T}
				\mathbb{E} \left[\frac 1{m N^{\gamma} T} \sum_{i=1}^{m} \norm{ \sum_{t=1}^{T}  \mathcal{F}_{(3)t} \mathcal{E}_{(3)t i\cdot}^\prime  }^2 \right]\nonumber\\
				&\leq
				\frac{r M_U^2 C_{\mathcal{FE}}}{m N^{4-\gamma} T}
				=
				O_p \left(\frac{1}{mTN^{4-\gamma}}\right).  \label{eq:MB_eq_85}
			\end{align}
			Therefore, term $a$ is $O_p (\frac{1}{\sqrt{mT}N^{2-\gamma/2}})$.
			
			For term $b$, for any $i=1, \dots, m$, because of Lemma \ref{lemma:MB_Lemma_2}\ref{lemma:MB_Lemma_2_i}
			\begin{align}
				\norm{\frac{1}{m N^2 T}  \sum_{t=1}^{T} \mathcal{E}_{(3)ti\cdot} \mathcal{F}_{(3)t}' \sum_{j=1}^{m}  u_j u_j'} &=
				\norm{ \frac{1}{m N^2 T}  \sum_{t=1}^{T} \mathcal{E}_{(3)ti\cdot} \mathcal{F}_{(3)t}' (U' U)}\nonumber\\
				&\leq \norm{ \frac{1}{N^2 T}  \sum_{t=1}^{T} \mathcal{E}_{(3)ti\cdot} \mathcal{F}_{(3)t}'}
				\norm{ \frac{U}{\sqrt{m}}}^2\nonumber\\
				&\leq \norm{ \frac{1}{N^2 T}  \sum_{t=1}^{T} \mathcal{E}_{(3)ti\cdot} \mathcal{F}_{(3)t}'} M_U^2.
				\label{eq:MB_eq_90}
			\end{align}
			Then, by Assumption \ref{as:MB_Indep} with $m=1$, 
			\begin{align}
				\norm{ \frac{1}{N^2 T}  \sum_{t=1}^{T} \mathcal{E}_{(3)ti\cdot} \mathcal{F}_{(3)t}'}
				&= \frac{\sqrt{m}}{N^{2-\gamma/2} \sqrt{T}}
				\norm{ \frac{1}{ \sqrt{mT} N^{\gamma/2}}  \sum_{t=1}^{T} \mathcal{E}_{(3)ti\cdot} \mathcal{F}_{(3)t}'}\nonumber \\
				&=   \frac{\sqrt{m}}{N^{2-\gamma/2} \sqrt{T}} 	O_p \left( \frac{1}{\sqrt{m}}\right).
				\label{eq:MB_eq_91}
			\end{align}
			By substituting  \eqref{eq:MB_eq_91} into \eqref{eq:MB_eq_90}, we prove that term $b$ is $O_p (\frac{1}{\sqrt{T} N^{2-\gamma/2}})$

			For term $c$, for any $i=1, \dots, m$, because of Assumption \ref{as:common_component}\ref{as:common_component_i}, 
			\begin{align}
				\norm{\frac{1}{m N^2 T}  \sum_{t=1}^{T} \sum_{j=1}^{m} \mathcal{E}_{(3)ti\cdot} \mathcal{E}_{(3)tj\cdot}'  u_j'} 
				&=
				\left\{  \sum_{t=1}^{T}\sum_{j=1}^{m}  \left( \frac{1}{m N^2 T} 
				\sum_{t=1}^{T} \sum_{j=1}^{m} \mathcal{E}_{(3)ti\cdot} \mathcal{E}_{(3)tj\cdot}'  U_{jk}
				\right)^2 \right\}^{1/2}\nonumber\\
				&\leq \sqrt{r} M_U \abs{ \frac{1}{m N^2 T} \sum_{t=1}^{T} \sum_{j=1}^{m}  \mathcal{E}_{(3)ti\cdot} \mathcal{E}_{(3)tj\cdot}' }\nonumber \\
				&\leq 
				\sqrt{r} M_U   
				\abs{\frac{1}{m N^2 T} \sum_{t=1}^{T} \sum_{j=1}^{m} \mathbb{E} \left[ \mathcal{E}_{(3)ti\cdot} \mathcal{E}_{(3)tj\cdot}' \right]  }\nonumber  \\
				&+
				\sqrt{r} M_U  
				\abs{ \frac{1}{m N^2 T} \sum_{t=1}^{T} \sum_{j=1}^{m} 
					\left\{	\mathcal{E}_{(3)ti\cdot} \mathcal{E}_{(3)tj\cdot}' - \mathbb{E} \left[ \mathcal{E}_{(3)ti\cdot} \mathcal{E}_{(3)tj\cdot}' \right] 
					\right\}  }.\nonumber 
			\end{align}
			By Assumption \ref{as:idiosyncratic_component}\ref{as:idiosyncratic_component_ii}, 
			\begin{align}
				\abs{\frac{1}{m N^2 T} \sum_{t=1}^{T} \sum_{j=1}^{m} \mathbb{E} \left[ \mathcal{E}_{(3)ti\cdot} \mathcal{E}_{(3)tj\cdot}' \right]  } 
				&\leq
				\frac{1}{m N^2 T} \sum_{t=1}^{T} \sum_{j=1}^{m} \abs{\mathbb{E} \left[ \mathcal{E}_{(3)ti\cdot} \mathcal{E}_{(3)tj\cdot}' \right]  } 
				\nonumber\\
				&\leq
				\frac{1}{m N^2 T}  \sum_{t=1}^{T} \sum_{j=1}^{m} M_{ij} N^{\gamma} 
				\leq \frac{M_{\mathcal{E}}}{mN^{2-\gamma}}.\nonumber
			\end{align}
			Moreover, by Assumption \ref{as:idiosyncratic_component}\ref{as:idiosyncratic_component_v} 
			\begin{equation}
				\mathbb{E} \left[  
				\abs{\frac{1}{m N^2 T} \sum_{t=1}^{T} \sum_{j=1}^{m} 
					\left\{	\mathcal{E}_{(3)ti\cdot} \mathcal{E}_{(3)tj\cdot}' - \mathbb{E} \left[ \mathcal{E}_{(3)ti\cdot} \mathcal{E}_{(3)tj\cdot}' \right] 
					\right\}}^2
				\right]
				\leq
				\frac{C_{\mathcal{E}}}{mN^{4-\gamma}T}.\nonumber
			\end{equation}
			Hence, term $c$ is $O_p \left(  \frac{1}{mN^{2-\gamma}} , \frac{1}{\sqrt{mT} N^{2-\gamma/2}}  \right)$.
			
			Given Lemma \ref{lemma:MB_Proposition_3}\ref{lemma:MB_Proposition_3_ii}, the terms $d$, $e$, and $f$ are dominated, since they are similar to $a$, $b$, and $c$, but with $(\widehat{u}'_j-u_j'J)$ replacing $u_j'$. This completes the proof of part \ref{lemma:MB_proposition_2_i}.
			
			Second, consider part \ref{lemma:MB_proposition_2_ii}.
			From \eqref{eq:Uhat_UHhat}
			\begin{align}
				\norm{\frac{ \widehat{U} - U \widehat{H} }{\sqrt{m}}}
				\leq&
				\left(
				\norm{\frac{\mathcal{F}_{(3)} \mathcal{E}_{(3)}' }{\sqrt{m}N^2 T}}
				\norm{\frac{U}{\sqrt{m}}}^2
				+
				\norm{\frac{\mathcal{E}_{(3)} \mathcal{F}_{(3)}' }{\sqrt{m}N^2 T}}
				\norm{\frac{U'U}{m}}
				+
				\norm{\frac{\mathcal{E}_{(3)} \mathcal{E}_{(3)}' }{mN^2 T}}
				\norm{\frac{U}{\sqrt{m}}} 
				\right)
				\norm{J} \norm{\left(\frac{\widehat{M}^{\mathcal{W}}}{mN^2} \right)^{-1}}\nonumber\\
				&+
				\left(
				\norm{\frac{\mathcal{F}_{(3)} \mathcal{E}_{(3)}' }{\sqrt{m}N^2 T}}
				\norm{\frac{U}{\sqrt{m}}}
				+
				\norm{\frac{\mathcal{E}_{(3)} \mathcal{F}_{(3)}' }{\sqrt{m}N^2 T}}
				\norm{\frac{U}{\sqrt{m}}}
				+
				\norm{\frac{\mathcal{E}_{(3)} \mathcal{E}_{(3)}' }{mN^2 T}}
				\right)
				\norm{\frac{ \widehat{U} - U J }{\sqrt{m}}} \norm{\left(\frac{\widehat{M}^{\mathcal{W}}}{mN^2} \right)^{-1}}\nonumber\\
				=&
				O_p \left(\max \left(
				\frac{1}{\sqrt{T}N^{2-\gamma/2}}, 
				\frac{1}{mN^{2-\gamma}}
				\right) \right),\nonumber
			\end{align}
			following from Lemma \ref{lemma:MB_Lemma_2}\ref{lemma:MB_Lemma_2_i}, \ref{lemma:MB_Proposition_3}, \ref{lemma:MB_Lemma_3}\ref{lemma:MB_Lemma_3_i}, \ref{lemma:MB_Lemma_3}\ref{lemma:MB_Lemma_3_ii}, \ref{lemma:MB_Lemma_3}\ref{lemma:MB_Lemma_3_iii}, \ref{lemma:MB_Lemma_7}\ref{lemma:MB_Lemma_7_iv} and $\norm{J} = O(1)$.
			This completes the proof.
		\end{proof}
		\begin{lemma}\label{lemma:MB_Lemma_11}
			Under Assumptions \ref{as:common_component}-\ref{as:MB_Assumption_9}, as $m, N, T \to \infty$, $ \norm{\widehat{H}} = O(1)$.
		\end{lemma}
		
		\begin{proof}
			By Lemmas \ref{lemma:MB_Lemma_2}, \ref{lemma:MB_Lemma_7} and \ref{lemma:MB_proposition_2}
			\begin{align}
				\norm{\widehat{H}} 
				&\leq
				\norm{\frac{\mathcal{F}_{(3)} \mathcal{F}_{(3)}'}{N^2 T}} \norm{\frac{U'\widehat{U}}{m}}
				\norm{\left( \frac{\widehat{M}^{\mathcal{W}}}{m N^2} \right)^{-1} }
				\leq
				\norm{\frac{\mathcal{F}_{(3)}}{N \sqrt{T}}}^2 \norm{\frac{U}{\sqrt{m}}}
				\norm{\frac{\widehat U}{\sqrt{m}}}
				\norm{\left( \frac{\widehat{M}^{\mathcal{W}}}{m N^2} \right)^{-1} }\nonumber \\
				&\leq 
				\norm{\frac{\mathcal{F}_{(3)}}{N \sqrt{T}}}^2 \norm{\frac{U}{\sqrt{m}}}
				\norm{J}
				\norm{\left( \frac{\widehat{M}^{\mathcal{W}}}{m N^2} \right)^{-1} }
				+
				\norm{\frac{\mathcal{F}_{(3)}}{N \sqrt{T}}}^2 \norm{\frac{U}{\sqrt{m}}}
				\norm{\frac{ \widehat{U} - U \widehat{H} }{\sqrt{m}}}
				\norm{\left( \frac{\widehat{M}^{\mathcal{W}}}{m N^2} \right)^{-1} }\nonumber \\
				&= O_p(1) + O_p \left(\max \left(
				\frac{1}{\sqrt{T}N^{2-\gamma/2}}, 
				\frac{1}{mN^{2-\gamma}}
				\right) \right).\nonumber
			\end{align}
		\end{proof}
		\begin{lemma}\label{lemma:H_J}
			Under Assumptions \ref{as:common_component}-\ref{as:MB_Assumption_9}, as $m,N,T \to \infty$
			\begin{enumerate}[label=(\roman*)]
				\item  $ \norm{\widehat{H} -J} = O_p\left(\frac{1}{\xi}\right) = o_p \left( \max \left( \frac{1}{\sqrt{T}N^{2-\gamma/2}}, \frac{1}{\sqrt mN^{2-\gamma/2}} \right)\right)$,\label{lemma:H_J_i}
				\item $\norm{\widehat{H}^{-1} -J} =  O_p\left(\frac{1}{\xi}\right) = o_p \left( \max \left( \frac{1}{\sqrt{T}N^{2-\gamma/2}}, \frac{1}{\sqrt mN^{1-\gamma/2}} \right)\right)$,
				\label{lemma:H_J_ii}
			\end{enumerate}
			where $\xi = 
			\min \left(
			\sqrt{mT} N^{2-\gamma/2},
			N^{3-\gamma/2} T, 
			m \sqrt{T} N^{2-\gamma}, 
			m N^{2-\gamma}
			\right)
			$.
		\end{lemma}

		\begin{proof}
			For part \ref{lemma:H_J_i}, using \eqref{eq:est_loadings} we have
			\begin{align}
				\frac{\widehat{U}' U \widehat{H}}{m}
				&=
				\frac{\widehat{U}' \widehat{U}}{m}
				+
				\frac{\widehat{U}' (U \widehat{H}- \widehat{U} ) }{m}\nonumber\\
				&= \frac{\widehat{M}^{\mathcal{W}}}{mN^2} 
				+ 
				\frac{\widehat{U}' (U \widehat{H}- \widehat{U} ) }{m} .\nonumber
			\end{align}
			Now, 
			\begin{equation}
				\frac{\widehat{U}' (U \widehat{H}- \widehat{U} ) }{m}
				=
				\frac{   (\widehat{U} - U \widehat{H} )' U \widehat{H} }{m} + \frac{(\widehat{U} - U \widehat{H} )' (\widehat{U} - U  \widehat{H} )}{m}.\nonumber
			\end{equation}
			Then, \begin{align}
				\norm{ \frac{\widehat{U}' (U \widehat{H}- \widehat{U} ) }{m} }
				&\leq \norm{  \frac{   (\widehat{U} - U \widehat{H} )' U \widehat{H} }{m}} 
				+ \norm{ \frac{(\widehat{U} - U \widehat{H} )' (\widehat{U} - U  \widehat{H} )}{m} }= I + II.\nonumber
			\end{align}
			First, consider $I$.
			From \eqref{eq:Uhat_UHhat}
			\begin{align}
				I =&  \frac{   (\widehat{U} - U \widehat{H} )' U \widehat{H} }{m}\nonumber\\
				=& \left(  \frac{\widehat{M}^{\mathcal{W}}}{mN^2}  \right)^{-1} J \frac{U'U}{m} \frac{\mathcal{F}_{(3)} \mathcal{E}_{(3)}' U \widehat{H}}{m N^2 T}
				+ \left(  \frac{\widehat{M}^{\mathcal{W}}}{mN^2}  \right)^{-1} J  	\frac{U' \mathcal{E}_{(3)} \mathcal{F}_{(3)}' }{m N^2 T}\frac{U'U \widehat{H}}{m}
				+  \left(  \frac{\widehat{M}^{\mathcal{W}}}{mN^2}  \right)^{-1} J \frac{U' \mathcal{E}_{(3)} \mathcal{E}_{(3)}' U \widehat{H}}{m^2 N^2 T}  \nonumber  \\
				&+  \left(  \frac{\widehat{M}^{\mathcal{W}}}{mN^2}  \right)^{-1} \frac{(\widehat{U} - UJ)'}{\sqrt{m}}  \frac{ \mathcal{E}_{(3)} \mathcal{F}_{(3)}' U'}{mN^2T} \frac{U \widehat{H}}{\sqrt{m}} 
				+  \left(  \frac{\widehat{M}^{\mathcal{W}}}{mN^2}  \right)^{-1} \frac{(\widehat{U} - UJ)'}{\sqrt{m}}  \frac{U  \mathcal{F}_{(3)} \mathcal{E}_{(3)}'}{mN^2T} \frac{U \widehat{H}}{\sqrt{m}}\nonumber \\
				&+ \left(  \frac{\widehat{M}^{\mathcal{W}}}{mN^2}  \right)^{-1} \frac{(\widehat{U} - UJ)'}{\sqrt{m}}  \frac{\mathcal{E}_{(3)} \mathcal{E}_{(3)}'}{mN^2T} \frac{U \widehat{H}}{\sqrt{m}}\nonumber \\
				=& I_a + I_b + I_c + I_d + I_e + I_f.\nonumber
			\end{align}
			Then, given \eqref{eq:MB_eq_85}, 
			\begin{equation}\label{eq:MB_eq_127}
				\mathbb{E} \left[ \norm{ \frac{\mathcal{F}_{(3)} \mathcal{E}_{(3)}' U}{m N^2 T}}^2  \right]
				= O \left(  \frac{1}{mTN^{4-\gamma}}  \right).
			\end{equation}	
			Therefore, by Assumption \ref{as:common_component}(i), Lemma \ref{lemma:MB_Lemma_7}(iv), Lemma \ref{lemma:MB_Lemma_11} and using $\norm{J} = O(1)$ and \eqref{eq:MB_eq_127}, 
			we get
			\begin{gather}
				\norm{I_a}  \leq 
				\norm{\left(  \frac{\widehat{M}^{\mathcal{W}}}{mN^2}  \right)^{-1}} \
				\norm{J} \  \norm{\frac{U'U}{m}} \
				\norm{ \frac{\mathcal{F}_{(3)} \mathcal{E}_{(3)}'U }{m N^2 T}}  \
				\norm{\widehat{H}} 
				= O_p \left(  \frac{1}{\sqrt{mT} N^{2-\gamma/2}}  \right) \nonumber \\
				\norm{I_b}  \leq 
				\norm{\left(  \frac{\widehat{M}^{\mathcal{W}}}{mN^2}  \right)^{-1}} \ 
				\norm{J} \
				\norm{ \frac{U' \mathcal{E}_{(3)} \mathcal{F}_{(3)}'}{m N^2 T}}  \
				\norm{\frac{U'U}{m}} \
				\norm{\widehat{H}} 
				= O_p \left(  \frac{1}{\sqrt{mT} N^{2-\gamma/2}}  \right) .\nonumber
			\end{gather}
			Moreover, because of Lemma \ref{lemma:MB_Lemma_2}\ref{lemma:MB_Lemma_2_i}, \ref{lemma:MB_Lemma_3}\ref{lemma:MB_Lemma_3_ii}, 
			Lemma \ref{lemma:MB_Lemma_7}\ref{lemma:MB_Lemma_7_iv}, 
			Lemma \ref{lemma:MB_Lemma_11} and  $\norm{J} = O(1)$, 
			\begin{equation}
				\norm{I_c} 
				\leq 
				\norm{\left(  \frac{\widehat{M}^{\mathcal{W}}}{mN^2}  \right)^{-1}} \ 
				\norm{J} \ 
				\norm{\frac{U}{\sqrt{m}}} \
				\norm{   \frac{U'\mathcal{E}_{(3)} \mathcal{E}_{(3)}'}{m^{3/2} N^2 T} } \
				\norm{\widehat{H}} 
				= O_p \left( \max \left(  \frac{1}{\sqrt{mT}N^{2-\gamma/2}}, \frac{1}{mN^{2-\gamma}} \right)  \right).\nonumber
			\end{equation}
			Similarly, because of Lemma \ref{lemma:MB_Proposition_3}, Lemma \ref{lemma:MB_Lemma_2}\ref{lemma:MB_Lemma_2_i},  \ref{lemma:MB_Lemma_3}\ref{lemma:MB_Lemma_3_i} and \ref{lemma:MB_Lemma_3_iii}, Lemma \ref{lemma:MB_Lemma_7}\ref{lemma:MB_Lemma_7_iv} and \ref{lemma:MB_Lemma_11}, 
			\begin{align}
				\norm{I_d} \leq&
				\norm{\left(  \frac{\widehat{M}^{\mathcal{W}}}{mN^2}  \right)^{-1}} \
				\norm{\frac{\widehat{U} - UJ}{\sqrt{m}}}  \
				\norm{\frac{ \mathcal{E}_{(3)} \mathcal{F}_{(3)}' }{\sqrt{m} N^2T}}  \
				\norm{\frac{U'U}{m}} \
				\norm{\widehat{H}} \nonumber\\
				&= O_p(1) \ O_p \left(\max \left( \frac{1}{N \sqrt{T}}, \frac{1}{m N^{2-\gamma}} \right) \right) \
				O_p\left( \frac{1}{ \sqrt{T} N^{2-\gamma/2}}  \right)  O_p(1) O_p(1)\nonumber\\
				&= O_p \left(\max \left( 
				\frac{1}{N^{3-\gamma/2} T}, 
				\frac{1}{m \sqrt{T} N^{2-\gamma}} 
				\right) \right),\nonumber\\
				\norm{I_e} \leq&
				\norm{\left(  \frac{\widehat{M}^{\mathcal{W}}}{mN^2}  \right)^{-1}} \
				\norm{\frac{\widehat{U} - UJ}{\sqrt{m}}}  \
				\norm{\frac{ \mathcal{E}_{(3)} \mathcal{F}_{(3)}' }{\sqrt{m} N^2T}}  \
				\norm{\frac{U}{\sqrt{m}}}^2 \
				\norm{\widehat{H}}\nonumber \\
				&= O_p \left(\max \left( 
				\frac{1}{N^{3-\gamma/2} T}, 
				\frac{1}{m \sqrt{T} N^{2-\gamma}} 
				\right) \right),\nonumber
			\end{align}
			and, because of Lemma \ref{lemma:MB_Proposition_3}, Lemma \ref{lemma:MB_Lemma_2}\ref{lemma:MB_Lemma_2_i},  \ref{lemma:MB_Lemma_3}\ref{lemma:MB_Lemma_3_ii}, Lemma \ref{lemma:MB_Lemma_7}\ref{lemma:MB_Lemma_7_iv} and \ref{lemma:MB_Lemma_11}	
			\begin{align}
				\norm{I_f} \leq&
				\norm{\left(  \frac{\widehat{M}^{\mathcal{W}}}{mN^2}  \right)^{-1}} \
				\norm{\frac{\widehat{U} - UJ}{\sqrt{m}}}  \
				\norm{\frac{ \mathcal{E}_{(3)} \mathcal{E}_{(3)}' }{m N^2T}}  \
				\norm{\frac{U}{\sqrt{m}}} \
				\norm{\widehat{H}}\nonumber \\
				&= O_p(1) \ O_p \left(\max \left( \frac{1}{N \sqrt{T}}, \frac{1}{m N^{2-\gamma}} \right) \right) \
				O_p\left( \frac{1}{ \sqrt{mT} N^{2-\gamma/2}}  \right)  O_p(1) O_p(1)\nonumber\\
				&= O_p \left(\max \left( 
				\frac{1}{\sqrt{m} N^{3-\gamma/2} T}, 
				\frac{1}{m^{3/2} \sqrt{T} N^{2-\gamma}} 
				\right) \right).\nonumber
			\end{align}
			Therefore, 
			\begin{equation}\label{eq:MB_eq_126}
				\norm{I} = O_p \left( 
				\max \left(
				\frac{1}{\sqrt{mT} N^{2-\gamma/2}},
				\frac{1}{N^{3-\gamma/2} T}, 
				\frac{1}{m \sqrt{T} N^{2-\gamma}}
				\right)
				\right).
			\end{equation}
			Second, consider $II$. From Lemma \ref{lemma:MB_proposition_2}\ref{lemma:MB_proposition_2_ii}
			\begin{equation}
				\norm{II} \leq \frac{1}{m} \norm{\widehat{U}- U \widehat{H}}^2
				= O_p \left( 
				\max \left(
				\frac{1}{N^{4-\gamma} T},
				\frac{1}{m^2 N^{4-2\gamma}}, 
				\right)
				\right).\nonumber
			\end{equation}
			Therefore,
			\begin{equation}\label{eq:MB_eq_125}
				\norm{ \frac{   \widehat{U}' \left(U - \widehat{U} \widehat{H} \right) }{m}  }
				\leq \norm{I} + \norm{II}
				= O_p \left( \frac{1}{\xi}\right).
			\end{equation}
			Recall that $\xi = 
			\min \left(
			\sqrt{mT} N^{2-\gamma/2},
			N^{3-\gamma/2} T, 
			m \sqrt{T} N^{2-\gamma}, 
			m N^{2-\gamma}
			\right)
			$.
			Because of \eqref{eq:MB_eq_125}, using \eqref{eq:est_loadings}, 
			\begin{align}
				\frac{\widehat{U}' U \widehat{H}}{m}
				&= 	\frac{\widehat{M}^{\mathcal{W}}}{mN^2} 
				+ O_p \left(\frac{1 }{\xi}\right).\nonumber
			\end{align}
			Or, equivalently, 
			\begin{align}
				\left( \frac{\widehat{M}^{\mathcal{W}}}{mN^2} \right)^{-1} 
				\frac{\widehat{U}' {U}}{m}
				=
				\widehat{H}^{-1}
				+ O_p \left(\frac{1 }{\xi}\right).\nonumber
			\end{align}
			Using this  in \eqref{eq:HDEF_PD} and using Assumption \ref{as:MB_Assumption_9} we have
			\begin{equation}
				\widehat{H}' 
				=
				\left( \frac{\widehat{M}^{\mathcal{W}}}{mN^2} \right)^{-1} 
				\frac{\widehat{U}' {U}}{m}
				=
				\widehat{H}^{-1}
				+  O_p \left(\frac{1 }{\xi}\right).\nonumber
			\end{equation}
			Therefore, as $m, T, N \to \infty$, $\widehat{H}$ is an $r \times r$ orthogonal matrix, thus it has eigenvalues $\pm 1$.
			
			Moreover, because of \eqref{eq:MB_eq_126}
			\begin{align}
				\frac{\widehat{U}' U }{m}
				=
				\frac{(\widehat{U}' - U \widehat{H} +  U \widehat{H} )' U }{m}
				&=  \frac{\widehat{H}' \widehat{U}' U }{m} +  O_p \left(\frac{1 }{\xi}\right).\label{eq:above}
			\end{align}
			And from \eqref{eq:above} it follows that
			\begin{align}
				\left( \frac{\widehat{M}^{\mathcal{W}}}{mN^2} \right) 
				\widehat{H}'
				=
				\widehat{H}'
				\frac{U' U}{m}
				+  O_p \left(\frac{1 }{\xi}\right).\nonumber
			\end{align}
			So, because of (eq. above), as $m, N, T \to \infty$, the columns of $\widehat{H}$ are the eigenvectors of $\frac{U' U}{m}$ with eigenvalues $\frac{\widehat{M}^{\mathcal{W}}}{mN^2} $.
			The eigenvectors are normalized since $\widehat{H}$  is orthogonal,  as $m, N, T \to \infty$.
			Moreover, under Assumption \ref{as:MB_Assumption_9} \ref{as:MB_Assumption_9_i}, $\frac{U' U}{m}$ is diagonal so,  as $m, N, T \to \infty$, $\widehat{H}$ must be diagonal with  eigenvalues $\pm 1$. 
			This proves part (i).
			Part (ii) follows from the fact that $\widehat{H}$ is orthogonal, as $m, N, T\to \infty$. This completes the proof.
		\end{proof}

		\begin{lemma}\label{lemma:VhatCNAR}
			Under Assumptions  \ref{as:common_component}-\ref{as:X_nu_v}, \ref{as:X_nu_app}, and \ref{as:subgauss}, if $\sqrt{ T}/ (mN^{2-\gamma}) \to 0$, as $m,N,T\to\infty$ 
			$$
			\norm{\widehat V^{-1}-V^{-1}}= O_p\left(\max\left(\frac 1{\sqrt N},\sqrt{\frac{\log N}{T}}\right)\right).
			$$
		\end{lemma}
		
		\begin{proof}
			First, we have
			\begin{align}
				\widehat \nu_t := y_t-\widehat X_t\widehat{\theta}^{\text{\tiny OLS}} &= y_t -\left(\widehat X_t-X_t\bar J+X_t\bar J \right)\left(\widehat{\theta}^{\text{\tiny OLS}}-\bar J\theta+\bar J\theta \right)\nonumber\\
				&= y_t - X_t\bar J\bar J\theta -  X_t\bar J\left(\widehat{\theta}^{\text{\tiny OLS}}-\bar J\theta\right)-\left(\widehat X_t-X_t\bar J\right)\bar J\theta -\left(\widehat X_t-X_t\bar J\right)\left(\widehat{\theta}^{\text{\tiny OLS}}-\bar J\theta\right)\nonumber\\
				&= \nu_t -  X_t\bar J\left(\widehat{\theta}^{\text{\tiny OLS}}-\bar J\theta\right)-\left(\widehat X_t-X_t\bar J\right)\bar J\theta -\left(\widehat X_t-X_t\bar J\right)\left(\widehat{\theta}^{\text{\tiny OLS}}-\bar J\theta\right)\nonumber\\
				& = \nu_t +\delta_t+\eta_t+\zeta_t.\label{eq:nuhat_th4}
			\end{align}
			Moreover, 
			\begin{align}
				\norm{\frac{\widehat X_t-X_t\bar J}{\sqrt N}}&=\frac 1{\sqrt N}\norm{\left( \frac{1}{N} \left( y_{t-1} \otimes  I_N  \right)' \left( \widehat{\mathcal{F}}_{(3)t-1} - J  \mathcal{F}_{(3)t-1} \right)', 0_{N \times 2N}   \right)'}\nonumber\\
				& \le \norm{\frac{y_{t-1}}{\sqrt N}} \norm{\frac{ \widehat{\mathcal{F}}_{(3)t-1} - J  \mathcal{F}_{(3)t-1}}{N}}  = O_p\left(\max\left(\frac 1{N^{3-\gamma/2}T},\frac 1{\sqrt m N^{1-\gamma/2}}\right)\right),\label{eq:Xhat_th4}
			\end{align}
			because of Theorem \ref{theorem:CLT_factors}\ref{theorem:CLT_factors_i} and since 
			\[
			\mathbb E\left[\norm{\frac{y_{t-1}}{\sqrt N}}^2\right] =\frac 1N\sum_{i=1}^N \mathbb E[y_{it}^2] = O(1),
			\]
			because of stationarity (see Proposition \ref{prop:station}). Likewise $\norm{X_t}= O_p(\sqrt N)$. Obviously $\norm{\theta}=O(1)$ and $\norm {\bar J}=1$. Thus, because of Theorem \ref{th:CLT_FNAR} and \eqref{eq:Xhat_th4}:
			\begin{equation}\label{eq:deltaeta_th4}
				\norm{\frac {\delta_t}{\sqrt N}} = O_p\left(\frac 1{\sqrt T}\right), \quad \norm{\frac {\eta_t}{\sqrt N}} = O_p\left(\max\left(\frac 1{N^{3-\gamma/2}T},\frac 1{\sqrt m N^{1-\gamma/2}}\right)\right).
			\end{equation}
			From \eqref{eq:nuhat_th4} and \eqref{eq:deltaeta_th4},  it follows that,
			\[
			\norm{\frac{\widehat \nu_t-\nu_t}{\sqrt N}} = O_p\left(\max\left(\frac 1{\sqrt T},\frac 1{N^{3-\gamma/2}T},\frac 1{\sqrt m N^{1-\gamma/2}}\right)\right).
			\]
			Now, from \eqref{eq:nuhat_th4} and \eqref{eq:moderror}, we also have that
			\begin{equation}\label{eq:eq:deltaeta2_th4}
				\widehat \nu_t = \Lambda G_t +\epsilon_t + \delta_t+\eta_t+\zeta_t.
			\end{equation}
			Recall, the PC estimators $\widehat \Lambda$ and $\widehat G_t$ or $\widetilde \Lambda$ and $\widetilde G_t$ defined in \citet{barigozzi2022} or \citet{bai2003inferential}, depending on the choice of the sample covariance matrix (see \eqref{eq:PCAerrors}) and for simplicity of notation write  $\widehat \Lambda$ and $\widehat G_t$ to indicate both sets of estimators. Then, since 
			\[
			\norm{\eta_t} = o_p(\norm{\delta_t})\;\text{ and }\; \norm{\zeta_t} = o_p(\norm{\delta_t}),
			\]
			because  $\sqrt{ T}/ (mN^{2-\gamma}) \to 0$, from \citet[Proposition 3]{barigozzi2022} or \citet[Proposition 1]{bai2020simpler}, we have
			\begin{align}\label{eq:Lhat_MB_V}
				\norm{\frac{\widehat\Lambda-\Lambda \mathfrak J}{\sqrt N}} = O_p\left(\max\left(\frac1 {N},\frac 1{\sqrt T}\right)\right),
			\end{align}
			and for any given $i=1,\ldots, N$ (see \citet[Proposition 4]{barigozzi2022} or \citet[Theorem 2]{bai2003inferential})
			\begin{align}\label{eq:Lhati_MB_V}
				\norm{\widehat\Lambda_{i\cdot}-\Lambda_{i\cdot} \mathfrak J} = O_p\left(\max\left(\frac1 {N},\frac 1{\sqrt T}\right)\right),
			\end{align}
			where $\mathfrak J$ is a $q\times q$ diagonal matrix with diagonal entries $\pm 1$.
			Moreover,   from \citet[Proposition 6]{barigozzi2022} or \citet[Theorem 1]{bai2003inferential},  for any given $t=1,\ldots, T$, we have
			\begin{align}\label{eq:Ghat_MB_V}
				\norm{\widehat G_{t}- \mathfrak JG_{t}} = O_p\left(\max\left(\frac1 {\sqrt N},\frac 1{\sqrt T}\right)\right).
			\end{align}
			Notice the term $1/\sqrt T$, coming from $\delta_t$, which is slower than the usual $1/T$ due to estimation of $\nu_t$. By steps analogous to \citet[Lemma A.1]{baiandng2006}, from \eqref{eq:Lhat_MB_V} and \eqref{eq:Ghat_MB_V}, we get
			\begin{align}\label{eq:Ghat2_MB_V}
				\frac 1T\sum_{t=1}^T \norm{\widehat G_{t}- \mathfrak JG_{t}}^2 = O_p\left(\max\left(\frac1 { N},\frac 1{ T}\right)\right).
			\end{align}
			And, from \eqref{eq:Lhati_MB_V} and \citet[Corollary B.13]{chenetal2020cnar},
			\begin{align}\label{eq:Lhatmax_MB_V}
				\max_{i=1,\ldots, N}\norm{\widehat\Lambda_{i\cdot}-\Lambda_{i\cdot} \mathfrak J} = O_p\left(\max\left(\frac1 {\sqrt N},\sqrt{\frac {\log N}{T}}\right)\right),
			\end{align}
			where the $\sqrt{\log N}$ terms is due to Assumption \ref{as:subgauss} of sub-Gaussian tails and it appears when taking the max over $i=1,\ldots, N$ in \eqref{eq:Lhati_MB_V} and applying the Bonferroni inequality.
			
			Let, $\widehat{\epsilon}_{it}= \widehat{\nu}_{it}- \widehat\Lambda_{i\cdot}\widehat G_t$ and recall that $\epsilon_{it}=\nu_{it}-\Lambda_{i\cdot}G_t$. Hence, from \eqref{eq:eq:deltaeta2_th4}
			\begin{equation}\label{eq:epshat}
				\abs{\widehat{\epsilon}_{it}-\epsilon_{it}}= \abs{\Lambda_{i\cdot}G_t-\widehat\Lambda_{i\cdot}\widehat G_t+\delta_{it}+\eta_{it}+\zeta_{it}}.
			\end{equation}
			From \eqref{eq:epshat} it follows that (see also \citealp[Corollary B.14]{chenetal2020cnar}):
			\begin{align}
				\max_{i=1,\ldots, N}\frac 1T\sum_{t=1}^T \abs{\widehat{\epsilon}_{it}-\epsilon_{it}}^2\le&\, \max_{i=1,\ldots, N}\frac 2T\sum_{t=1}^T\abs{\Lambda_{i\cdot}G_t-\widehat\Lambda_{i\cdot}\widehat G_t}^2\nonumber\\
				&+\max_{i=1,\ldots, N}\frac 2T\sum_{t=1}^T\abs{\delta_{it}}^2+\max_{i=1,\ldots, N}\frac 2T\sum_{t=1}^T\abs{\eta_{it}}^2+\max_{i=1,\ldots, N}\frac 2T\sum_{t=1}^T\abs{\zeta_{it}}^2\nonumber\\
				&\le 8 \max_{i=1,\ldots, N}\norm{\Lambda_{i\cdot}\mathfrak J} \frac 1T\sum_{t=1}^T \norm{\widehat G_{t}- \mathfrak JG_{t}}^2+ 8 \max_{i=1,\ldots, N}\norm{\widehat\Lambda_{i\cdot}-\Lambda_{i\cdot} \mathfrak J} \frac 1T\sum_{t=1}^T \norm{\mathfrak J G_t}^2\nonumber\\
				&+\max_{i=1,\ldots, N}\frac 2T\sum_{t=1}^T\abs{\delta_{it}}^2+\max_{i=1,\ldots, N}\frac 2T\sum_{t=1}^T\abs{\eta_{it}}^2+\max_{i=1,\ldots, N}\frac 2T\sum_{t=1}^T\abs{\zeta_{it}}^2\nonumber \\
				=&\,O_p\left(\max\left(\frac1 { N},\frac {\log N}{T}\right)\right)+O_p\left(\max\left(\frac 1{ T},\frac {\log N}{N^{6-\gamma}T^2},\frac {\log N}{ m^2 N^{4-\gamma}},\frac{\log N}{mN^{2-\gamma}T}\right)\right)\nonumber\\
				=&\, O_p\left(\max\left(\frac1 { N},\frac {\log N}{T}\right)\right),\label{eq:coveps_th4}
			\end{align}
			because we assumed $\sqrt{ T}/ (mN^{2-\gamma}) \to 0$.
			Indeed, the first term  in the second last line of \eqref{eq:coveps_th4} is due to  \eqref{eq:Ghat2_MB_V} and \eqref{eq:Lhatmax_MB_V}. As for the second term in the second last line of \eqref{eq:coveps_th4}, we have		
			\[
			\max_{i=1,\ldots, N}\frac 2T\sum_{t=1}^T\abs{\delta_{it}}^2\le \max_{i=1,\ldots, N}\frac 2T\sum_{t=1}^T \norm{X_{i\cdot t}}^2 \norm{\bar J\left(\widehat{\theta}^{\text{\tiny OLS}}-\bar J\theta\right)}^2 = O_p\left(\frac 1T\right) O_p\left(1+\sqrt{\frac{\log N}{T}}\right),
			\]
			by Theorem \ref{th:CLT_FNAR} and since, by Assumption \ref{as:subgauss}\ref{as:subgauss_i} and Bonferroni inequality, for any $s>0$ and all $j=1,\ldots, r+2$,
			\[
			\text{P}\left(\abs{\max_{i=1,\ldots, N}\frac 2T\sum_{t=1}^T X_{ij t}^2- \mathbb E[X_{ijt}^2]}>s \right)\le 2N \exp(-cTs^2),
			\]
			for some finite $c$ independent of $j$, $N$, and $T$. Similarly, by Assumptions \ref{as:X_nu_v} and \ref{as:subgauss}\ref{as:subgauss_ii} we have
			\[
			\max_{i=1,\ldots, N}\frac 2T\sum_{t=1}^T\abs{\eta_{it}}^2\le \max_{i=1,\ldots, N}\frac 2T\sum_{t=1}^T \norm{\widehat X_{i\cdot t}-X_{i\cdot t}}^2 \norm{\theta}^2 = O_p\left(\max\left(\frac {\log N}{N^{6-\gamma}T^2},\frac {\log N}{ m^2 N^{4-\gamma}},\frac{\log N}{mN^{2-\gamma}T}\right)\right),
			\]
			since the error in estimating $X_t$ is function of 
			\begin{equation}
				\frac 1{mN^2T}\sum_{t=1}^T\sum_{j=1}^m u_j \mathcal E_{(3)tji} \left(y_{t-1}\otimes\left\{y_{t-1}\otimes I_N\right\}' \frac{\mathcal E_{(3)t-1}'U}{mN}\right)_{i\cdot},
			\end{equation}
			see \eqref{eq:b2_prop1} and \eqref{eq:b2lead_prop1} in the proof of Proposition \ref{lemma:part2_lemma_1}.
			
			Moreover, (see also \citealp[Corollary B.14]{chenetal2020cnar}):
			\begin{align}
				\max_{i=1,\ldots, N}\frac 1T\sum_{t=1}^T \abs{\widehat{\epsilon}_{it}-\epsilon_{it}}\abs{\epsilon_{it}}\le&\, \max_{i=1,\ldots, N}\left(\frac 1T\sum_{t=1}^T \abs{\widehat{\epsilon}_{it}-\epsilon_{it}}^2\left\{\frac 1T\sum_{t=1}^T\epsilon_{it}^2-\mathbb E[\epsilon_{it}^2]\right\}\right)^{1/2}\nonumber\\
				&+ \max_{i=1,\ldots, N}\left(\frac 1T\sum_{t=1}^T \abs{\widehat{\epsilon}_{it}-\epsilon_{it}}^2\mathbb E[\epsilon_{it}^2]\right)^{1/2}\nonumber\\
				=&\, O_p\left(\max\left(\frac1 { \sqrt N},\sqrt{\frac {\log N}{T}}\right)\right)O_p\left(\frac 1{\sqrt T}\right)
				+O_p\left(\max\left(\frac1 { \sqrt N},\sqrt{\frac {\log N}{T}}\right)\right),\label{eq:coveps2_th4}
			\end{align}
			by Assumptions \ref{as:errors_nu}\ref{as:errors_nu_idio_i}, \ref{as:errors_nu}\ref{as:errors_nu_idio_iv}, and \ref{as:subgauss}\ref{as:subgauss_iii} and following the same reasoning as in \eqref{eq:coveps_th4}.
			
			Therefore, from \eqref{eq:coveps_th4} and \eqref{eq:coveps2_th4}, and Assumption \ref{as:errors_nu}\ref{as:errors_nu_idio_iv},
			\begin{align}
				\norm{\widehat S-S}&\le \max_{i=1,\ldots, N}\frac 1T\sum_{t=1}^T \abs{\widehat{\epsilon}_{it}-\epsilon_{it}}^2+2\max_{i=1,\ldots, N}\frac 1T\sum_{t=1}^T \abs{\widehat{\epsilon}_{it}-\epsilon_{it}}\abs{\epsilon_{it}}+\max_{i=1,\ldots, N}\abs{\frac 1T\sum_{t=1}^T \left\{\epsilon_{it}^2-\mathbb E[\epsilon_{it}^2]\right\}}\nonumber\\
				&= O_p\left(\max\left(\frac1 { N},\frac {\log N}{T}\right)\right)+O_p\left(\max\left(\frac1 { \sqrt N},\sqrt{\frac {\log N}{T}}\right)\right)+O_p\left(\frac 1{\sqrt T}\right)\nonumber\\
				&= O_p\left(\max\left(\frac1 { \sqrt N},\sqrt{\frac {\log N}{T}}\right)\right).\label{eq:Shat_th4}
			\end{align}
			And from \eqref{eq:Shat_th4} and Assumption \ref{as:errors_nu}\ref{as:errors_nu_idio_i}
			\begin{align}\label{eq:Shatinv_th4}
				\norm{\widehat S^{-1}-S^{-1}}&\le \norm{S^{-1}} \norm{\widehat S-S}\norm {\widehat S^{-1}}\nonumber\\ 
				&= \frac 1{\min_{i=1,\ldots, N} \mathbb E[\epsilon_{it}^2]} O_p\left(\max\left(\frac1 { \sqrt N},\sqrt{\frac {\log N}{T}}\right)\right)
				\frac 1{\min_{i=1,\ldots, N} \frac 1T\sum_{t=1}^T\epsilon_{it}^2}\nonumber\\
				&\le \frac 1{\underline M_\epsilon}O_p\left(\max\left(\frac1 { \sqrt N},\sqrt{\frac {\log N}{T}}\right)\right) \frac 1{\underline M_\epsilon+O_p\left(\max\left(\frac1 { \sqrt N},\sqrt{\frac {\log N}{T}}\right)\right)}\nonumber\\
				& = O_p\left(\max\left(\frac1 { \sqrt N},\sqrt{\frac {\log N}{T}}\right)\right).
			\end{align}
			Thus, from \eqref{eq:Lhat_MB_V} and \eqref{eq:Shatinv_th4}
			\begin{align}\label{eq:LSL_th4}
				\norm{\frac{\widehat \Lambda'\widehat S^{-1}\widehat \Lambda-\Lambda'S^{-1}\Lambda}{N}} = O_p\left(\max\left(\frac1 { \sqrt N},\sqrt{\frac {\log N}{T}}\right)\right).
			\end{align}
			Furthermore, recalling the definition of $\widehat V^{-1}$ in \eqref{eq:invVsherman} 
			\begin{equation}
				\widehat V^{-1}=\widehat S^{-1}-\widehat S^{-1} \frac{\widehat\Lambda}{\sqrt N}\left(\frac {I_N}N+ \frac{\widehat\Lambda'\widehat S^{-1}\widehat \Lambda}N\right)^{-1} \frac{\widehat\Lambda'}{\sqrt N}\widehat  S^{-1},\nonumber
			\end{equation}
			and since we can always write
			\begin{equation}
				V^{-1}= S^{-1}- S^{-1} \frac{\Lambda}{\sqrt N}\left(\frac {I_N}N+ \frac{\Lambda' S^{-1} \Lambda}N\right)^{-1} \frac{\Lambda'}{\sqrt N}  S^{-1},\nonumber
			\end{equation}
			we have
			\begin{align}
				\norm{\widehat V^{-1}-V^{-1}} = O_p\left(\max\left(\frac1 { \sqrt N},\sqrt{\frac {\log N}{T}}\right)\right),
			\end{align}
			because of \eqref{eq:Lhat_MB_V}, \eqref{eq:Shatinv_th4}, and \eqref{eq:LSL_th4}. This completes the proof.
		\end{proof}
		\begin{lemma}\label{lem:mancavasololui}
			Under Assumptions \ref{as:common_component}-\ref{as:X_nu_v}, as $N,T\to\infty$, if $\sqrt N/T\to 0$, 
			\begin{align}
				\frac 1{\sqrt{NT}}\sum_{t=1}^T \widehat W_t^{*'}\epsilon_t=&\,\frac 1{\sqrt{NT}}\sum_{t=1}^T W_t'\epsilon_t+\sqrt{\frac NT}\xi^*_{NT}\nonumber\\
				&+\sqrt NO_p(\Vert\widehat{\theta}^*-\bar J\theta \Vert^2)+O_p(\Vert\widehat{\theta}^*-\bar J\theta \Vert)+\sqrt NO_p\left(\max\left(\frac 1N, \frac 1T\right)\right),\nonumber
			\end{align}
			with
			\[
			\xi^*_{NT}:=-\frac 1{NT}\sum_{t=1}^T\bar J\left(X_t-\frac 1T\sum_{s=1}^T (G_t'G_s)X_s\right)'\Lambda\sum_{u=1}^T G_u\left(\frac 1N\sum_{i=1}^N\epsilon_{iu}\epsilon_{it} \right).
			\]
		\end{lemma}
		
		\begin{proof} First note that $M_{{\Lambda}}-M_{\widehat{\Lambda}^*}=P_{{\Lambda}}-P_{\widehat{\Lambda}^*}$, where 
			$P_{{\Lambda}}={\Lambda}({\Lambda}'{\Lambda})^{-1}{\Lambda}'$
			and
			$P_{\widehat{\Lambda}^*}=\widehat{\Lambda}^{*}(\widehat{\Lambda}^{*'}\widehat{\Lambda}^*)^{-1}\widehat{\Lambda}^{*'}$, then
			\begin{align}
				\frac 1{\sqrt {NT}}\sum_{t=1}^TX_t'(M_{{\Lambda}}-M_{\widehat{\Lambda}^*})\epsilon_t=&\,\frac 1{\sqrt {NT}}\sum_{t=1}^TX_t'\widehat{\Lambda}^{*}(\widehat{\Lambda}^{*'}\widehat{\Lambda}^*)^{-1}\widehat{\Lambda}^{*'}\epsilon_t-\frac 1{\sqrt {NT}}\sum_{t=1}^TX_t'{\Lambda}({\Lambda}'{\Lambda})^{-1}{\Lambda}'\epsilon_t\nonumber\\
				=&\, \frac 1{\sqrt {NT}}\sum_{t=1}^TX_t'\left(\widehat{\Lambda}^{*}-\Lambda J\right)\left(J\Lambda'\Lambda J\right)^{-1}J\Lambda'\epsilon_t\nonumber\\
				&+ \frac 1{\sqrt {NT}}\sum_{t=1}^TX_t'\left(\widehat{\Lambda}^{*}-\Lambda J\right)\left(J\Lambda'\Lambda J\right)^{-1}
				\left(\widehat{\Lambda}^{*}-\Lambda J\right)'\epsilon_t\nonumber\\
				&+\frac 1{\sqrt {NT}}\sum_{t=1}^TX_t'\Lambda J\left(J\Lambda'\Lambda J\right)^{-1}
				\left(\widehat{\Lambda}^{*}-\Lambda J\right)'\epsilon_t\nonumber\\
				&+ \frac 1{\sqrt {NT}}\sum_{t=1}^TX_t'\left(\widehat{\Lambda}^{*}-\Lambda J\right)\left(\widehat{\Lambda}^{*'}\widehat{\Lambda}^*-J\Lambda'\Lambda J\right)^{-1}
				J\Lambda'\epsilon_t\nonumber\\
				&+\frac 1{\sqrt {NT}}\sum_{t=1}^TX_t'\left(\widehat{\Lambda}^{*}-\Lambda J\right)\left(\widehat{\Lambda}^{*'}\widehat{\Lambda}^*-J\Lambda'\Lambda J\right)^{-1}
				\left(\widehat{\Lambda}^{*}-\Lambda J\right)'\epsilon_t\nonumber\\
				&+ \frac 1{\sqrt {NT}}\sum_{t=1}^TX_t'\Lambda J\left(\widehat{\Lambda}^{*'}\widehat{\Lambda}^*-J\Lambda'\Lambda J\right)^{-1}
				\left(\widehat{\Lambda}^{*}-\Lambda J\right)'\epsilon_t\nonumber\\
				&+ \frac 1{\sqrt {NT}}\sum_{t=1}^TX_t'\Lambda J\left(\widehat{\Lambda}^{*'}\widehat{\Lambda}^*-J\Lambda'\Lambda J\right)^{-1}
				J\Lambda'\epsilon_t\nonumber\\
				=:&\, a+b+c+d+e+f+g.\label{alfabeto}
			\end{align}		
			Then, since $\left(\widehat{\lambda}_i^{*'}-\lambda_i' J\right)\left({J\Lambda'\Lambda J}\right)^{-1}J\lambda_j$ is a scalar,
			\begin{align}
				a&=\frac 1{\sqrt {NT}}\sum_{t=1}^T\sum_{i=1}^N X_{it}\left(\widehat{\lambda}_i^{*'}-\lambda_i' J\right)\left(\frac{J\Lambda'\Lambda J}N\right)^{-1}J\frac 1N\sum_{j=1}^N \lambda_j\epsilon_{jt}\nonumber\\
				&= \frac 1N \sum_{i=1}^N \left(\widehat{\lambda}_i^{*'}-\lambda_i' J\right)\left(\frac{J\Lambda'\Lambda J}N\right)^{-1}J\left(\frac 1{\sqrt {NT}}\sum_{t=1}^T\sum_{j=1}^N \lambda_jX_{it}\epsilon_{jt}\right),\nonumber
			\end{align}
			and, therefore,
			\begin{align}
				\Vert a\Vert \le&\, \left[\frac 1N\sum_{i=1}^N \left\Vert \widehat{\lambda}_i^{*'}-\lambda_i' J \right\Vert^2 \right]^{1/2}\, \left\Vert \left(\frac{J\Lambda'\Lambda J}N\right)^{-1}\right\Vert\,\left\Vert J\right\Vert\left[\frac 1N\sum_{i=1}^N \left\Vert
				\left(\frac 1{\sqrt {NT}}\sum_{t=1}^T\sum_{j=1}^N \lambda_jX_{it}\epsilon_{jt}
				\right)
				\right\Vert^2
				\right]^{1/2}\nonumber\\
				&=\left[O_p\left(\Vert\widehat{\theta}^*-\bar J\theta\Vert\right)+ O_p\left(\max\left(\frac 1{\sqrt N},\frac 1{\sqrt T}\right)\right)\right] O_p(1),
			\end{align}
			because of \eqref{A1Bai09}, Assumption \ref{as:errors_nu}\ref{as:errors_nu_common_i},  and \citet[Lemma 3]{barigozzi2022}.
			Similarly,
			\begin{align}
				b&= \frac 1{\sqrt {NT}}\sum_{t=1}^T\sum_{i=1}^N X_{it}\left(\widehat{\lambda}_i^{*'}-\lambda_i' J\right)\left(\frac{J\Lambda'\Lambda J}N\right)^{-1}\frac 1N
				\sum_{j=1}^N\left(\widehat{\lambda}_j^{*}-J\lambda_j\right)\epsilon_{jt}\nonumber\\
				&=\sqrt N\frac 1{N^2}\sum_{i=1}^N\sum_{j=1}^N
				\left(\widehat{\lambda}_i^{*'}-\lambda_i' J\right)\left(\frac{J\Lambda'\Lambda J}N\right)^{-1}
				\left(\widehat{\lambda}_j^{*}-J\lambda_j\right)\left(\frac 1{\sqrt {T}} \sum_{t=1}^T X_{it}\epsilon_{jt}\right).\nonumber
			\end{align}
			Therefore,
			\begin{align}
				\Vert b\Vert \le&\, \sqrt N \left(\frac 1N\sum_{i=1}^N \left\Vert\widehat{\lambda}_i^{*'}-\lambda_i' J\right\Vert^2\right) \left\Vert \left(\frac{J\Lambda'\Lambda J}N\right)^{-1}\right\Vert
				\left(\frac 1{N^2}\sum_{i=1}^N\sum_{j=1}^N\left\Vert \frac 1{\sqrt {T}} \sum_{t=1}^T X_{it}\epsilon_{jt}\right\Vert^2\right)^{1/2}\nonumber\\
				&=\sqrt N\left[O_p\left(\Vert\widehat{\theta}^*-\bar J\theta\Vert^2\right)+ O_p\left(\max\left(\frac 1{ N},\frac 1{ T}\right)\right)\right]O_p(1),
			\end{align}
			because of \eqref{A1Bai09}, Assumption \ref{as:errors_nu}\ref{as:errors_nu_common_i},  and \citet[Lemma 2]{barigozzi2022}.
			Then, consider
			\begin{align}
				c=&\,\frac 1{\sqrt {NT}}\sum_{t=1}^T X_{t}'\Lambda J\left(\frac{J\Lambda'\Lambda J}N\right)^{-1}
				\frac 1N	\left(\widehat{\Lambda}^{*}-\Lambda J\right)'\epsilon_{t}\nonumber\\
				=&\, \frac{\sqrt {NT}} T\left\{\frac 1{T}\sum_{t=1}^T\sum_{s=1}^T \frac{X_t'\Lambda J}{N}\left(\frac{J\Lambda'\Lambda J}{N}\right)^{-1}G_s \left(\frac 1{N} \sum_{i=1}^N \epsilon_{it}\epsilon_{is}\right)\right\}\nonumber\\
				& + O_p\left(\Vert{\widehat{\theta}^*-\bar J\theta}\Vert\right)+\sqrt{\frac{N}{T}}O_p\left(\Vert{\widehat{\theta}^*-\bar J\theta}\Vert\right)+{\sqrt N}O_p\left(\max\left(\frac 1N,\frac 1T\right)\right)\nonumber\\
				=&\, \sqrt {\frac NT}\psi_{NT} + O_p\left(\Vert{\widehat{\theta}^*-\bar J\theta}\Vert\right)+\sqrt{\frac{N}{T}}O_p\left(\Vert{\widehat{\theta}^*-\bar J\theta}\Vert\right)+{\sqrt N}O_p\left(\max\left(\frac 1N,\frac 1T\right)\right).\nonumber
			\end{align}
			And it easily seen that $\Vert \psi_{NT}\Vert=O_p(1)$ so  $\sqrt{\frac NT}\psi_{NT} $ dominates $\sqrt{\frac{N}{T}}O_p\left(\Vert{\widehat{\theta}^*-\bar J\theta}\Vert\right)$, since $\Vert{\widehat{\theta}^*-\bar J\theta}\Vert=o_p(1)$ as shown in \eqref{bello} in the proof of Lemma \ref{118}\ref{118_i}. Finally, $d$, $e$, and $f$ are dominated by $a$, $b$, and $c$, respectively, and $g$ behaves as $a$. 
			
			Therefore, from \eqref{alfabeto} and Lemma \ref{108}\ref{108_iv} we have
			\begin{align}
				\frac 1{\sqrt {NT}}&\,\sum_{t=1}^TX_t'(M_{{\Lambda}}-M_{\widehat{\Lambda}^*})\epsilon_t=\frac{\sqrt {NT}} T\psi_{NT}\label{eq:bari}\\
				&+O_p\left(\Vert\widehat{\theta}^*-\bar J\theta\Vert\right)+ O_p\left(\max\left(\frac 1{\sqrt N},\frac 1{\sqrt T}\right)\right)\nonumber\\
				&+\sqrt N\left[O_p\left(\Vert\widehat{\theta}^*-\bar J\theta\Vert^2\right)+ O_p\left(\max\left(\frac 1{ N},\frac 1{ T}\right)\right)\right]\nonumber\\
				&+ O_p\left(\Vert{\widehat{\theta}^*-\bar J\theta}\Vert\right)
				+{\sqrt N}O_p\left(\max\left(\frac 1N,\frac 1T\right)\right)\nonumber\\
				=&\, \frac{\sqrt {NT}} T\psi_{NT}+O_p\left(\Vert\widehat{\theta}^*-\bar J\theta\Vert\right)+\sqrt N O_p\left(\Vert\widehat{\theta}^*-\bar J\theta\Vert^2\right)+
				\sqrt NO_p\left(\max\left(\frac 1{ N},\frac 1{ T}\right)\right),\nonumber
			\end{align}
			where
			\[
			\psi_{NT} := \frac 1{T}\sum_{t=1}^T\sum_{s=1}^T \frac{X_t'\Lambda J}{N}\left(\frac{J\Lambda'\Lambda J}{N}\right)^{-1}G_s \left(\frac 1{N} \sum_{i=1}^N \epsilon_{it}\epsilon_{is}\right).
			\]
			
			Let now $V_t:=\frac 1T\sum_{t=1}^T(G_s'G_t) X_t$. Then, replacing $X_t$ with $V_t$ in \eqref{alfabeto} we have
			\begin{align}
				\frac 1{\sqrt {NT}}\sum_{t=1}^T&V_t'(M_{{\Lambda}}-M_{\widehat{\Lambda}^*})\epsilon_t\label{eq:lecce}\\
				=&\, \frac{\sqrt {NT}} T\psi_{NT}^*+O_p\left(\Vert\widehat{\theta}^*-\bar J\theta\Vert\right)+\sqrt N O_p\left(\Vert\widehat{\theta}^*-\bar J\theta\Vert^2\right)+
				\sqrt NO_p\left(\max\left(\frac 1{ N},\frac 1{ T}\right)\right),\nonumber
			\end{align}
			where
			\[
			\psi_{NT}^* :=- \frac 1{T}\sum_{t=1}^T\sum_{s=1}^T \frac{V_t'\Lambda J}{N}\left(\frac{J\Lambda'\Lambda J}{N}\right)^{-1}G_s \left(\frac 1{N} \sum_{i=1}^N \epsilon_{it}\epsilon_{is}\right).
			\]
			By combining \eqref{eq:bari} and \eqref{eq:lecce} we complete the proof.
		\end{proof}
		
		\begin{lemma}\label{118}
			Let $D(\widehat{\Lambda}^*):=\frac 1{NT} \sum_{t=1}^T \bar J\widehat{W}_{t}^{*'} \widehat{W}_{t}^{*}\bar J$ and 
			$D({\Lambda}):=\frac 1{NT} \sum_{t=1}^T \bar J{W}_{t}^{'} {W}_{t}\bar J$.
			Under Assumptions \ref{as:common_component}-\ref{as:X_nu_v}, as $N,T\to\infty$,	
			\begin{enumerate}[label=(\roman*)]
				\item $\norm{D(\widehat{\Lambda}^*)^{-1}-D({\Lambda})^{-1}}=o_p(1)$.
				\label{118_i}
				\item $\sqrt{\frac TN}\norm{D(\widehat{\Lambda}^*)^{-1}-D({\Lambda})^{-1}}=o_p(1)$ if $\sqrt T/N\to 0$.
				\label{118_ii}
				\item $\sqrt{\frac NT}\norm{D(\widehat{\Lambda}^*)^{-1}-D({\Lambda})^{-1}}=o_p(1)$ if $\sqrt N/T\to 0$.
				\label{118_iii}
			\end{enumerate}	
		\end{lemma}
		
		\begin{proof}
			For part \ref{118_i}, notice that
			\begin{align}
				D(\widehat{\Lambda}^*)-D({\Lambda})=&\, \frac 1{NT}\sum_{t=1}^TX_t'(M_{\widehat{\Lambda}^*}-M_{\Lambda})X_t\nonumber\\
				&-\frac 1N\left[\frac 1{T^2}\sum_{t=1}^T\sum_{s=1}^TX_t'(M_{\widehat{\Lambda}^*}-M_{\Lambda})X_s (G_t'G_s)\right].\nonumber
			\end{align}
			Then the proof follows from the fact that by \eqref{A1Bai09} and \citet[Proposition 5]{barigozzi2022}
			\begin{align}\label{proiettori}
				\norm{M_{\widehat{\Lambda}^*}-M_{\Lambda}}^2=2 \text{tr}\left(I_q-\frac{\widehat{\Lambda}^{*'} \Lambda(\Lambda'\Lambda)^{-1}\Lambda'\widehat{\Lambda}^{*'}}{N}\right) = O_p\left(\Vert{\widehat{\theta}^*-\bar J\theta}\Vert\right) + O_p\left(\max\left(\frac 1{N},\frac 1T\right)\right).
			\end{align}
			which follows from
			\begin{align}
				\frac 1N\Lambda'(\widehat{\Lambda}^*-\Lambda J)&=\frac{\Lambda'\widehat{\Lambda}^*}{N}-\frac{\Lambda'{\Lambda}J}{N}= O_p\left(\Vert{\widehat{\theta}^*-\bar J\theta}\Vert\right) + O_p\left(\max\left(\frac 1{N},\frac 1T\right)\right),\nonumber\\
				\frac 1N\widehat{\Lambda}^{*'}(\widehat{\Lambda}^*-\Lambda J)&=\frac{\widehat{\Lambda}^{*'}\widehat{\Lambda}^*}{N}-\frac{\widehat{\Lambda}^{*'}\widehat{\Lambda}^{*}}{N}= O_p\left(\Vert{\widehat{\theta}^*-\bar J\theta}\Vert\right) + O_p\left(\max\left(\frac 1{N},\frac 1T\right)\right).\nonumber
			\end{align}
			Therefore,
			\begin{align}
				\norm{D(\widehat{\Lambda}^*)-D({\Lambda})}=O_p\left(\Vert{\widehat{\theta}^*-\bar J\theta}\Vert\right) + O_p\left(\max\left(\frac 1{N},\frac 1T\right)\right).\label{figo}
			\end{align}
			This proves part \ref{118_i}.
			
			For part \ref{118_ii}, from \eqref{uffissima} and the second step \eqref{toronto} (which does not use this lemma), we have
			\begin{align}
				\sqrt{NT}\left(\widehat\theta^*-\bar J\theta\right)&=\left(D(\widehat{\Lambda}^*)\right)^{-1} \left\{\frac 1{\sqrt{NT}}\sum_{t=1}^T \bar J{W}_t^{'}\epsilon_t\right.\nonumber\\
				&-\sqrt{\frac TN}\left[\frac {1}{NT}\sum_{t=1}^T \bar JX_t'M_{\widehat{\Lambda}^*}\left(\frac 1T\sum_{s=1}^T
				\mathbb E[\epsilon_s\epsilon_s']\right)\widehat{\Lambda}^*\left(\frac{\Lambda'\widehat{\Lambda}^*}{N}\right)^{-1}G_t\right]\nonumber\\
				&\left.-\sqrt{\frac NT} \left[\frac 1{NT}\sum_{t=1}^T\bar J\left(X_t-\frac 1T\sum_{s=1}^T (G_t'G_s)X_s\right)'\Lambda\left(\frac{\Lambda'\Lambda}{N}\right)^{-1}\sum_{u=1}^T G_u\left(\frac 1N\sum_{i=1}^N\epsilon_{iu}\epsilon_{it} \right)
				\right]
				\right\}+o_p(1)\nonumber\\
				&= \left(D(\widehat{\Lambda}^*)\right)^{-1}\frac 1{\sqrt{NT}}\sum_{t=1}^T \bar J{W}_t^{'}\epsilon_t+\sqrt{\frac TN} \zeta_{NT} +\sqrt {\frac NT} \xi_{NT}+o_p(1),\label{cor1Bai}
			\end{align}
			where 
			\begin{align}
				\zeta_{NT}:=-\left(D(\widehat{\Lambda}^*)\right)^{-1}\frac {1}{NT}\sum_{t=1}^T \bar JX_t'M_{\widehat{\Lambda}^*}\left(\frac 1T\sum_{s=1}^T
				\mathbb E[\epsilon_s\epsilon_s']\right)\widehat{\Lambda}^*\left(\frac{\Lambda'\widehat{\Lambda}^*}{N}\right)^{-1}G_t,\label{sos}
			\end{align}
			and
			\begin{align}
				\xi_{NT}&:=-\left(D(\widehat{\Lambda}^*)\right)^{-1}
				\frac 1{NT}\sum_{t=1}^T\bar J\left(X_t-\frac 1T\sum_{s=1}^T (G_t'G_s)X_s\right)'\Lambda\left(\frac{\Lambda'\Lambda}{N}\right)^{-1}\sum_{u=1}^T G_u\left(\frac 1N\sum_{i=1}^N\epsilon_{iu}\epsilon_{it}\right)\label{aiuto}.
			\end{align}
			It is easily seen that $\Vert \xi_{NT}\Vert=O_p(1)$ and $\Vert \zeta_{NT}\Vert=O_p(1)$. Moreover, by Assumptions \ref{as:Z}\ref{as:Z_iii} and \ref{as:Z}\ref{as:Z_iv}, we also have $\left(D(\widehat{\Lambda}^*)\right)^{-1}\frac 1{\sqrt{NT}}\sum_{t=1}^T \bar J{W}_t^{'}\epsilon_t=O_p(1)$.
			Therefore, from \eqref{cor1Bai}
			\[
			\sqrt{NT}\left(\widehat\theta^*-\bar J\theta\right)=O_p\left(\sqrt{\frac TN}\right)+O_p\left(\sqrt{\frac NT}\right).
			\]
			Hence,
			\begin{align}
				\left(\widehat\theta^*-\bar J\theta\right)=O_p\left(\max\left(\frac 1N,\frac 1T\right)\right).\label{bello}
			\end{align}
			By \eqref{figo} and \eqref{bello} we have
			\[
			\norm{D(\widehat{\Lambda}^*)-D({\Lambda})} = O_p\left(\max\left(\frac 1{\sqrt N},\frac 1{\sqrt T}\right)\right).
			\]
			The proof of parts \ref{118_ii} and \ref{118_iii} follows immediately. This completes the proof.
		\end{proof}
		
		\begin{lemma}\label{rimini}
			Under Assumptions \ref{as:common_component}-\ref{as:X_nu_v}, as $N,T\to\infty$,
			\begin{enumerate}[label=(\roman*)]
				\item if $\sqrt T/N\to 0$,
				\[
				\sqrt {\frac TN}\norm{\zeta_{NT}-\zeta_{NT}^0}= o_p(1),
				\]
				where 
				\[
				\zeta_{NT}^0:=- \left(D({\Lambda})\right)^{-1}\frac {1}{NT}\sum_{t=1}^T \bar JX_t'M_{{\Lambda}}\left(\frac 1T\sum_{s=1}^T
				\mathbb E[\epsilon_s\epsilon_s']\right){\Lambda}\left(\frac{\Lambda'{\Lambda}}{N}\right)^{-1}G_t,
				\]
				and
				$\zeta_{NT}$ is defined in \eqref{sos} in the proof of Lemma \ref{118}.
				\label{rimini_i}
				\item if $\sqrt N/T\to 0$,
				\[
				\sqrt {\frac NT}\norm{\xi_{NT}-\xi_{NT}^0}= o_p(1),
				\]
				where  
				\[
				\xi_{NT}^0 := - \left(D({\Lambda})\right)^{-1}\frac 1{NT}\sum_{t=1}^T\bar J\left(X_t-\frac 1T\sum_{s=1}^T (G_t'G_s)X_s\right)'\Lambda\left(\frac{\Lambda'\Lambda}{N}\right)^{-1}\sum_{u=1}^T G_u\left(\frac 1N\sum_{i=1}^N\mathbb E[\epsilon_{iu}\epsilon_{it}]\right),
				\]
				and $\xi_{NT}$ is defined in \eqref{aiuto} in the proof of Lemma \ref{118}.
				\label{rimini_ii}
			\end{enumerate}
		\end{lemma}
		
		\begin{proof}
			For part \ref{rimini_i}, we have 
			\begin{align}
				\left\Vert \frac {1}{NT}\sum_{t=1}^T \bar JX_t'M_{\widehat{\Lambda}^*}\left(\frac 1T\sum_{s=1}^T
				\mathbb E[\epsilon_s\epsilon_s']\right)\widehat{\Lambda}^*\left(\frac{\Lambda'\widehat{\Lambda}^*}{N}\right)^{-1}G_t\right.&-\left.
				\frac {1}{NT}\sum_{t=1}^T \bar JX_t'M_{{\Lambda}}\left(\frac 1T\sum_{s=1}^T
				\mathbb E[\epsilon_s\epsilon_s']\right){\Lambda}\left(\frac{\Lambda'{\Lambda}}{N}\right)^{-1}G_t\right\Vert\nonumber\\
				&= O_p\left(\max\left(\frac 1{\sqrt N},\frac 1{\sqrt T}\right)\right),\nonumber
			\end{align}		
			by \eqref{proiettori} and \eqref{bello} in the proof of Lemma \ref{118}. Thus, since $\sqrt {\frac TN}O_p\left(\max\left(\frac 1{\sqrt N},\frac 1{\sqrt T}\right)\right)= o_p(1)$ if $\sqrt T/N\to 0$, by using  Lemma \ref{118}\ref{118_iii}, we prove part \ref{rimini_i}.
			
			For part \ref{rimini_ii}, by Lemma \ref{118}\ref{118_iii}, if $\sqrt N/T\to 0$, we have
			\begin{align}
				\sqrt {\frac NT}\left({\xi_{NT}-\xi_{NT}^0}\right) = \sqrt {\frac NT}\left(\left(D(\widehat{\Lambda}^*)\right)^{-1}\xi_{NT}^*-\xi_{NT}^0\right)= \sqrt {\frac NT}\left(\left(D({\Lambda})\right)^{-1}\xi_{NT}^*-\xi_{NT}^0\right)+o_p(1),\label{riccione}
			\end{align}
			where $\xi_{NT}^*$ is defined in Lemma \ref{108}, and since $\Vert \xi_{NT}^*\Vert=O_p(1)$. Moreover, let
			\[
			A_{tu}:= G_u'\otimes \left[\left(X_t-\frac 1T\sum_{s=1}^T (G_t'G_s)X_s\right)'\frac{\Lambda}N\right], 
			\]
			which is such that $\norm{A_{tu}}=O_p(1)$. Then, since $\norm{\left(\frac{\Lambda'\Lambda}{N}\right)^{-1}}=O_p(1)$ by Assumption \ref{as:errors_nu}\ref{as:errors_nu_common_i}, we have
			\begin{align}
				\left(D({\Lambda})\right)^{-1}\xi_{NT}^*-\xi_{NT}^0&=-\left(D({\Lambda})\right)^{-1}\frac 1T \sum_{t=1}^T\sum_{u=1}^T A_{tu}\left\{\frac 1N\sum_{i=1}^N \epsilon_{iu}\epsilon_{it}-\mathbb E[\epsilon_{iu}\epsilon_{it}]\right\}\text{vec}\left(\left(\frac{\Lambda'\Lambda}{N}\right)^{-1}\right)\nonumber\\
				&=O_p\left(\frac 1{\sqrt N}\right).\label{riccione2}
			\end{align}		
			by Assumption \ref{as:errors_nu}\ref{as:errors_nu_idio_iv}. Thus, by substituing \eqref{riccione2} into \eqref{riccione}, we have
			$\sqrt {\frac NT}\left({\xi_{NT}-\xi_{NT}^0}\right)= O_p\left(\frac 1{\sqrt T}\right)$. This completes the proof.
		\end{proof}
		
		\begin{lemma}\label{108}
			Under Assumptions \ref{as:common_component}-\ref{as:X_nu_v}, as $N,T\to\infty$,
			\begin{enumerate}[label=(\roman*)]
				\item $\norm{\frac 1N (\widehat{\Lambda}^*-\Lambda J)\epsilon_t}= \frac 1{\sqrt N}O_p(\Vert{\widehat{\theta}^*-\bar J\theta}\Vert)+O_p\left(\max\left(\frac 1N,\frac 1T\right)\right)$.
				\label{108_i}
				\item $\norm{\frac 1{N\sqrt T}\sum_{t=1}^T (\widehat{\Lambda}^*-\Lambda J)\epsilon_t}= \frac 1{\sqrt N}O_p(\Vert{\widehat{\theta}^*-\bar J\theta}\Vert)
				+\frac 1{\sqrt T}O_p(\Vert{\widehat{\theta}^*-\bar J\theta}\Vert)+O_p\left(\frac 1{\sqrt T}\right)
				+O_p\left(\max\left(\frac 1N,\frac 1T\right)\right)$.
				\label{108_ii}
				\item $\norm{\frac 1{N T}\sum_{t=1}^TG_t' (\widehat{\Lambda}^*-\Lambda J)\epsilon_t}= \frac 1{\sqrt {NT}}O_p(\Vert{\widehat{\theta}^*-\bar J\theta}\Vert)
				+\frac 1{T}O_p(\Vert{\widehat{\theta}^*-\bar J\theta}\Vert)
				+O_p\left(\frac 1{T}\right)
				+\frac 1{\sqrt T}O_p\left(\max\left(\frac 1N,\frac 1T\right)\right)$.
				\label{108_iii}
				\item
				\begin{align}
					&\norm{
						\frac 1{NT}\sum_{t=1}^T \frac{X_t'\Lambda J}{N}\left(\frac{J\Lambda'\Lambda J}{N}\right)^{-1}(\widehat{\Lambda}^*-\Lambda J)\epsilon_t-
						\frac 1{T^2}\sum_{t=1}^T\sum_{s=1}^T \frac{X_t'\Lambda J}{N}\left(\frac{J\Lambda'\Lambda J}{N}\right)^{-1}G_s \left(\frac 1{N} \sum_{i=1}^N \epsilon_{it}\epsilon_{is}\right)
					}\nonumber\\
					&=\frac 1{\sqrt {NT}}O_p(\Vert{\widehat{\theta}^*-\bar J\theta}\Vert)
					+\frac 1{T}O_p(\Vert{\widehat{\theta}^*-\bar J\theta}\Vert)
					+\frac 1{\sqrt T}O_p\left(\max\left(\frac 1N,\frac 1T\right)\right).\nonumber
				\end{align}
				\label{108_iv}
			\end{enumerate}
		\end{lemma}
		
		\begin{proof}
			Part \ref{108_i} follows from \eqref{A1Bai09} and \citet[Proposition 5]{barigozzi2022}.
			
			For part \ref{108_ii}, using \eqref{8indiani} and \eqref{indianata}, we have
			\begin{align}
				\frac 1{N \sqrt T}\sum_{t=1}^TG_t' (\widehat{\Lambda}^*-\Lambda J)\epsilon_t&= \frac 1{N\sqrt T}\sum_{t=1}^T \left(\frac{\widehat{\Lambda}^{*'}\Lambda}{N}\right)^{-1}
				\left\{\sum_{j=1}^8 I_j\right\}' \epsilon_t + o_p\left(\frac 1{\sqrt {NT}}\right)\nonumber\\
				&=:A_1+A_2+A_3+A_4+A_5+A_6+A_7+A_8+ o_p\left(\frac 1{\sqrt {NT}}\right).\label{ottoA}
			\end{align}
			For $A_1$ in \eqref{ottoA}, we have
			\begin{align}
				\norm{A_1}&\le \frac 1{\sqrt N} \norm{ \left(\frac{\widehat{\Lambda}^{*'}\Lambda}{N}\right)^{-1}}
				\left(
				\frac 1T\sum_{t=1}^T \norm{\sum_{s=1}^T \frac{\widehat{\Lambda}^{*'}X_s}{N}}\bar J(\bar J\theta-\widehat{\theta}^*)(\bar J\theta-\widehat{\theta}^*)'\bar J\frac{X_s'\epsilon_t}{\sqrt{NT}}
				\right)\nonumber\\
				&\le \frac 1{\sqrt N} \norm{ \left(\frac{\widehat{\Lambda}^{*'}\Lambda}{N}\right)^{-1}} \frac 1T\sum_{t=1}^T \norm{\frac{\widehat{\Lambda}^*}{\sqrt N}}\, \norm{\frac 1{\sqrt{NT}}\sum_{s=1}^T \bar JX_s'\epsilon_t}\,\frac{\norm{X_s}}{\sqrt N} \norm{\bar J\theta-\widehat{\theta}^*}^2\nonumber\\
				&= \frac 1{\sqrt N} O_p\left(\norm{\bar J\theta-\widehat{\theta}^*}\right)O_p(1),\label{A1agosto}
			\end{align}
			by \eqref{jsm}, \eqref{indiano1}, and Assumption \ref{as:Z}\ref{as:Z_iii}. 	Similarly, for $A_2$ in \eqref{ottoA}, we have
			\begin{align}
				A_2&=\frac 1{N T}\frac 1{\sqrt T} \sum_{t=1}^T \sum_{s=1}^T G_s (\bar J\theta-\widehat{\theta}^*)'\bar JX_s'\epsilon_t.\nonumber
			\end{align}
			Thus,
			\begin{align}
				\norm{A_2}=\frac 1{\sqrt N} O_p\left(\norm{\bar J\theta-\widehat{\theta}^*}\right).\label{A2agosto}
			\end{align}
			And by the same arguments, for $A_3$ in \eqref{ottoA}, we have
			\begin{align}
				\norm{A_3}&\le\frac 1{\sqrt N} \norm{ \left(\frac{\widehat{\Lambda}^{*'}\Lambda}{N}\right)^{-1}}\left(\frac 1T\sum_{t=1}^T \norm{
					\sum_{s=1}^T \frac{\widehat{\Lambda}^{*'}\epsilon_s}{N} (\bar J\theta-\widehat{\theta}^*)'\bar J\frac{X_s'\epsilon_t}{\sqrt{NT}}
				}\right)\nonumber\\
				&\le \frac 1{\sqrt N} \norm{ \left(\frac{\widehat{\Lambda}^{*'}\Lambda}{N}\right)^{-1}}\frac 1T \sum_{t=1}^T \norm{\frac{\widehat{\Lambda}^*}{\sqrt N}}\,
				\norm{\frac 1{\sqrt{NT}}\sum_{s=1}^T \bar J X_s'\epsilon_t}\, \frac{\norm{\epsilon_s}}{\sqrt N} \norm{\bar J\theta-\widehat{\theta}^*}\nonumber\\
				&= \frac 1{\sqrt N} O_p\left(\norm{\bar J\theta-\widehat{\theta}^*}\right)O_p(1),\label{A3agosto}
			\end{align}
			by \citet[Lemma 2]{barigozzi2022}. 
			
			For $A_4$ in \eqref{ottoA}, we have
			\begin{align}
				A_4&= \frac 1{N\sqrt T} \frac 1{NT} \sum_{t=1}^T\sum_{s=1}^T\left(\frac{\widehat{\Lambda}^{*'}\Lambda}{N}\right)^{-1} \widehat{\Lambda}^{*'} X_s \bar J(\bar J\theta-\widehat{\theta}^*)G_s'\Lambda'\epsilon_t\nonumber\\
				&= \frac 1{\sqrt N}  \left(\frac{\widehat{\Lambda}^{*'}\Lambda}{N}\right)^{-1}\left(\frac1 T \sum_{s=1}^T \frac{\widehat{\Lambda}^{*'} X_s\bar J}{N}\right)
				(\bar J\theta-\widehat{\theta}^*)\left(\frac 1{\sqrt {NT}}\sum_{t=1}^T G_s'\Lambda'\epsilon_t\right).
				\nonumber
			\end{align}
			Thus, by Assumption \ref{as:X_nu}\ref{as:X_nu_ii}, \eqref{jsm} and \citet[Lemma 3]{barigozzi2022},
			\begin{align}
				\norm{A_4}&=\frac 1{\sqrt N} O_p\left(\norm{\bar J\theta-\widehat{\theta}^*}\right).\label{A4agosto}
			\end{align}
			For $A_5$ in \eqref{ottoA}, we have
			\begin{align}
				A_5&=\frac 1{N\sqrt T} \sum_{t=1}^T\left(\frac{\widehat{\Lambda}^{*'}\Lambda}{N}\right)^{-1} \frac 1{NT}\sum_{s=1}^T \widehat{\Lambda}^{*'} X_s\bar J (\bar J\theta-\widehat{\theta}^*)\epsilon_s'\epsilon_t\nonumber\\
				&= \frac 1{\sqrt T} \left(\frac{\widehat{\Lambda}^{*'}\Lambda}{N}\right)^{-1}\left\{\frac 1{\sqrt T} \sum_{s=1}^T\frac{\widehat{\Lambda}^{*'} X_s\bar J}{N}  (\bar J\theta-\widehat{\theta}^*)\frac 1N\sum_{k=1}^N\epsilon_{ks}\right\} \frac 1{\sqrt T}\sum_{t=1}^T \epsilon_{kt}.\nonumber
			\end{align}
			Thus, by Assumptions \ref{as:errors_nu}\ref{as:errors_nu_idio_ii} and \ref{as:X_nu}\ref{as:X_nu_ii}, and \eqref{jsm}
			\begin{align}
				\norm{A_5}&=\frac 1{\sqrt T} O_p\left(\norm{\bar J\theta-\widehat{\theta}^*}\right).\label{A5agosto}
			\end{align}
			For $A_6$ in \eqref{ottoA}, we have
			\begin{align}
				A_6=&\, \frac 1{N\sqrt T}\sum_{t=1}^T \left(\frac{\widehat{\Lambda}^{*'}\Lambda}{N}\right)^{-1}\frac 1{NT} \sum_{s=1}^T \widehat{\Lambda}^{*'}\epsilon_s G_s'\Lambda'\epsilon_t\nonumber\\
				=&\,\frac 1{N^2T}\frac 1{\sqrt T} \sum_{s=1}^T \left(\frac{\widehat{\Lambda}^{*'}\Lambda}{N}\right)^{-1} J \Lambda'\epsilon_sG_s'
				\sum_{t=1}^T \Lambda'\epsilon_t+ \frac 1{N^2T}\frac 1{\sqrt T} \sum_{s=1}^T \left(\frac{\widehat{\Lambda}^{*'}\Lambda}{N}\right)^{-1} (\widehat{\Lambda}^{*'}-J \Lambda')\epsilon_sG_s'
				\sum_{t=1}^T \Lambda'\epsilon_t\nonumber\\
				=:&\, A_{6,1}+A_{6,2}.\label{A6agosto0}
			\end{align}
			Then, by \eqref{jsm} and \citet[Lemma 3]{barigozzi2022},
			\begin{align}
				A_{6,1}&=\frac 1{N\sqrt T} \left(\frac{\widehat{\Lambda}^{*'}\Lambda}{N}\right)^{-1}J\left(\frac1 {\sqrt{NT}}\sum_{s=1}^T \Lambda' \epsilon_s G_s'\right) \left(\frac 1{\sqrt{NT}}\sum_{t=1}^T\Lambda'\epsilon_t \right)=O_p\left(\frac 1{N\sqrt T}\right).\label{A61agosto}
			\end{align}
			Similarly, by \eqref{jsm}, \eqref{A1Bai09}, and \citet[Lemma 3]{barigozzi2022},
			\begin{align}
				A_{6,2}&=\frac 1{N\sqrt T} \left(\frac{\widehat{\Lambda}^{*'}\Lambda}{N}\right)^{-1}\left(\frac1 {\sqrt{NT}}\sum_{s=1}^T (\widehat{\Lambda}^{*'}-J\Lambda') \epsilon_s G_s'\right) \left(\frac 1{\sqrt{NT}}\sum_{t=1}^T\Lambda'\epsilon_t \right)\nonumber\\
				&=\frac 1{\sqrt N}\left\{O_p\left(\widehat{\theta}^*-\bar J\theta\right) + O_p\left(\max\left(\frac 1{\sqrt N},\frac 1{\sqrt T}\right)\right) \right\}\nonumber\\
				&=
				\frac 1{\sqrt N}O_p\left(\widehat{\theta}^*-\bar J\theta\right)+ O_p\left(\max\left(\frac 1{ N},\frac 1{ T}\right)\right).\label{A62agosto}
			\end{align}
			By substituting \eqref{A61agosto} and \eqref{A62agosto} into \eqref{A6agosto0}
			\begin{align}
				\norm{A_6}&=\frac 1{\sqrt N}O_p\left(\widehat{\theta}^*-\bar J\theta\right)+ O_p\left(\max\left(\frac 1{ N},\frac 1{ T}\right)\right).\label{A6agosto}
			\end{align}
			For $A_7$ in \eqref{ottoA}, we have
			\begin{align}
				A_7&=\frac 1{N\sqrt T}\frac 1T \sum_{t=1}^T\sum_{s=1}^T G_s\epsilon_s'\epsilon_t= \frac 1{\sqrt T} \left(\frac 1{\sqrt{NT}}\sum_{s=1}^T G_s \epsilon_s'\right)
				\left(\frac 1{\sqrt{NT}}\sum_{t=1}^T  \epsilon_t\right).\label{A7agosto0}
			\end{align}
			Thus, by \citet[Lemmas 2 and 3]{barigozzi2022},
			\begin{align}
				\norm{A_7}=O_p\left(\frac 1{\sqrt T}\right).\label{A7agosto}
			\end{align}
			Finally, for $A_8$ in \eqref{ottoA}, we have
			\begin{align}
				A_8=&\, \frac 1{N^2 T} \frac 1{\sqrt T} \sum_{t=1}^T\sum_{s=1}^T   \left(\frac{\widehat{\Lambda}^{*'}\Lambda}{N}\right)^{-1} \widehat{\Lambda}^{*'} \epsilon_s\epsilon_s'\epsilon_t\nonumber\\
				=&\,\frac 1{N^2 T} \frac 1{\sqrt T} \sum_{t=1}^T\sum_{s=1}^T   \left(\frac{\widehat{\Lambda}^{*'}\Lambda}{N}\right)^{-1} J {\Lambda}'\epsilon_s\epsilon_s'\epsilon_t\nonumber\\
				&+ \frac 1{N^2 T} \frac 1{\sqrt T} \sum_{t=1}^T\sum_{s=1}^T   \left(\frac{\widehat{\Lambda}^{*'}\Lambda}{N}\right)^{-1} (\widehat{\Lambda}^{*'} -J\Lambda')\epsilon_s\epsilon_s'\epsilon_t\nonumber\\
				=:&\, A_{8,1}+A_{8,2}.\label{A8agosto0}
			\end{align}
			Then, by \eqref{jsm}, \citet[Lemma 3]{barigozzi2022}, and Assumptions \ref{as:errors_nu}\ref{as:errors_nu_idio_ii} and \ref{as:errors_nu}\ref{as:errors_nu_idio_iv},
			\begin{align}
				A_{8,1}=&\, \frac 1{NT} \sum_{s=1}^T  \left(\frac{\widehat{\Lambda}^{*'}\Lambda}{N}\right)^{-1}\left\{
				\left(\frac 1{\sqrt N} J\Lambda'\epsilon_s\right)
				\left(\frac1{\sqrt{NT}} \sum_{t=1}^T \epsilon_s'\epsilon_t-\mathbb E[\epsilon_s'\epsilon_t]\right)
				\right\}\nonumber\\
				&+ \frac 1{NT} \sum_{s=1}^T  \left(\frac{\widehat{\Lambda}^{*'}\Lambda}{N}\right)^{-1}\left\{
				\left(\frac 1{\sqrt N} J\Lambda'\epsilon_s\right)
				\left(\frac1{\sqrt{NT}} \sum_{t=1}^T \mathbb E[\epsilon_s'\epsilon_t]\right)
				\right\}\nonumber\\
				=\,&O_p\left(\frac 1N\right)+O_p\left(\frac 1{\sqrt{NT}}\right).\label{A81agosto}
			\end{align}
			Similarly,
			\begin{align}
				A_{8,2}=&\,\frac 1{\sqrt N}\frac 1T\sum_{s=1}^T \frac 1{\sqrt{N T}}\sum_{t=1}^T \left(\frac{\widehat{\Lambda}^{*'}\Lambda}{N}\right)^{-1} \frac{(\widehat{\Lambda}^{*'} -J\Lambda' )\epsilon_s}N \left\{\epsilon_s'\epsilon_t-\mathbb E[\epsilon_s'\epsilon_t]\right\}\nonumber\\
				&+\frac 1{\sqrt N}\frac 1T\sum_{s=1}^T \frac 1{\sqrt{N T}}\sum_{t=1}^T \left(\frac{\widehat{\Lambda}^{*'}\Lambda}{N}\right)^{-1} \frac{(\widehat{\Lambda}^{*'} -J\Lambda' )\epsilon_s}N \mathbb E[\epsilon_s'\epsilon_t]\nonumber\\
				=:&\, A_{8,2,1}+A_{8,2,2}.\label{A82agosto}
			\end{align}
			By \eqref{jsm}, \eqref{A1Bai09}, and \citet[Lemmas 2 and 3]{barigozzi2022},
			\begin{align}
				\norm{A_{8,2,1}}&\le \frac 1{\sqrt N} \norm{\left(\frac{\widehat{\Lambda}^{*'}\Lambda}{N}\right)^{-1}} \left(\frac 1T\sum_{s=1}^T\norm{\frac 1{\sqrt{NT}}\sum_{t=1}^T\left\{\epsilon_s'\epsilon_t-\mathbb E[\epsilon_s'\epsilon_t]\right\} }^2\right)^{1/2}\left(\frac 1T\sum_{s=1}^T \frac{\norm{\epsilon_s}^2}{N}\right)^{1/2}\frac{\norm{\widehat{\Lambda}^{*} -\Lambda J }}{\sqrt N}\nonumber\\
				&=\frac 1{\sqrt N}\left\{O_p\left(\widehat{\theta}^*-\bar J\theta\right) + O_p\left(\max\left(\frac 1{\sqrt N},\frac 1{\sqrt T}\right)\right) \right\}\nonumber\\
				&=
				\frac 1{\sqrt N}O_p\left(\widehat{\theta}^*-\bar J\theta\right)+ O_p\left(\max\left(\frac 1{ N},\frac 1{ T}\right)\right).\label{A821agosto}
			\end{align}
			And, by the same arguments, and Assumption \ref{as:errors_nu}\ref{as:errors_nu_idio_ii}
			\begin{align}
				\norm{A_{8,2,2}}&\le \frac 1{\sqrt T} \frac{\norm{\widehat{\Lambda}^{*} -\Lambda J }}{\sqrt N}  \norm{\left(\frac{\widehat{\Lambda}^{*'}\Lambda}{N}\right)^{-1}}\frac 1T\sum_{t=1}^T\sum_{s=1}^T \mathbb E[\epsilon_s'\epsilon_t]\frac{\Vert \epsilon_s\Vert}{\sqrt N}\nonumber\\
				&=\frac 1{\sqrt T}\left\{O_p\left(\widehat{\theta}^*-\bar J\theta\right) + O_p\left(\max\left(\frac 1{\sqrt N},\frac 1{\sqrt T}\right)\right) \right\}\nonumber\\
				&=
				\frac 1{\sqrt T}O_p\left(\widehat{\theta}^*-\bar J\theta\right)+ O_p\left(\max\left(\frac 1{ N},\frac 1{ T}\right)\right).\label{A822agosto}
			\end{align}
			Therefore, from \eqref{A821agosto} and \eqref{A822agosto}, we have
			\begin{align}
				\norm{A_{8,2}}= \frac 1{\sqrt N}O_p\left(\widehat{\theta}^*-\bar J\theta\right)+\frac 1{\sqrt T}O_p\left(\widehat{\theta}^*-\bar J\theta\right)+ O_p\left(\max\left(\frac 1{ N},\frac 1{ T}\right)\right).\label{A82agosto}
			\end{align}
			By substituting \eqref{A81agosto} and \eqref{A82agosto} into \eqref{A8agosto0} we obtain:
			\begin{align}
				\norm{A_{8}}=  \frac 1{\sqrt N}O_p\left(\widehat{\theta}^*-\bar J\theta\right)+\frac 1{\sqrt T}O_p\left(\widehat{\theta}^*-\bar J\theta\right)+ O_p\left(\max\left(\frac 1{ N},\frac 1{ T}\right)\right). \label{A8agosto}
			\end{align}
			
			Summing up, by substituting \eqref{A1agosto}, \eqref{A2agosto}, \eqref{A3agosto}, \eqref{A4agosto}, \eqref{A5agosto}, \eqref{A6agosto}, \eqref{A7agosto}, and \eqref{A8agosto} into \eqref{ottoA}, we prove part \ref{108_ii}.
			
			For part \ref{108_iii} just divide part \ref{108_ii} by $\sqrt T$ and
			notice that $\Vert G_t\Vert = O_p(1)$ because of Assumption \ref{as:errors_nu}\ref{as:errors_nu_common_ii}.
			
			For part  \ref{108_iv} just replace in part \ref{108_iii} $G_t$ with $X_t'\Lambda J(J\Lambda'\Lambda J)^{-1}$ which is also $O_p(1)$ and the second term on the right-hand-side of the statement comes from the $O_p(T^{-1})$ term in part \ref{108_iii} (see also the expression of $A_7$ in \eqref{A7agosto0}).
		\end{proof}

		
		\section{Monte Carlo simulations}\label{app:MC_appendix}
		
		\subsection{Data generating process}\label{app:MC_appendix_setup}

		We assume $r=1$ and $q=1$, i.e., one network factor and one node factor.
		We then generate artificial time series of $y_t$ and $\mathcal{W}_t$, for $t=1, \dots, T$, according to the three model equations:
		\begin{align*}
			y_t &= \beta {\frac{F_{t-1}}{N}}  y_{t-1} + \rho y_{t-1}+ \alpha+ \nu_t \\
			\nu_{t} &=  \Lambda G_t+\epsilon_{t}\\
			\mathcal{W}_t &= \mathcal{F}_t \times_3 U + \mathcal{E}_t
		\end{align*}
		where $\mathcal{F}_t$ here has size $N \times N \times 1$, $U$ is a column vector of length $m$, $G_t$ is a scalar and $\Lambda$ is the $N$-dimensional column vector of node-specific factor loadings.
		
		The Monte Carlo (MC) simulations are implemented as follows.  We consider $N \in \{10,20,50,100,200\}$, $m \in \{20,50,100\}$, and $T \in \{50,80,100\}$. 
		First of all, we fix the values of FNAR parameters $\beta=0.5$, $\rho=0.3$, and $\alpha=0.2$.
		Also, for each value of the pair ($N ,m$), we randomly generate the entries of the loading	vectors $U$ and $\Lambda$ once (and independently)
		from $\mathcal{N}(1,1)$, and then we keep them fixed across MC iterations (see \citealt{chenetal2020cnar}).
		
		All results are based on $S=500$ iterations. Then, at each MC iteration we simulate the model according to the following.
		
		\begin{itemize}	
			
			\item For all $t=1,\ldots,T$, we generate the entries of the network factor matrix $F_{1,t}$ from $\mathcal{N}(0,1)$, except for the diagonal elements, which are set to zero to ensure the interpretation of $F_{1,t}$ as a network.
			
			\item
			For the multi-network of idiosyncratic terms $\mathcal{E}_t$, which is $N\times N\times m$, we consider two cases.
			\begin{enumerate}[label=\Roman*]
				\item $\mathcal{E}_t$ has zero autocorrelation and zero cross-layer correlation.
				\item $\mathcal{E}_t$ is serially and cross-layer  correlated. 
				Specifically, we generate $\mathcal{E}_t$ according to the equation 
				$$
				\mathcal{E}_{i j \cdot, t} = \rho_{\mathcal{E}} \mathcal{E}_{i j \cdot, t-1} + \mathcal{N}(0_{m}, \Sigma_{\mathcal{E}}^{(m)}), \quad i,j=1,\ldots, N.
				$$ 
				Where $\rho_{\mathcal{E}}=0.5$ and the $(i,j)$-th element of
				$\Sigma_{\mathcal{E}}^{(m)}$  is given by $\tau_{\mathcal{E}}^{\vert i-j \vert }$ if $\vert i-j \vert < d $, and 0 otherwise.
				We set $\tau_{\mathcal{E}}$ to 0.5 and $d$ to 5.
			\end{enumerate}
			In both case I and case II, we assume no cross-correlation in the first two dimensions of $\mathcal{E}_t$.
			Also, we set all diagonal elements of each $N \times N$ layer $\mathcal{E}_{\cdot \cdot k, t}$ to zero, $k=1, \dots, m$ and $t = 1, \dots, T$.

			\item 	To keep the noise-to-signal ratio in check, we choose to rescale the simulated elements of $\mathcal{E}_t$ so that the (iteration-specific) sample variance across all the elements in the $N \times N \times m \times T$ tensor of idiosyncratic components $\mathcal{E}$ is equal to half the sample variance across all the elements in the $N \times N \times m \times T$ tensor of common components $\mathcal{F} \times_3 U$.

			\item We simulate the node factor $G_t$ according  to $G_t = \rho_G G_{t-1} + \mathcal{N}(0, 1)$ where $\rho_G= 0.2$. This allows us to study the iterated estimator under serially correlated FNAR errors, i.e., when the OLS and GLS are not consistent.
			
			\item The $N$ node-specific idiosyncratic terms in vector $\epsilon_{t}$ are generated from independent $\mathcal{N}(0, 1)$. 
			Then, to control the noise-to-signal ratio in the FNAR errors $\nu_t$, we rescale the simulated elements of $\epsilon_{t}$ so that the sample variance of $\text{vec}((\epsilon_1 \cdots \epsilon_T))$ is half the sample variance of  $\text{vec}((\Lambda G_1 \cdots \Lambda G_T))$.

			\item We simulate $y_t$ and $\mathcal{W}_t$ according to the previous steps.
		\end{itemize}	
		
		Next, at each iteration we re-estimate 
		$\mathcal{F}_t$, $U$, $\beta$, $\rho$ and $\alpha$ on the simulated $y_t$ and $\mathcal{W}_t$.
		When estimating $\mathcal{F}_t$ and $U$, we implement a small-$N$ adjustment to our tensor principal component estimator. In particular, recall that the $N \times N$ network matrices (slices) composing the $N \times N \times m$ weight tensor $\mathcal{W}_{t}$ have zero diagonal elements. Accordingly, the number of elements of each network matrix that are effectively used to calculate the inner product $\widehat{\Gamma}^{\mathcal{W}}$, used for PC estimation,
		is not $N^2$, but $N^2 - N$.
		Of course, this is not an issue for large $N$, and asymptotic results for $N \to \infty$ are derived by simply considering $N^2$ elements in each network matrix, since the term $N$ is dominated by $N^2$ and is therefore asymptotically negligible.
		However, for small values of $N$, estimation improves by taking into account the presence of zeros along the diagonals.
		For this reason, in the MC simulations, we introduce the following adjustment to estimated loadings and factors, replacing $N^2$ with $N^2-N$:
		\begin{equation*}
			\widehat{U} :=  \widehat{V}^{\mathcal{W}}( \widehat{M}^{\mathcal{W}})^{1/2} (N^2-N)^{-1/2}, \qquad 
			\widehat{\mathcal{F}}_{t} :=
			\mathcal{W}_{t} \times_3 [ (N^2-N) ( \widehat{M}^{\mathcal{W}} )^{-1} \widehat{U}' ].	 
		\end{equation*}
		
		To estimate the FNAR parameters, we use the \citet{bai2009panel} iterative estimator, given the serial correlation assumed in the node-specific factor $G_t$.
		Recall that factors, loadings and the FNAR parameter $\beta$ are consistently estimated only up to a sign. To address this issue,  we change the signs of $\widehat{U}$ to ensure that $\widehat{U}' U $ is positive, then $\widehat{\mathcal{F}}_t$ and $\widehat{\beta}$ are adjusted accordingly.

		For	different values of $T$, $N$, and $m$, we report the MC root mean squared errors of the estimates, both in absolute terms (\textit{RMSE}) and relative to the true parameter values (\textit{ReRMSE}). 
		Let us define $\mathcal{F}^{mc}$ as the $N \times N \times 1 \times T \times S$ tensor storing all simulated tensors $\mathcal{F}_t$ for all time periods $T$ and all $S$ iterations, and let us define $\widehat{\mathcal{F}}^{mc}$ analogously for factor estimates.
		Also, let us define $\widehat{U}^{mc}$ as the $N \times 1 \times S$ tensor storing all estimated loadings $\widehat{U}$ for all $S$ iterations, and $U^{mc}$ the tensor of same dimension obtained by replicating the true $U$   $S$ times. 
		Then, in the case of network factors and loadings,
		the  \textit{RMSE} and relative \textit{RMSE} are calculated as 
		$ \Vert \widehat{\mathcal{F}}^{mc} - \mathcal{F}^{mc} \Vert_F $, $ \vert \vert \widehat{U}^{mc} - U^{mc} \Vert_F$, 
		$ \Vert \widehat{\mathcal{F}}^{mc} - \mathcal{F}^{mc} \Vert_F  / \Vert \mathcal{F}^{mc} \vert \vert_F $
		and $ \Vert \widehat{U}^{mc} - U^{mc} \Vert_F   / \Vert U^{mc} \Vert_F $, respectively. Likewise, we compute the \textit{RMSE} and relative \textit{RMSE} for the elements of the estimated FNAR coefficients, averaged over all $S$ iterations and when considering the estimator $\widehat{\theta}^\dag$ (the results for the elements of $\widehat{\theta}^*$ are almost identical).	Finally, we also report the MC distribution of the estimated network effect $\widehat{\beta}$. 
		
		Results for case I are below, while results for case II are in Section \ref{sec:simulations}, except those for $T=80$ which are below.

		\subsection{RMSEs and Histograms - case I}\label{app:MC_appendix_resA}

\begin{landscape}
			\begin{figure}[H]
				\centering
				\caption{Monte Carlo histograms of the network effect $\widehat{\beta}$ - $T=10$, case I}
				\begin{subfigure}{.24\textwidth}
					\centering
					\includegraphics[width=\linewidth]{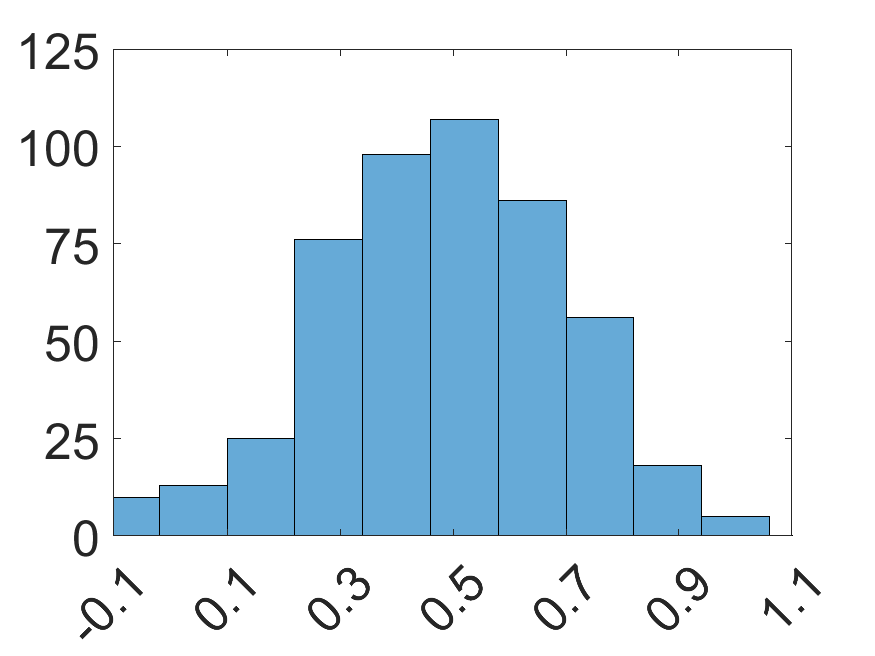}
					\caption{$N=10, m=20$}
				\end{subfigure}%
				\hfill
				\begin{subfigure}{.24\textwidth}
					\centering
					\includegraphics[width=\linewidth]{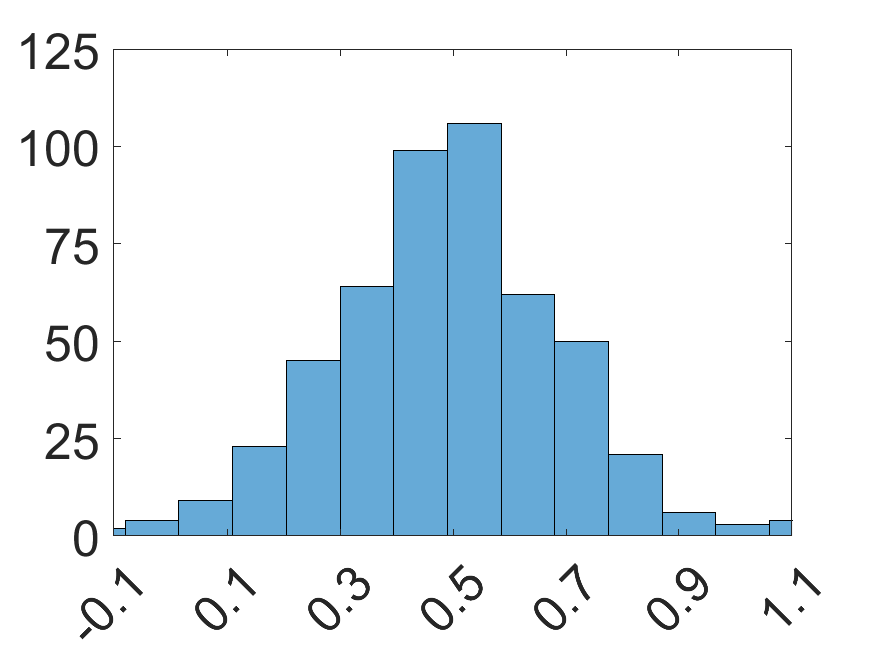}
					\caption{$N=20, m=20$}
				\end{subfigure}
				\hfill
				\begin{subfigure}{.24\textwidth}
					\centering
					\includegraphics[width=\linewidth]{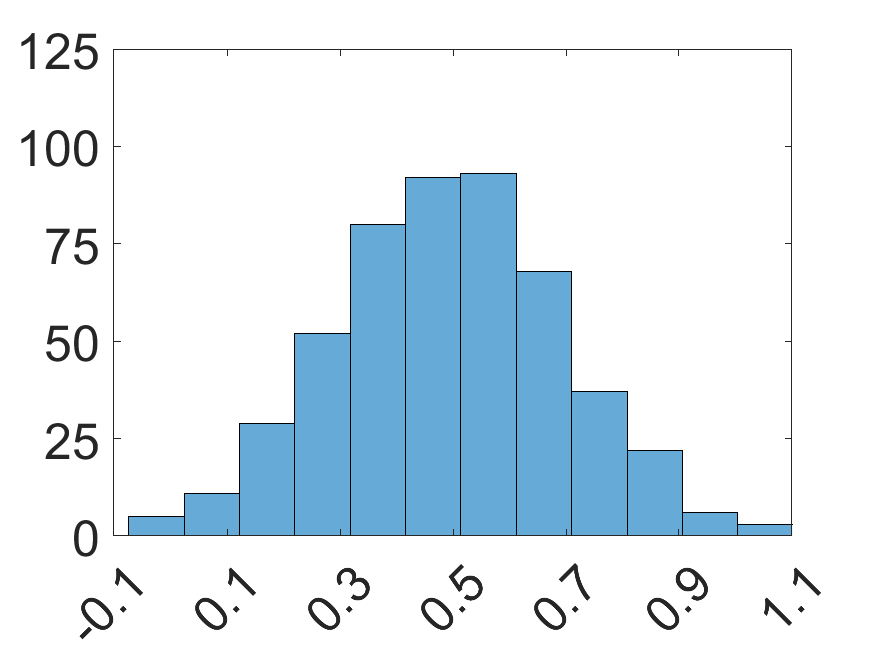}
					\caption{$N=50, m=20$}
				\end{subfigure}
				\hfill
				\begin{subfigure}{.24\textwidth}
					\centering
					\includegraphics[width=\linewidth]{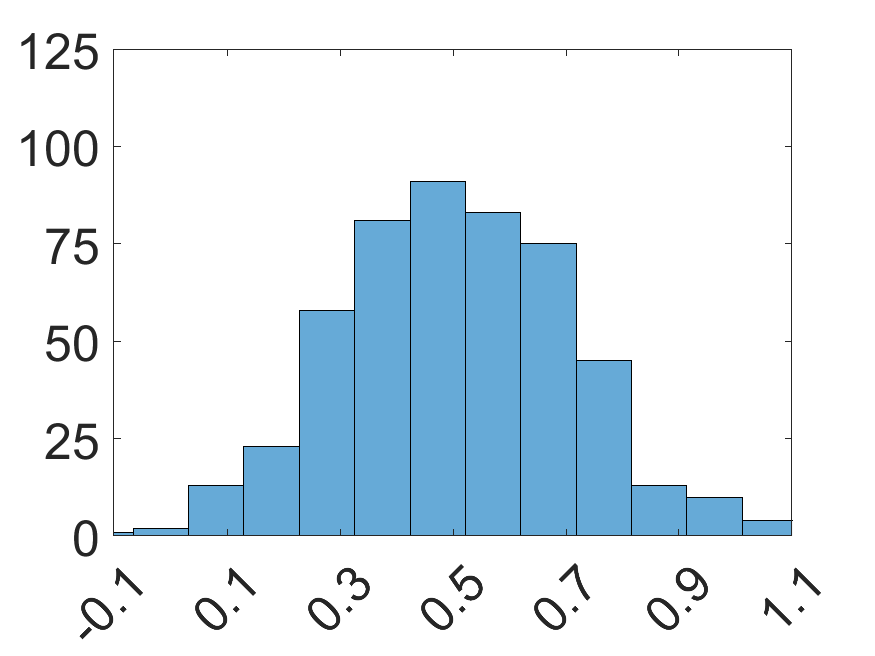}
					\caption{$N=100, m=20$}
				\end{subfigure}
				\hfill
				\begin{subfigure}{.24\textwidth}
					\centering
					\includegraphics[width=\linewidth]{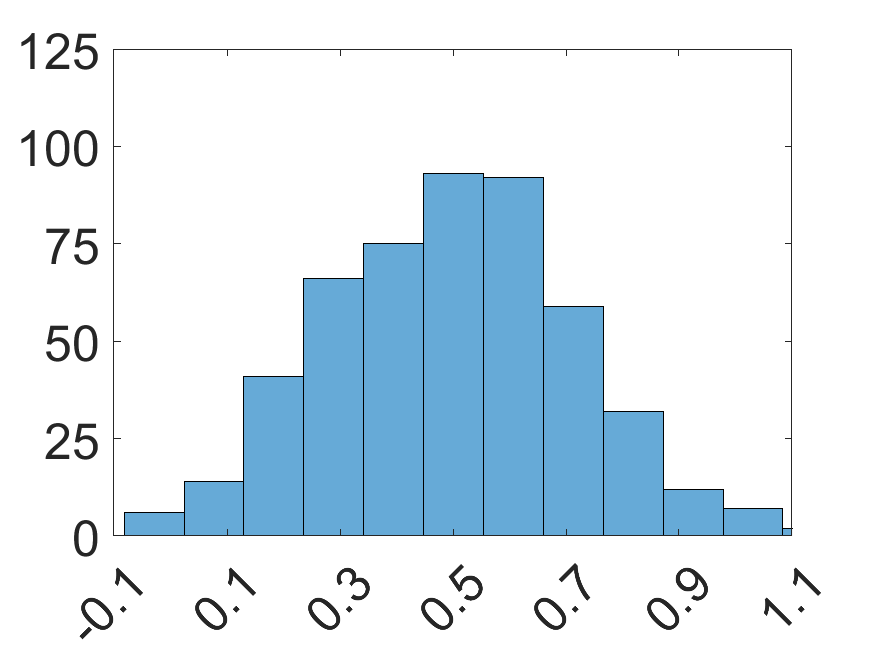}
					\caption{$N=200, m=20$}
				\end{subfigure}
				
				\begin{subfigure}{.24\textwidth}
					\centering
					\includegraphics[width=\linewidth]{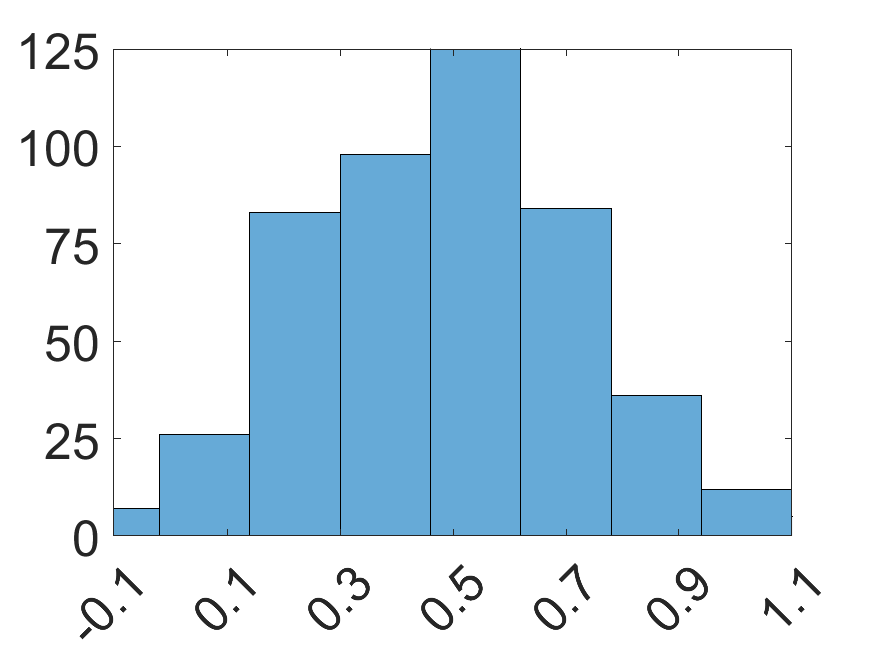}
					\caption{$N=10, m=50$}
				\end{subfigure}%
				\hfill
				\begin{subfigure}{.24\textwidth}
					\centering
					\includegraphics[width=\linewidth]{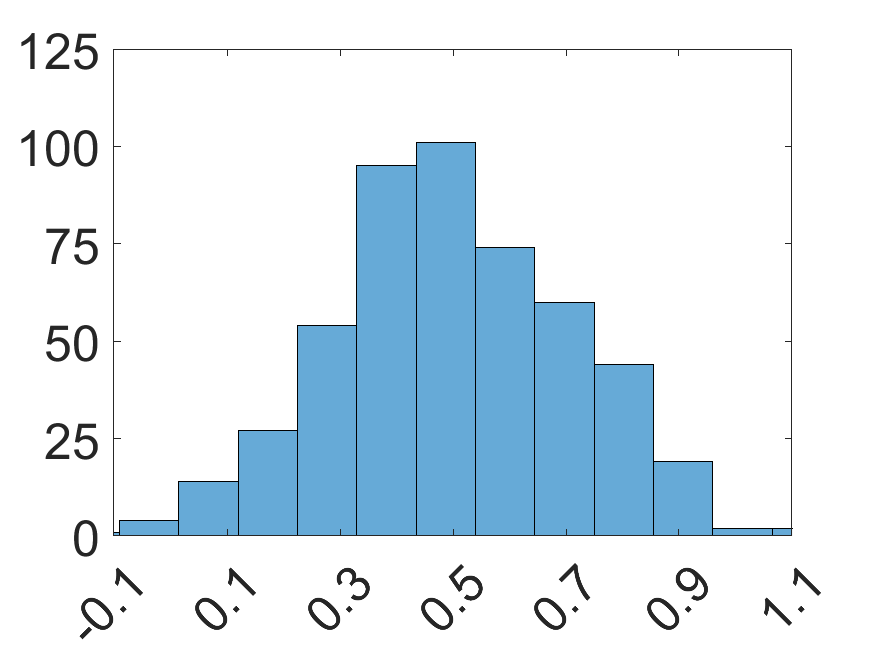}
					\caption{$N=20, m=50$}
				\end{subfigure}
				\hfill
				\begin{subfigure}{.24\textwidth}
					\centering
					\includegraphics[width=\linewidth]{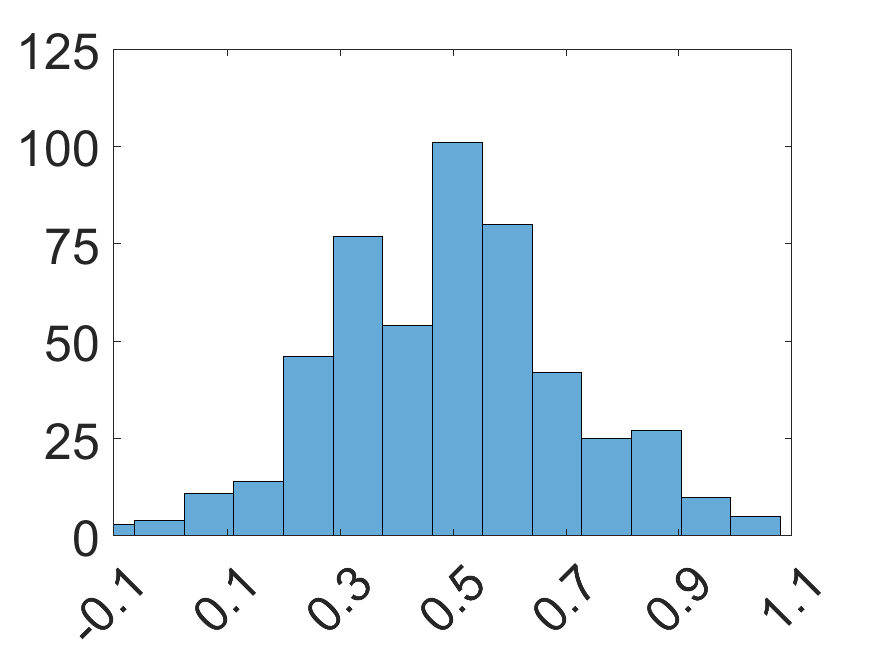}
					\caption{$N=50, m=50$}
				\end{subfigure}
				\hfill
				\begin{subfigure}{.24\textwidth}
					\centering
					\includegraphics[width=\linewidth]{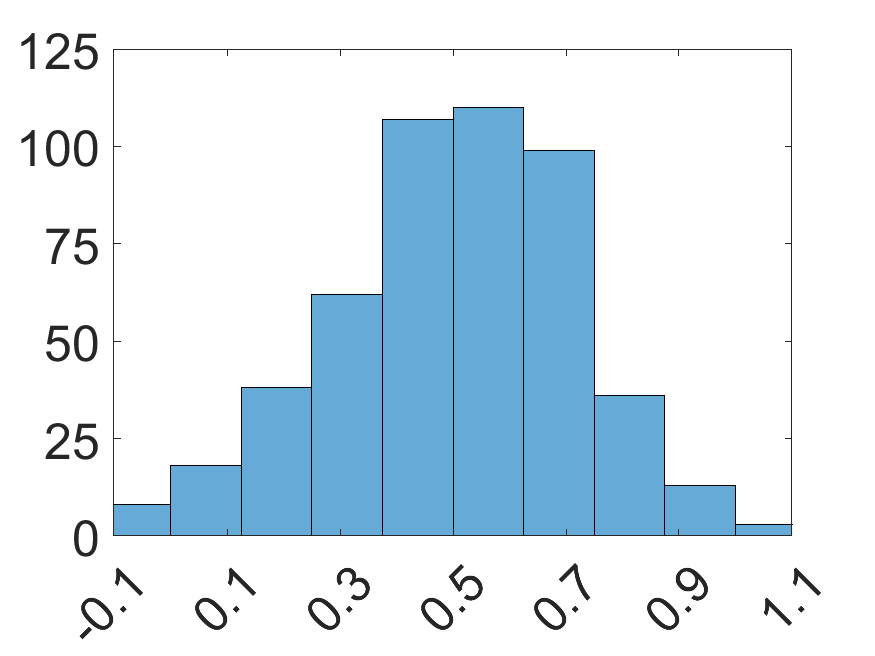}
					\caption{$N=100, m=50$}
				\end{subfigure}
				\hfill
				\begin{subfigure}{.24\textwidth}
					\centering
					\includegraphics[width=\linewidth]{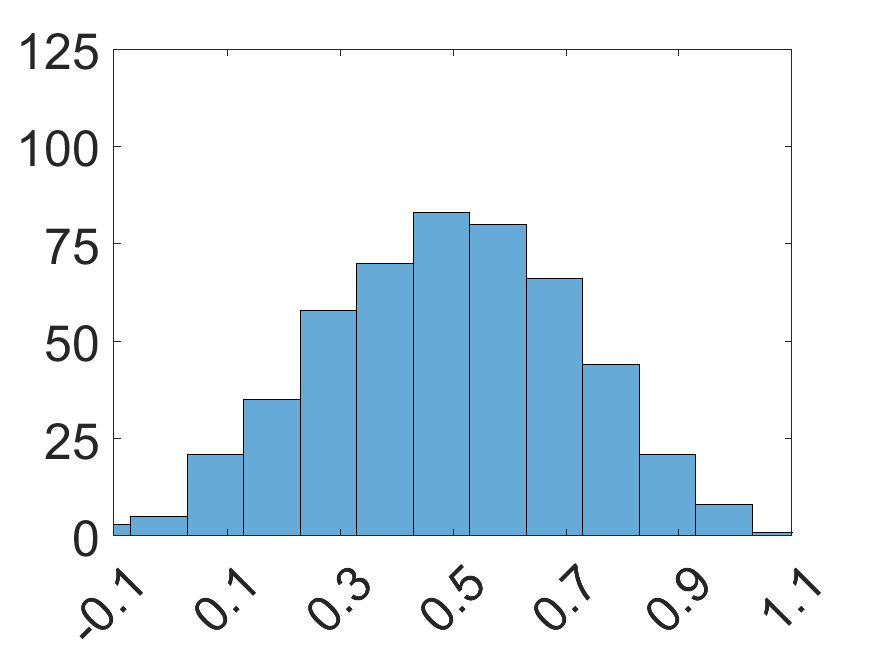}
					\caption{$N=200, m=50$}
				\end{subfigure}
				
				\begin{subfigure}{.24\textwidth}
					\centering
					\includegraphics[width=\linewidth]{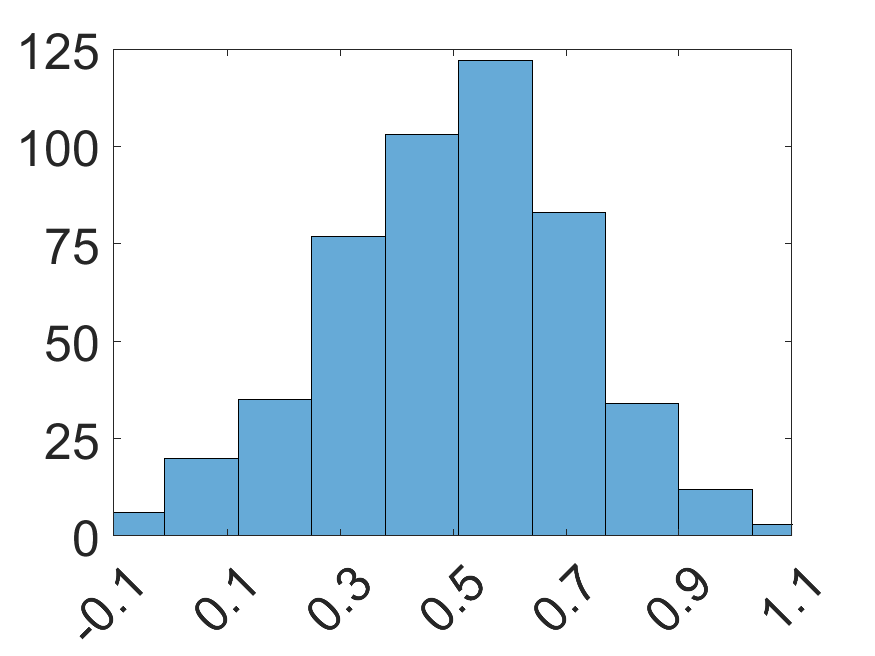}
					\caption{$N=10, m=100$}
				\end{subfigure}%
				\hfill
				\begin{subfigure}{.24\textwidth}
					\centering
					\includegraphics[width=\linewidth]{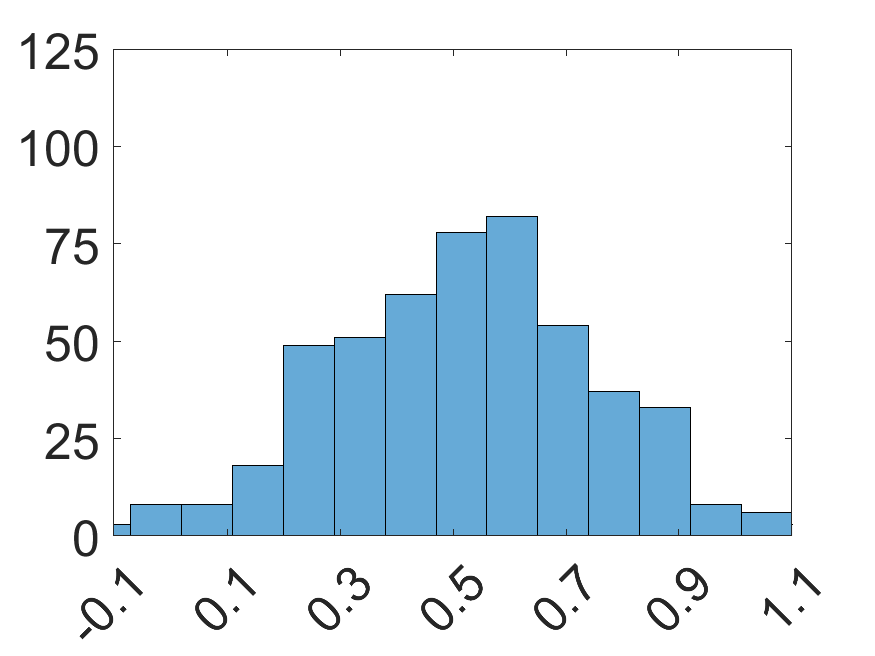}
					\caption{$N=20, m=100$}
				\end{subfigure}
				\hfill
				\begin{subfigure}{.24\textwidth}
					\centering
					\includegraphics[width=\linewidth]{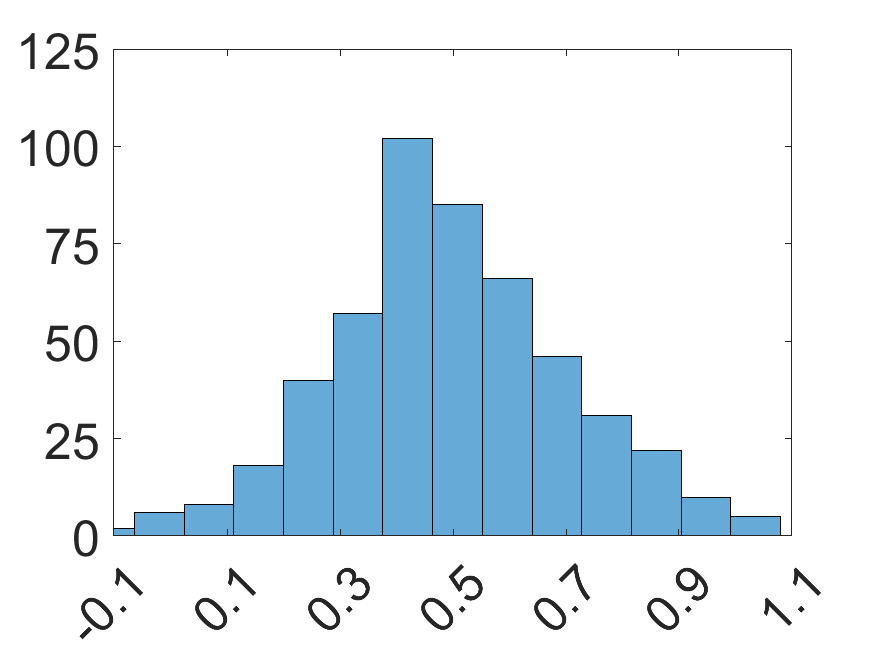}
					\caption{$N=50, m=100$}
				\end{subfigure}
				\hfill
				\begin{subfigure}{.24\textwidth}
					\centering
					\includegraphics[width=\linewidth]{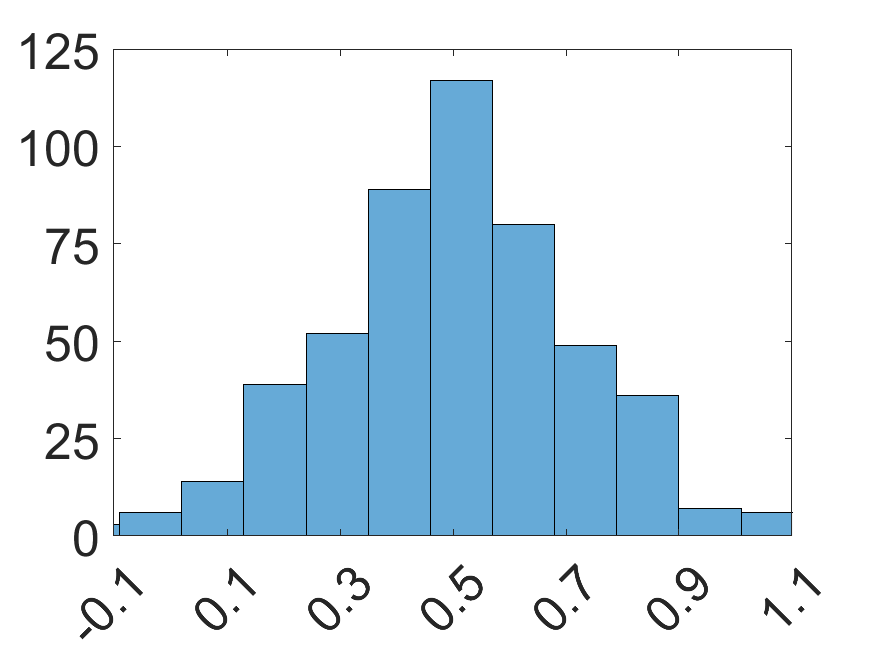}
					\caption{$N=100, m=100$}
				\end{subfigure}
				\hfill
				\begin{subfigure}{.24\textwidth}
					\centering
					\includegraphics[width=\linewidth]{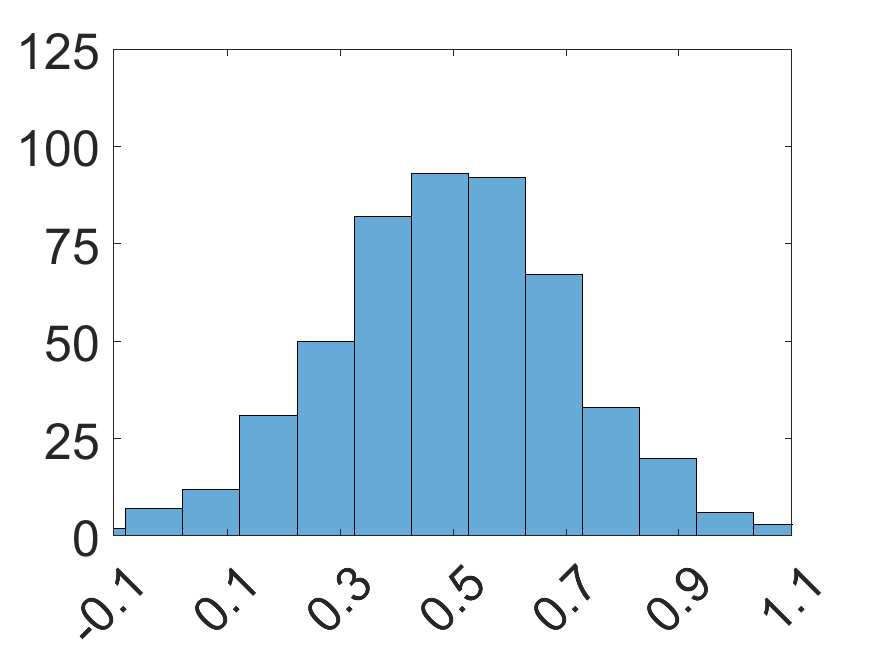}
					\caption{$N=200, m=100$}
				\end{subfigure}
				\subcaption*{\textit{Note:} the true value is $\beta= 0.5$.}
				\label{fig:MC_hist_beta_CaseI_T10}
			\end{figure}
		\end{landscape}

\begin{landscape}
			\begin{figure}[H]
				\centering
				\caption{Monte Carlo histograms of the network effect $\widehat{\beta}$ - $T=50$, case I}
				\begin{subfigure}{.24\textwidth}
					\centering
					\includegraphics[width=\linewidth]{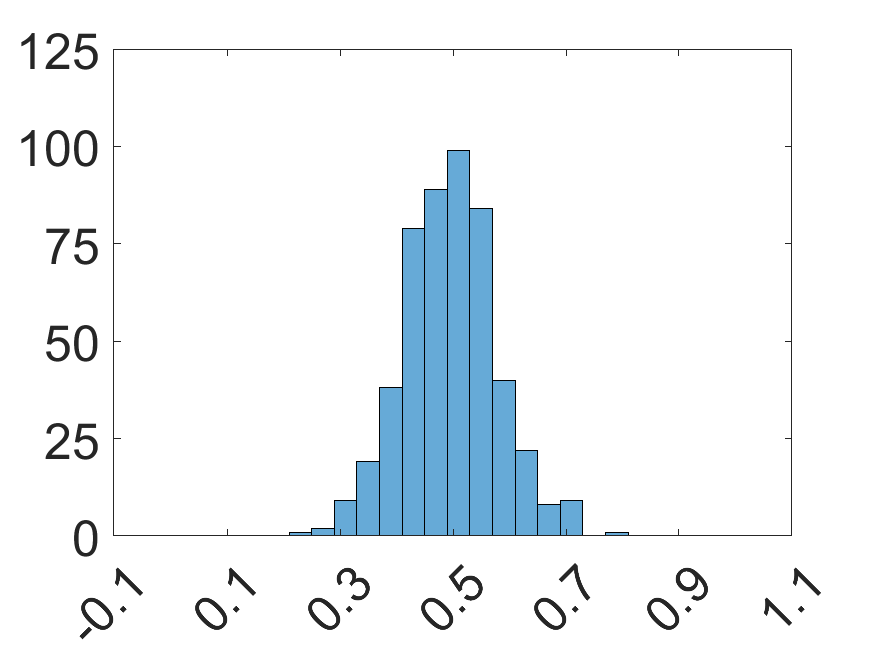}
					\caption{$N=10, m=20$}
				\end{subfigure}%
				\hfill
				\begin{subfigure}{.24\textwidth}
					\centering
					\includegraphics[width=\linewidth]{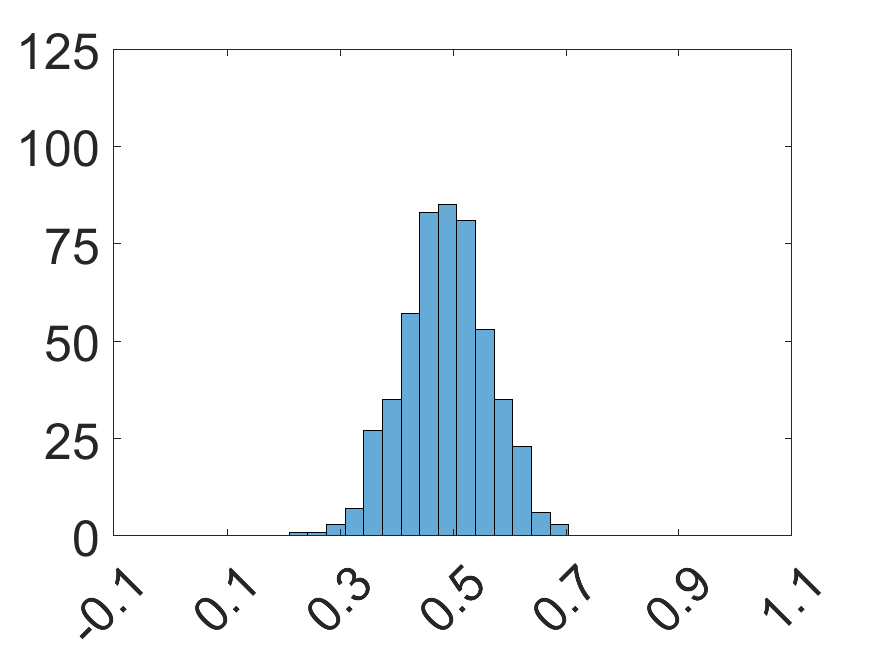}
					\caption{$N=20, m=20$}
				\end{subfigure}
				\hfill
				\begin{subfigure}{.24\textwidth}
					\centering
					\includegraphics[width=\linewidth]{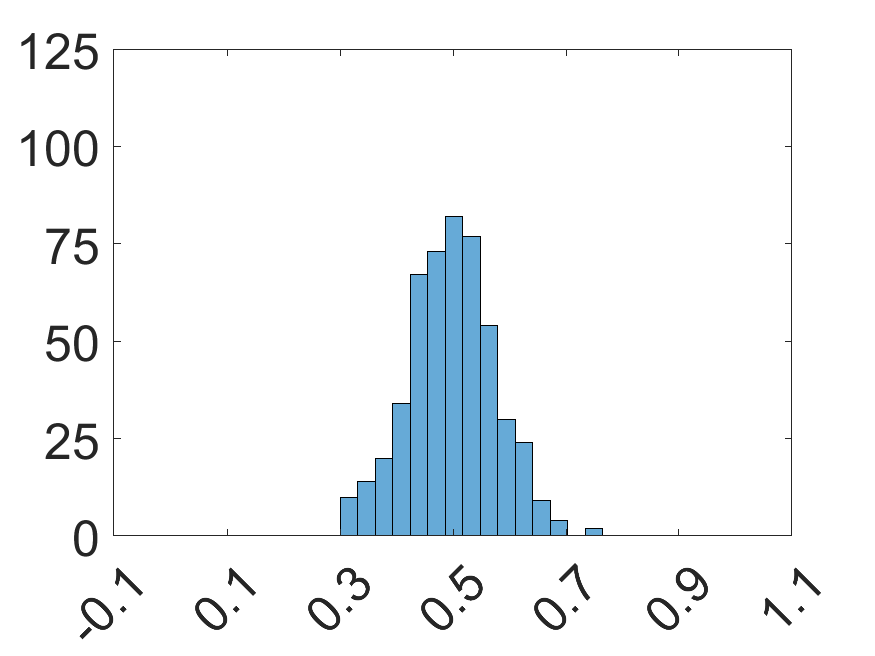}
					\caption{$N=50, m=20$}
				\end{subfigure}
				\hfill
				\begin{subfigure}{.24\textwidth}
					\centering
					\includegraphics[width=\linewidth]{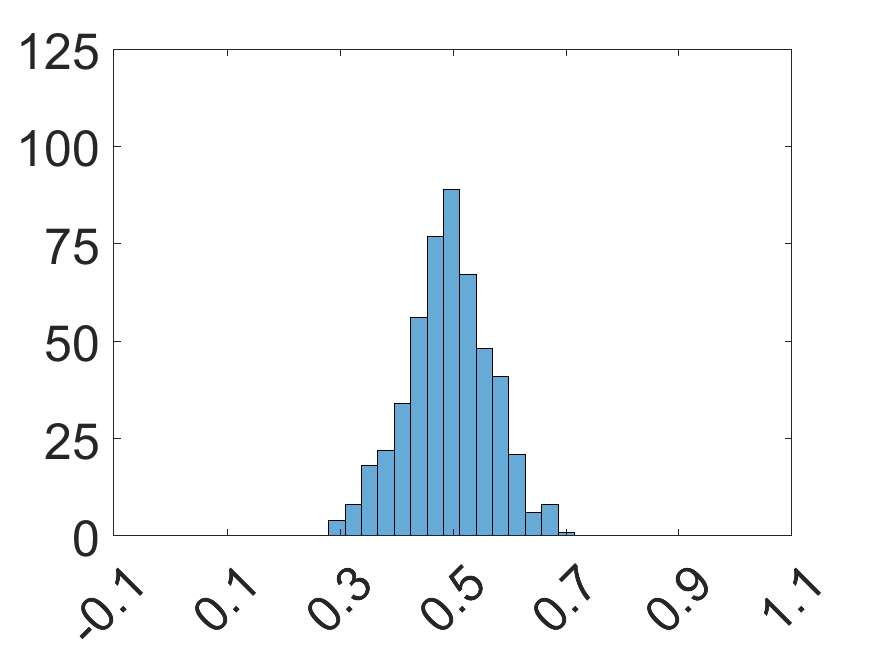}
					\caption{$N=100, m=20$}
				\end{subfigure}
				\hfill
				\begin{subfigure}{.24\textwidth}
					\centering
					\includegraphics[width=\linewidth]{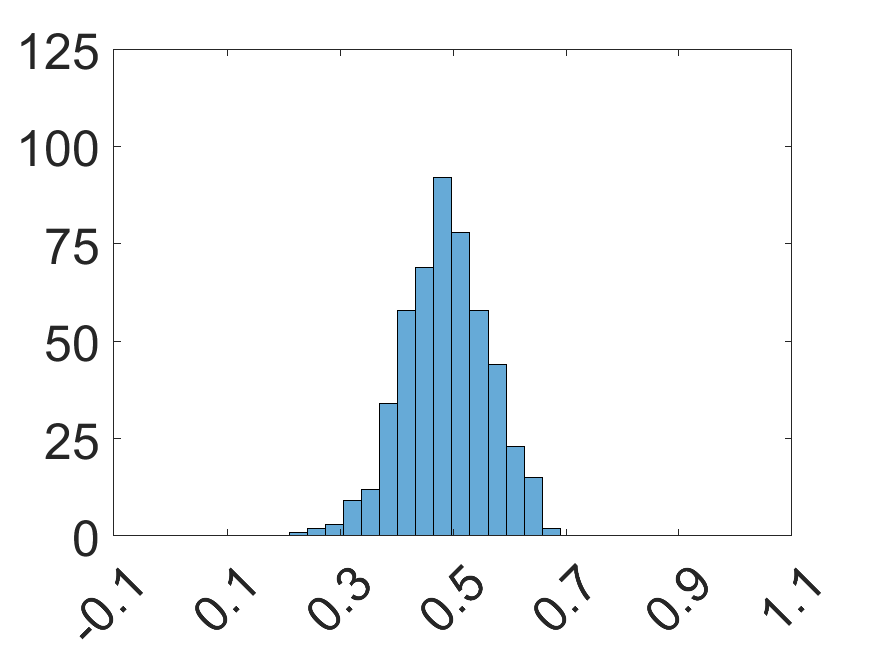}
					\caption{$N=200, m=20$}
				\end{subfigure}
				
				\begin{subfigure}{.24\textwidth}
					\centering
					\includegraphics[width=\linewidth]{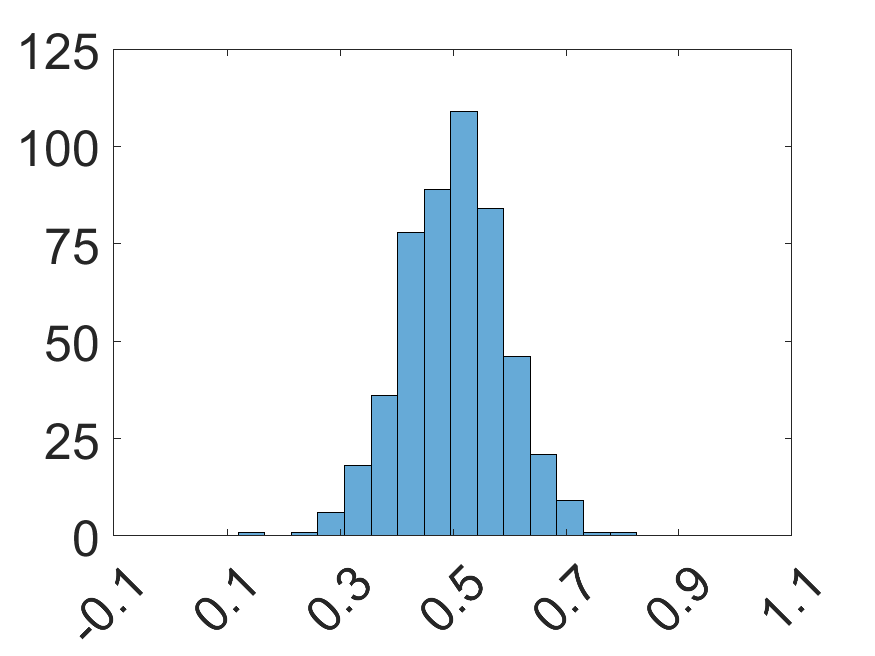}
					\caption{$N=10, m=50$}
				\end{subfigure}%
				\hfill
				\begin{subfigure}{.24\textwidth}
					\centering
					\includegraphics[width=\linewidth]{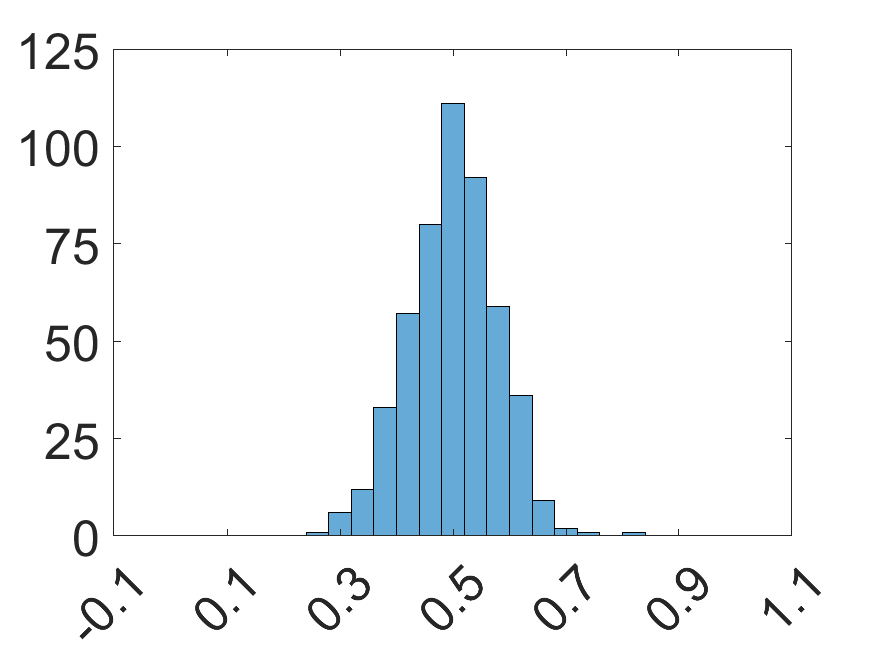}
					\caption{$N=20, m=50$}
				\end{subfigure}
				\hfill
				\begin{subfigure}{.24\textwidth}
					\centering
					\includegraphics[width=\linewidth]{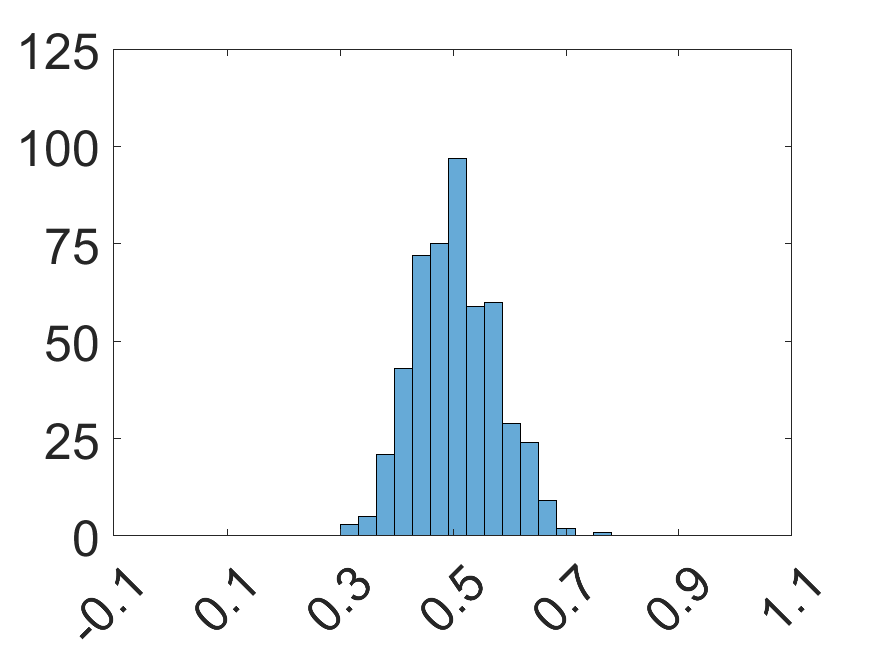}
					\caption{$N=50, m=50$}
				\end{subfigure}
				\hfill
				\begin{subfigure}{.24\textwidth}
					\centering
					\includegraphics[width=\linewidth]{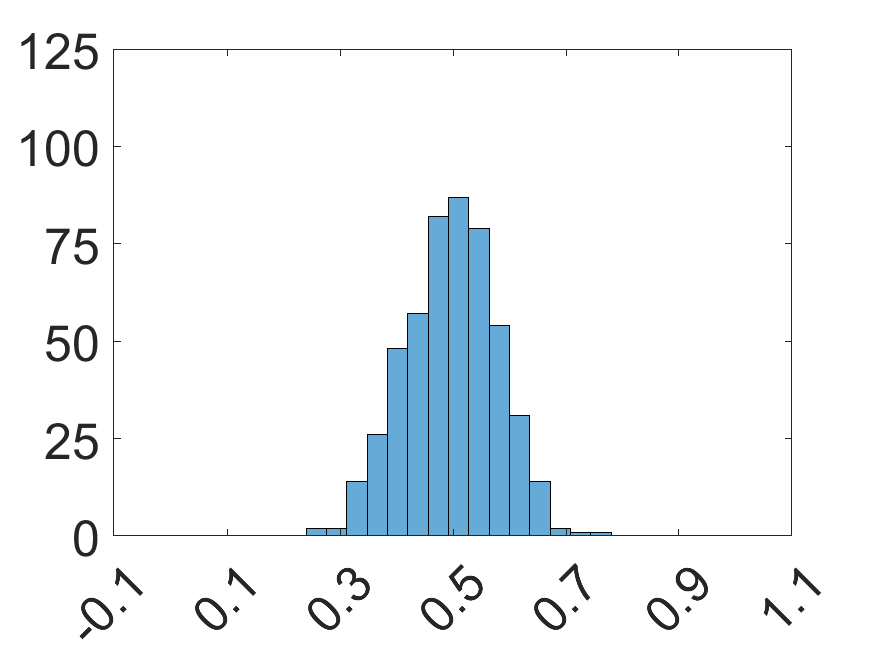}
					\caption{$N=100, m=50$}
				\end{subfigure}
				\hfill
				\begin{subfigure}{.24\textwidth}
					\centering
					\includegraphics[width=\linewidth]{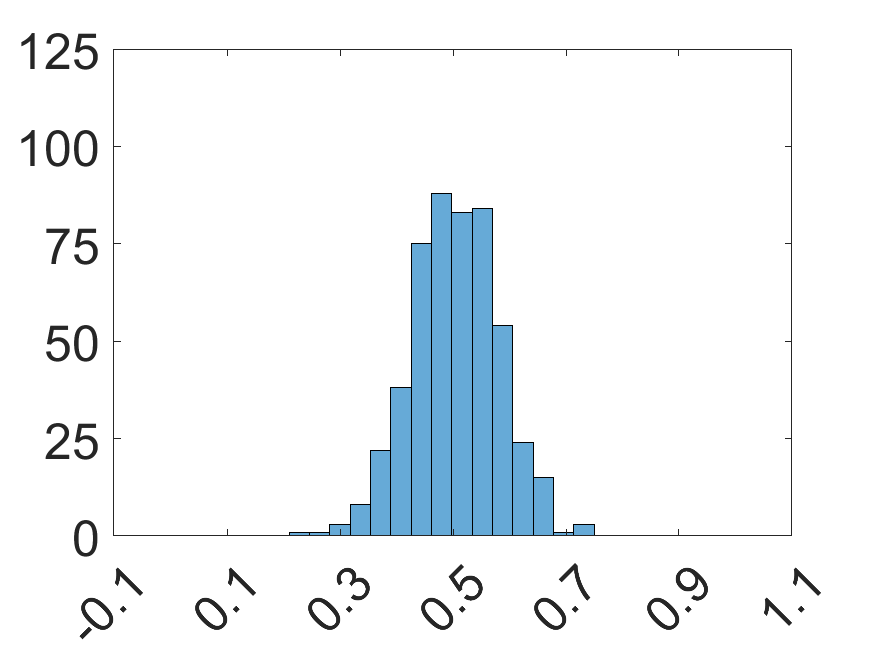}
					\caption{$N=200, m=50$}
				\end{subfigure}
				
				\begin{subfigure}{.24\textwidth}
					\centering
					\includegraphics[width=\linewidth]{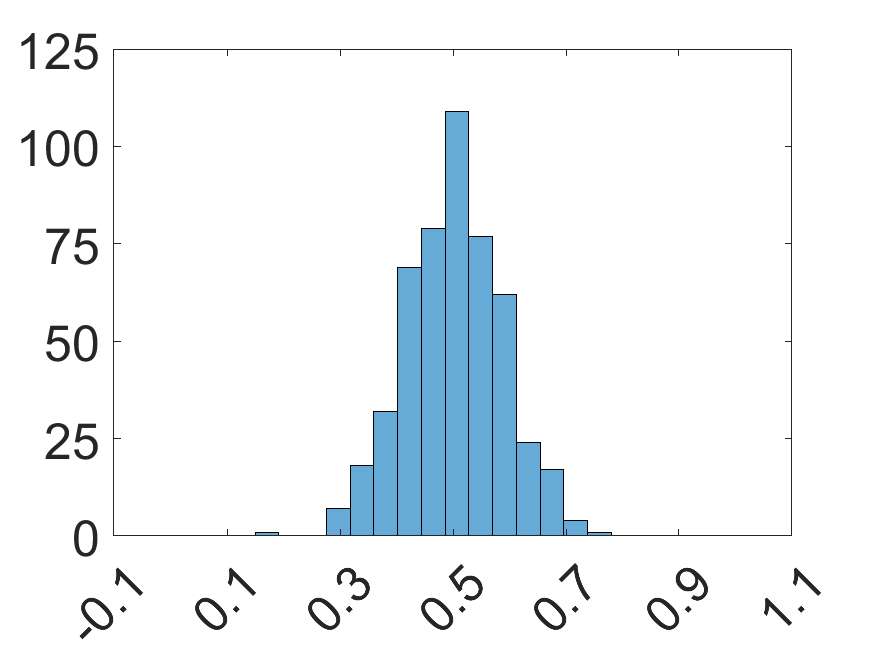}
					\caption{$N=10, m=100$}
				\end{subfigure}%
				\hfill
				\begin{subfigure}{.24\textwidth}
					\centering
					\includegraphics[width=\linewidth]{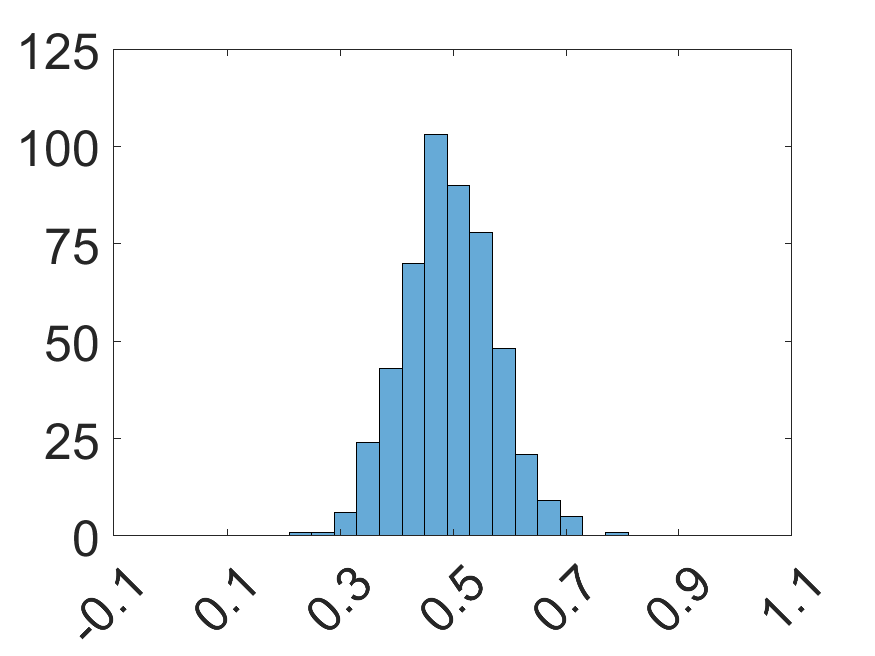}
					\caption{$N=20, m=100$}
				\end{subfigure}
				\hfill
				\begin{subfigure}{.24\textwidth}
					\centering
					\includegraphics[width=\linewidth]{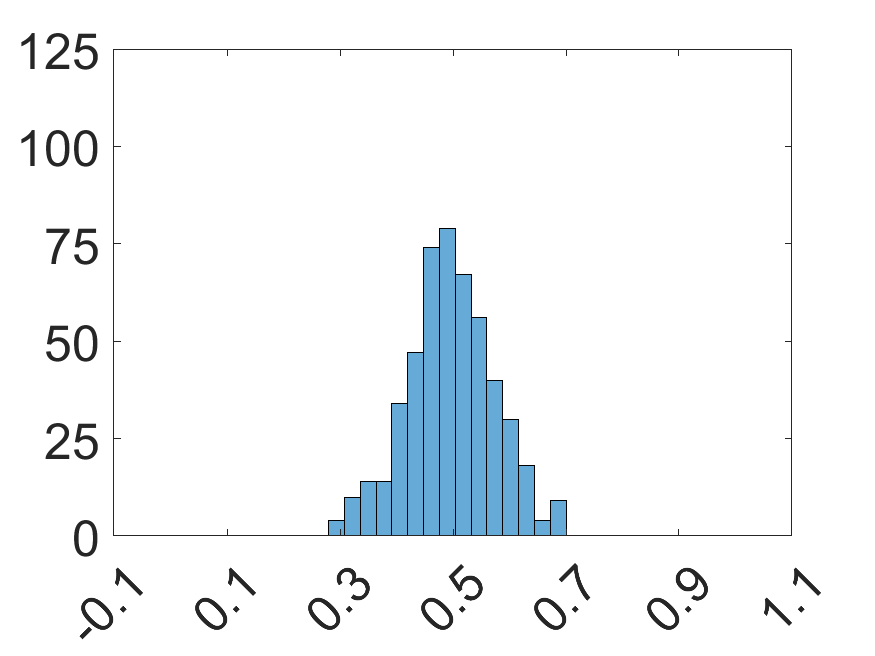}
					\caption{$N=50, m=100$}
				\end{subfigure}
				\hfill
				\begin{subfigure}{.24\textwidth}
					\centering
					\includegraphics[width=\linewidth]{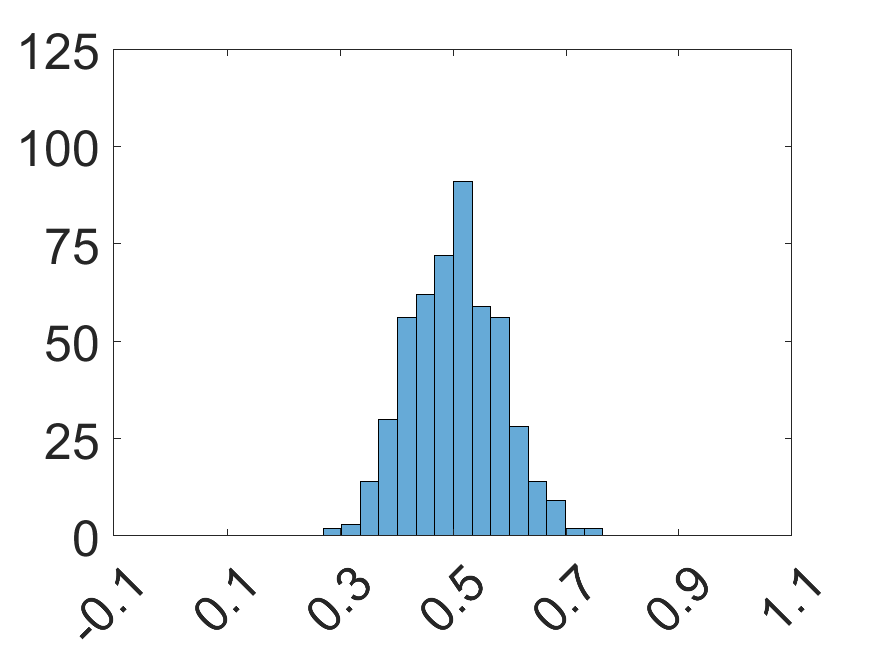}
					\caption{$N=100, m=100$}
				\end{subfigure}
				\hfill
				\begin{subfigure}{.24\textwidth}
					\centering
					\includegraphics[width=\linewidth]{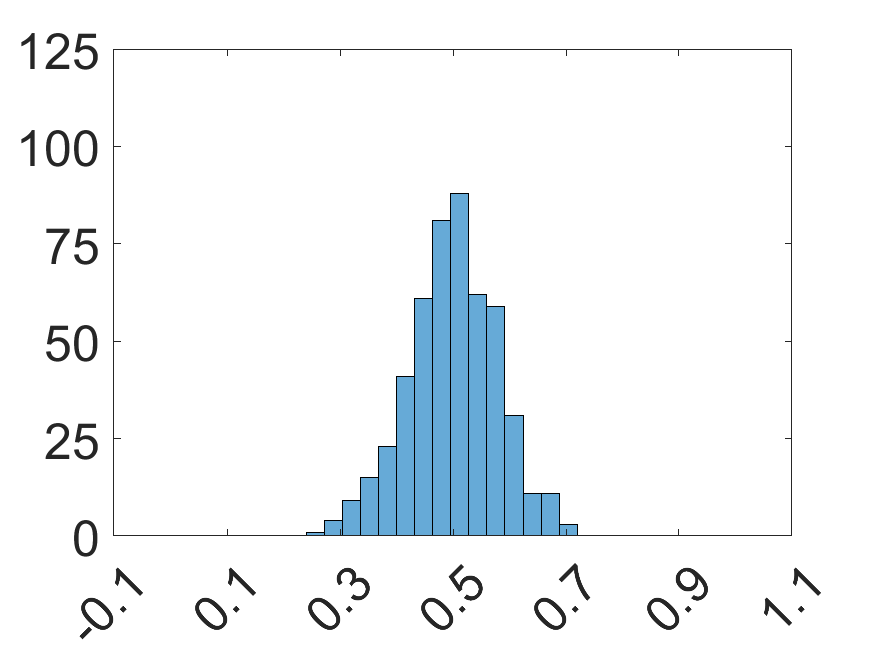}
					\caption{$N=200, m=100$}
				\end{subfigure}
				\subcaption*{\textit{Note:} the true value is $\beta= 0.5$.}
				\label{fig:MC_hist_beta_CaseI_T50}
			\end{figure}
		\end{landscape}

\begin{landscape}		
			\begin{figure}[H]
				\centering
				\caption{Monte Carlo histograms of the network effect $\widehat{\beta}$ - $T=100$, case I}
				\begin{subfigure}{.24\textwidth}
					\centering
					\includegraphics[width=\linewidth]{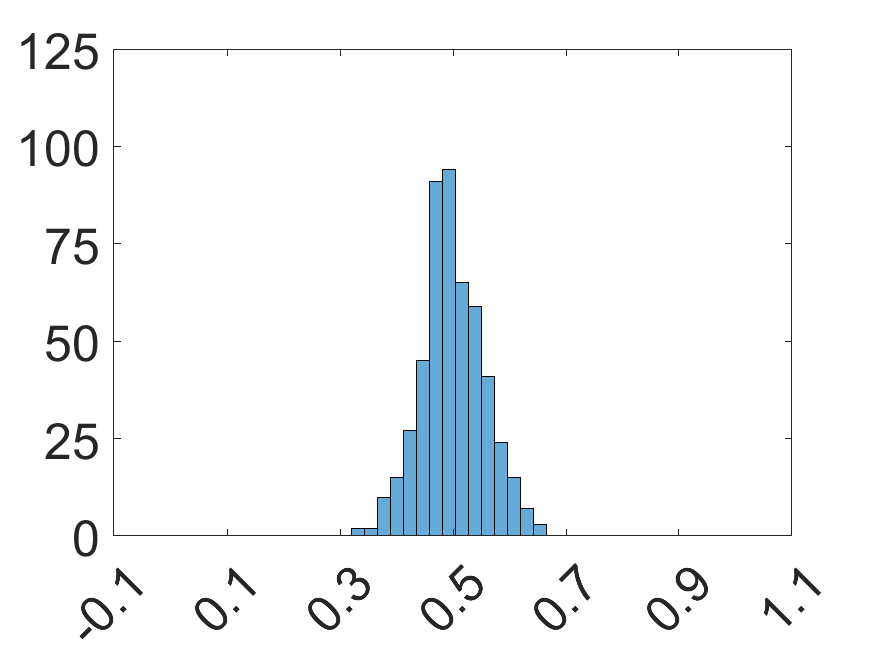}
					\caption{$N=10, m=20$}
				\end{subfigure}%
				\hfill
				\begin{subfigure}{.24\textwidth}
					\centering
					\includegraphics[width=\linewidth]{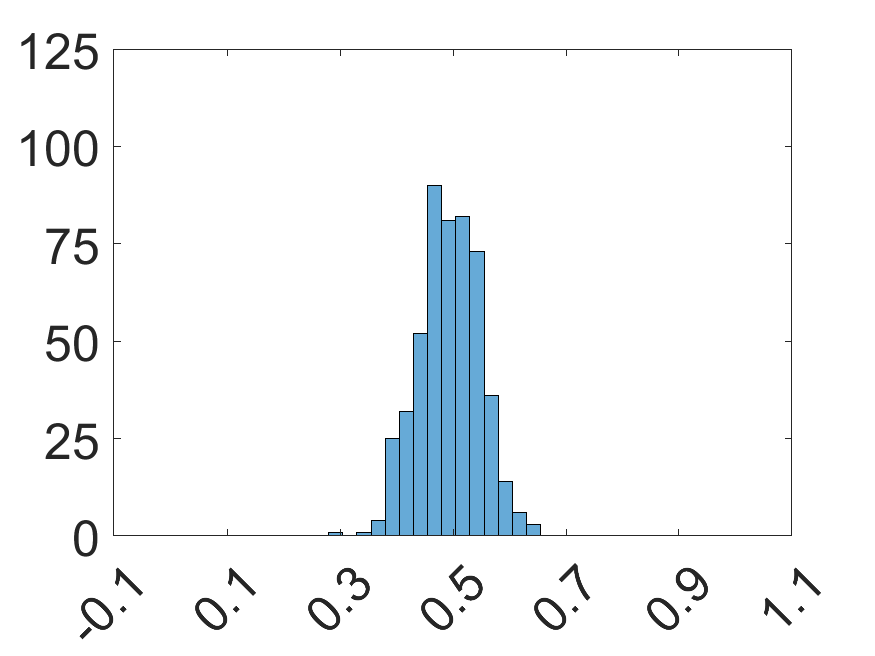}
					\caption{$N=20, m=20$}
				\end{subfigure}
				\hfill
				\begin{subfigure}{.24\textwidth}
					\centering
					\includegraphics[width=\linewidth]{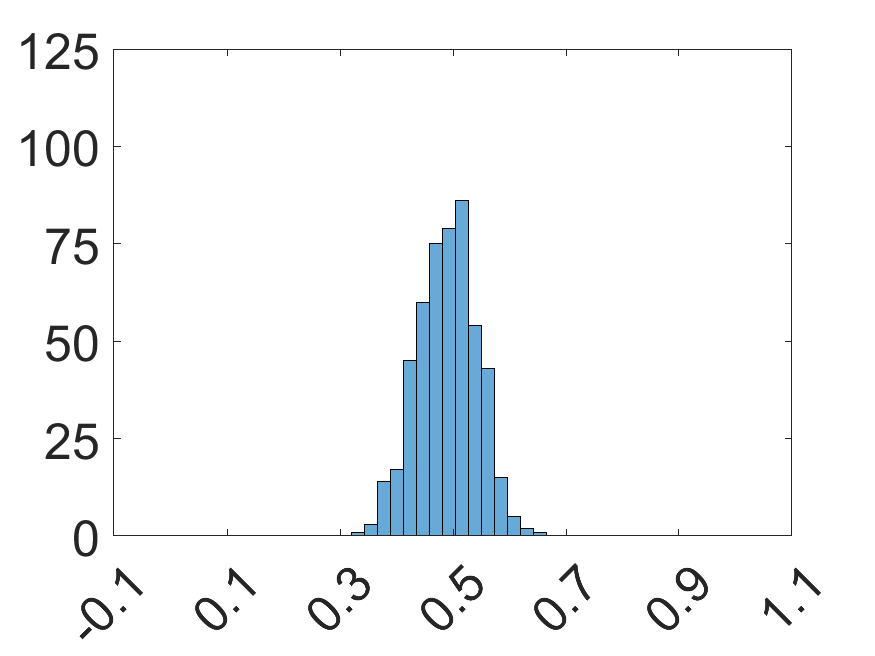}
					\caption{$N=50, m=20$}
				\end{subfigure}
				\hfill
				\begin{subfigure}{.24\textwidth}
					\centering
					\includegraphics[width=\linewidth]{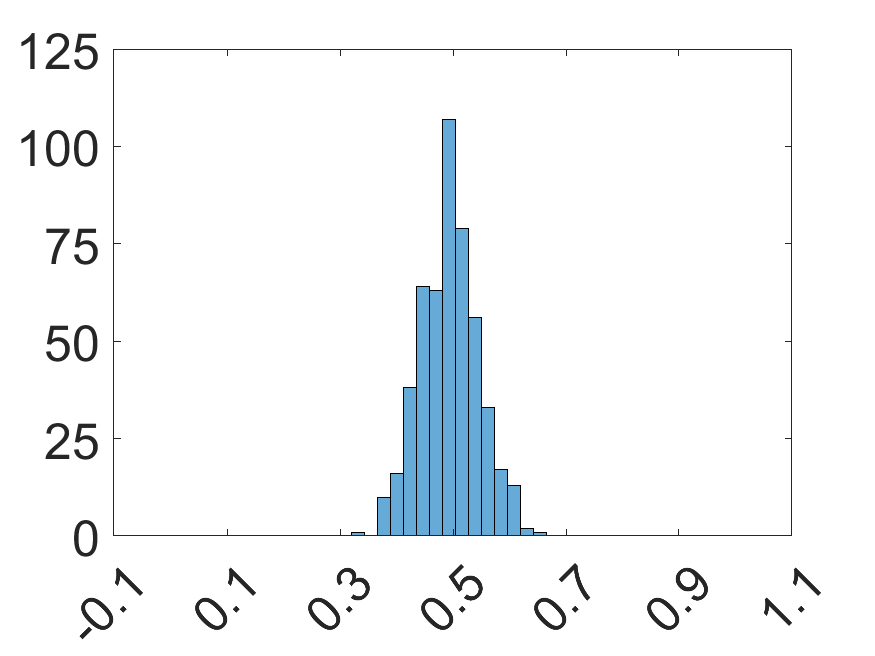}
					\caption{$N=100, m=20$}
				\end{subfigure}		
				\hfill
				\begin{subfigure}{.24\textwidth}
					\centering
					\includegraphics[width=\linewidth]{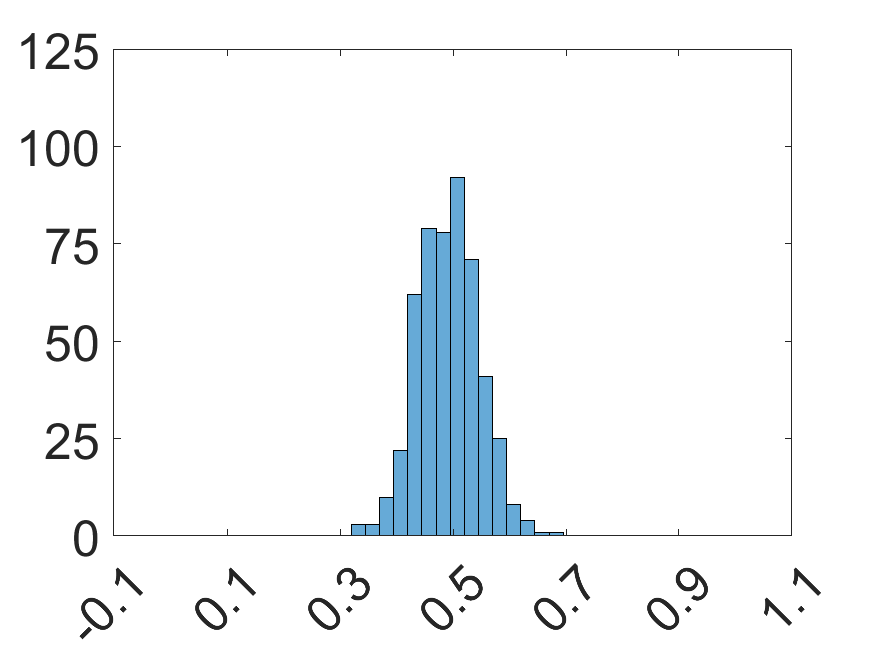}
					\caption{$N=200, m=20$}
				\end{subfigure}	
				
				\begin{subfigure}{.24\textwidth}
					\centering
					\includegraphics[width=\linewidth]{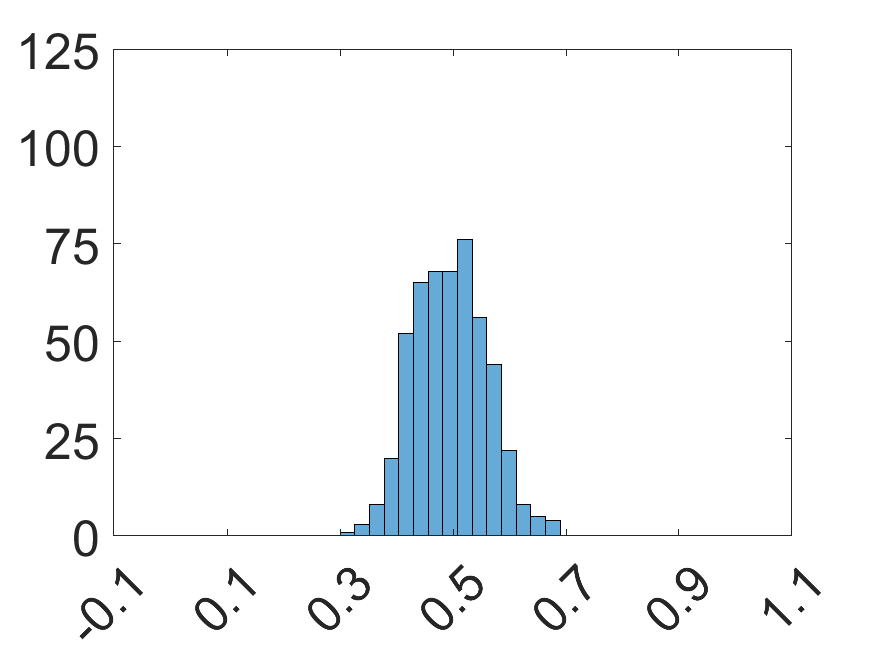}
					\caption{$N=10, m=50$}
				\end{subfigure}%
				\hfill
				\begin{subfigure}{.24\textwidth}
					\centering
					\includegraphics[width=\linewidth]{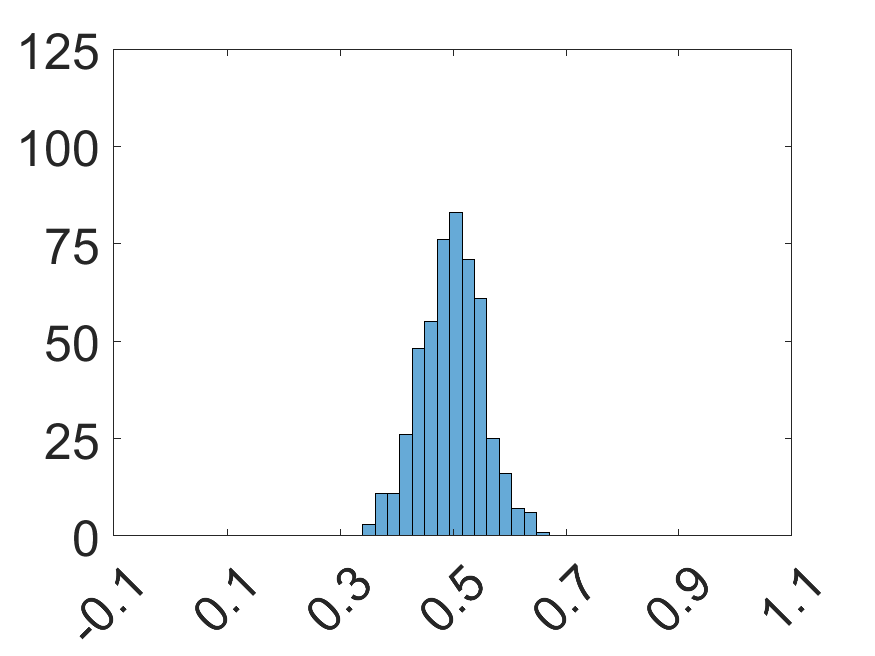}
					\caption{$N=20, m=50$}
				\end{subfigure}
				\hfill
				\begin{subfigure}{.24\textwidth}
					\centering
					\includegraphics[width=\linewidth]{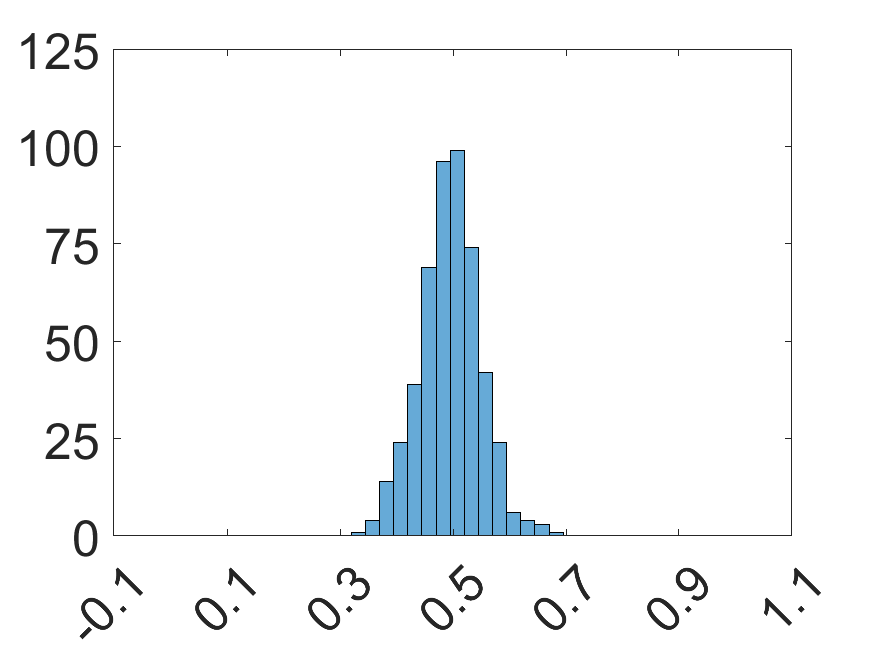}
					\caption{$N=50, m=50$}
				\end{subfigure}
				\hfill
				\begin{subfigure}{.24\textwidth}
					\centering
					\includegraphics[width=\linewidth]{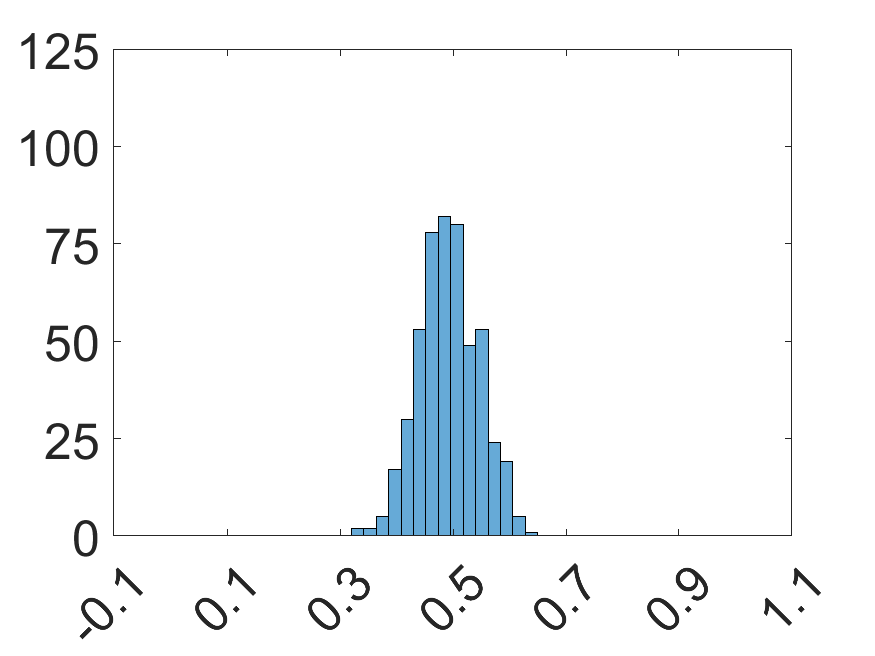}
					\caption{$N=100, m=50$}
				\end{subfigure}
				\hfill
				\begin{subfigure}{.24\textwidth}
					\centering
					\includegraphics[width=\linewidth]{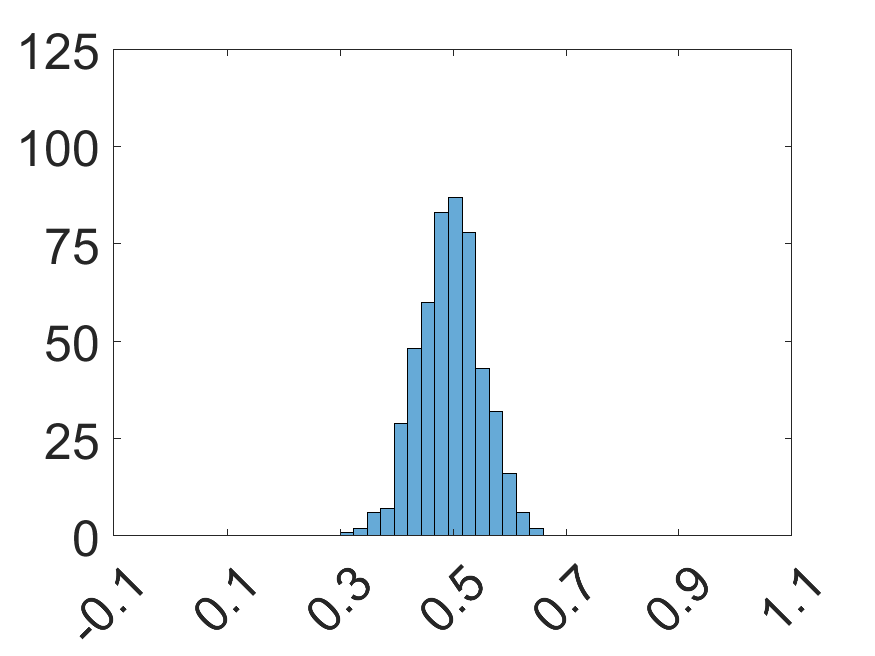}
					\caption{$N=200, m=50$}
				\end{subfigure}
				
				\begin{subfigure}{.24\textwidth}
					\centering
					\includegraphics[width=\linewidth]{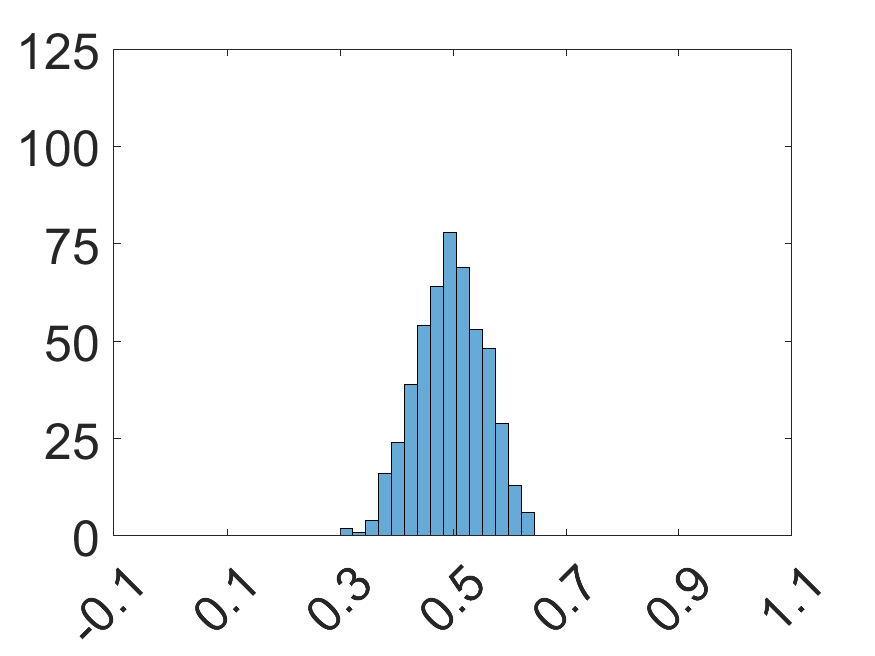}
					\caption{$N=10, m=100$}
				\end{subfigure}%
				\hfill
				\begin{subfigure}{.24\textwidth}
					\centering
					\includegraphics[width=\linewidth]{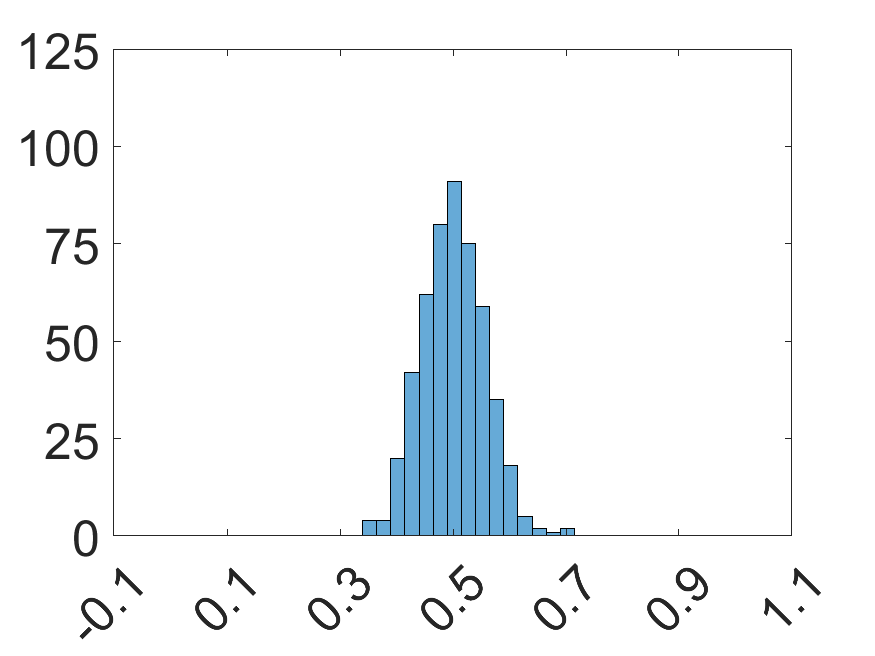}
					\caption{$N=20, m=100$}
				\end{subfigure}
				\hfill
				\begin{subfigure}{.24\textwidth}
					\centering
					\includegraphics[width=\linewidth]{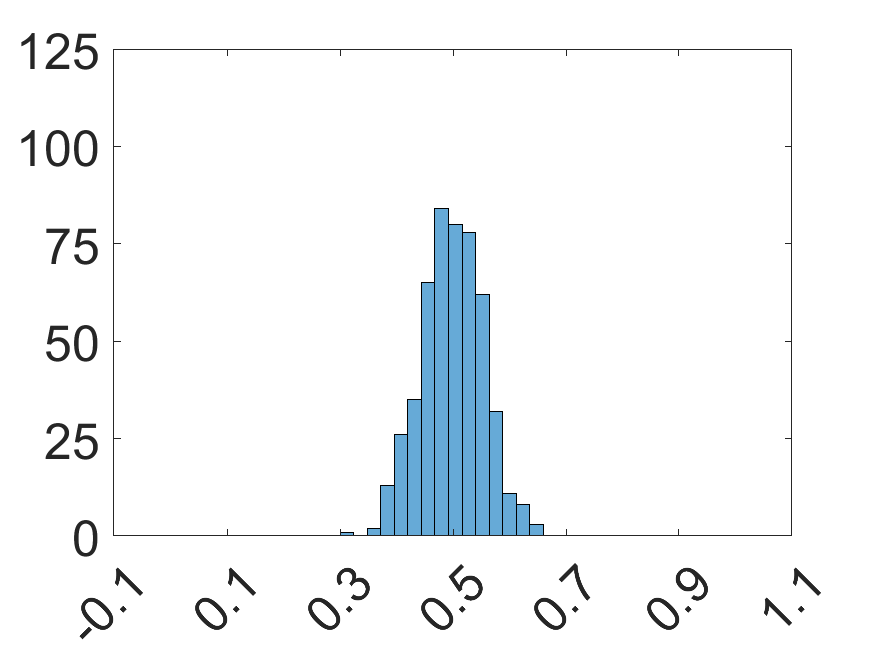}
					\caption{$N=50, m=100$}
				\end{subfigure}
				\hfill
				\begin{subfigure}{.24\textwidth}
					\centering
					\includegraphics[width=\linewidth]{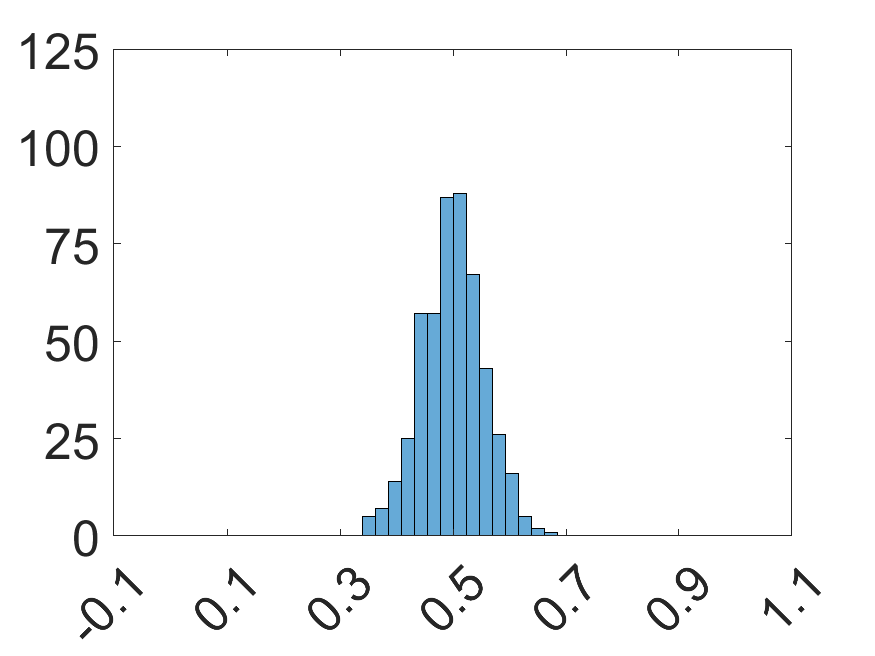}
					\caption{$N=100, m=100$}
				\end{subfigure}
				\hfill
				\begin{subfigure}{.24\textwidth}
					\centering
					\includegraphics[width=\linewidth]{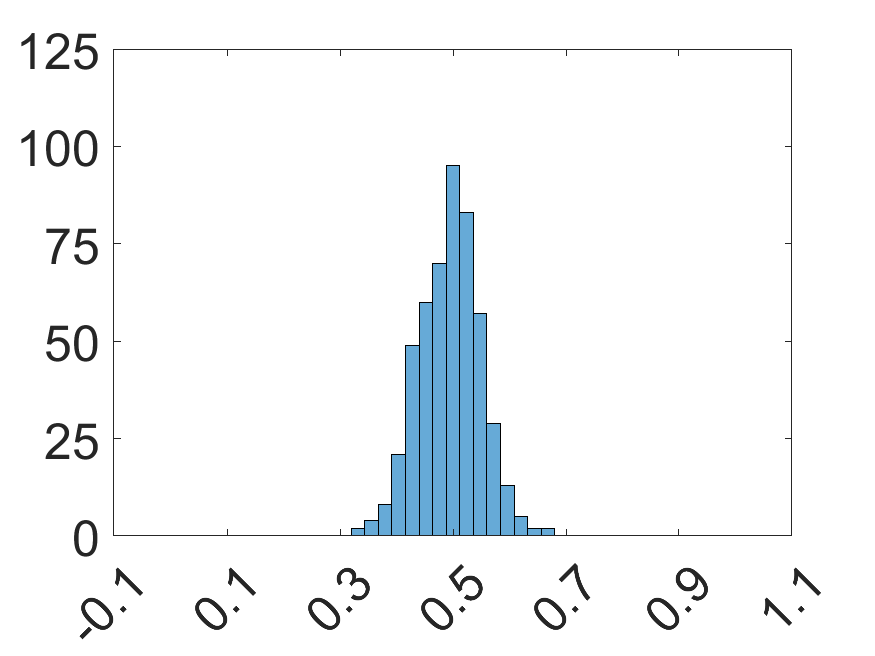}
					\caption{$N=200, m=100$}
				\end{subfigure}
				\subcaption*{\textit{Note:} the true value is $\beta= 0.5$.}
				\label{fig:MC_hist_beta_CaseI_T100}
			\end{figure}
		\end{landscape}

		Tables \ref{tab:MC_RMSE_Case1_T10}, \ref{tab:MC_RMSE_Case1_T50}, \ref{tab:MC_RMSE_Case1_T80} and \ref{tab:MC_RMSE_Case1_T100} report the {RMSE} and {ReRMSE} of parameter estimates for $T=10$, $T=50$, $T=80$ and $T=100$, respectively, under case I. 
		Figures \ref{fig:MC_hist_beta_CaseI_T10}, \ref{fig:MC_hist_beta_CaseI_T50} and \ref{fig:MC_hist_beta_CaseI_T100} show the histograms of the MC estimates of $\beta$, for different combinations of $N,m,T$, under case I.

		\begin{table}[H]
			\centering
			\scriptsize{
				\caption{Monte Carlo RMSEs - $T=10$, case I}\label{tab:MC_RMSE_Case1_T10}
				\scalebox{1}[1]{
					\begin{tabular}{ll | cc|cc|cc|cc|cc }
						\hline \hline
						&& \multicolumn{2}{|c|}{$N = 10$} & \multicolumn{2}{c|}{$N = 20$} & \multicolumn{2}{c|}{$N = 50$}& \multicolumn{2}{c|}{$N = 100$} & \multicolumn{2}{c}{$N = 200$} \\
						&& RMSE        & ReRMSE       & RMSE        & ReRMSE       & RMSE        & ReRMSE   & RMSE  & ReRMSE  & RMSE  & ReRMSE  \\
						\hline
						\multirow{5}{*}{$m = 20$}  & $\beta$     & 0.234 & 46.7\%  & 0.205 & 41.0\% & 0.205 & 41.1\% & 0.203 & 40.6\% & 0.221 & 44.1\% \\
						& $\rho$        & 0.133 & 44.3\%  & 0.096 & 32.0\% & 0.068 & 22.7\% & 0.056 & 18.7\% & 0.050 & 16.7\% \\
						& $\alpha$      & 0.213 & 106.7\% & 0.116 & 58.2\% & 0.082 & 40.8\% & 0.054 & 27.1\% & 0.043 & 21.5\% \\
						& $\mathcal{F}$ & 0.151 & 15.9\%  & 0.153 & 15.7\% & 0.155 & 15.7\% & 0.156 & 15.7\% & 0.156 & 15.7\% \\
						& $U$           & 0.078 & 3.6\%   & 0.046 & 2.1\%  & 0.031 & 1.4\%  & 0.028 & 1.3\%  & 0.028 & 1.3\%  \\
						\hline
						\multirow{5}{*}{$m = 50$}  & $\beta$     & 0.264 & 52.8\%  & 0.220 & 43.9\% & 0.213 & 42.5\% & 0.230 & 46.0\% & 0.233 & 46.7\% \\
						& $\rho$        & 0.144 & 48.0\%  & 0.103 & 34.4\% & 0.071 & 23.5\% & 0.055 & 18.3\% & 0.046 & 15.2\% \\
						& $\alpha$      & 0.171 & 85.5\%  & 0.122 & 61.2\% & 0.074 & 37.0\% & 0.050 & 25.2\% & 0.039 & 19.3\% \\
						& $\mathcal{F}$ & 0.097 & 10.2\%  & 0.098 & 10.0\% & 0.099 & 10.0\% & 0.099 & 10.0\% & 0.099 & 10.0\% \\
						& $U$           & 0.059 & 3.3\%   & 0.031 & 1.7\%  & 0.014 & 0.8\%  & 0.011 & 0.6\%  & 0.010 & 0.5\%  \\
						\hline
						\multirow{5}{*}{$m = 100$} & $\beta$     & 0.235 & 46.9\%  & 0.233 & 46.7\% & 0.210 & 42.1\% & 0.223 & 44.7\% & 0.215 & 43.0\% \\
						& $\rho$        & 0.144 & 47.9\%  & 0.105 & 35.1\% & 0.073 & 24.3\% & 0.057 & 19.1\% & 0.051 & 16.8\% \\
						& $\alpha$      & 0.234 & 117.1\% & 0.113 & 56.6\% & 0.069 & 34.4\% & 0.062 & 31.1\% & 0.043 & 21.4\% \\
						& $\mathcal{F}$ & 0.070 & 7.4\%   & 0.070 & 7.2\%  & 0.070 & 7.1\%  & 0.070 & 7.1\%  & 0.070 & 7.1\%  \\
						& $U$           & 0.053 & 3.3\%   & 0.026 & 1.6\%  & 0.011 & 0.7\%  & 0.007 & 0.4\%  & 0.005 & 0.3\%   \\
						\hline \hline
					\end{tabular}  	
				}
			}
		\end{table}

		\begin{table}[H]
			\centering
			\scriptsize{
				\caption{Monte Carlo RMSEs - $T=50$, case I}\label{tab:MC_RMSE_Case1_T50}
				\scalebox{1}[1]{
					\begin{tabular}{ll | cc|cc|cc|cc|cc }
						\hline \hline
						&& \multicolumn{2}{|c|}{$N = 10$} & \multicolumn{2}{c|}{$N = 20$} & \multicolumn{2}{c|}{$N = 50$}& \multicolumn{2}{c|}{$N = 100$} & \multicolumn{2}{c}{$N = 200$} \\
						&& RMSE        & ReRMSE       & RMSE        & ReRMSE       & RMSE        & ReRMSE   & RMSE  & ReRMSE  & RMSE  & ReRMSE  \\
						\hline
						\multirow{5}{*}{$m = 20$}  & $\beta$     & 0.083 & 16.6\% & 0.078 & 15.5\% & 0.077 & 15.4\% & 0.075 & 14.9\% & 0.076 & 15.3\% \\
						& $\rho$        & 0.044 & 14.8\% & 0.031 & 10.2\% & 0.021 & 7.1\%  & 0.015 & 5.0\%  & 0.011 & 3.6\%  \\
						& $\alpha$      & 0.083 & 41.6\% & 0.043 & 21.5\% & 0.029 & 14.4\% & 0.020 & 10.1\% & 0.014 & 7.2\%  \\
						& $\mathcal{F}$ & 0.149 & 15.7\% & 0.153 & 15.7\% & 0.155 & 15.7\% & 0.156 & 15.7\% & 0.156 & 15.7\% \\
						& $U$           & 0.043 & 1.9\%  & 0.032 & 1.4\%  & 0.028 & 1.3\%  & 0.028 & 1.3\%  & 0.027 & 1.2\%  \\
						\hline
						\multirow{5}{*}{$m = 50$}  & $\beta$     & 0.089 & 17.9\% & 0.078 & 15.6\% & 0.073 & 14.6\% & 0.080 & 16.1\% & 0.077 & 15.4\% \\
						& $\rho$        & 0.047 & 15.7\% & 0.032 & 10.7\% & 0.020 & 6.7\%  & 0.014 & 4.8\%  & 0.010 & 3.5\%  \\
						& $\alpha$      & 0.068 & 34.1\% & 0.045 & 22.4\% & 0.030 & 15.2\% & 0.019 & 9.4\%  & 0.015 & 7.4\%  \\
						& $\mathcal{F}$ & 0.095 & 10.0\% & 0.097 & 10.0\% & 0.099 & 10.0\% & 0.099 & 10.0\% & 0.099 & 10.0\% \\
						& $U$           & 0.028 & 1.6\%  & 0.016 & 0.9\%  & 0.010 & 0.6\%  & 0.009 & 0.5\%  & 0.009 & 0.5\%  \\
						\hline
						\multirow{5}{*}{$m = 100$} & $\beta$     & 0.085 & 17.1\% & 0.081 & 16.1\% & 0.078 & 15.6\% & 0.079 & 15.7\% & 0.078 & 15.7\% \\
						& $\rho$        & 0.047 & 15.6\% & 0.031 & 10.3\% & 0.021 & 7.0\%  & 0.015 & 5.0\%  & 0.011 & 3.6\%  \\
						& $\alpha$      & 0.087 & 43.6\% & 0.044 & 22.1\% & 0.027 & 13.6\% & 0.019 & 9.6\%  & 0.014 & 7.0\%  \\
						& $\mathcal{F}$ & 0.068 & 7.1\%  & 0.069 & 7.1\%  & 0.070 & 7.1\%  & 0.070 & 7.1\%  & 0.070 & 7.1\%  \\
						& $U$           & 0.024 & 1.5\%  & 0.012 & 0.8\%  & 0.006 & 0.4\%  & 0.005 & 0.3\%  & 0.004 & 0.3\%  \\
						\hline \hline
					\end{tabular}  	
				}
			}
		\end{table}

		\begin{table}[H]
			\centering
			\scriptsize{
				\caption{Monte Carlo RMSEs - $T=80$, case I}\label{tab:MC_RMSE_Case1_T80}
				\scalebox{1}[1]{
					\begin{tabular}{ll | cc|cc|cc|cc|cc }
						\hline \hline
						&& \multicolumn{2}{|c|}{$N = 10$} & \multicolumn{2}{c|}{$N = 20$} & \multicolumn{2}{c|}{$N = 50$}& \multicolumn{2}{c|}{$N = 100$} & \multicolumn{2}{c}{$N = 200$} \\
						&& RMSE        & ReRMSE       & RMSE        & ReRMSE       & RMSE        & ReRMSE   & RMSE  & ReRMSE  & RMSE  & ReRMSE  \\
						\hline
						\multirow{5}{*}{$m = 20$}  & $\beta$     & 0.060 & 12.1\% & 0.064 & 12.7\% & 5.7\%  & 11.5\% & 0.061 & 12.1\% &  0.061    &  12.3\%      \\
						& $\rho$        & 0.034 & 11.5\% & 0.025 & 8.2\%  & 1.6\%  & 5.4\%  & 0.011 & 3.7\%  &  0.008    &  2.7\%      \\
						& $\alpha$      & 0.065 & 32.4\% & 0.033 & 16.3\% & 2.3\%  & 11.4\% & 0.017 & 8.3\%  &  0.012    &  6.0\%      \\
						& $\mathcal{F}$ & 0.149 & 15.7\% & 0.153 & 15.7\% & 15.5\% & 15.7\% & 0.156 & 15.7\% & 0.156     &  15.7\%      \\
						& $U$           & 0.038 & 1.7\%  & 0.030 & 1.4\%  & 2.8\%  & 1.3\%  & 0.028 & 1.3\%  & 0.027     &  1.2\%      \\
						\hline
						\multirow{5}{*}{$m = 50$}  & $\beta$     & 0.068 & 13.6\% & 0.058 & 11.6\% & 6.3\%  & 12.7\% & 0.061 & 12.2\% & 0.06 & 12.4\% \\
						& $\rho$        & 0.036 & 11.9\% & 0.024 & 7.8\%  & 1.6\%  & 5.2\%  & 0.011 & 3.8\%  & 0.01 & 2.7\%  \\
						& $\alpha$      & 0.057 & 28.6\% & 0.036 & 18.1\% & 2.2\%  & 11.2\% & 0.014 & 7.1\%  & 0.01 & 5.8\%  \\
						& $\mathcal{F}$ & 0.095 & 10.0\% & 0.097 & 10.0\% & 9.9\%  & 10.0\% & 0.099 & 10.0\% & 0.10 & 10.0\% \\
						& $U$           & 0.023 & 1.3\%  & 0.013 & 0.7\%  & 1.0\%  & 0.6\%  & 0.009 & 0.5\%  & 0.01 & 0.5\%  \\
						\hline
						\multirow{5}{*}{$m = 100$} & $\beta$     & 0.068 & 13.7\% & 0.060 & 12.1\% & 6.7\%  & 13.4\% & 0.062 & 12.5\% & 0.06 & 12.0\% \\
						& $\rho$        & 0.036 & 12.1\% & 0.025 & 8.2\%  & 1.6\%  & 5.3\%  & 0.011 & 3.7\%  & 0.01 & 2.9\%  \\
						& $\alpha$      & 0.064 & 32.0\% & 0.036 & 18.2\% & 2.1\%  & 10.4\% & 0.016 & 8.1\%  & 0.01 & 5.7\%  \\
						& $\mathcal{F}$ & 0.067 & 7.1\%  & 0.069 & 7.1\%  & 7.0\%  & 7.1\%  & 0.070 & 7.1\%  & 0.07 & 7.1\%  \\
						& $U$           & 0.019 & 1.2\%  & 0.010 & 0.6\%  & 0.5\%  & 0.3\%  & 0.005 & 0.3\%  & 0.00 & 0.3\%    \\
						\hline \hline
					\end{tabular}   	
				}
			}
		\end{table}

		\begin{table}[H]
			\centering
			\scriptsize{
				\caption{Monte Carlo RMSEs - $T=100$, case I}\label{tab:MC_RMSE_Case1_T100}
				\scalebox{1}[1]{
					\begin{tabular}{ll | cc|cc|cc|cc|cc }
						\hline \hline
						&& \multicolumn{2}{|c|}{$N = 10$} & \multicolumn{2}{c|}{$N = 20$} & \multicolumn{2}{c|}{$N = 50$}& \multicolumn{2}{c|}{$N = 100$} & \multicolumn{2}{c}{$N = 200$} \\
						&& RMSE        & ReRMSE       & RMSE        & ReRMSE       & RMSE        & ReRMSE   & RMSE  & ReRMSE   & RMSE  & ReRMSE   \\
						\hline
						\multirow{5}{*}{$m = 20$}  & $\beta$     & 0.055 & 11.0\% & 0.055 & 11.0\% & 0.053 & 10.7\% & 0.052 & 10.4\% & 0.055 & 11.0\% \\
						& $\rho$        & 0.031 & 10.2\% & 0.022 & 7.2\%  & 0.014 & 4.6\%  & 0.010 & 3.3\%  & 0.008 & 2.5\%  \\
						& $\alpha$      & 0.056 & 28.1\% & 0.029 & 14.4\% & 0.020 & 10.1\% & 0.014 & 7.2\%  & 0.010 & 5.0\%  \\
						& $\mathcal{F}$ & 0.149 & 15.7\% & 0.153 & 15.7\% & 0.155 & 15.7\% & 0.156 & 15.7\% & 0.156 & 15.7\% \\
						& $U$           & 0.036 & 1.6\%  & 0.029 & 1.3\%  & 0.028 & 1.3\%  & 0.028 & 1.3\%  & 0.027 & 1.2\%  \\
						\hline
						\multirow{5}{*}{$m = 50$}  & $\beta$     & 0.064 & 12.9\% & 0.055 & 11.0\% & 0.054 & 10.8\% & 0.053 & 10.6\% & 0.056 & 11.2\% \\
						& $\rho$        & 0.031 & 10.3\% & 0.021 & 7.0\%  & 0.014 & 4.7\%  & 0.010 & 3.3\%  & 0.008 & 2.5\%  \\
						& $\alpha$      & 0.048 & 24.0\% & 0.032 & 15.8\% & 0.019 & 9.6\%  & 0.014 & 6.9\%  & 0.010 & 5.2\%  \\
						& $\mathcal{F}$ & 0.095 & 10.0\% & 0.097 & 10.0\% & 0.099 & 10.0\% & 0.099 & 10.0\% & 0.099 & 10.0\% \\
						& $U$           & 0.021 & 1.2\%  & 0.013 & 0.7\%  & 0.010 & 0.5\%  & 0.009 & 0.5\%  & 0.009 & 0.5\%  \\
						\hline
						\multirow{5}{*}{$m = 100$} & $\beta$     & 0.060 & 12.1\% & 0.056 & 11.3\% & 0.054 & 10.8\% & 0.055 & 10.9\% & 0.055 & 11.0\% \\
						& $\rho$        & 0.030 & 10.0\% & 0.022 & 7.4\%  & 0.014 & 4.5\%  & 0.010 & 3.4\%  & 0.007 & 2.5\%  \\
						& $\alpha$      & 0.059 & 29.3\% & 0.030 & 15.1\% & 0.018 & 8.8\%  & 0.014 & 7.1\%  & 0.010 & 4.9\%  \\
						& $\mathcal{F}$ & 0.067 & 7.1\%  & 0.069 & 7.1\%  & 0.070 & 7.1\%  & 0.070 & 7.1\%  & 0.070 & 7.1\%  \\
						& $U$           & 0.017 & 1.1\%  & 0.009 & 0.6\%  & 0.005 & 0.3\%  & 0.004 & 0.3\%  & 0.004 & 0.3\%   \\
						\hline \hline
					\end{tabular}   	
				}
			}
		\end{table}

		\begin{landscape}		
			\begin{figure}[H]
				\centering
				\caption{Monte Carlo histograms of the network effect $\widehat{\beta}$ - $T=80$, case I}
				\begin{subfigure}{.24\textwidth}
					\centering
					\includegraphics[width=\linewidth]{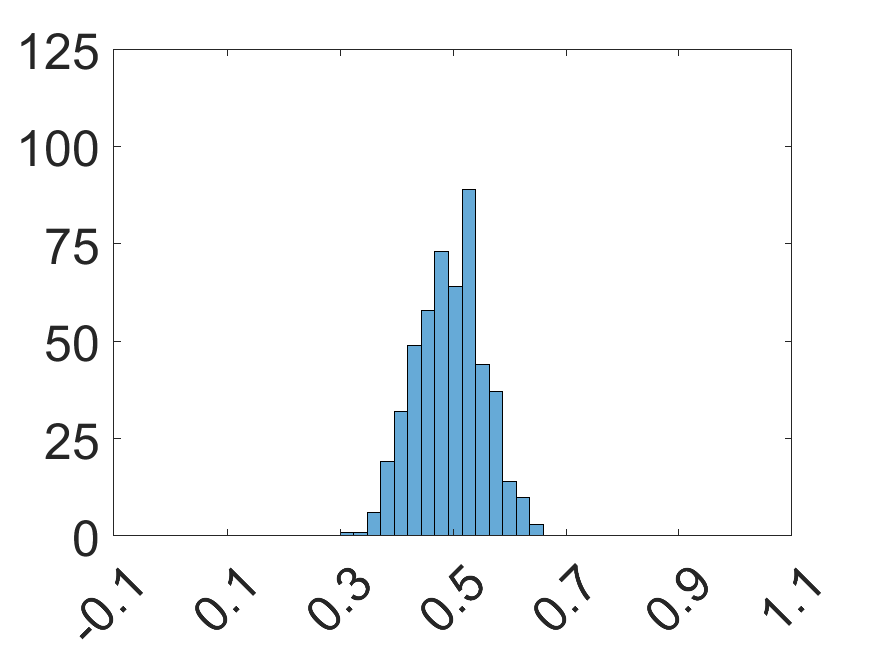}
					\caption{$N=10, m=20$}
				\end{subfigure}%
				\hfill
				\begin{subfigure}{.24\textwidth}
					\centering
					\includegraphics[width=\linewidth]{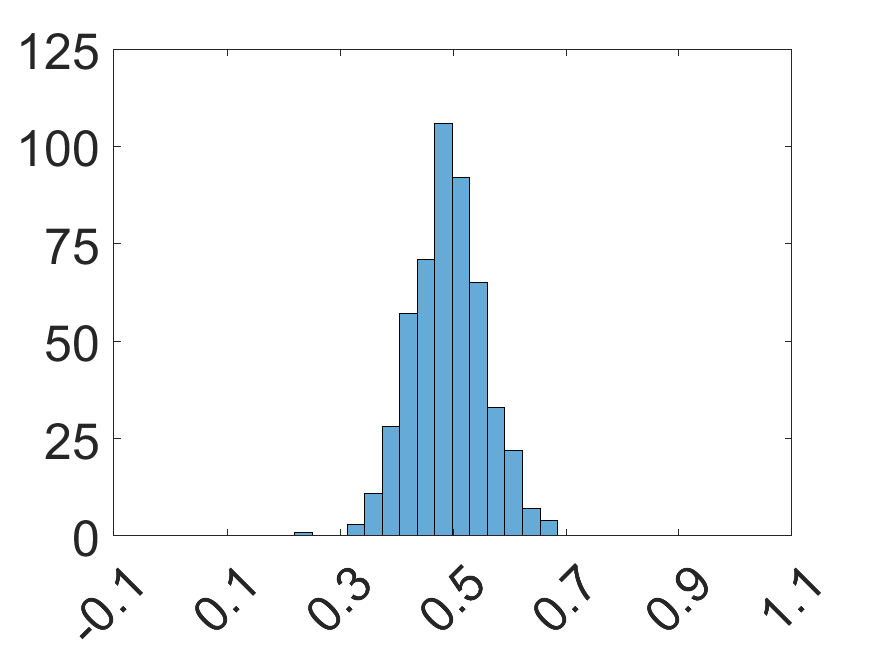}
					\caption{$N=20, m=20$}
				\end{subfigure}
				\hfill
				\begin{subfigure}{.24\textwidth}
					\centering
					\includegraphics[width=\linewidth]{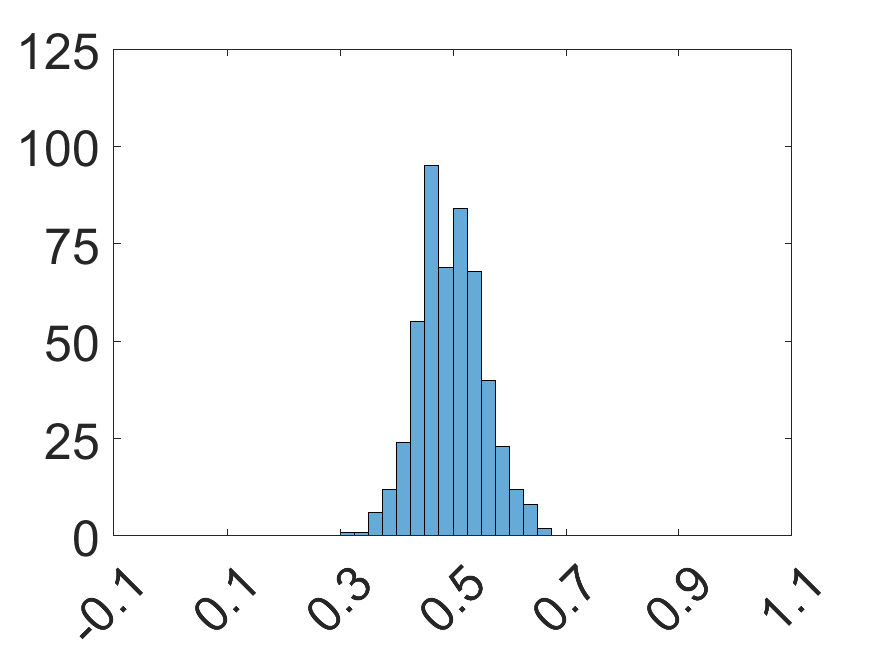}
					\caption{$N=50, m=20$}
				\end{subfigure}
				\hfill
				\begin{subfigure}{.24\textwidth}
					\centering
					\includegraphics[width=\linewidth]{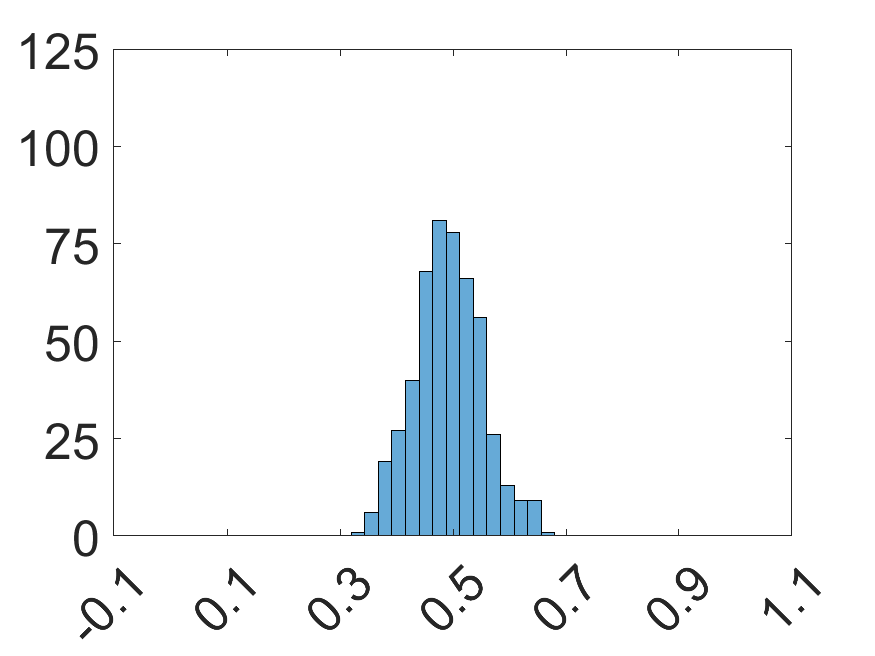}
					\caption{$N=100, m=20$}
				\end{subfigure}	
				\hfill
				\begin{subfigure}{.24\textwidth}
					\centering
					\includegraphics[width=\linewidth]{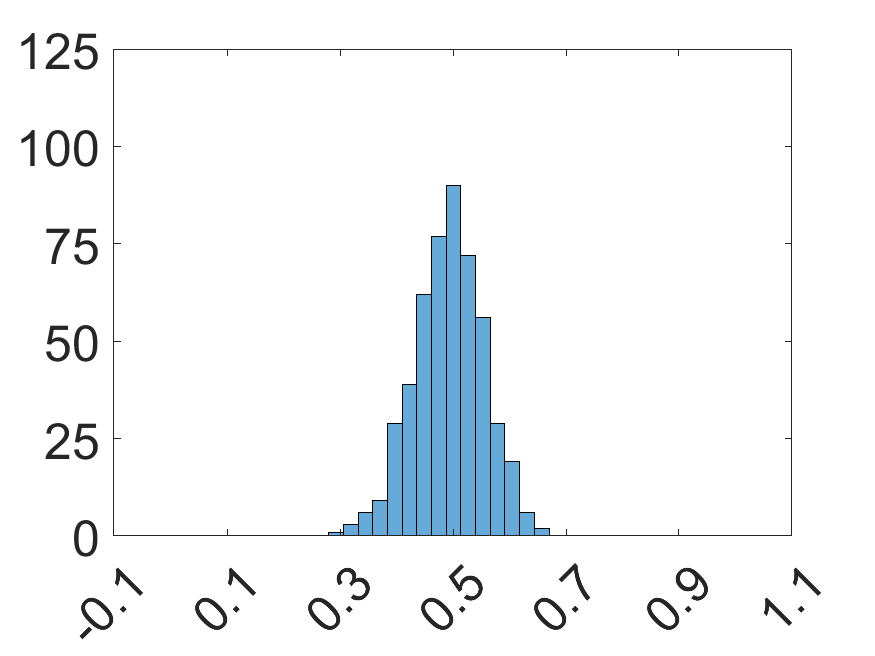}
					\caption{$N=200, m=20$}
				\end{subfigure}		
				
				\begin{subfigure}{.24\textwidth}
					\centering
					\includegraphics[width=\linewidth]{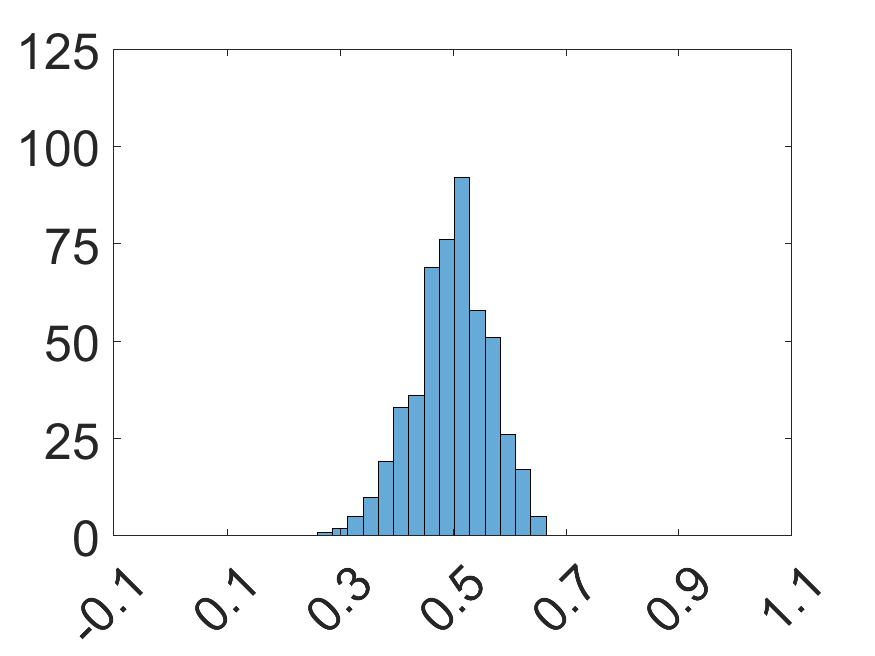}
					\caption{$N=10, m=50$}
				\end{subfigure}%
				\hfill
				\begin{subfigure}{.24\textwidth}
					\centering
					\includegraphics[width=\linewidth]{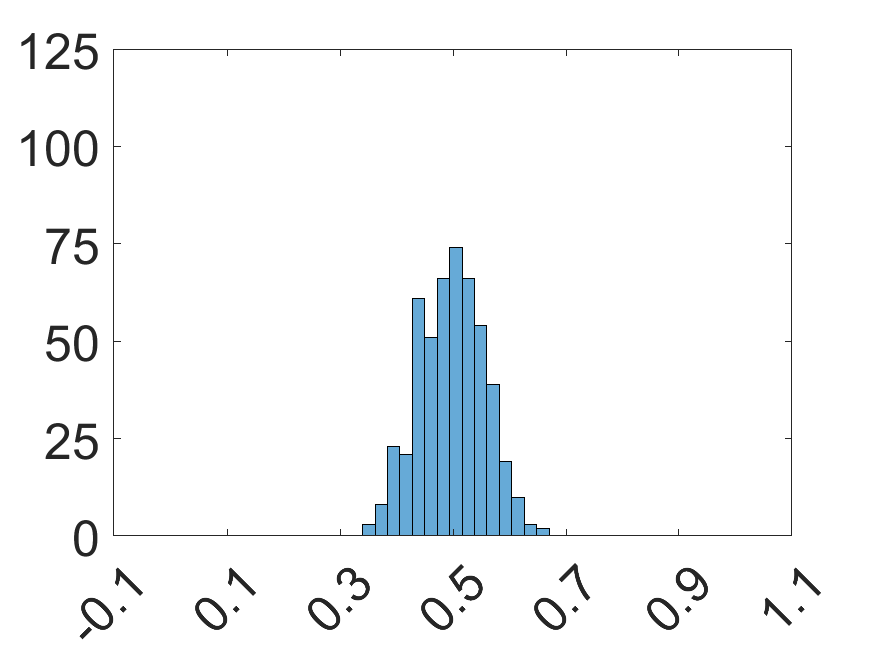}
					\caption{$N=20, m=50$}
				\end{subfigure}
				\hfill
				\begin{subfigure}{.24\textwidth}
					\centering
					\includegraphics[width=\linewidth]{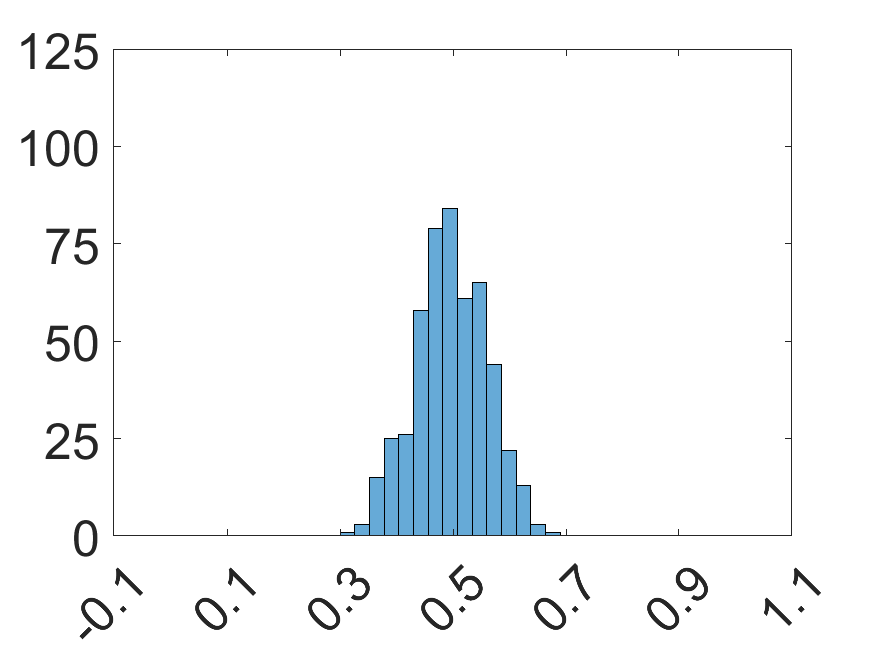}
					\caption{$N=50, m=50$}
				\end{subfigure}
				\hfill
				\begin{subfigure}{.24\textwidth}
					\centering
					\includegraphics[width=\linewidth]{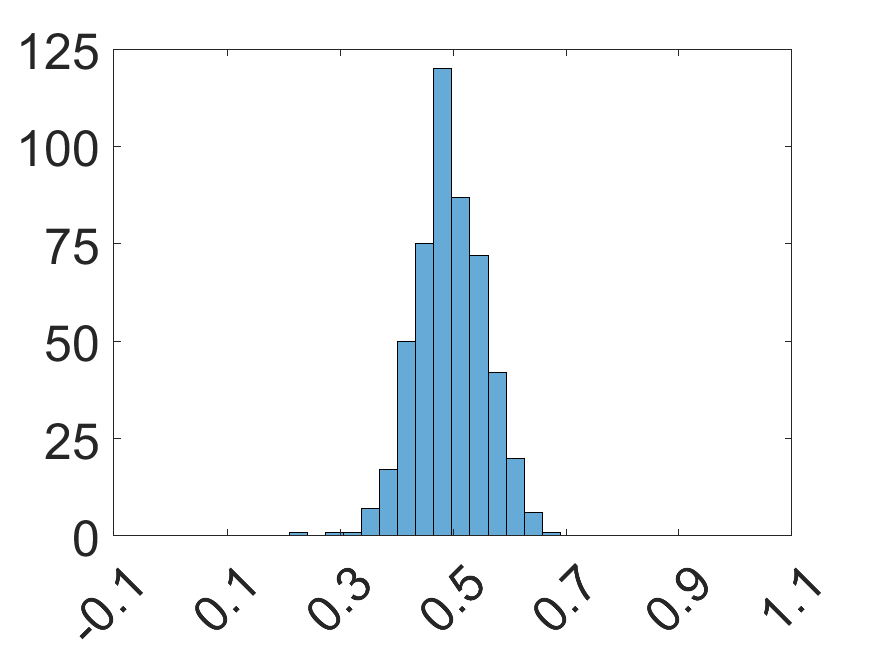}
					\caption{$N=100, m=50$}
				\end{subfigure}
				\hfill
				\begin{subfigure}{.24\textwidth}
					\centering
					\includegraphics[width=\linewidth]{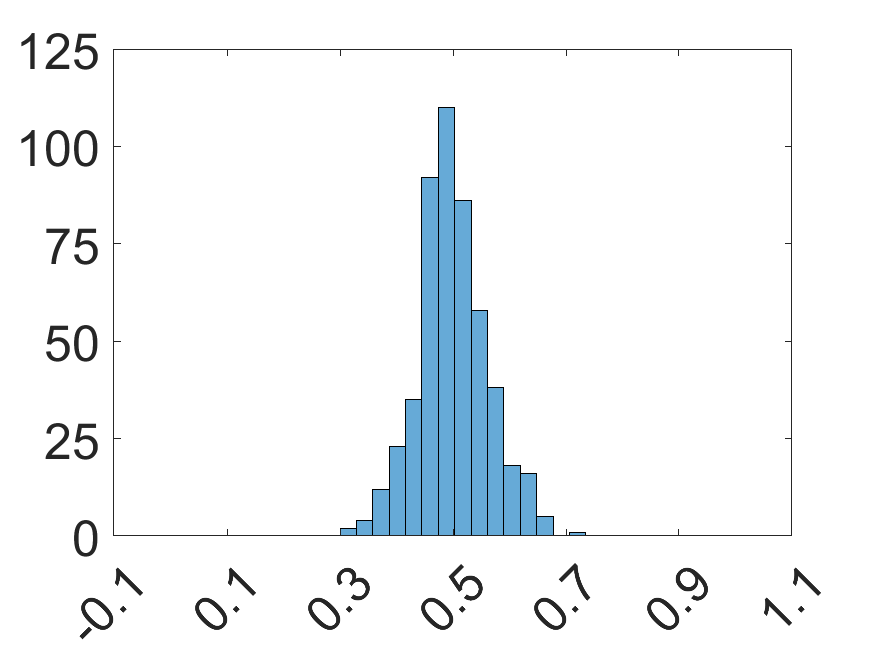}
					\caption{$N=200, m=50$}
				\end{subfigure}
				
				\begin{subfigure}{.24\textwidth}
					\centering
					\includegraphics[width=\linewidth]{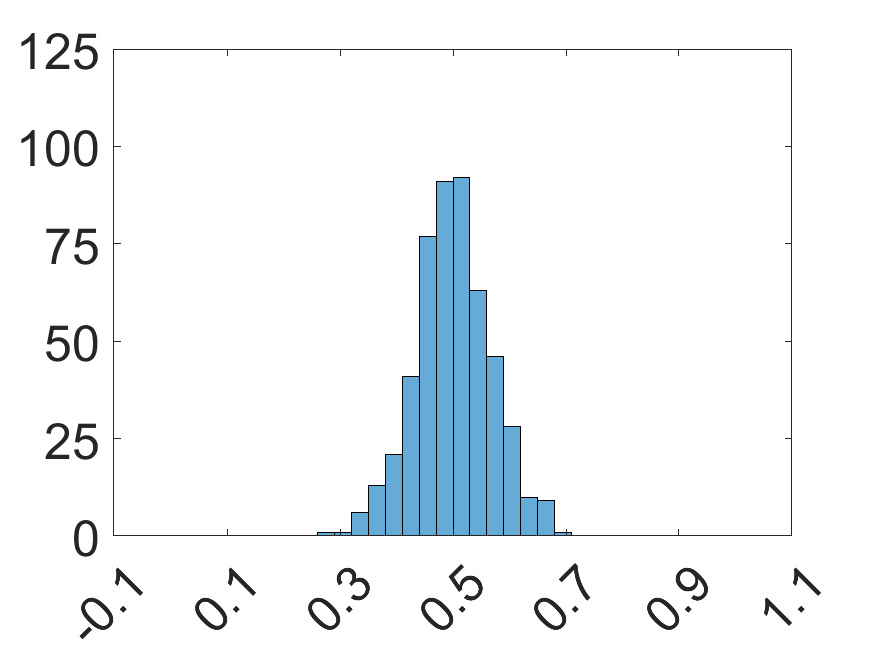}
					\caption{$N=10, m=100$}
				\end{subfigure}%
				\hfill
				\begin{subfigure}{.24\textwidth}
					\centering
					\includegraphics[width=\linewidth]{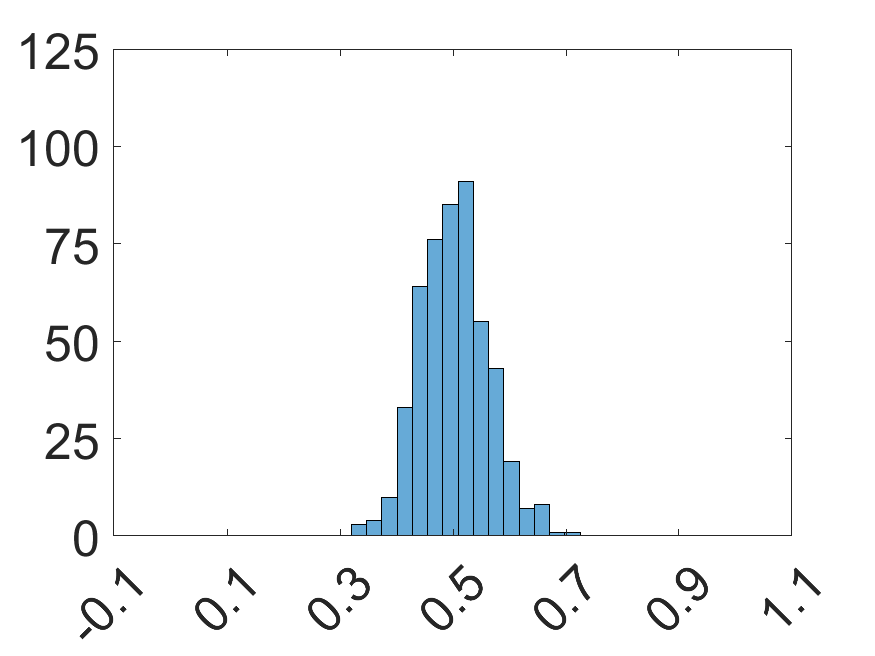}
					\caption{$N=20, m=100$}
				\end{subfigure}
				\hfill
				\begin{subfigure}{.24\textwidth}
					\centering
					\includegraphics[width=\linewidth]{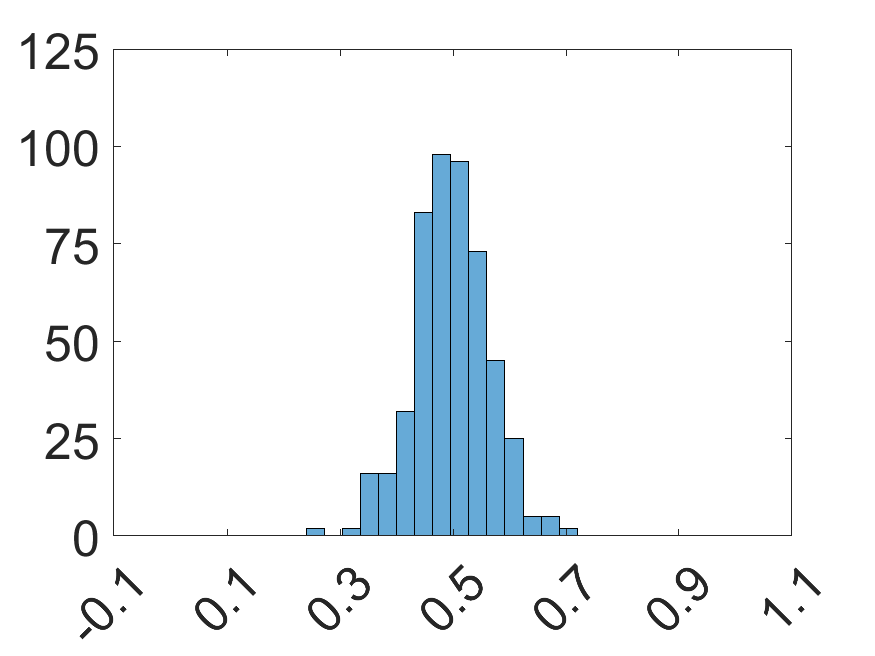}
					\caption{$N=50, m=100$}
				\end{subfigure}
				\hfill
				\begin{subfigure}{.24\textwidth}
					\centering
					\includegraphics[width=\linewidth]{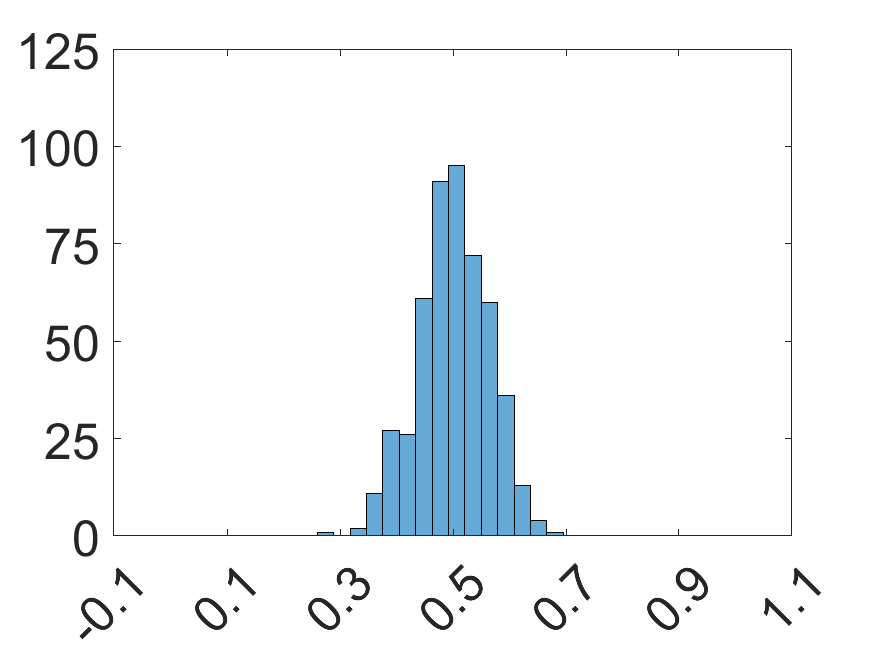}
					\caption{$N=100, m=100$}
				\end{subfigure}
				\hfill
				\begin{subfigure}{.24\textwidth}
					\centering
					\includegraphics[width=\linewidth]{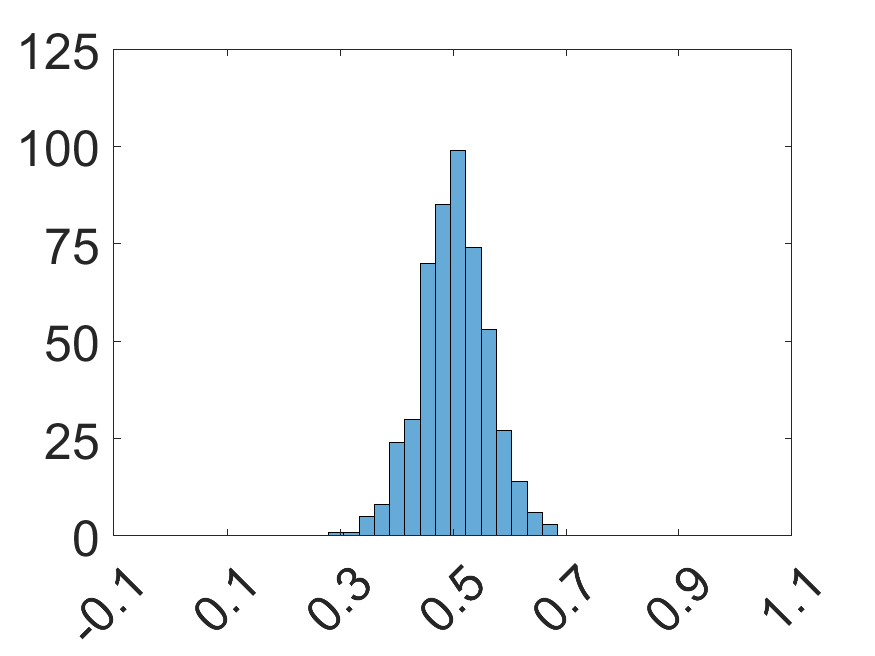}
					\caption{$N=200, m=100$}
				\end{subfigure}
				\subcaption*{\textit{Note:} the true value is $\beta= 0.5$.}
				\label{fig:MC_hist_beta_CaseI_T80}
			\end{figure}
		\end{landscape}

		\subsection{RMSEs and Histograms - case II}\label{app:MC_appendix_res}

\begin{landscape}
			\begin{figure}[H]
				\centering
				\caption{Monte Carlo histograms of the network effect $\widehat{\beta}$ - $T=50$, case II}
				\begin{subfigure}{.24\textwidth}
					\centering
					\includegraphics[width=\linewidth]{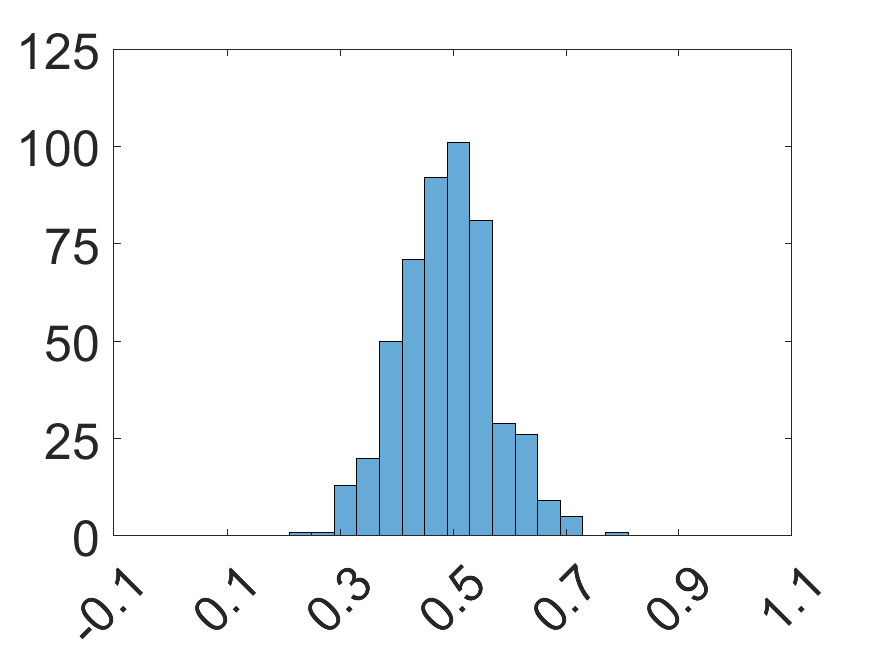}
					\caption{$N=10, m=20$}
				\end{subfigure}%
				\hfill
				\begin{subfigure}{.24\textwidth}
					\centering
					\includegraphics[width=\linewidth]{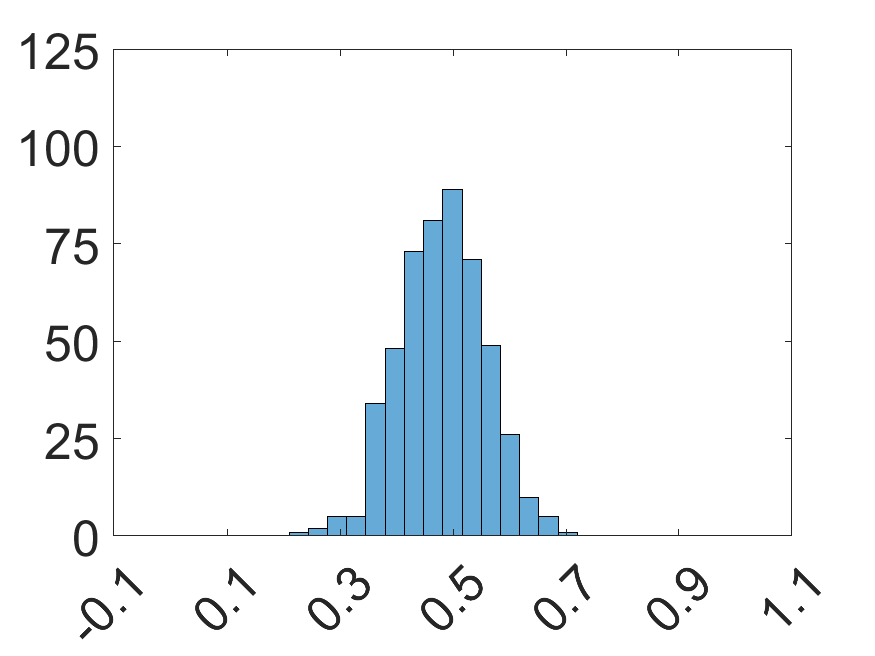}
					\caption{$N=20, m=20$}
				\end{subfigure}
				\hfill
				\begin{subfigure}{.24\textwidth}
					\centering
					\includegraphics[width=\linewidth]{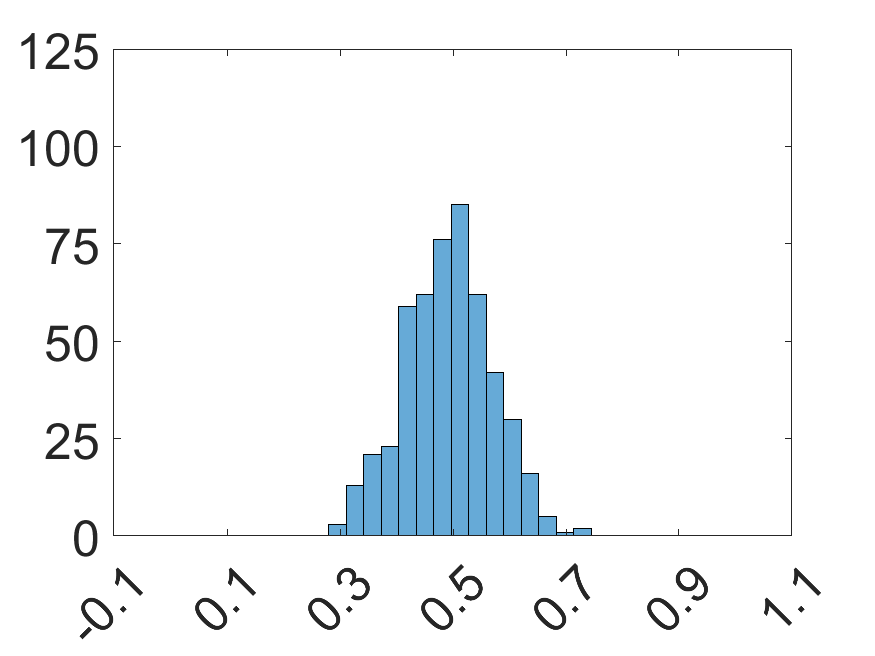}
					\caption{$N=50, m=20$}
				\end{subfigure}
				\hfill
				\begin{subfigure}{.24\textwidth}
					\centering
					\includegraphics[width=\linewidth]{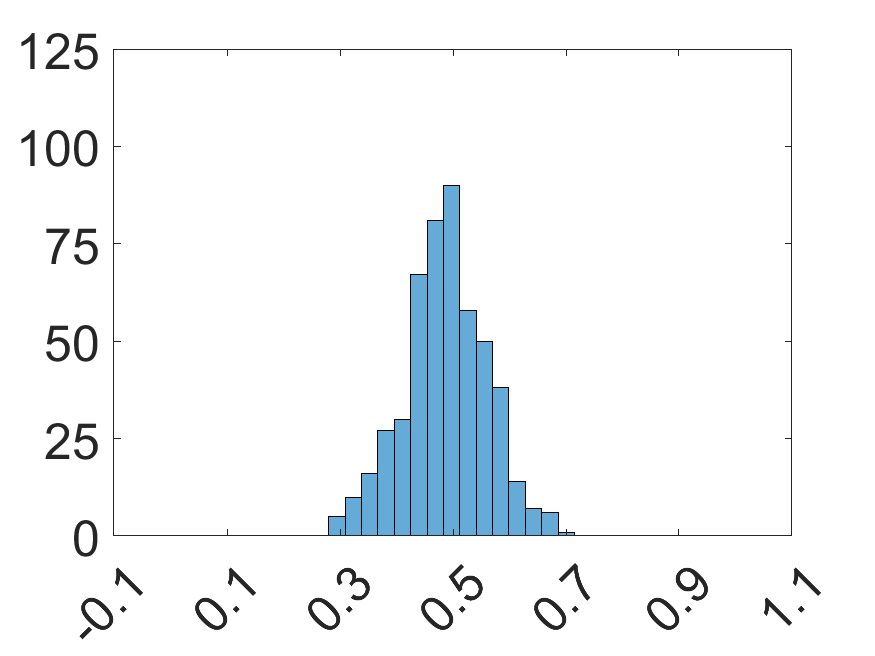}
					\caption{$N=100, m=20$}
				\end{subfigure}
				\hfill
				\begin{subfigure}{.24\textwidth}
					\centering
					\includegraphics[width=\linewidth]{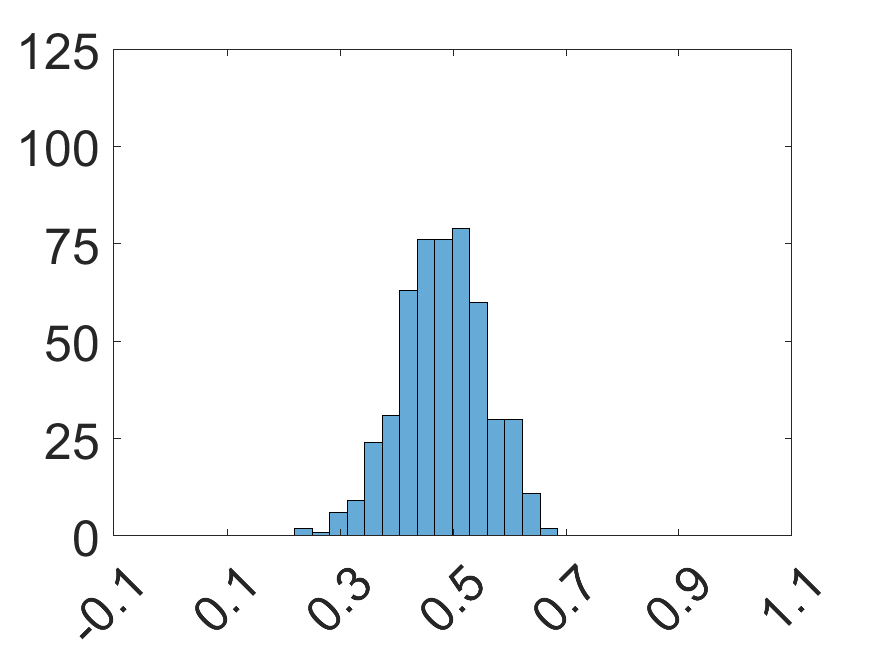}
					\caption{$N=200, m=20$}
				\end{subfigure}	
				
				\begin{subfigure}{.24\textwidth}
					\centering
					\includegraphics[width=\linewidth]{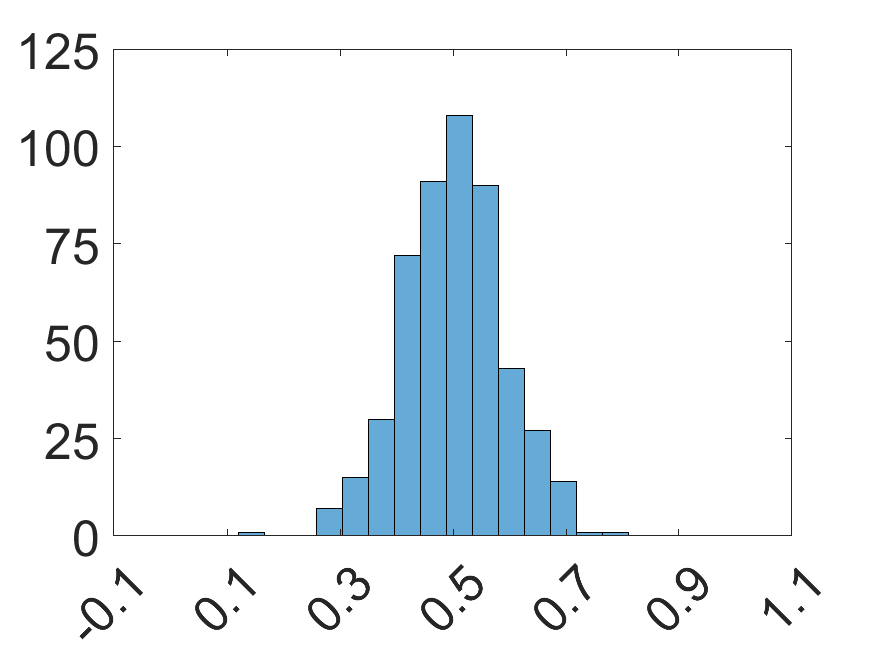}
					\caption{$N=10, m=50$}
				\end{subfigure}%
				\hfill
				\begin{subfigure}{.24\textwidth}
					\centering
					\includegraphics[width=\linewidth]{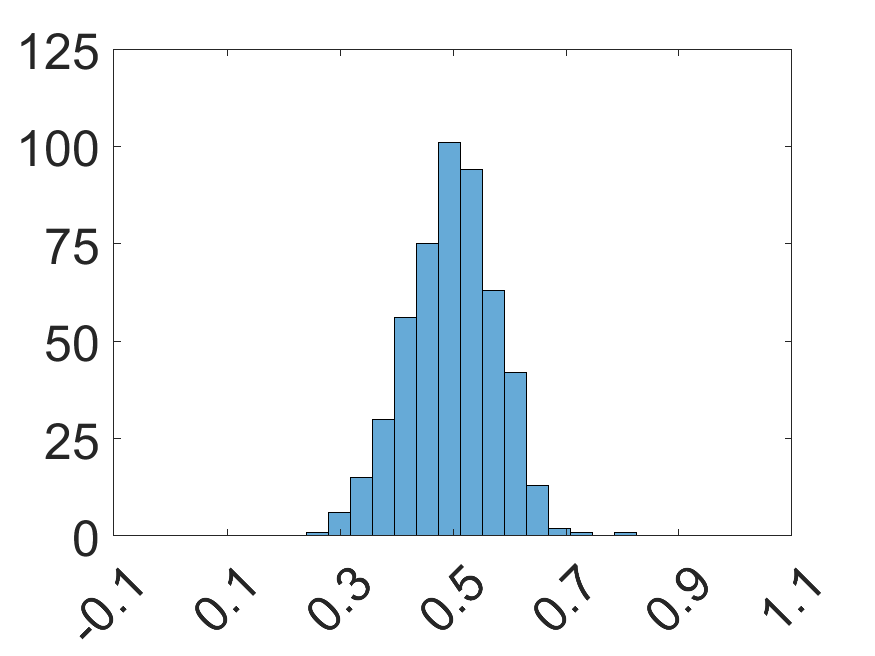}
					\caption{$N=20, m=50$}
				\end{subfigure}
				\hfill
				\begin{subfigure}{.24\textwidth}
					\centering
					\includegraphics[width=\linewidth]{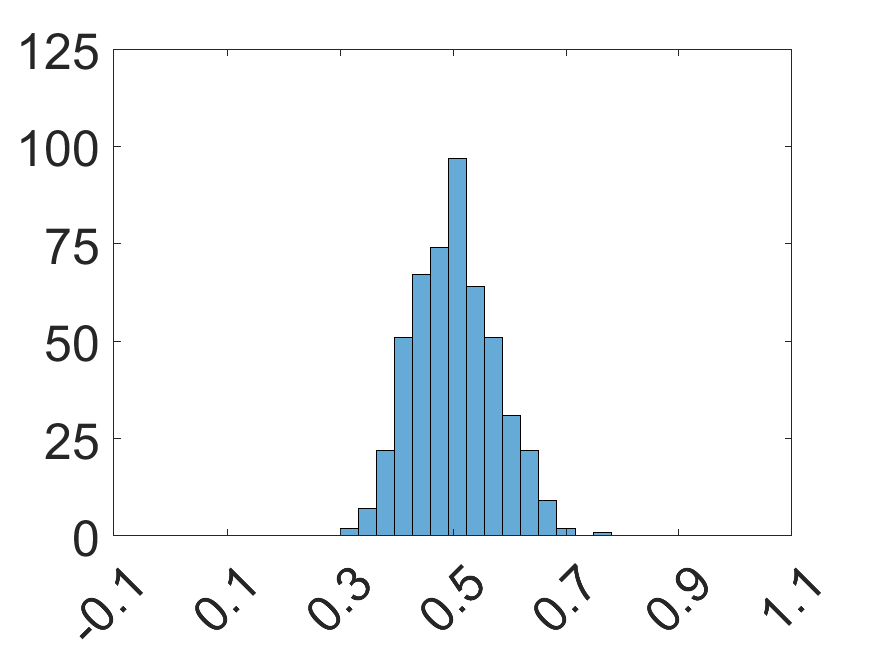}
					\caption{$N=50, m=50$}
				\end{subfigure}
				\hfill
				\begin{subfigure}{.24\textwidth}
					\centering
					\includegraphics[width=\linewidth]{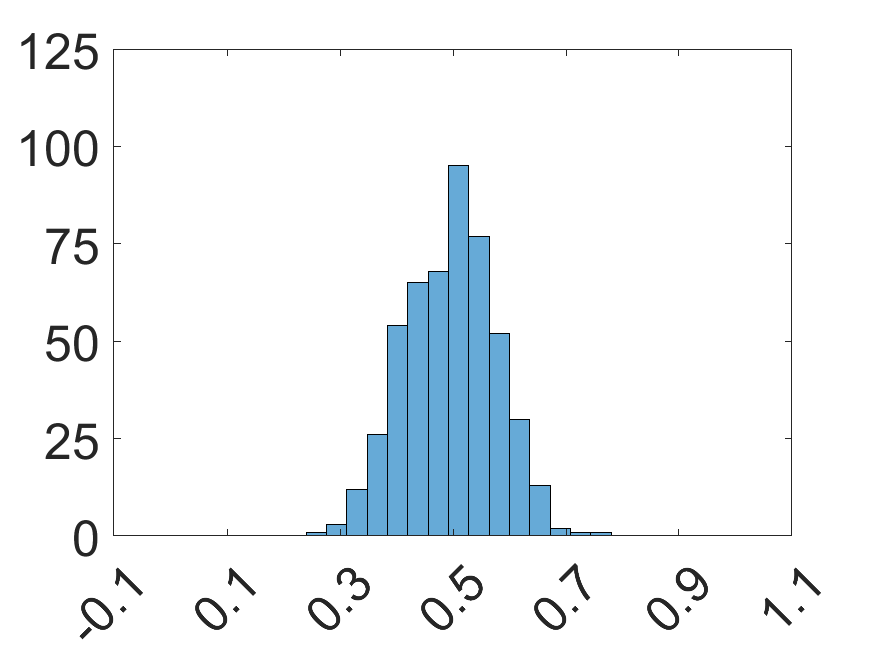}
					\caption{$N=100, m=50$}
				\end{subfigure}
				\hfill
				\begin{subfigure}{.24\textwidth}
					\centering
					\includegraphics[width=\linewidth]{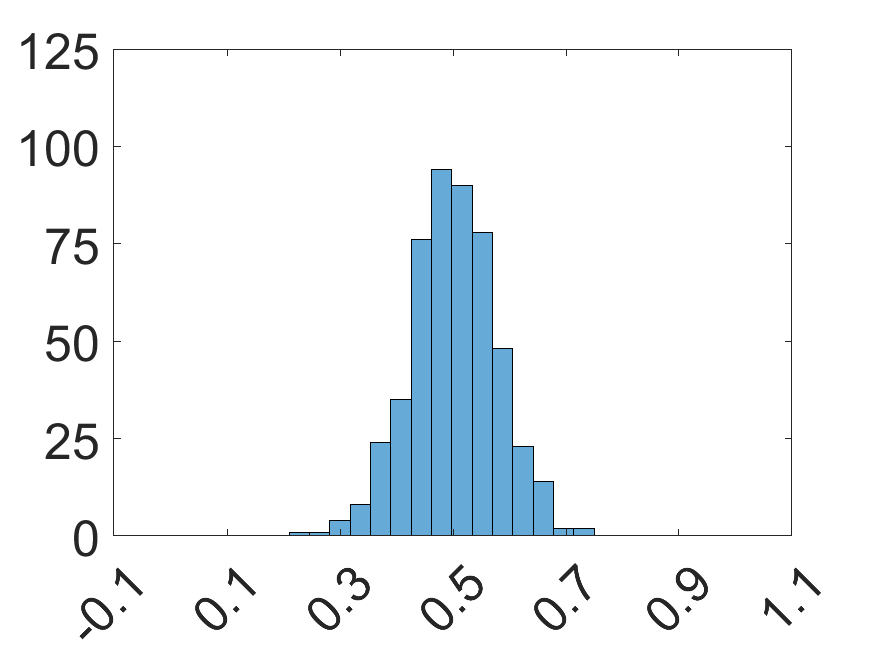}
					\caption{$N=200, m=50$}
				\end{subfigure}
				
				\begin{subfigure}{.24\textwidth}
					\centering
					\includegraphics[width=\linewidth]{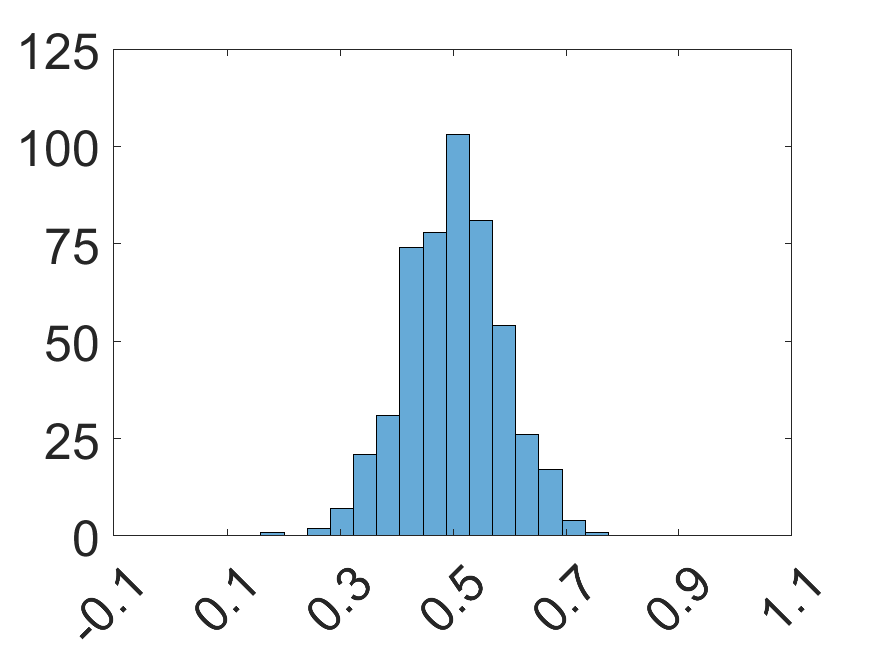}
					\caption{$N=10, m=100$}
				\end{subfigure}%
				\hfill
				\begin{subfigure}{.24\textwidth}
					\centering
					\includegraphics[width=\linewidth]{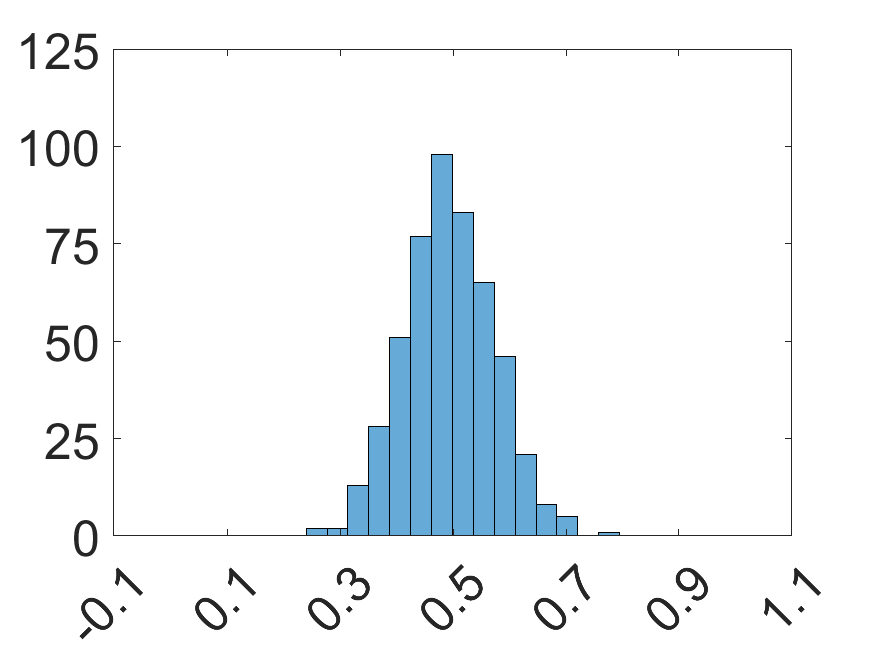}
					\caption{$N=20, m=100$}
				\end{subfigure}
				\hfill
				\begin{subfigure}{.24\textwidth}
					\centering
					\includegraphics[width=\linewidth]{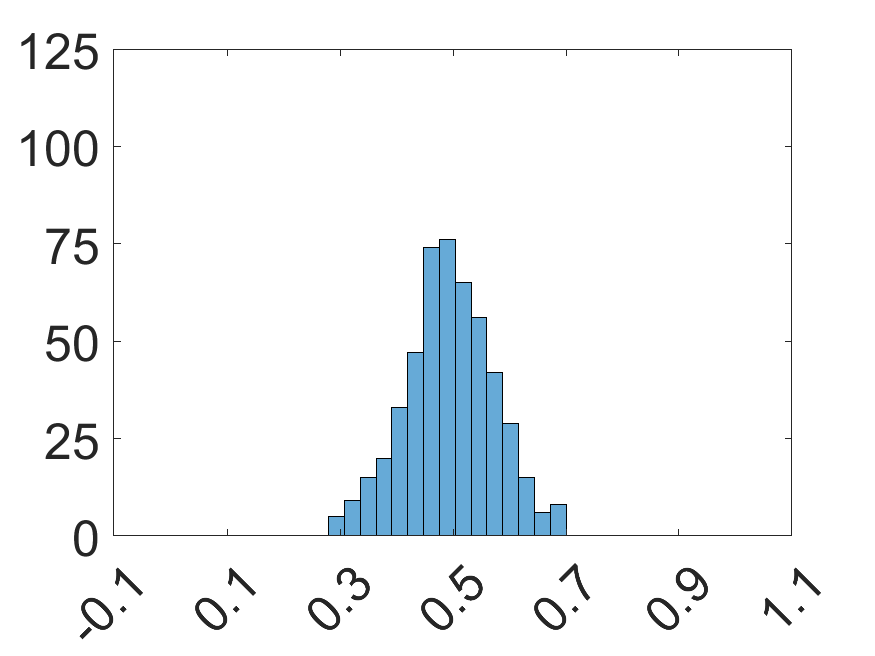}
					\caption{$N=50, m=100$}
				\end{subfigure}
				\hfill
				\begin{subfigure}{.24\textwidth}
					\centering
					\includegraphics[width=\linewidth]{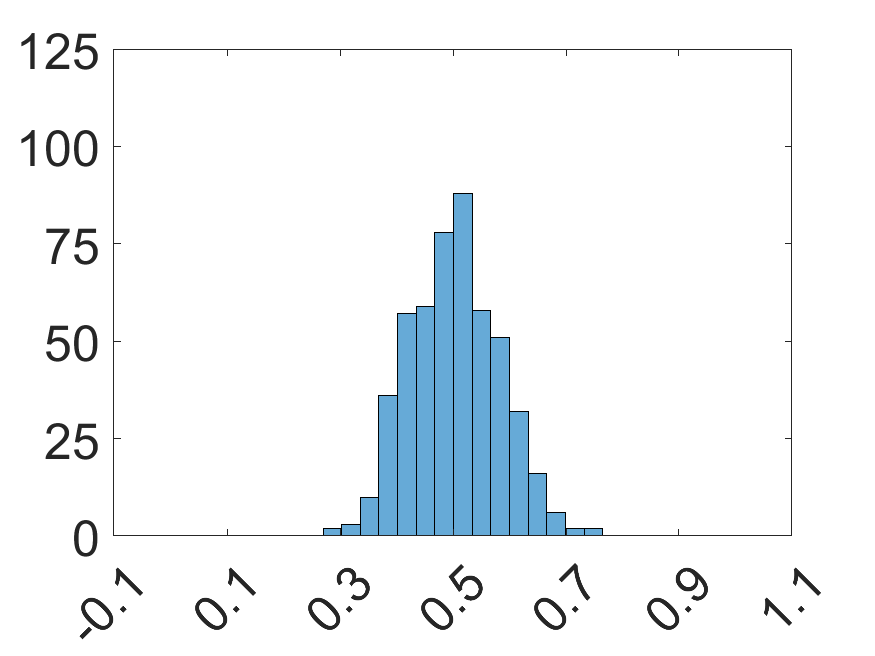}
					\caption{$N=100, m=100$}
				\end{subfigure}		
				\hfill
				\begin{subfigure}{.24\textwidth}
					\centering
					\includegraphics[width=\linewidth]{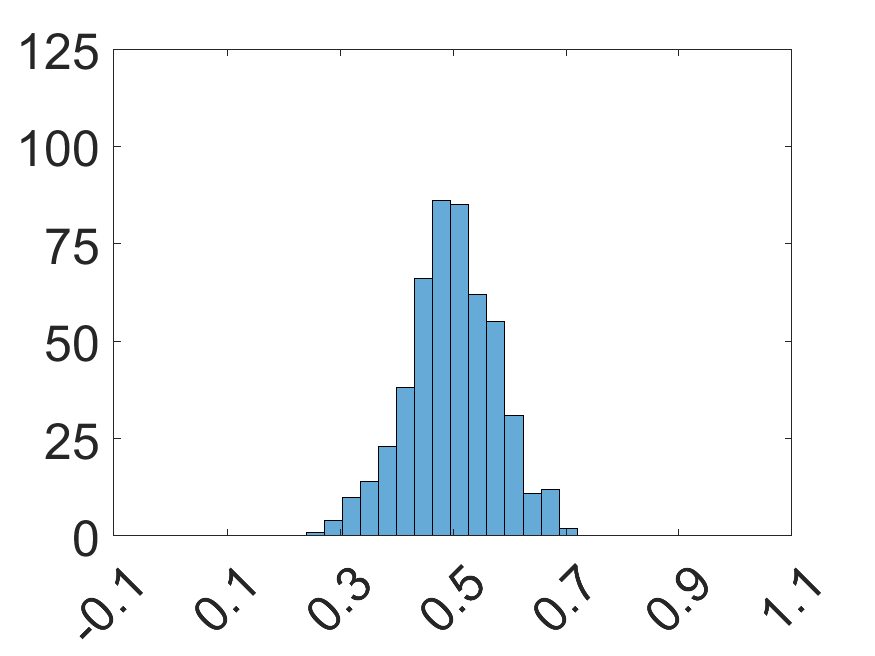}
					\caption{$N=200, m=100$}
				\end{subfigure}
				\subcaption*{\textit{Note:} the true value is $\beta= 0.5$. }
				\label{fig:MC_hist_beta_CaseII_T50}
			\end{figure}
		\end{landscape}

\begin{landscape}			
			\begin{figure}[H]
				\centering
				\caption{Monte Carlo histograms of the network effect $\widehat{\beta}$ - $T=100$, case II}
				\begin{subfigure}{.24\textwidth}
					\centering
					\includegraphics[width=\linewidth]{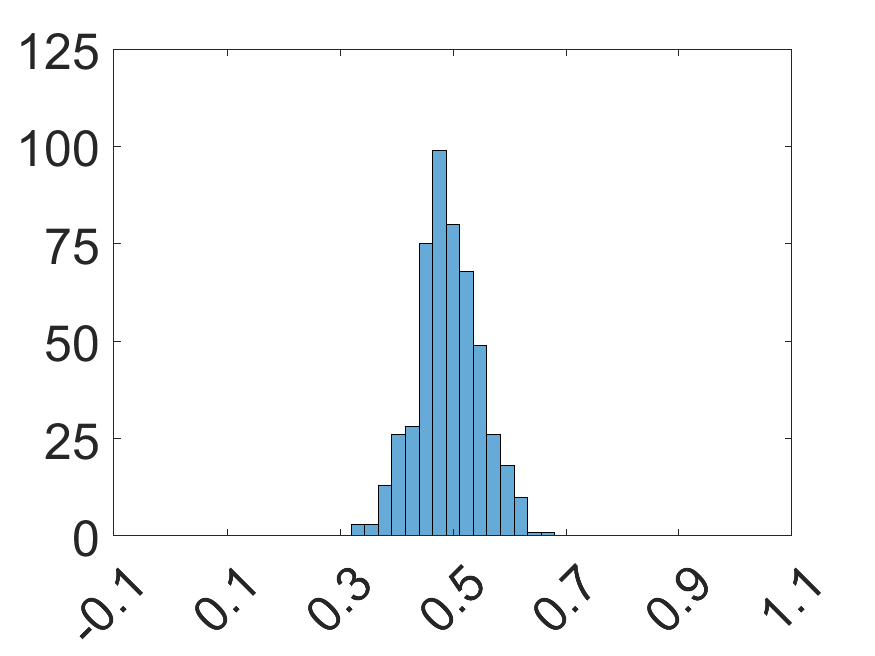}
					\caption{$N=10, m=20$}
				\end{subfigure}%
				\hfill
				\begin{subfigure}{.24\textwidth}
					\centering
					\includegraphics[width=\linewidth]{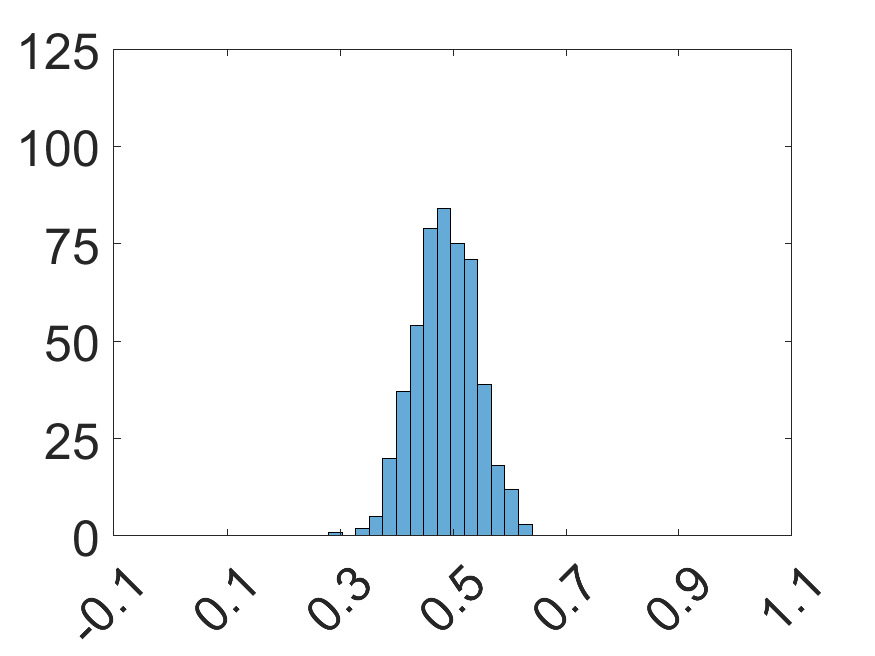}
					\caption{$N=20, m=20$}
				\end{subfigure}
				\hfill
				\begin{subfigure}{.24\textwidth}
					\centering
					\includegraphics[width=\linewidth]{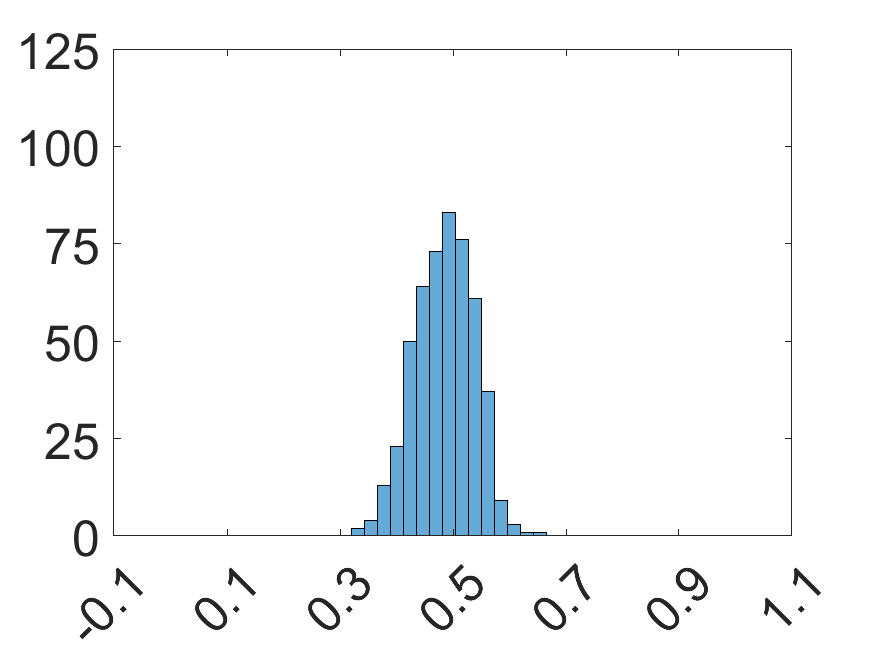}
					\caption{$N=50, m=20$}
				\end{subfigure}
				\hfill
				\begin{subfigure}{.24\textwidth}
					\centering
					\includegraphics[width=\linewidth]{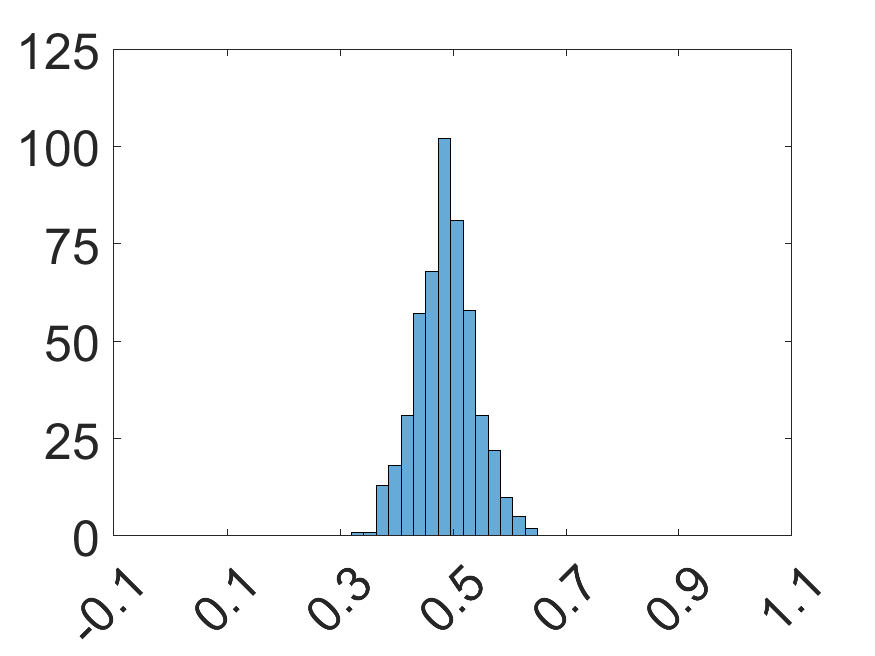}
					\caption{$N=100, m=20$}
				\end{subfigure}
				\hfill
				\begin{subfigure}{.24\textwidth}
					\centering
					\includegraphics[width=\linewidth]{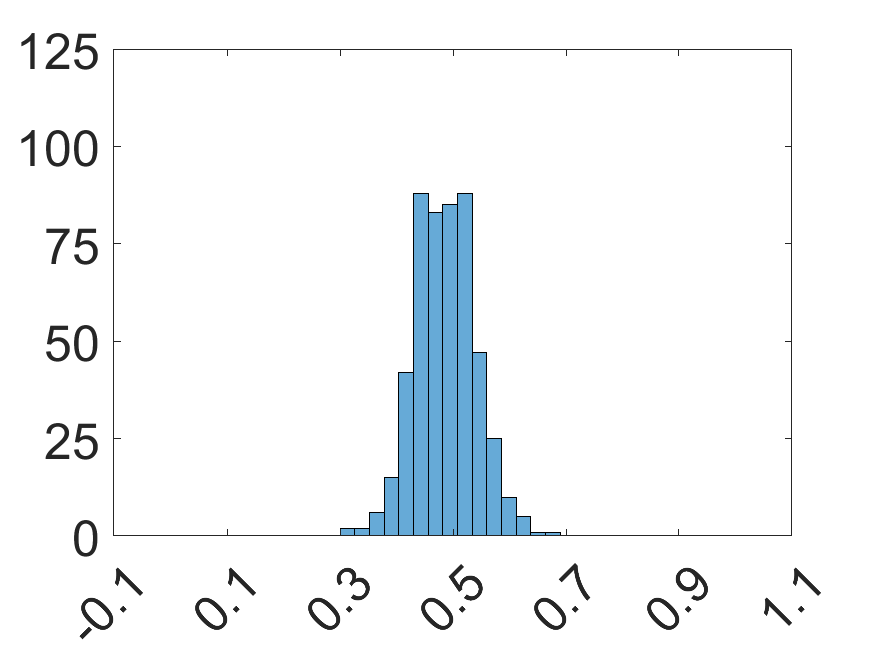}
					\caption{$N=200, m=20$}
				\end{subfigure}
				
				\begin{subfigure}{.24\textwidth}
					\centering
					\includegraphics[width=\linewidth]{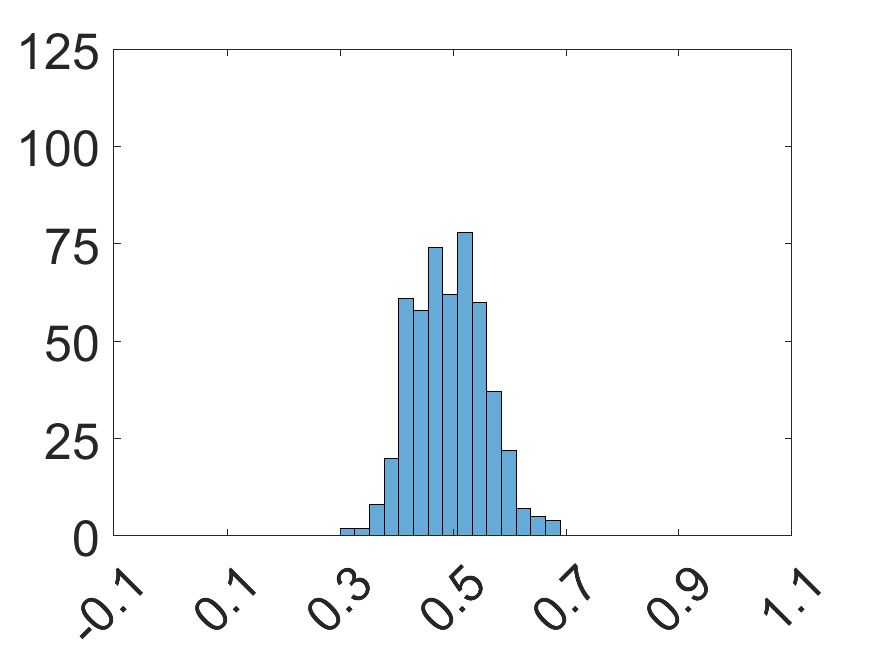}
					\caption{$N=10, m=50$}
				\end{subfigure}%
				\hfill
				\begin{subfigure}{.24\textwidth}
					\centering
					\includegraphics[width=\linewidth]{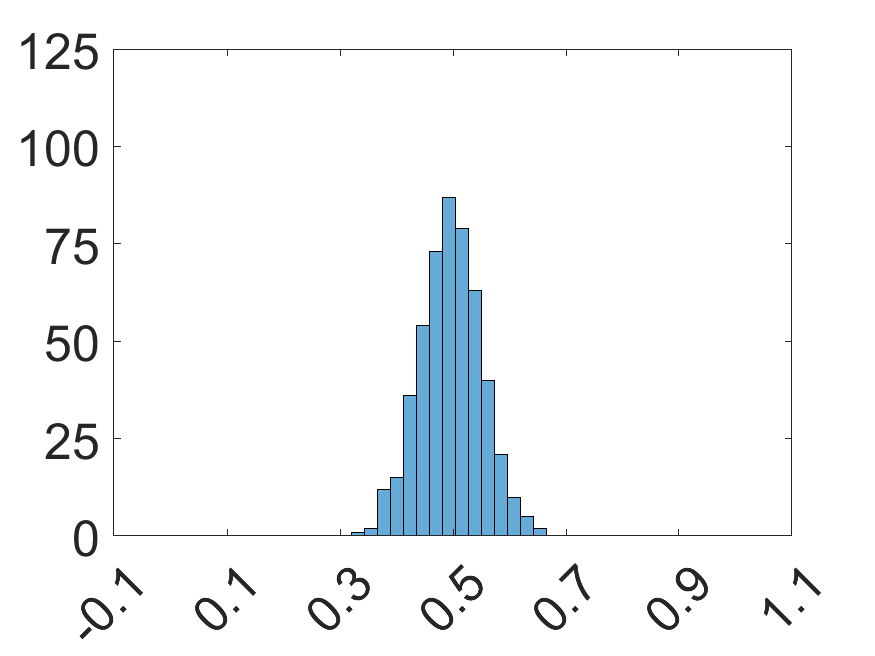}
					\caption{$N=20, m=50$}
				\end{subfigure}
				\hfill
				\begin{subfigure}{.24\textwidth}
					\centering
					\includegraphics[width=\linewidth]{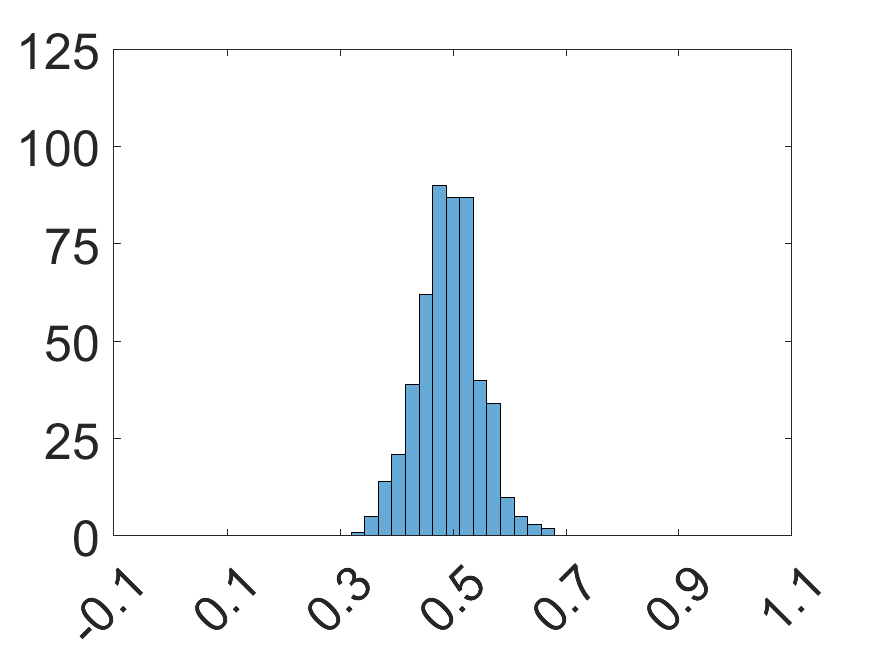}
					\caption{$N=50, m=50$}
				\end{subfigure}
				\hfill
				\begin{subfigure}{.24\textwidth}
					\centering
					\includegraphics[width=\linewidth]{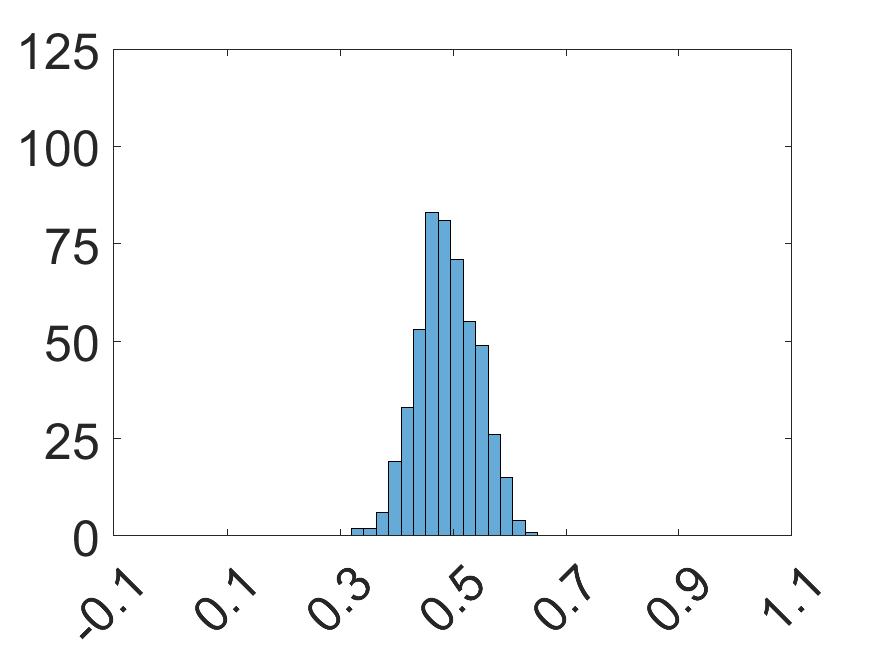}
					\caption{$N=100, m=50$}
				\end{subfigure}
				\hfill
				\begin{subfigure}{.24\textwidth}
					\centering
					\includegraphics[width=\linewidth]{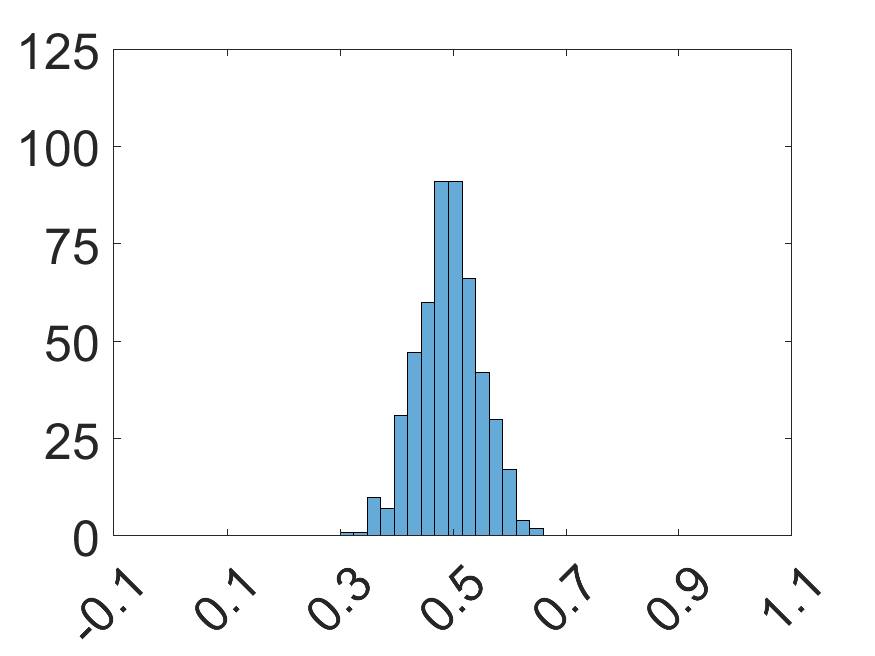}
					\caption{$N=200, m=50$}
				\end{subfigure}
				
				\begin{subfigure}{.24\textwidth}
					\centering
					\includegraphics[width=\linewidth]{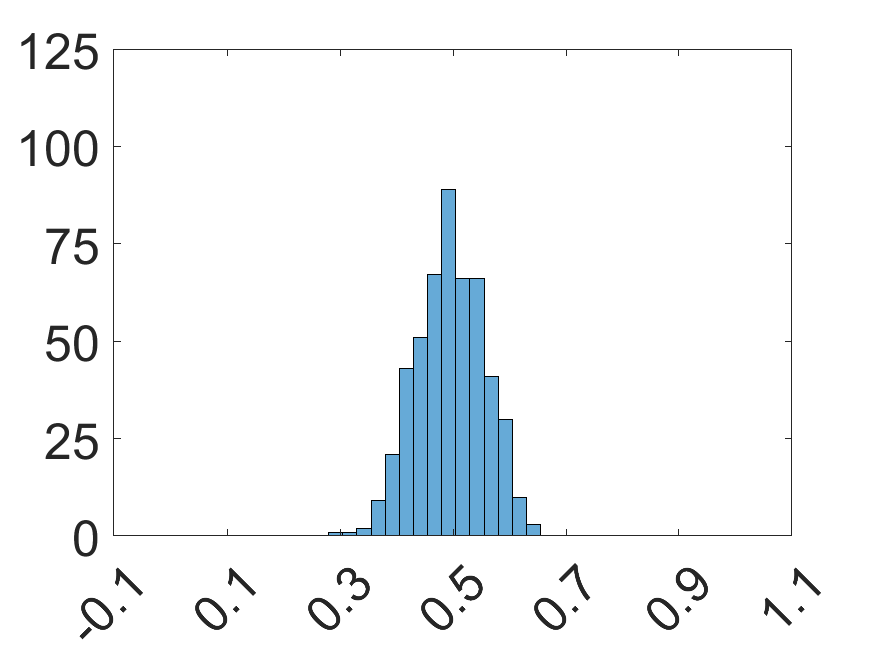}
					\caption{$N=10, m=100$}
				\end{subfigure}%
				\hfill
				\begin{subfigure}{.24\textwidth}
					\centering
					\includegraphics[width=\linewidth]{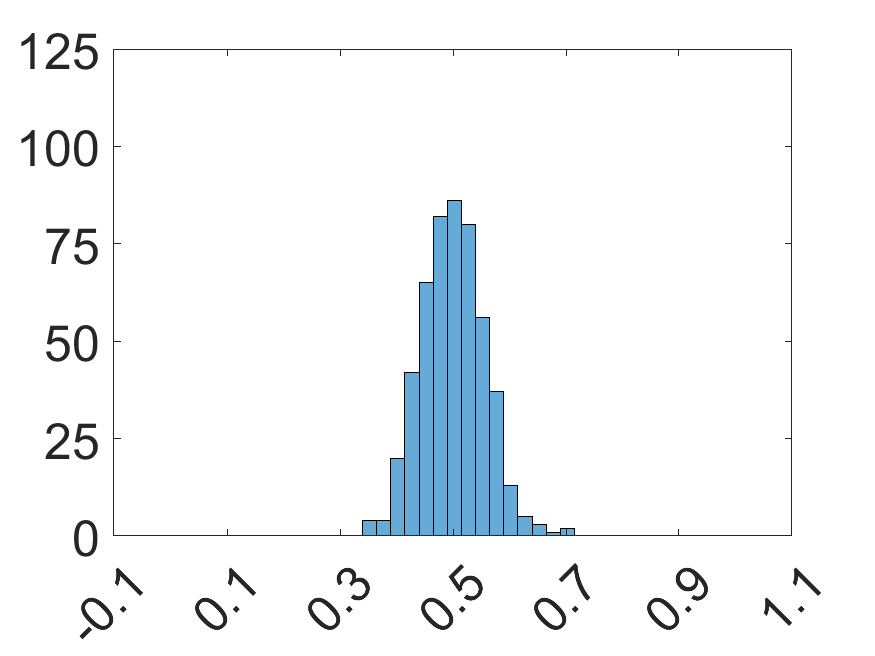}
					\caption{$N=20, m=100$}
				\end{subfigure}
				\hfill
				\begin{subfigure}{.24\textwidth}
					\centering
					\includegraphics[width=\linewidth]{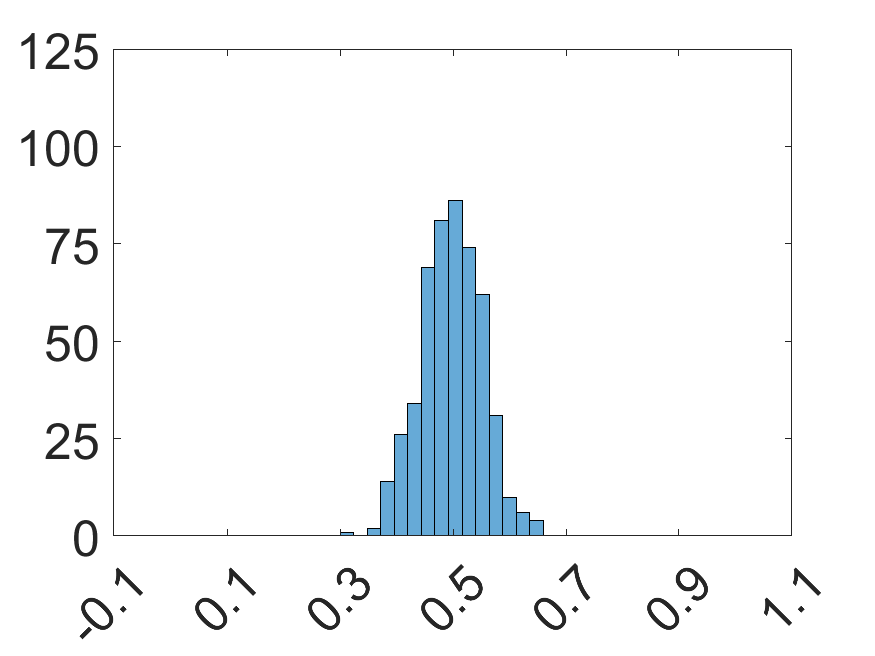}
					\caption{$N=50, m=100$}
				\end{subfigure}
				\hfill
				\begin{subfigure}{.24\textwidth}
					\centering
					\includegraphics[width=\linewidth]{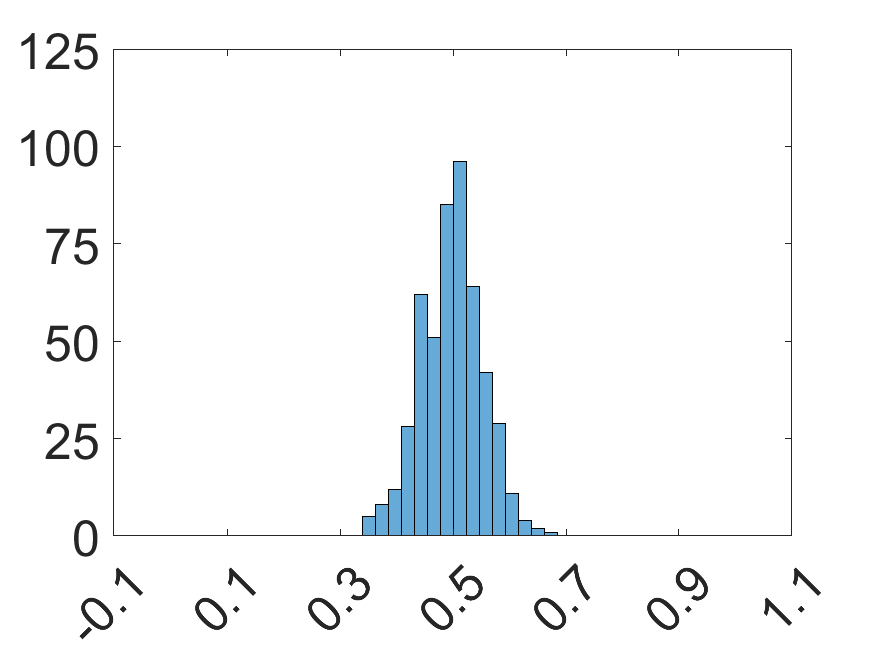}
					\caption{$N=100, m=100$}
				\end{subfigure}
				\hfill
				\begin{subfigure}{.24\textwidth}
					\centering
					\includegraphics[width=\linewidth]{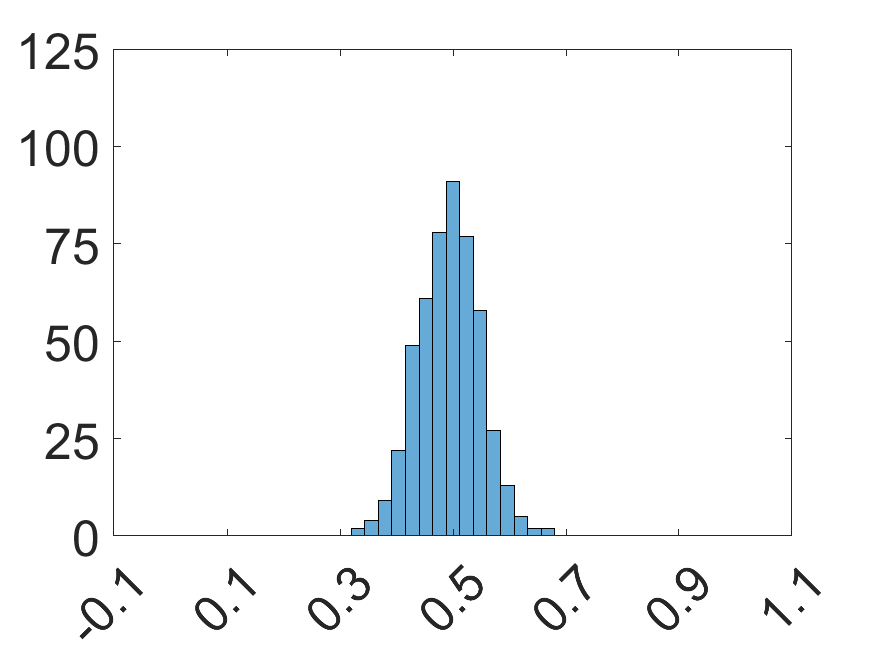}
					\caption{$N=200, m=100$}
				\end{subfigure}	
				\subcaption*{\textit{Note:} the true value is $\beta= 0.5$.}
				\label{fig:MC_hist_beta_CaseII_T100}
			\end{figure}
		\end{landscape}

		Tables \ref{tab:MC_RMSE_Case2_T10} and \ref{tab:MC_RMSE_Case2_T80} report the {RMSE} and {ReRMSE} of parameter estimates for $T=10$ and $T=80$, under case II. The results for $T=50$ and $T=100$ are reported in Section \ref{sec:simulations} of the paper. 
		Figures \ref{fig:MC_hist_beta_CaseII_T50} and \ref{fig:MC_hist_beta_CaseII_T100} show the histograms of the MC estimates of $\beta$, for different combinations of $N,m,T$, under case II.

		\begin{table}[H]
			\centering
			\scriptsize{
				\caption{Monte Carlo RMSEs - $T=10$, case II}\label{tab:MC_RMSE_Case2_T10}
				\scalebox{1}[1]{
					\begin{tabular}{ll | cc|cc|cc|cc|cc }
						\hline \hline
						&& \multicolumn{2}{|c|}{$N = 10$} & \multicolumn{2}{c|}{$N = 20$} & \multicolumn{2}{c|}{$N = 50$}& \multicolumn{2}{c|}{$N = 100$} & \multicolumn{2}{c}{$N = 200$}\\
						&& RMSE        & ReRMSE       & RMSE        & ReRMSE       & RMSE        & ReRMSE   & RMSE  & ReRMSE   & RMSE  & ReRMSE  \\
						\hline
						\multirow{5}{*}{$m = 20$}  & $\beta$     & 0.229 & 45.8\%  & 0.207 & 41.5\% & 0.208 & 41.6\% & 0.204 & 40.8\% & 0.220 & 44.0\% \\
						& $\rho$        & 0.133 & 44.3\%  & 0.096 & 32.0\% & 0.068 & 22.8\% & 0.056 & 18.7\% & 0.050 & 16.7\% \\
						& $\alpha$      & 0.214 & 107.0\% & 0.116 & 58.2\% & 0.082 & 40.9\% & 0.054 & 27.2\% & 0.043 & 21.5\% \\
						& $\mathcal{F}$ & 0.207 & 21.8\%  & 0.212 & 21.7\% & 0.214 & 21.7\% & 0.215 & 21.7\% & 0.216 & 21.7\% \\
						& $U$           & 0.108 & 4.9\%   & 0.086 & 3.9\%  & 0.079 & 3.6\%  & 0.078 & 3.5\%  & 0.078 & 3.5\%  \\
						\hline
						\multirow{5}{*}{$m = 50$}  & $\beta$     & 0.265 & 53.0\%  & 0.221 & 44.2\% & 0.212 & 42.5\% & 0.230 & 46.0\% & 0.234 & 46.8\% \\
						& $\rho$        & 0.144 & 48.0\%  & 0.103 & 34.4\% & 0.071 & 23.5\% & 0.055 & 18.2\% & 0.046 & 15.2\% \\
						& $\alpha$      & 0.171 & 85.4\%  & 0.122 & 61.1\% & 0.074 & 36.9\% & 0.050 & 25.2\% & 0.039 & 19.3\% \\
						& $\mathcal{F}$ & 0.133 & 14.0\%  & 0.135 & 13.9\% & 0.137 & 13.9\% & 0.138 & 13.9\% & 0.138 & 13.9\% \\
						& $U$           & 0.065 & 3.6\%   & 0.039 & 2.2\%  & 0.028 & 1.6\%  & 0.027 & 1.5\%  & 0.026 & 1.5\%  \\
						\hline
						\multirow{5}{*}{$m = 100$} & $\beta$     & 0.235 & 47.0\%  & 0.235 & 46.9\% & 0.211 & 42.1\% & 0.223 & 44.6\% & 0.215 & 43.0\% \\
						& $\rho$        & 0.144 & 48.0\%  & 0.106 & 35.2\% & 0.073 & 24.3\% & 0.057 & 19.1\% & 0.051 & 16.8\% \\
						& $\alpha$      & 0.235 & 117.4\% & 0.113 & 56.6\% & 0.069 & 34.4\% & 0.062 & 31.1\% & 0.043 & 21.4\% \\
						& $\mathcal{F}$ & 0.095 & 10.0\%  & 0.096 & 9.9\%  & 0.097 & 9.8\%  & 0.098 & 9.8\%  & 0.098 & 9.8\%  \\
						& $U$           & 0.054 & 3.4\%   & 0.029 & 1.8\%  & 0.016 & 1.0\%  & 0.013 & 0.8\%  & 0.012 & 0.7\%  \\
						\hline \hline
					\end{tabular}   	
				}
			}
		\end{table}

		\begin{table}[H]
			\centering
			\scriptsize{
				\caption{Monte Carlo RMSEs - $T=80$, case II}\label{tab:MC_RMSE_Case2_T80}
				\scalebox{1}[1]{
					\begin{tabular}{ll | cc|cc|cc|cc|cc }
						\hline \hline
						&& \multicolumn{2}{|c|}{$N = 10$} & \multicolumn{2}{c|}{$N = 20$} & \multicolumn{2}{c|}{$N = 50$}& \multicolumn{2}{c|}{$N = 100$} & \multicolumn{2}{c}{$N = 200$}\\
						&& RMSE        & ReRMSE       & RMSE        & ReRMSE       & RMSE        & ReRMSE   & RMSE  & ReRMSE  & RMSE  & ReRMSE  \\
						\hline
						\multirow{5}{*}{$m = 20$}  & $\beta$     & 0.062 & 12.3\% & 0.065 & 12.9\% & 0.058 & 11.6\% & 0.062 & 12.4\% & 0.063 & 12.5\% \\
						& $\rho$        & 0.034 & 11.4\% & 0.025 & 8.2\%  & 0.016 & 5.4\%  & 0.011 & 3.7\%  & 0.008 & 2.7\%  \\
						& $\alpha$      & 0.065 & 32.5\% & 0.033 & 16.4\% & 0.023 & 11.4\% & 0.017 & 8.3\%  & 0.012 & 6.0\%  \\
						& $\mathcal{F}$ & 0.206 & 21.7\% & 0.211 & 21.7\% & 0.214 & 21.7\% & 0.216 & 21.7\% & 0.216 & 21.7\% \\
						& $U$           & 0.083 & 3.7\%  & 0.079 & 3.6\%  & 0.078 & 3.5\%  & 0.078 & 3.5\%  & 0.078 & 3.5\%  \\
						\hline
						\multirow{5}{*}{$m = 50$}  & $\beta$     & 0.068 & 13.7\% & 0.058 & 11.6\% & 0.063 & 12.7\% & 0.061 & 12.3\% & 0.062 & 12.4\% \\
						& $\rho$        & 0.036 & 11.9\% & 0.023 & 7.8\%  & 0.016 & 5.2\%  & 0.011 & 3.8\%  & 0.008 & 2.7\%  \\
						& $\alpha$      & 0.057 & 28.7\% & 0.036 & 18.1\% & 0.022 & 11.2\% & 0.014 & 7.1\%  & 0.012 & 5.8\%  \\
						& $\mathcal{F}$ & 0.132 & 13.9\% & 0.135 & 13.9\% & 0.137 & 13.9\% & 0.138 & 13.9\% & 0.138 & 13.9\% \\
						& $U$           & 0.034 & 1.9\%  & 0.028 & 1.6\%  & 0.026 & 1.5\%  & 0.026 & 1.5\%  & 0.026 & 1.5\%  \\
						\hline
						\multirow{5}{*}{$m = 100$} & $\beta$     & 0.068 & 13.7\% & 0.060 & 12.1\% & 0.067 & 13.4\% & 0.062 & 12.4\% & 0.060 & 12.0\% \\
						& $\rho$        & 0.036 & 12.1\% & 0.025 & 8.2\%  & 0.016 & 5.3\%  & 0.011 & 3.7\%  & 0.009 & 2.9\%  \\
						& $\alpha$      & 0.064 & 32.0\% & 0.036 & 18.2\% & 0.021 & 10.4\% & 0.016 & 8.1\%  & 0.011 & 5.7\%  \\
						& $\mathcal{F}$ & 0.093 & 9.8\%  & 0.096 & 9.8\%  & 0.097 & 9.8\%  & 0.098 & 9.8\%  & 0.098 & 9.8\%  \\
						& $U$           & 0.022 & 1.4\%  & 0.015 & 0.9\%  & 0.012 & 0.8\%  & 0.012 & 0.7\%  & 0.012 & 0.7\%  \\
						\hline \hline
					\end{tabular}  	 	
				}
			}
		\end{table}

		\begin{landscape}
			\begin{figure}[H]
				\centering
				\caption{Monte Carlo histograms of the network effect $\widehat{\beta}$ - $T=10$, case II}
				\begin{subfigure}{.24\textwidth}
					\centering
					\includegraphics[width=\linewidth]{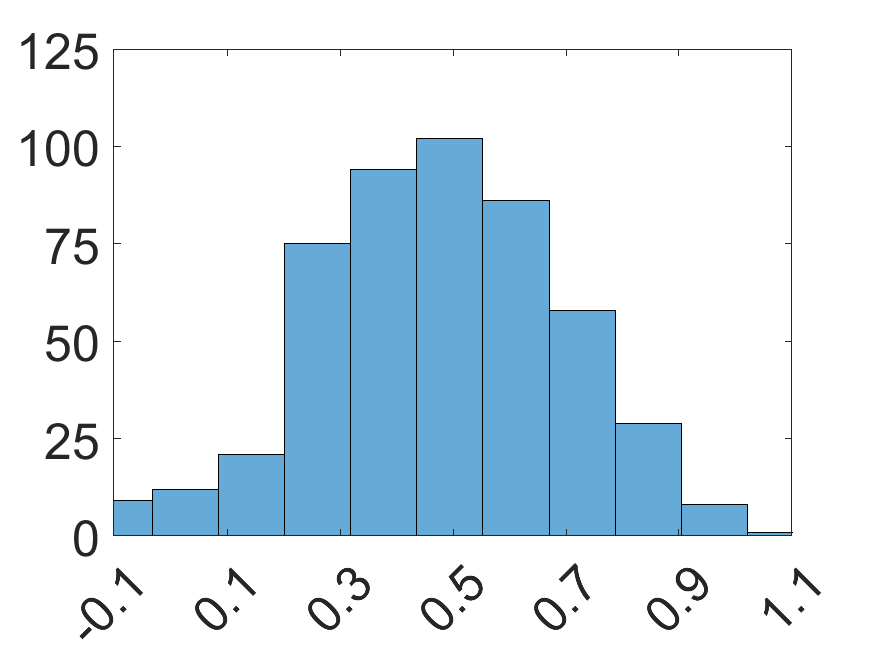}
					\caption{$N=10, m=20$}
				\end{subfigure}%
				\hfill
				\begin{subfigure}{.24\textwidth}
					\centering
					\includegraphics[width=\linewidth]{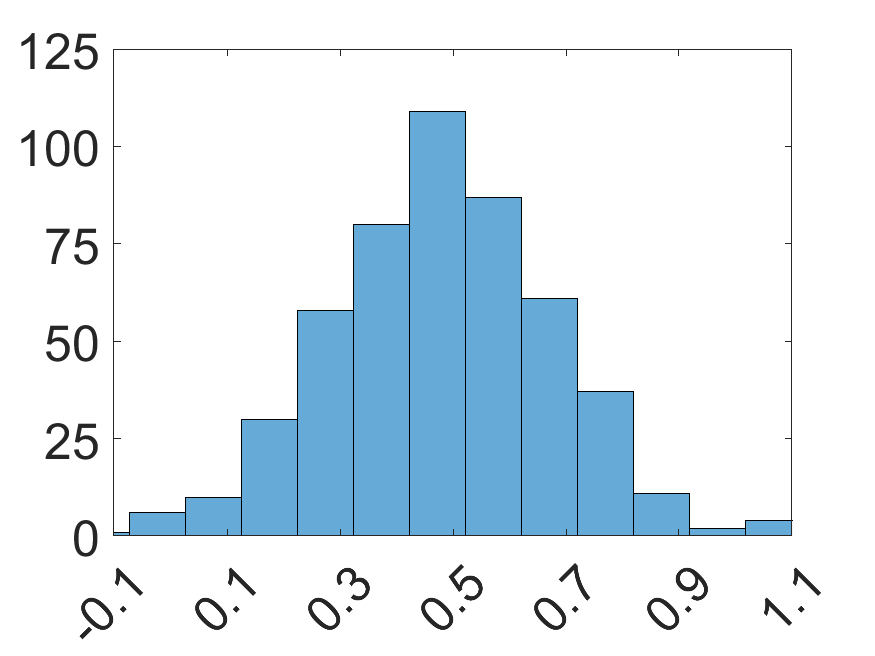}
					\caption{$N=20, m=20$}
				\end{subfigure}
				\hfill
				\begin{subfigure}{.24\textwidth}
					\centering
					\includegraphics[width=\linewidth]{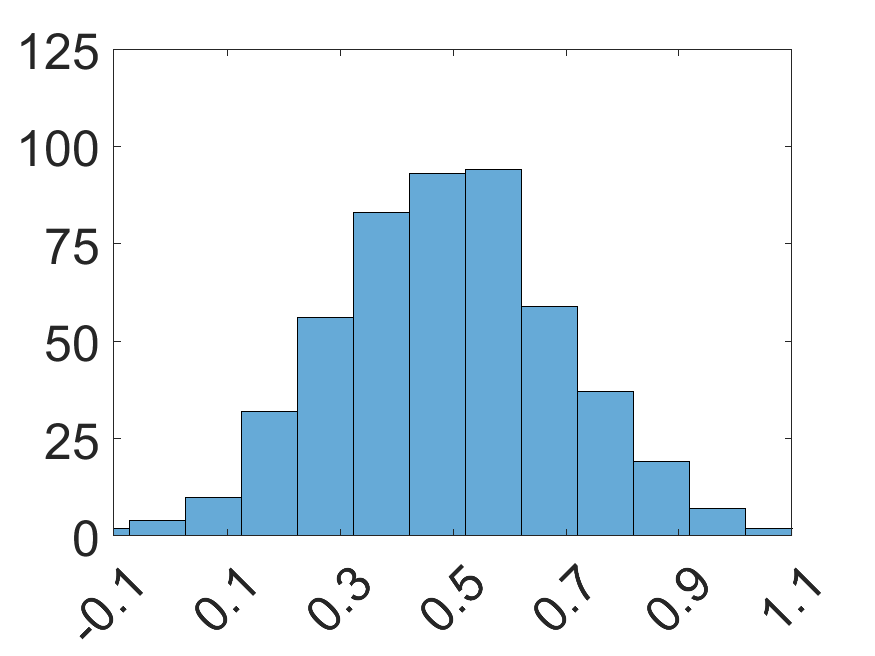}
					\caption{$N=50, m=20$}
				\end{subfigure}
				\hfill
				\begin{subfigure}{.24\textwidth}
					\centering
					\includegraphics[width=\linewidth]{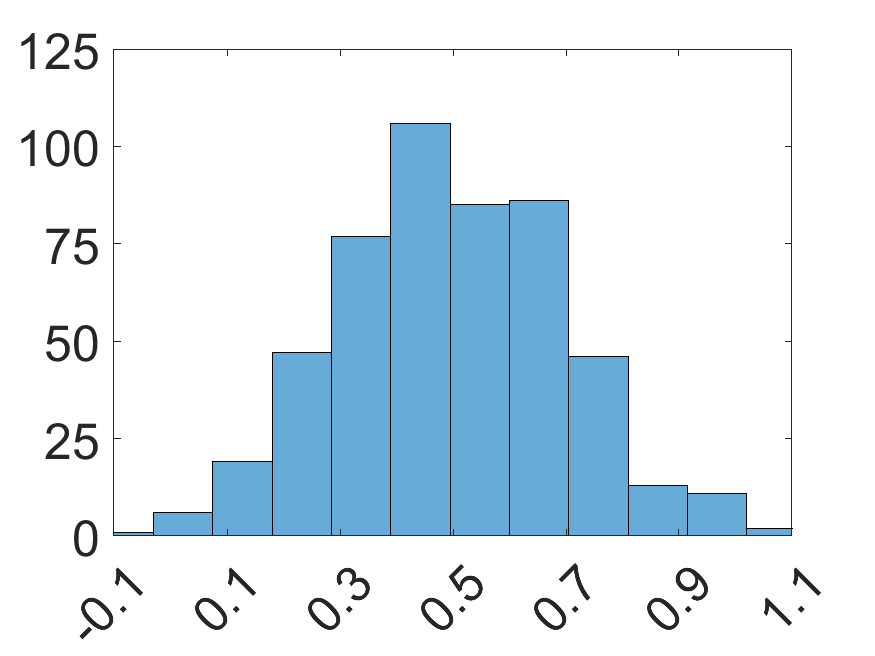}
					\caption{$N=100, m=20$}
				\end{subfigure}
				\hfill
				\begin{subfigure}{.24\textwidth}
					\centering
					\includegraphics[width=\linewidth]{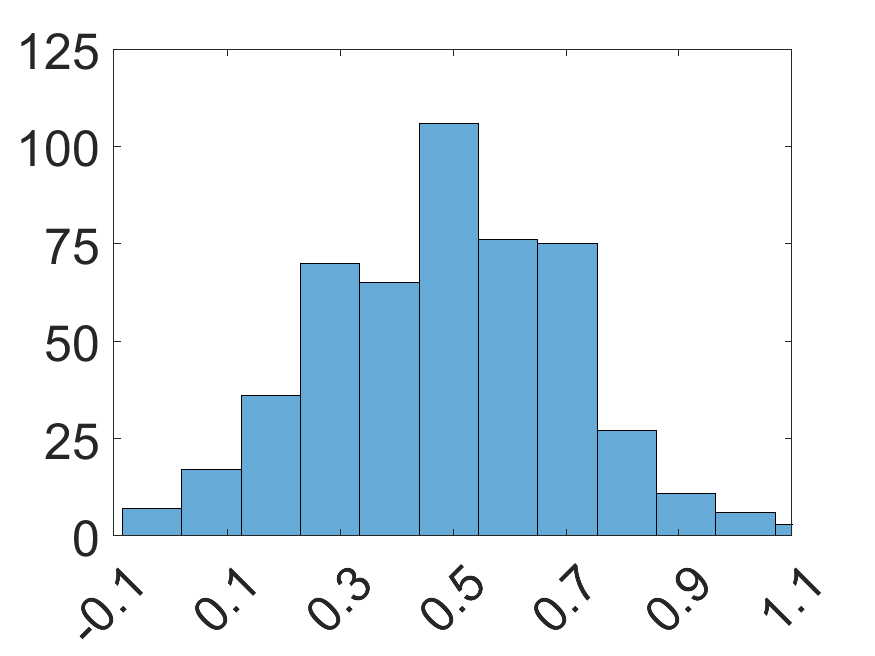}
					\caption{$N=200, m=20$}
				\end{subfigure}
				
				\begin{subfigure}{.24\textwidth}
					\centering
					\includegraphics[width=\linewidth]{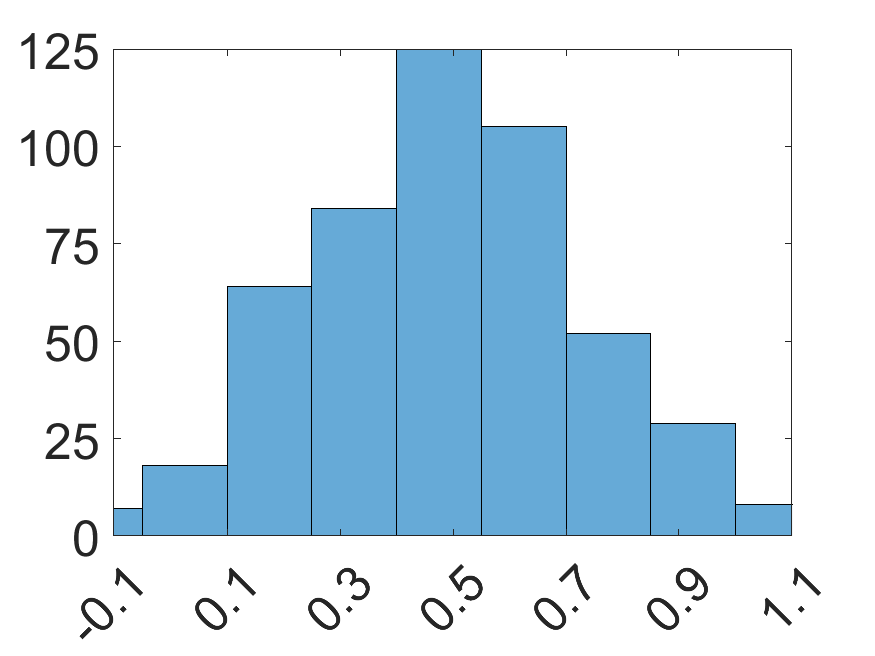}
					\caption{$N=10, m=50$}
				\end{subfigure}%
				\hfill
				\begin{subfigure}{.24\textwidth}
					\centering
					\includegraphics[width=\linewidth]{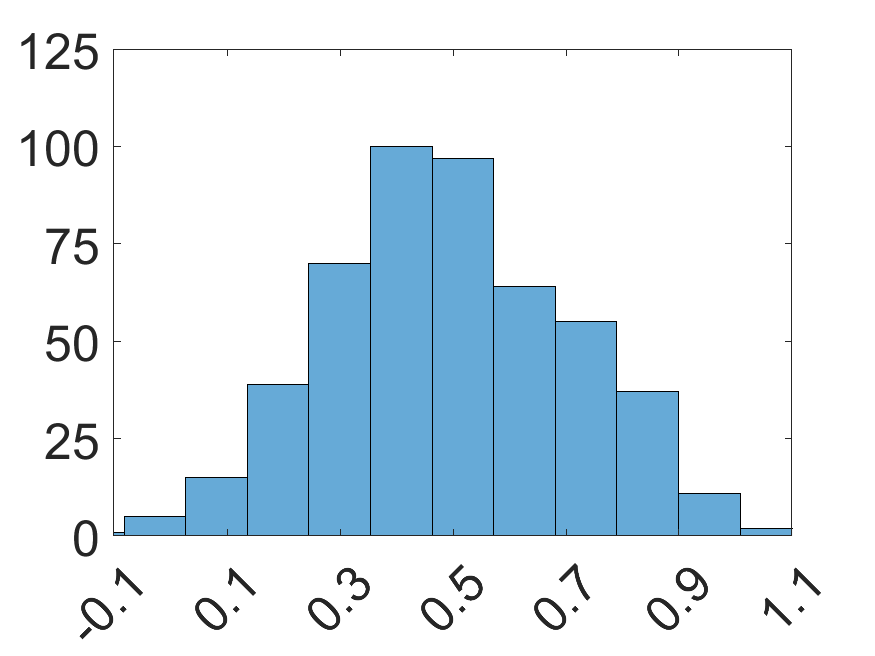}
					\caption{$N=20, m=50$}
				\end{subfigure}
				\hfill
				\begin{subfigure}{.24\textwidth}
					\centering
					\includegraphics[width=\linewidth]{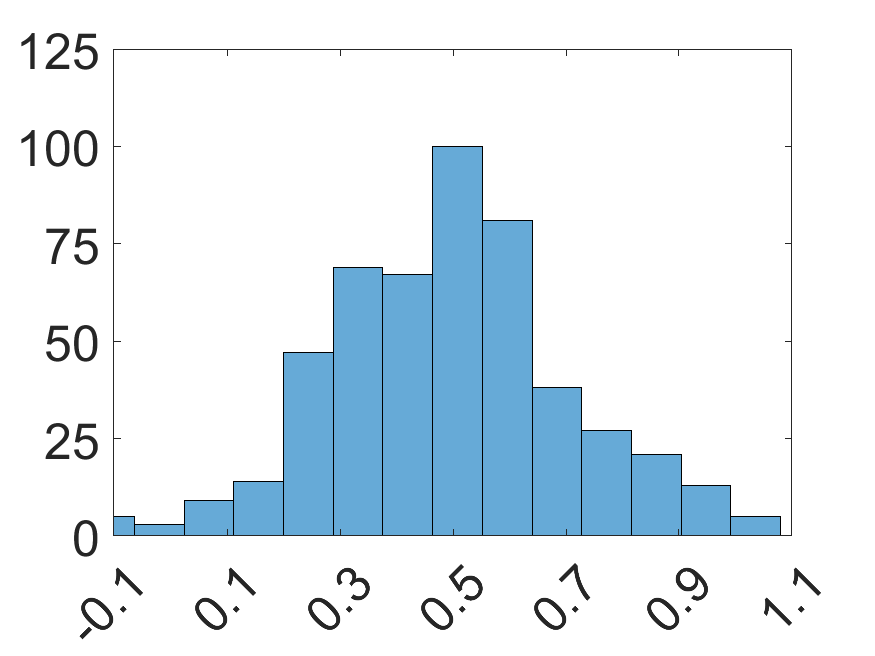}
					\caption{$N=50, m=50$}
				\end{subfigure}
				\hfill
				\begin{subfigure}{.24\textwidth}
					\centering
					\includegraphics[width=\linewidth]{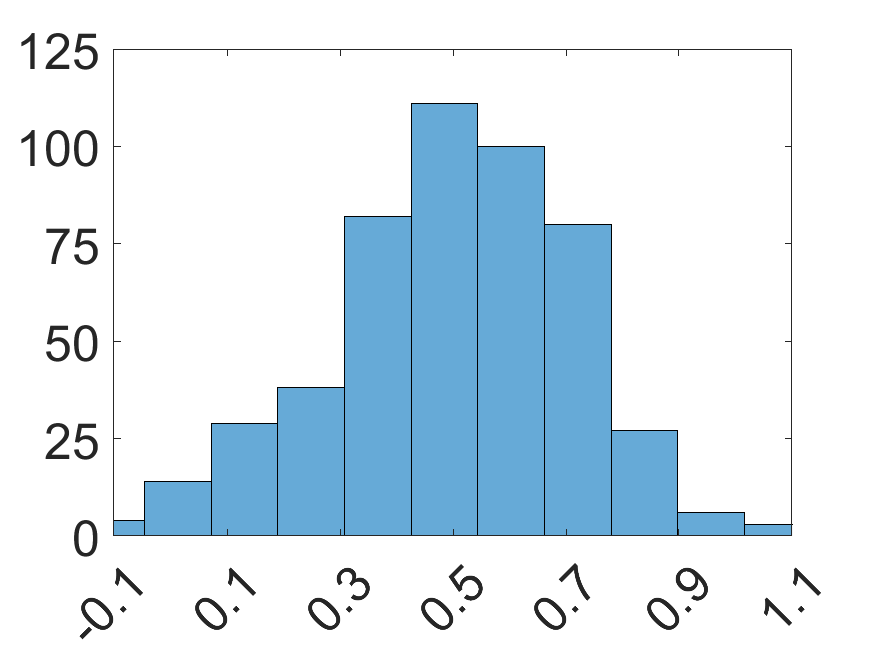}
					\caption{$N=100, m=50$}
				\end{subfigure}
				\hfill
				\begin{subfigure}{.24\textwidth}
					\centering
					\includegraphics[width=\linewidth]{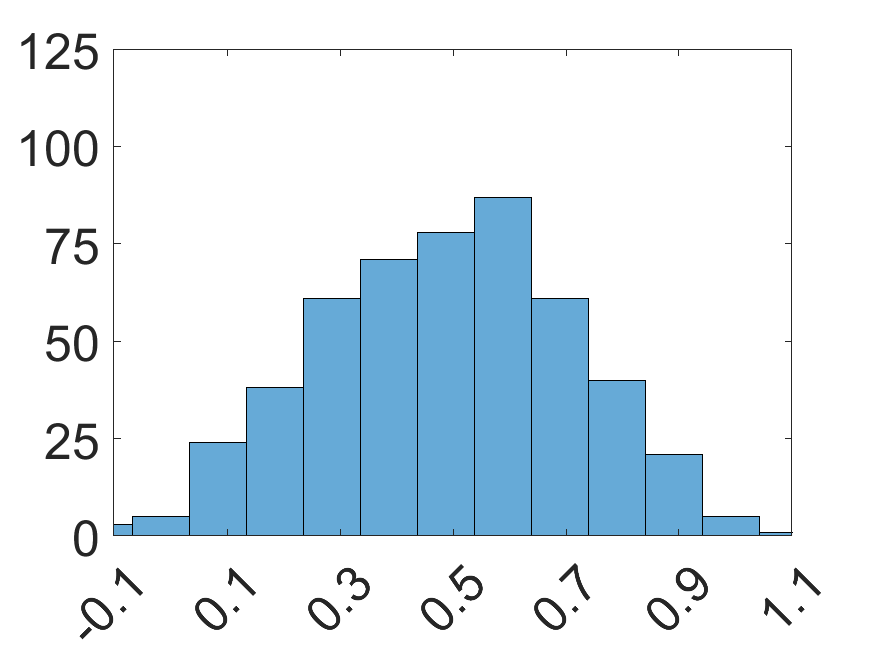}
					\caption{$N=200, m=50$}
				\end{subfigure}		
				
				\begin{subfigure}{.24\textwidth}
					\centering
					\includegraphics[width=\linewidth]{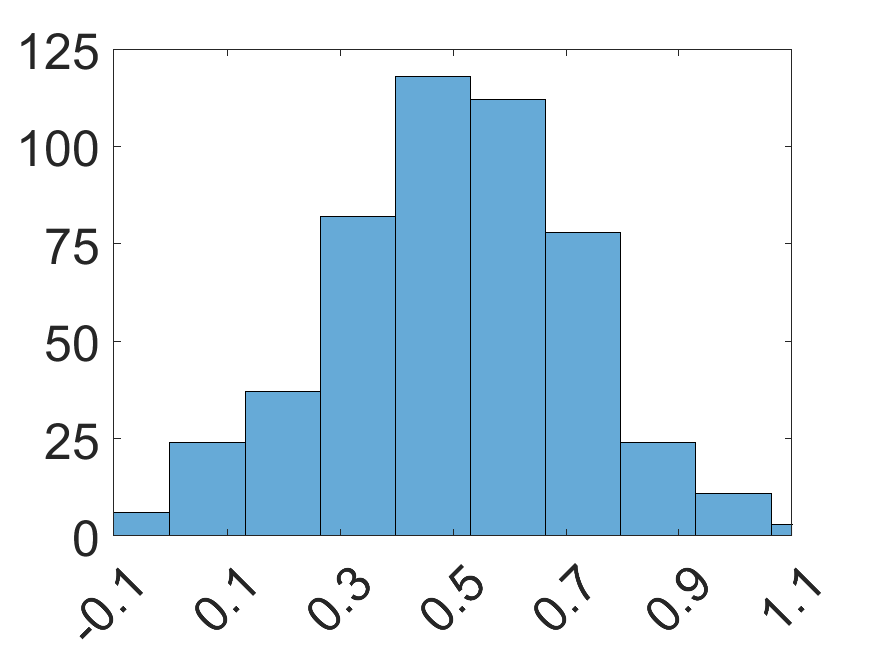}
					\caption{$N=10, m=100$}
				\end{subfigure}%
				\hfill
				\begin{subfigure}{.24\textwidth}
					\centering
					\includegraphics[width=\linewidth]{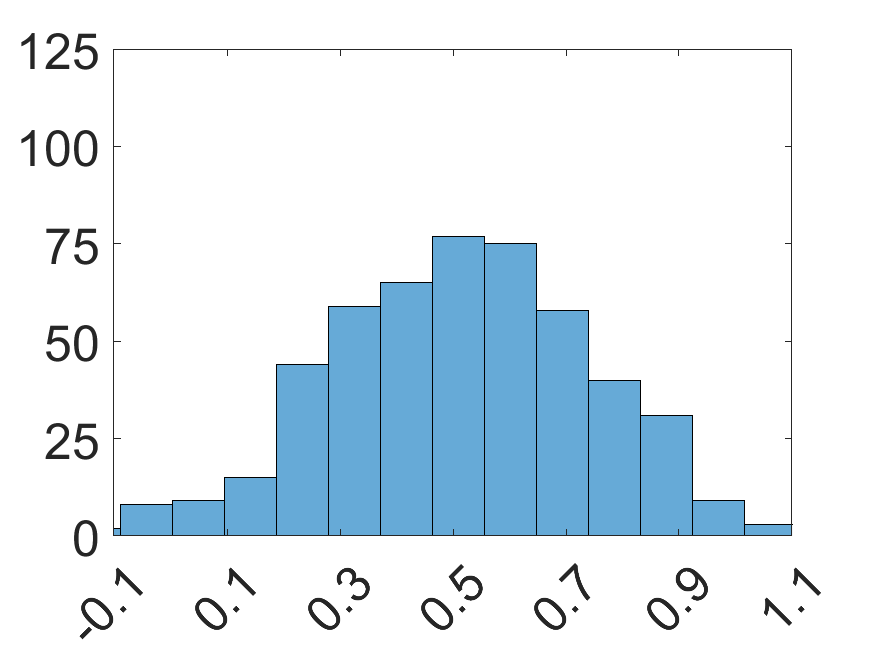}
					\caption{$N=20, m=100$}
				\end{subfigure}
				\hfill
				\begin{subfigure}{.24\textwidth}
					\centering
					\includegraphics[width=\linewidth]{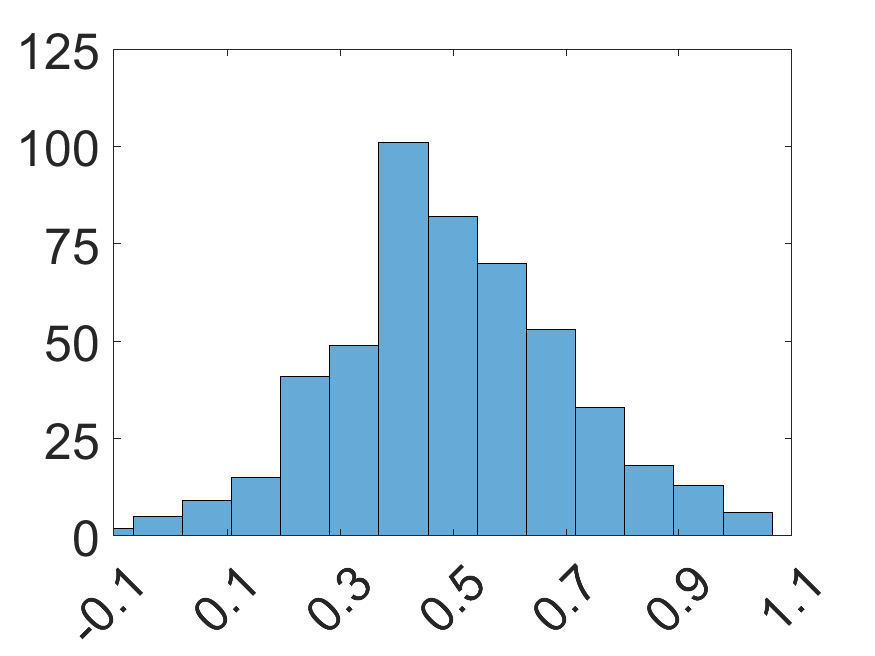}
					\caption{$N=50, m=100$}
				\end{subfigure}
				\hfill
				\begin{subfigure}{.24\textwidth}
					\centering
					\includegraphics[width=\linewidth]{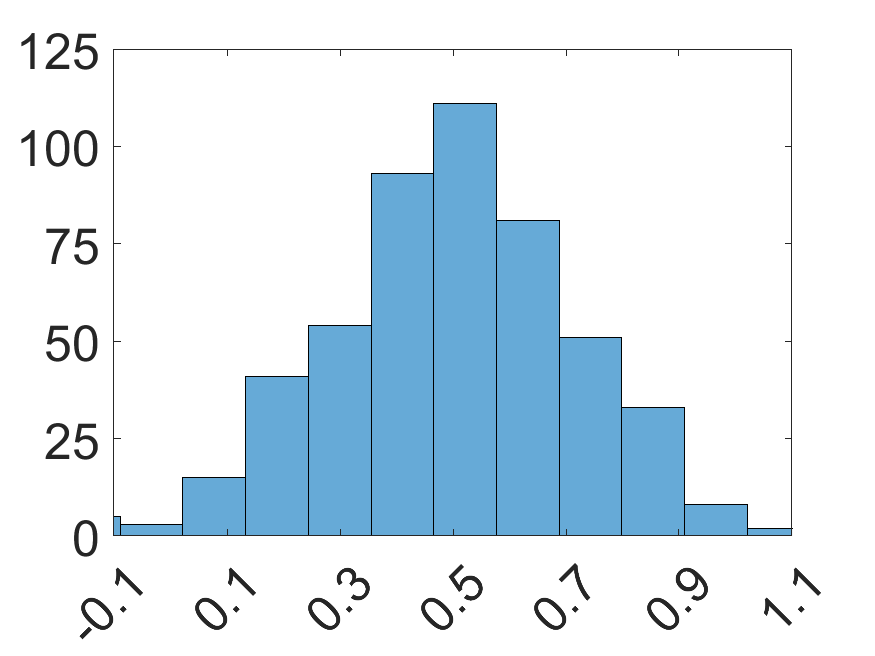}
					\caption{$N=100, m=100$}
				\end{subfigure}
				\hfill
				\begin{subfigure}{.24\textwidth}
					\centering
					\includegraphics[width=\linewidth]{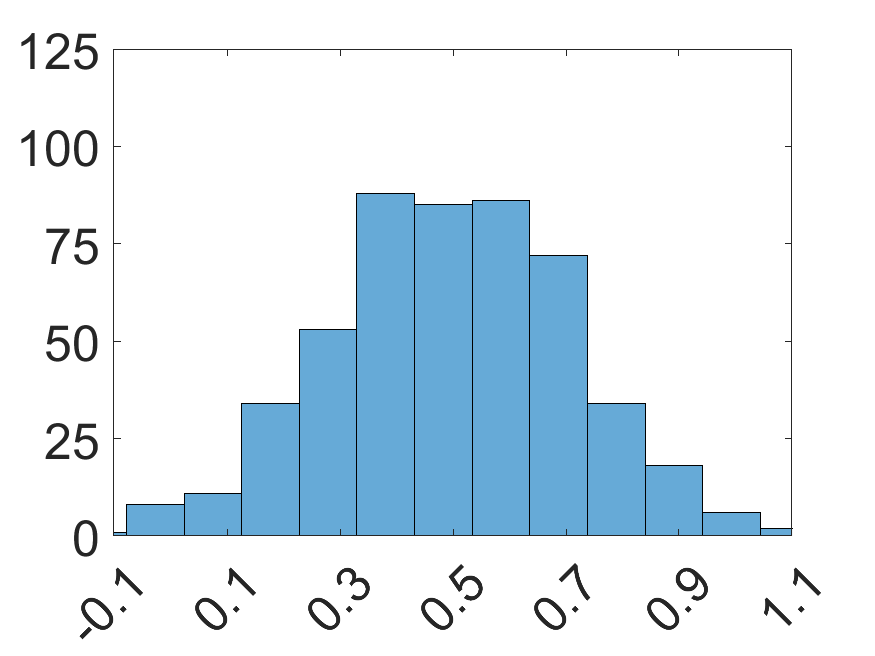}
					\caption{$N=200, m=100$}
				\end{subfigure}
				\subcaption*{\textit{Note:} the true value is $\beta= 0.5$. }
				\label{fig:MC_hist_beta_CaseII_T10}
			\end{figure}
		\end{landscape}

		\begin{landscape}			
			\begin{figure}[H]
				\centering
				\caption{Monte Carlo histograms of the network effect $\widehat{\beta}$ - $T=80$, case II}
				\begin{subfigure}{.24\textwidth}
					\centering
					\includegraphics[width=\linewidth]{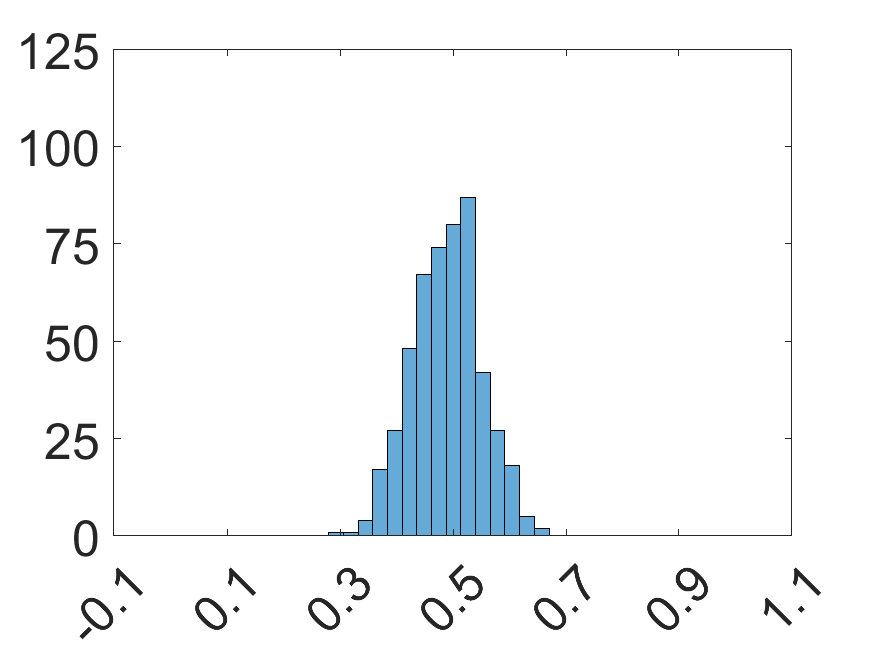}
					\caption{$N=10, m=20$}
				\end{subfigure}%
				\hfill
				\begin{subfigure}{.24\textwidth}
					\centering
					\includegraphics[width=\linewidth]{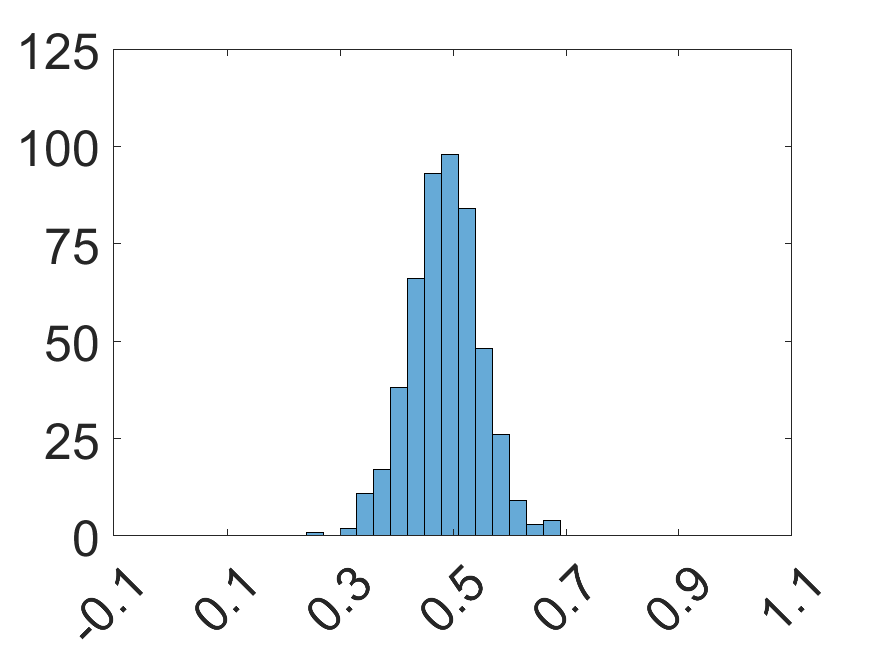}
					\caption{$N=20, m=20$}
				\end{subfigure}
				\hfill
				\begin{subfigure}{.24\textwidth}
					\centering
					\includegraphics[width=\linewidth]{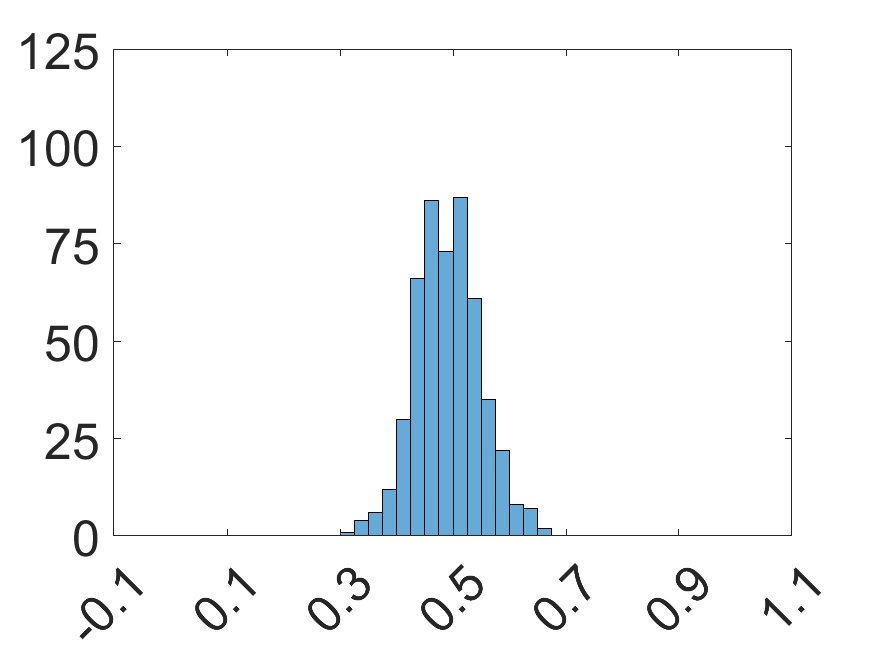}
					\caption{$N=50, m=20$}
				\end{subfigure}
				\hfill
				\begin{subfigure}{.24\textwidth}
					\centering
					\includegraphics[width=\linewidth]{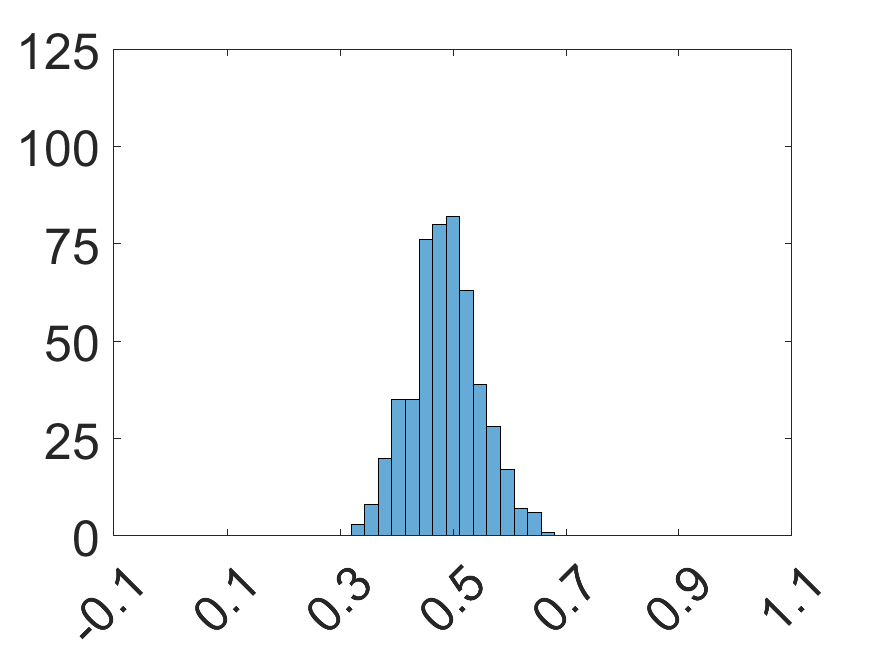}
					\caption{$N=100, m=20$}
				\end{subfigure}
				\hfill
				\begin{subfigure}{.24\textwidth}
					\centering
					\includegraphics[width=\linewidth]{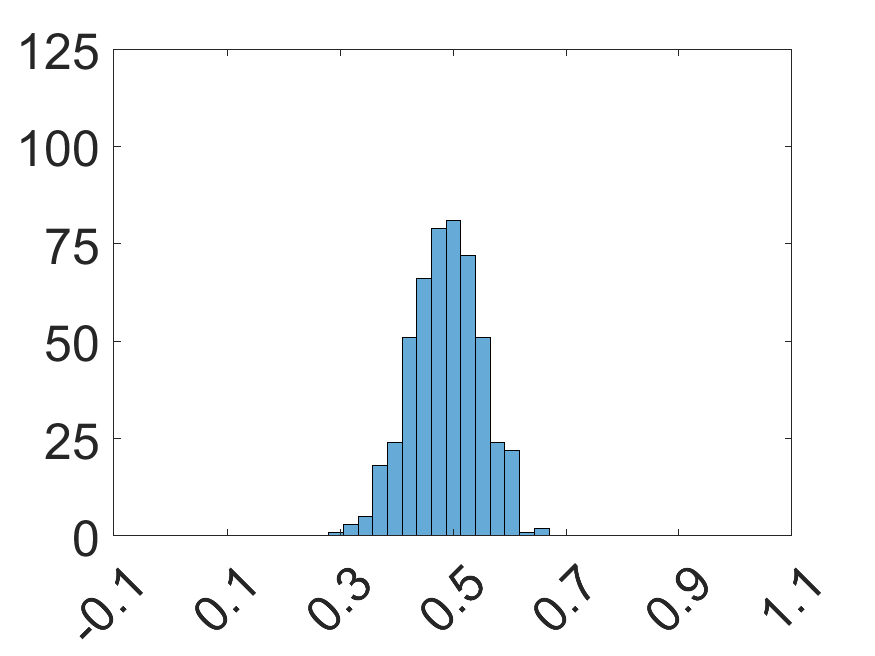}
					\caption{$N=200, m=20$}
				\end{subfigure}
				
				\begin{subfigure}{.24\textwidth}
					\centering
					\includegraphics[width=\linewidth]{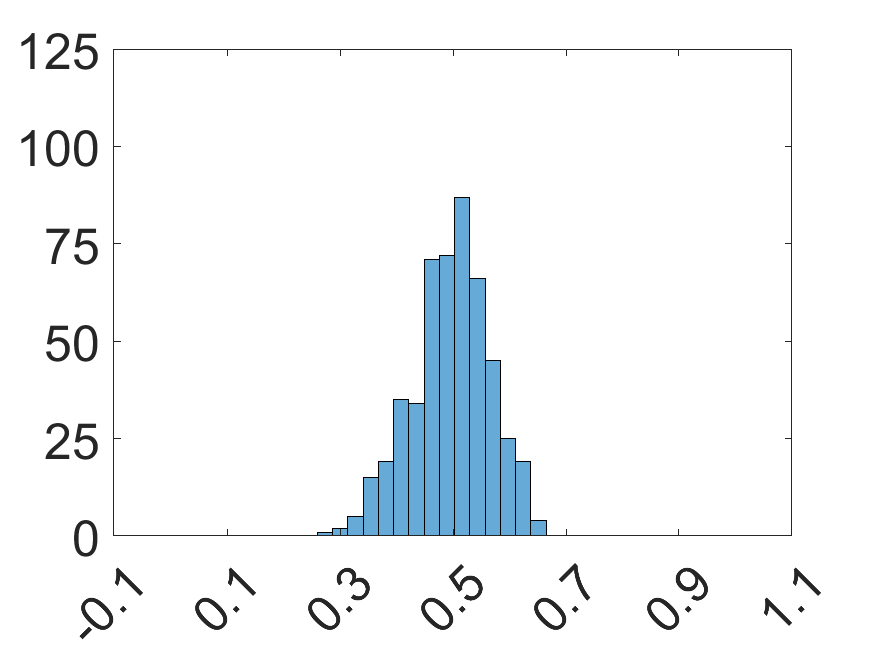}
					\caption{$N=10, m=50$}
				\end{subfigure}%
				\hfill
				\begin{subfigure}{.24\textwidth}
					\centering
					\includegraphics[width=\linewidth]{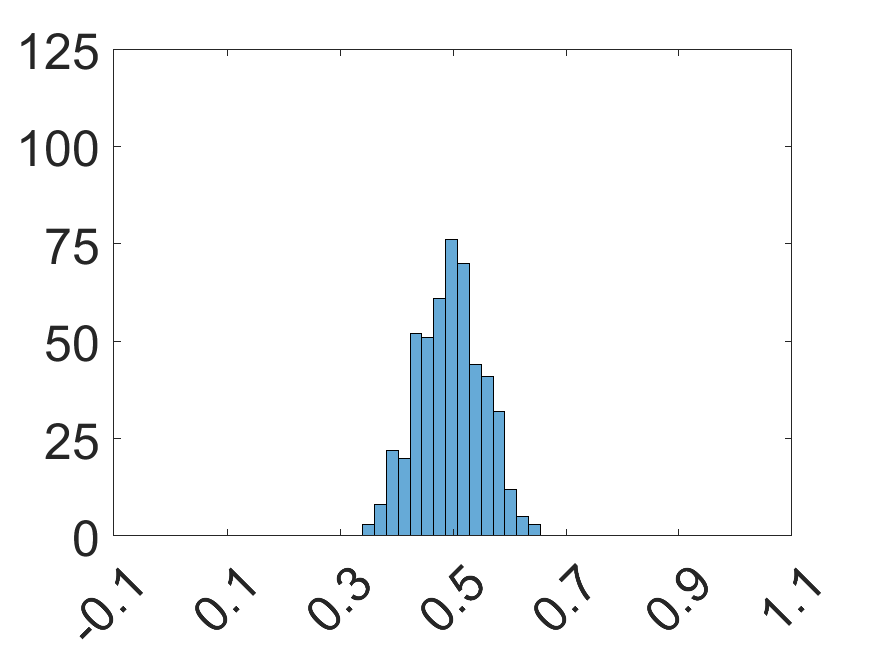}
					\caption{$N=20, m=50$}
				\end{subfigure}
				\hfill
				\begin{subfigure}{.24\textwidth}
					\centering
					\includegraphics[width=\linewidth]{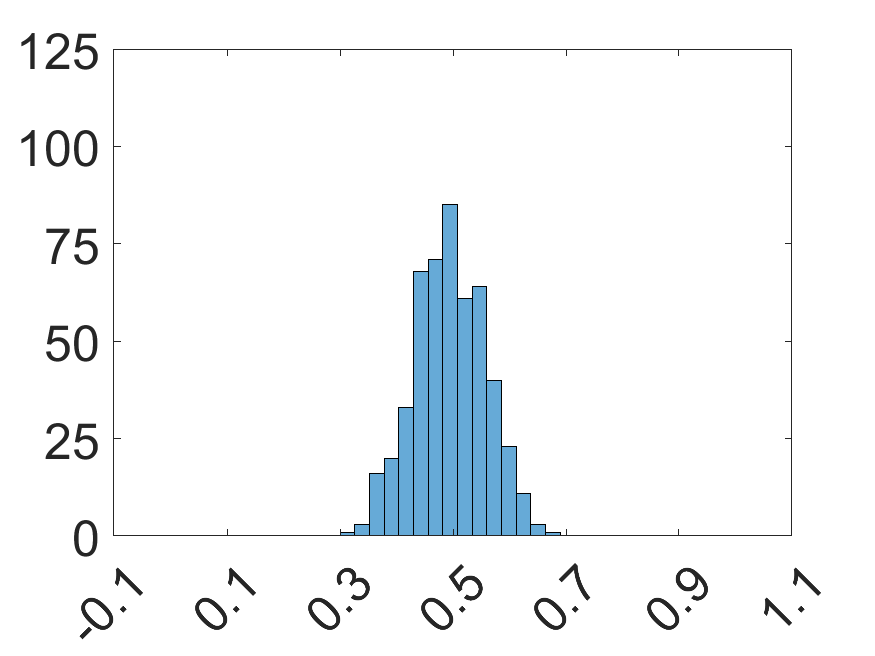}
					\caption{$N=50, m=50$}
				\end{subfigure}
				\hfill
				\begin{subfigure}{.24\textwidth}
					\centering
					\includegraphics[width=\linewidth]{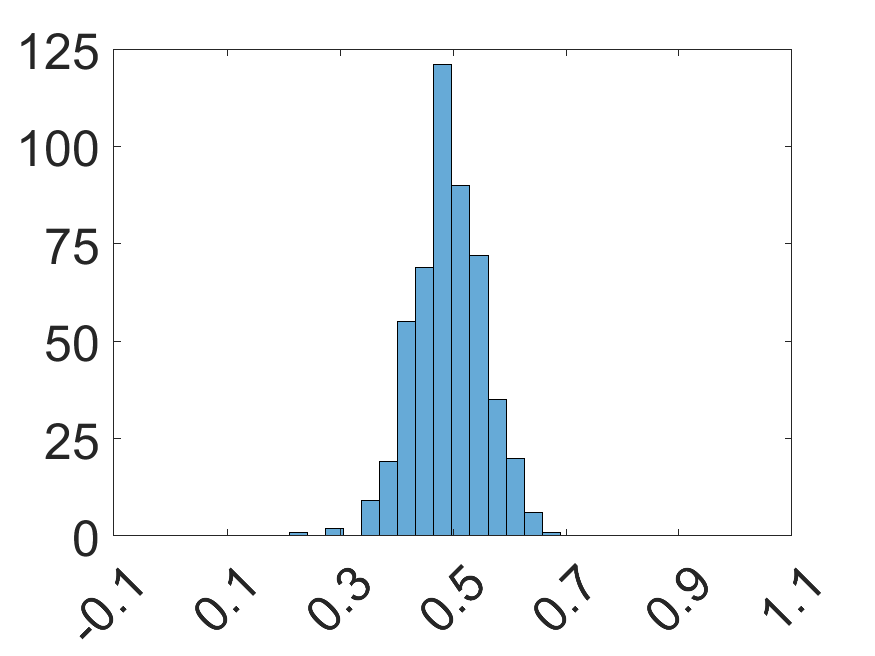}
					\caption{$N=100, m=50$}
				\end{subfigure}
				\hfill
				\begin{subfigure}{.24\textwidth}
					\centering
					\includegraphics[width=\linewidth]{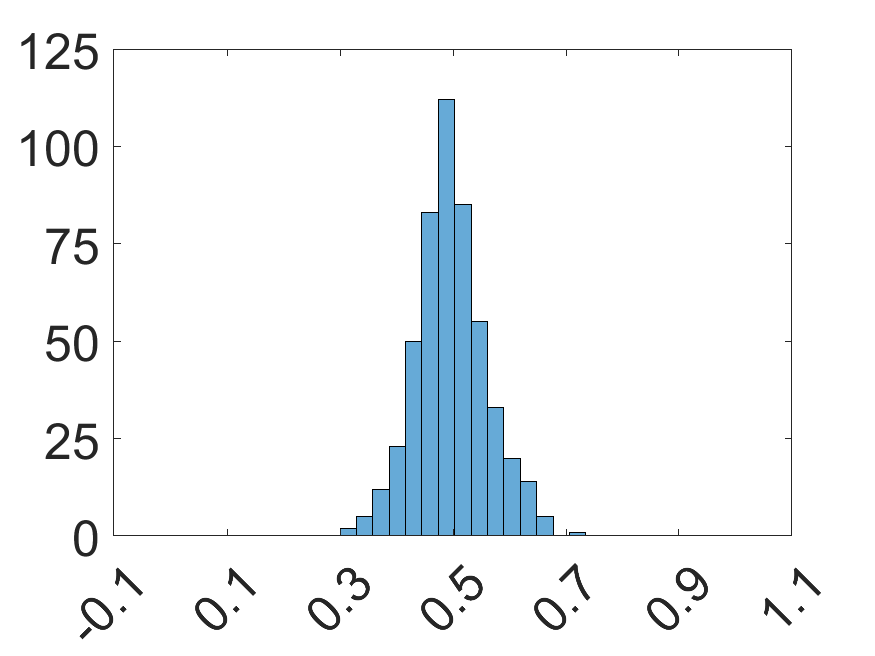}
					\caption{$N=200, m=50$}
				\end{subfigure}
				
				\begin{subfigure}{.24\textwidth}
					\centering
					\includegraphics[width=\linewidth]{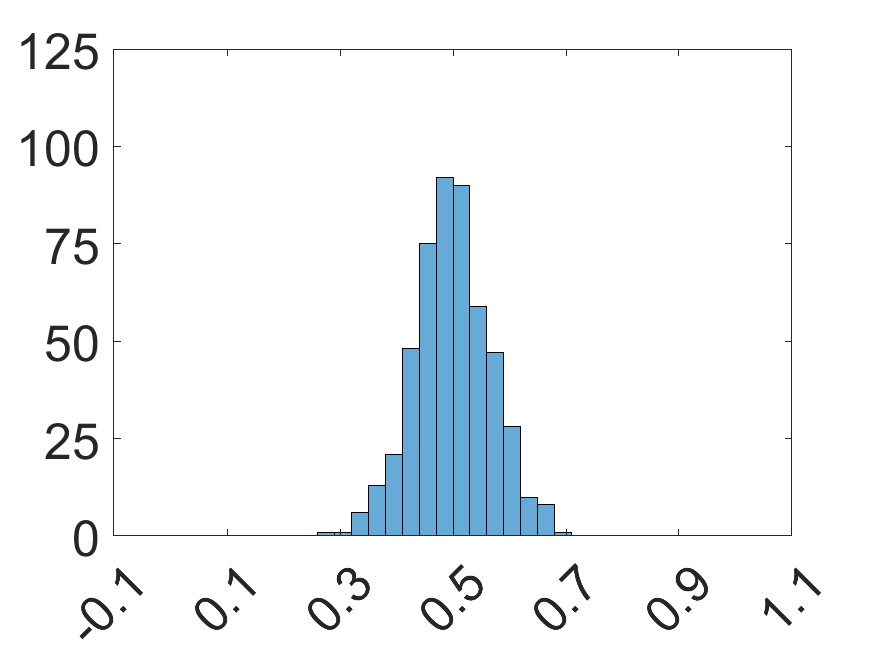}
					\caption{$N=10, m=100$}
				\end{subfigure}%
				\hfill
				\begin{subfigure}{.24\textwidth}
					\centering
					\includegraphics[width=\linewidth]{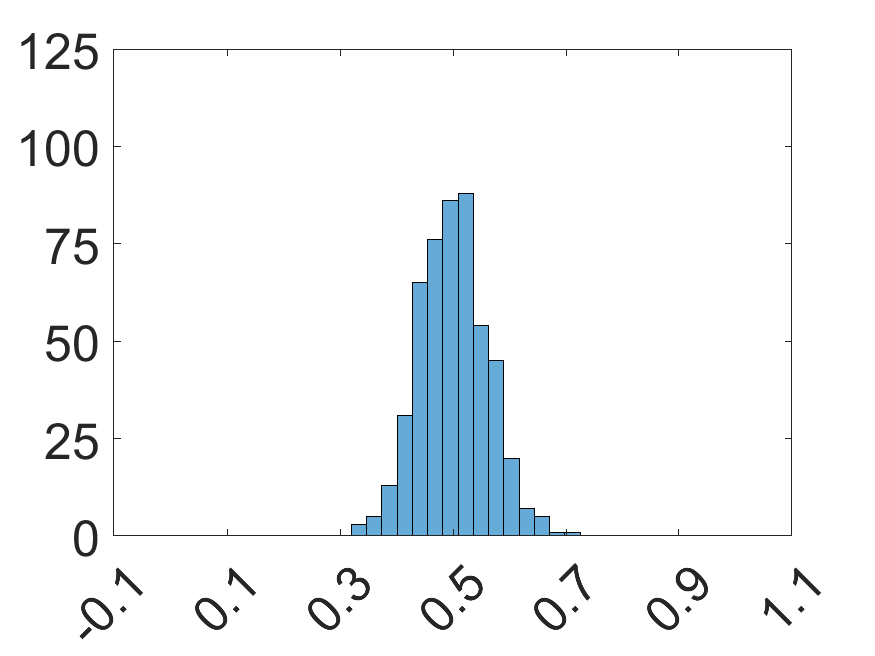}
					\caption{$N=20, m=100$}
				\end{subfigure}
				\hfill
				\begin{subfigure}{.24\textwidth}
					\centering
					\includegraphics[width=\linewidth]{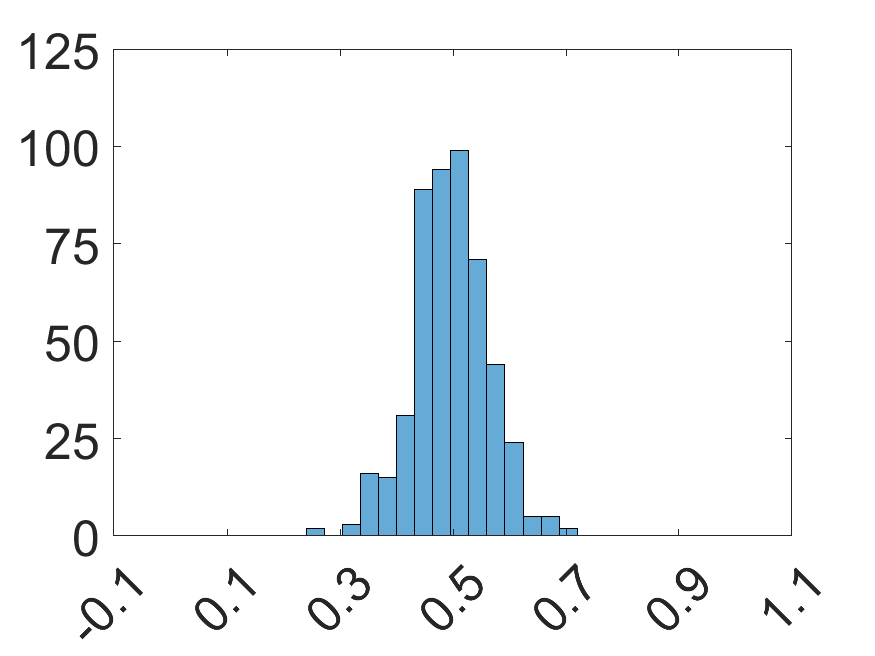}
					\caption{$N=50, m=100$}
				\end{subfigure}
				\hfill
				\begin{subfigure}{.24\textwidth}
					\centering
					\includegraphics[width=\linewidth]{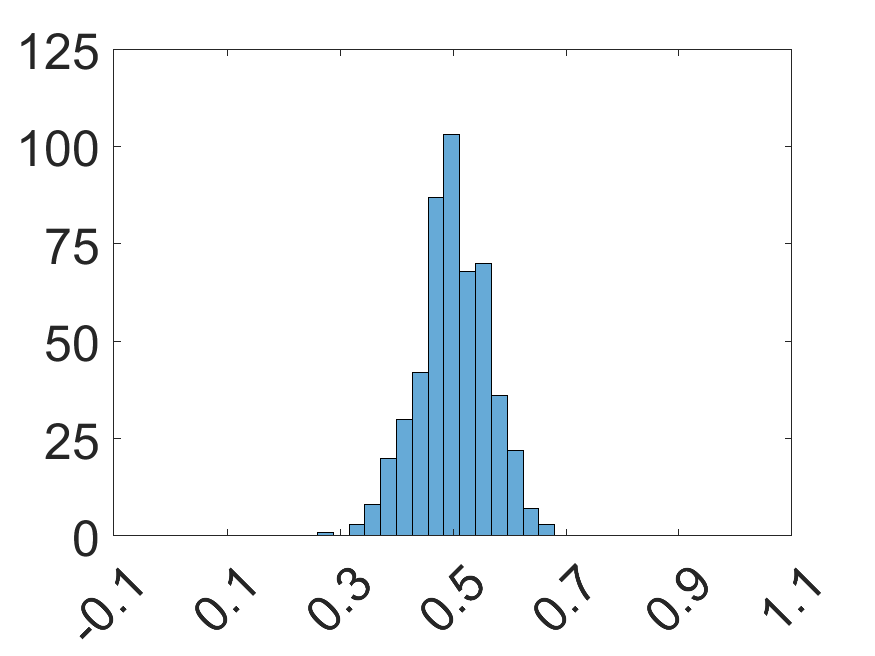}
					\caption{$N=100, m=100$}
				\end{subfigure}
				\hfill
				\begin{subfigure}{.24\textwidth}
					\centering
					\includegraphics[width=\linewidth]{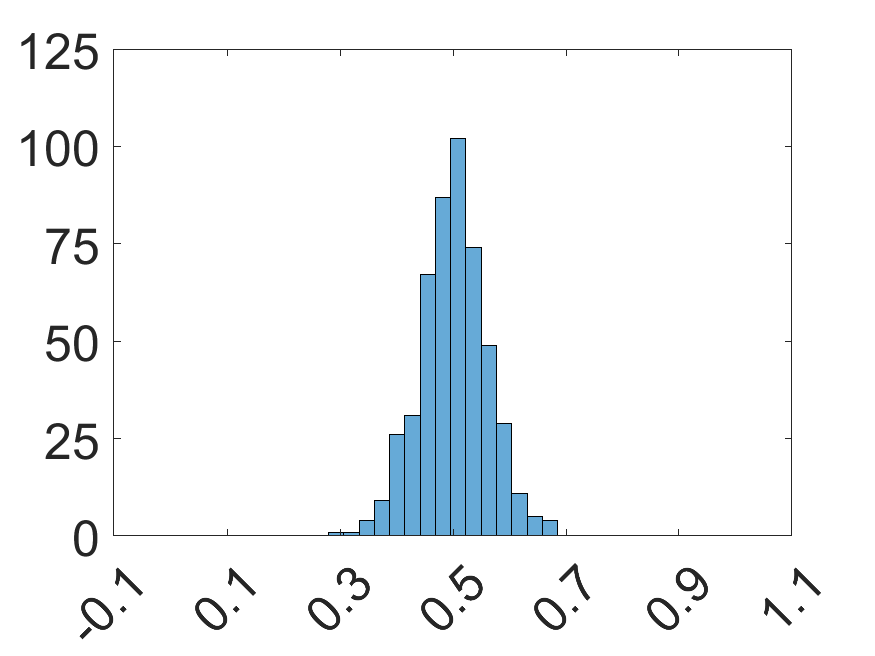}
					\caption{$N=200, m=100$}
				\end{subfigure}		
				
				\subcaption*{\textit{Note:} the true value is $\beta= 0.5$.}
				\label{fig:MC_hist_beta_CaseII_T80}
			\end{figure}
		\end{landscape}

		\subsection{Comparison of asymptotic standard errors}\label{app:SError}
		
		In this section, we compare the standard errors of the iterative estimators $\widehat {\theta}^{\dag}$ and $\widehat{\theta}^*$
		(see \eqref{eq:thetastar}). We compute these in two ways.

		First, we use their analytical expressions. Let $\widehat {\beta}^{\dag}$ and $\widehat{\beta}^*$  be the first elements of $\widehat {\theta}^{\dag}$ and $\widehat{\theta}^*$, respectively, and let
		$\widehat{\Sigma}(\theta^\dag)= \widehat{\Avar}\left[ \sqrt{NT} (\widehat{\theta}^{\dag} -\theta) \right]$ and
		$\widehat{\Sigma}(\theta^*)= \widehat{\Avar}\left[ \sqrt{NT} (\widehat{\theta}^{*} -\theta) \right]$
		be the estimated asymptotic covariance matrices as defined in  \eqref{eq:AVARtheta} (see Section \ref{app:MC_appendix_setup} above on how to fix the sign indeterminacy). Then, at each iteration of the MC exercise we compute: 
		$(NT)^{-1/2}\sqrt{[\widehat{\Sigma}(\theta^\dag)]_{11}}$ and $(NT)^{-1/2}\sqrt{[\widehat{\Sigma}(\theta^*)]_{11}}$, which are the estimators of the asymptotic standard errors of $\widehat {\beta}^{\dag}$ and $\widehat{\beta}^*$. Averages of these quantities over the $S$ iterations are in Table \ref{tab:SEbeta} (second and third column) for selected values of $m,N,$ and $T$.
		
		Second, we compute the MC standard errors of $\widehat {\beta}^{\dag}$ and $\widehat{\beta}^*$. Results are in Table \ref{tab:SEbeta} (fourth and fifth column) for selected values of $m,N,$ and $T$.

		\begin{table}[H]
			\centering
			\scriptsize{
				\caption{Standard errors of  $\widehat {\beta}^{\dag}$ and $\widehat{\beta}^*$ - $T=100$, $m=100$}\label{tab:SEbeta}
				\scalebox{1}[1]{
					\begin{tabular}{l | c c | c c }
						\hline \hline
						&\multicolumn{2}{c|}{Asymptotic s.e.}&\multicolumn{2}{c}{MC s.e.} \\
						$N= 50$ & case I & case II& case I & case II\\
						\hline
						$\widehat {\beta}^{\dag}$&0.0539&	0.0540&0.0540&	0.0540\\
						$\widehat {\beta}^{*}$&0.0539&	0.0540&0.0540&	0.0540\\
						\hline
						&\multicolumn{2}{c|}{Asymptotic s.e.}&\multicolumn{2}{c}{MC s.e.} \\
						$N= 100$ & case I & case II& case I & case II\\
						\hline
						$\widehat {\beta}^{\dag}$&0.0547&	0.0546&0.0541&	0.0541\\
						$\widehat {\beta}^{*}$&0.0547&	0.0546&0.0541&	0.0541\\
						\hline
						\hline
					\end{tabular}  	 	
				}
			}
		\end{table}

		\subsection{Comparison with TOPUP and TIPUP}\label{app:comparison_Chen_et_al}
		
		In this section, we compare the Monte Carlo performance of our tensor PC estimator with that of the two estimators proposed by \citet{chenyangzhang22}, TOPUP and TIPUP. These are designed for the case of idiosyncratic tensors with no serial autocorrelation. They are implemented considering different values of $h_0$, i.e., the maximum lag used in the computation of inner and outer products of the tensor data. In particular, we report results for $h_0 = 1$ and $h_0 = 10$.
		In Table \ref{tab:comparison_Chen_et_al}, we report the relative RMSE of the estimates of $U$ provided by TOPUP, TIPUP, and our PC-based approach, as well as the RMSE of the  TOPUP and TIPUP estimators with respect to ours.
		As expected, the TOPUP and PCA estimators behave very similarly when there is no autocorrelation in $\mathcal{E}$ (columns (a) and (b)), while the performance of TOPUP keeps deteriorating as the autocorrelation parameter $\rho_{\mathcal{E}}$ increases to 0.5 (columns (c) and (d)) and up to 0.9 (columns (e) and (f)). Our estimator has instead a similar performance in all considered cases, since by construction it is not affected by serial idiosyncratic autocorrelations, as long as the autocovariances are summable as prescribed by Assumption \ref{as:idiosyncratic_component}\ref{as:idiosyncratic_component_ii}. The TIPUP estimator seems to work poorly in all considered settings. 
		
		\begin{table}[H]
			\centering
			\caption{Comparison with TOPUP and TIPUP estimators ($N=10$, $T=100$)}
			\label{tab:comparison_Chen_et_al}
			\scriptsize{
				\begin{tabular}{l |c | cc | cc | cc}
					\hline \hline
					&           & \multicolumn{2}{c |}{$\rho_{\mathcal{E}}   = 0.0$} & \multicolumn{2}{c |}{$\rho_{\mathcal{E}}   = 0.5$} & \multicolumn{2}{c}{$\rho_{\mathcal{E}}   = 0.9$} \\
					&           & (a)                    & (b)                 & (c)                   & (d)      & (e)		&(f)            \\		
					&                            & ReRMSE                   & ratio to PC                & ReRMSE                   & ratio to PC                 & ReRMSE                  & ratio to PC              \\
					\hline
					PC            & \multirow{5}{*}{$m = 20$}  & 3.7\%  & 1.00  & 3.7\%  & 1.00  & 3.7\%   & 1.00  \\
					TOPUP $h0=1$  &                            & 2.7\%  & 0.74  & 3.3\%  & 0.91  & 10.2\%  & 2.75  \\
					TOPUP $h0=10$ &                            & 2.7\%  & 0.73  & 3.3\%  & 0.89  & 10.1\%  & 2.73  \\
					TIPUP $h0=1$  &                            & 58.8\% & 15.95 & 91.4\% & 24.75 & 92.5\%  & 24.88 \\
					TIPUP $h0=10$ &                            & 15.0\% & 4.08  & 46.3\% & 12.53 & 88.9\%  & 23.91 \\
					\hline
					PC            & \multirow{5}{*}{$m = 50$}  & 1.8\%  & 1.00  & 1.8\%  & 1.00  & 1.8\%   & 1.00  \\
					TOPUP $h0=1$  &                            & 1.4\%  & 0.77  & 1.7\%  & 0.94  & 5.4\%   & 2.94  \\
					TOPUP $h0=10$ &                            & 1.3\%  & 0.72  & 1.6\%  & 0.90  & 5.3\%   & 2.93  \\
					TIPUP $h0=1$  &                            & 57.7\% & 32.16 & 87.2\% & 48.54 & 100.7\% & 55.31 \\
					TIPUP $h0=10$ &                            & 14.8\% & 8.27  & 20.4\% & 11.37 & 81.0\%  & 44.52 \\
					\hline
					PC            & \multirow{5}{*}{$m = 100$} & 1.3\%  & 1.00  & 1.3\%  & 1.00  & 1.3\%   & 1.00  \\
					TOPUP $h0=1$  &                            & 1.0\%  & 0.81  & 1.2\%  & 0.98  & 3.9\%   & 3.05  \\
					TOPUP $h0=10$ &                            & 0.9\%  & 0.72  & 1.1\%  & 0.90  & 3.9\%   & 3.05  \\
					TIPUP $h0=1$  &                            & 57.5\% & 45.25 & 68.6\% & 53.97 & 91.3\%  & 70.80 \\
					TIPUP $h0=10$ &                            & 14.7\% & 11.56 & 16.3\% & 12.82 & 52.1\%  & 40.38  \\ 
					\hline 	\hline              
				\end{tabular}
			}
		\end{table}		
		
		\section{Empirical application: data and additional results}\label{app:data}
		
		In this section we provide additional information about the data used in the empirical application considered in the paper, as well as additional results. 
		In Section \ref{app:data Network layers} we provide details about how the network layers are built and about GDP data. 
		Additional empirical results are provided in Section \ref{app:additional results}.
		
		\subsection{Data}\label{app:data Network layers}
		
		\noindent\textsc{Countries.}
		The $N=24$ countries we consider are further divided into: advanced economies, which are:
		Australia (AUS), Belgium (BEL), Canada (CAN), France (FRA), Germany (DEU), Italy (ITA), Japan (JAP), South Korea (KOR), Netherlands (NLD), Norway (NOR), Spain (ESP), Sweden (SWE), Switzerland (CHE), United Kingdom (GBR), and United States (USA); and emerging economies, which are: Brazil (BRA), China (CHN),  Hong Kong (HKG),  India (IND), Indonesia (IDN), Mexico (MEX),  Saudi Arabia (SAU), South Africa (ZAF), and  Turkey (TUR), 
		
		\noindent\textsc{Real GDP.} For real GDP growth rates, we use the dataset compiled by \citet{mohaddesraissi20} to which we refer for details on the primary data sources and data adjustments. The dataset contains time series of log real GDP indices from 1979Q2 to 2019Q4, we take the first differences to calculate quarterly growth rates.
		For Hong Kong, which is not included in the dataset of \citet{mohaddesraissi20}, we retrieve data from the IMF's International Financial Statistics.

		\noindent\textsc{Network layers.}
		The full list of the $m=25$ network layers is reported in Table \ref{tab:layer_list}. Here we summarize the way we built our data.
		
		\begin{table}[t!]
			\centering
			\caption{Layers of the network of countries.}
			\begin{tabular}{cp{14cm}}
				\hline \hline
				number & layer  \\	
				\hline
				1 & \textbf{Trade in goods}: Food and live animals                                               \\
				2 & \textbf{Trade in goods}: Beverages and tobacco                                               \\
				3 & \textbf{Trade in goods}: Crude materials, inedible, except fuels                             \\
				4 & \textbf{Trade in goods}: Mineral fuels, lubricants and related materials                     \\
				5 & \textbf{Trade in goods}: Animal and vegetable oils, fats and waxes                           \\
				6 & \textbf{Trade in goods}: Chemicals and related products, n.e.s.                              \\
				7 & \textbf{Trade in goods}: Manufactured goods classified chiefly by material                   \\
				8 & \textbf{Trade in goods}: Machinery and transport equipment                                   \\
				9 & \textbf{Trade in goods}: Miscellaneous manufactured articles                                 \\
				\hline 
				10 & \textbf{Trade in services}: Transport \\
				11 & \textbf{Trade in services}: Travel \\
				12 & \textbf{Trade in services}: Construction \\
				13 & \textbf{Trade in services}: Insurance and pension services \\
				14 & \textbf{Trade in services}: Financial services \\
				15 & \textbf{Trade in services}: Charges for the use of intellectual property n.i.e. \\
				16 & \textbf{Trade in services}: Telecommunications, computer, and information services \\
				17 & \textbf{Trade in services}: Other business services \\
				18 & \textbf{Trade in services}: Personal, cultural, and recreational services \\
				19 & \textbf{Trade in services}: Government goods and services n.i.e. \\
				\hline 
				20 & \textbf{Financial assets and liabilities}: Equity\\
				21 & \textbf{Financial assets and liabilities}: Long-term debt\\
				22 & \textbf{Financial assets and liabilities}: Short-term debt\\
				23 &  \textbf{Financial assets and liabilities}: Flows of banks' assets and liabilities\\
				\hline
				24 & \textbf{Mergers and acquisitions}: Goods sectors (Agriculture, forestry and fishing; Mining and construction; Manufacturing)\\
				25 & \textbf{Mergers and acquisitions}: Services sectors (Transportation, communications and utilities; Wholesale and retail trade; Finance, insurance and real estate; Services; Public Administration)\\
				\hline \hline                        
			\end{tabular}
			\label{tab:layer_list}
		\end{table}


		\noindent\textit{Trade in goods and services.}
		We calculate trade weights for goods using trade data from the UN Comtrade Database and classifying products into 10 categories, based on the 1-digit Standard International Trade Classification (SITC, revisions 3-4).
		To facilitate interpretation of the results, we exclude unclassified commodities (SITC code 9).
		
		Trade weights for services are computed using data from the OECD-WTO Balanced Trade in Services Database (BPM6 edition). Services are classified into 10 categories, based on the Balance of Payments Services Classification (EBOPS, 2010 edition for the period 2005-2019 and 2002 edition for the period 2001-2004). Two categories of EBOPS 2010, namely ``SA" (manufacturing services on physical inputs owned by others) and ``SB" (maintenance and repair services n.i.e.), are excluded because they were absent in EBOPS 2002 and have a large number of missing values in EBOPS 2010.

		Given a product/service type representing the $k$-th layer of the network, the weight of country $j$ for country $i$ in layer $k$ at time $t$ (i.e., the element $ijk$ of the weight tensor $\mathcal{W}_t$) is calculated as:
		\begin{equation}\nonumber		
			\mathcal{W}_{ijk,t} = N  \frac{imports_{ij,k,t}+exports_{ij,k,t}}{\sum_{h=1}^{N}(imports_{ih,k,t}+exports_{ih,k,t})},
		\end{equation}
		where $imports_{ij,k,t}$ is the dollar amount of commodities of type $k$ imported by country $i$ from country $j$ at time $t$ and  $exports_{ij,k,t}$ is the amount of exports of $k$ from country $i$ to country $j$.

		\noindent\textit{Financial assets and liabilities.}		
		To construct the financial layers of the network, we first use data on outstanding bilateral assets/liabilities, classified into three categories of financial claims: equity, short-term debt and long-term debt. The data source is the IMF's Coordinated Portfolio Investment Survey (CPIS) database. The liabilities data that we use are derived from the assets data reported by counterparties. As explained by the IMF, while many countries report liabilities data directly, ``more reliable detailed cross border positions data can usually be collected on an economy's holdings of portfolio investment because the holder (creditor) will usually know what securities it holds. On the liabilities side, the issuer of a security (debtor) may not know the residency of the holder because the securities may be held by foreign custodians or other intermediaries. Using the assets data reported by CPIS participating economies, the IMF derives liabilities data for all economies (CPIS reporters as well as nonreporters); these data are termed {\it derived liabilities}'' (source: https://datahelp.imf.org/knowledgebase/articles/500647-why-are-coordinated-portfolio-investment-cpis-da).
		
		Letting $k$ identify a category of financial claims, the weight of country $j$ for country $i$ in layer $k$ is calculated as:
		\begin{equation}
			\mathcal{W}_{ijk,t}  = N  \frac{assets_{ij,k,t}+liabilities_{ij,k,t}}{\sum_{h=1}^{N}(assets_{ih,k,t}+liabilities_{ih,k,t})},\nonumber
		\end{equation}
		where $assets_{ij,k,t}$ is the stock of assets held by country $i$ and issued by country $j$, and $liabilities_{ij,k,t}$ is the stock of liabilities of country $i$ towards country $j$.  
		
		\noindent\textit{Flows of banks' assets and liabilities.}
		We use data on flows of banks' assets and liabilities, as measured in the Locational Banking Statistics by the Bank for International Settlements (BIS). We cannot calculate bilateral capital flows by simply taking the first differences of the IMF CPIS assets and liabilities. The reason is that financial positions are reported at their market values in the CPIS, so first differences also reflect changes in valuation and exchange rate movements. In its International Banking Statistics, the BIS reports changes in amounts outstanding adjusted for exchange-rate movements and breaks in reporting methodologies, in order to approximate flows. In the case of BIS banking data, we generally use data provided directly by country $i$. Five countries (China, India, Indonesia, Norway and Saudi Arabia) do not report data. For these countries, we use derived flows, i.e., data reported by their counterparties. As for bilateral flows among non-reporting countries, we assume them to be zero.
		
		The weight of country $j$ for country $i$ in layer $k$ is calculated as:
		\begin{equation}
			\mathcal{W}_{ijk,t} = N  \frac{|\Delta bank\_assets_{ij,t}|+|\Delta bank\_liabilities_{ij,t}|}{\sum_{h=1,h\neq i}^{N}(|\Delta bank\_assets_{ih,t}| + |\Delta bank\_liabilities_{ih,t}|)},\nonumber
		\end{equation}
		where $\Delta bank\_assets_{ij,t}$ is the annual change (at year-end) in outstanding assets held by banks located in country $i$ and issued by counterparties of any type located in country $j$, and  $ \allowbreak \Delta bank\_liabilities_{ij,t}$ is the annual change in liabilities of banks located in country $i$ and held by counterparties of any type located in country $j$.

		\noindent\textit{Mergers and acquisitions.}
		We collect all cross-border mergers and acquisitions (M\&A) deals involving companies located in two different countries and classify them in two macro-sectors of economic activity, ``goods sectors" and ``services sectors", using the Standard Industrial Classification (SIC). Data are from Bureau Van Dijk's Zephyr database (160,713 completed deals between the 24 countries considered over the period 2001-2019, corresponding to an average of around 15 deals for each pair of countries per year). We consider not only operations that are strictly speaking mergers or acquisitions, i.e., involving more than 50\% of the target firms' capital, but also capital increases and acquisitions of minority stakes. We classify the M\&A deals using the primary economic sector of the target companies. The rationale for this choice is that acquirors tend to be larger companies operating in a larger number of sectors than targets, so using the sectors of targets should provide a more accurate classification. 
		Finally, given the greater volatility of M\&A weights compared to the other types of weights (M\&A operations are rarer than transactions in goods, services and portfolio instruments), we smooth out the M\&A weights by taking 3-year moving averages.

		If $k$ identifies M\&A deals in a given sector, the weight of country $j$ for country $i$ is given by:
		\begin{equation}
			\mathcal{W}_{ijk,t} = N  \frac{m\&a_{ij,k,t}}{\sum_{h=1,h\neq i}^{N}(m\&a_{ih,k,t})}
		\end{equation}
		where $m\&a_{ij,k,t}$ denotes the dollar value of all mergers and acquisitions in sector $k$ at time $t$ involving a company located in country $i$ and a company located in country $j$. This includes both the deals in which country $i$'s companies are acquirors and the deals in which they are targets.

		\subsection{Additional results}\label{app:additional results}

		\subsubsection{Cosine similarity} 
		
		The main premise of our approach is that the different layers of the network are driven by common factor networks. As a preliminary analysis, it is therefore useful to assess the degree of similarity between the observed layers. As suggested by \citet{bargiglietal15}, we consider cosine similarity as a measure of layer similarity for weighted networks. 
		The cosine similarity ($C$) between two vectors $x$ and $y$ is their inner product normalized by the product of their norms, i.e.:
		$C(x,y) = \frac{x'y}{\norm{x}\,\norm{y}}. $
		To calculate the similarity between any two layers $h$ and $k$, we vectorize the matrices $\mathcal{W}_{\cdot\cdot,h,t}$ and $\mathcal{W}_{\cdot\cdot,k,t}$ for each $t$, then stack the resulting vectors for $t=1,\dots,T$ to obtain two column vectors of length $N^2T \times 1$ each. Results are presented in Table \ref{fig:similarity}.

		\begin{landscape}
			\begin{table}[t!]
				\vspace{50pt}
				\centering
				\caption{Cosine similarity between layers.}
				\scriptsize
				{
					\begin{tabular}{l|lllllllllllllllllllllllll}
						\hline \hline
						\textit{layer} & 1 & 2 & 3 & 4 & 5 & 6 & 7 & 8 & 9 & 10 & 11 & 12 & 13 & 14 & 15 & 16 & 17 & 18 & 19 & 20 & 21 &22 &23 &24 &25 \\
						\hline
						1 & & 0.75 & 0.79 & 0.74 & 0.70 & 0.87 & 0.87 & 0.85 & 0.80 & 0.80 & 0.83 & 0.74 & 0.71 & 0.71 & 0.67 & 0.78 & 0.77 & 0.74 & 0.64 & 0.62 & 0.71 & 0.54 & 0.56 & 0.54 & 0.58 \\
						2 & 0.75 & & 0.63 & 0.58 & 0.57 & 0.77 & 0.72 & 0.74 & 0.67 & 0.74 & 0.78 & 0.69 & 0.72 & 0.70 & 0.63 & 0.74 & 0.75 & 0.73 & 0.60 & 0.64 & 0.68 & 0.59 & 0.59 & 0.58 & 0.61 \\
						3 & 0.79 & 0.63 & & 0.68 & 0.65 & 0.76 & 0.82 & 0.80 & 0.78 & 0.68 & 0.71 & 0.61 & 0.57 & 0.54 & 0.51 & 0.63 & 0.62 & 0.61 & 0.50 & 0.48 & 0.53 & 0.40 & 0.45 & 0.48 & 0.49 \\
						4 & 0.74 & 0.58 & 0.68 & & 0.63 & 0.72 & 0.73 & 0.69 & 0.60 & 0.63 & 0.68 & 0.65 & 0.57 & 0.58 & 0.50 & 0.64 & 0.61 & 0.61 & 0.47 & 0.46 & 0.50 & 0.42 & 0.46 & 0.45 & 0.47 \\
						5 & 0.70 & 0.57 & 0.65 & 0.63 & & 0.64 & 0.65 & 0.61 & 0.57 & 0.56 & 0.60 & 0.57 & 0.49 & 0.48 & 0.44 & 0.55 & 0.53 & 0.53 & 0.43 & 0.40 & 0.47 & 0.36 & 0.37 & 0.40 & 0.43 \\
						6 & 0.87 & 0.77 & 0.76 & 0.72 & 0.64 & & 0.93 & 0.94 & 0.87 & 0.84 & 0.85 & 0.77 & 0.75 & 0.72 & 0.72 & 0.83 & 0.83 & 0.78 & 0.68 & 0.66 & 0.70 & 0.57 & 0.61 & 0.61 & 0.62 \\
						7 & 0.87 & 0.72 & 0.82 & 0.73 & 0.65 & 0.93 & & 0.93 & 0.89 & 0.80 & 0.84 & 0.74 & 0.70 & 0.67 & 0.62 & 0.77 & 0.75 & 0.71 & 0.61 & 0.57 & 0.65 & 0.53 & 0.56 & 0.55 & 0.57 \\
						8 & 0.85 & 0.74 & 0.80 & 0.69 & 0.61 & 0.94 & 0.93 & & 0.89 & 0.83 & 0.85 & 0.75 & 0.74 & 0.70 & 0.68 & 0.80 & 0.80 & 0.76 & 0.67 & 0.63 & 0.69 & 0.56 & 0.58 & 0.59 & 0.62 \\
						9 & 0.80 & 0.67 & 0.78 & 0.60 & 0.57 & 0.87 & 0.89 & 0.89 & & 0.78 & 0.80 & 0.69 & 0.68 & 0.66 & 0.65 & 0.74 & 0.75 & 0.70 & 0.64 & 0.62 & 0.65 & 0.52 & 0.56 & 0.54 & 0.57 \\
						10 & 0.80 & 0.74 & 0.68 & 0.63 & 0.56 & 0.84 & 0.80 & 0.83 & 0.78 & & 0.84 & 0.80 & 0.82 & 0.77 & 0.79 & 0.87 & 0.86 & 0.81 & 0.74 & 0.74 & 0.76 & 0.61 & 0.65 & 0.63 & 0.66 \\
						11 & 0.83 & 0.78 & 0.71 & 0.68 & 0.60 & 0.85 & 0.84 & 0.85 & 0.80 & 0.84 & & 0.80 & 0.78 & 0.78 & 0.70 & 0.83 & 0.83 & 0.83 & 0.68 & 0.70 & 0.75 & 0.65 & 0.65 & 0.61 & 0.66 \\
						12 & 0.74 & 0.69 & 0.61 & 0.65 & 0.57 & 0.77 & 0.74 & 0.75 & 0.69 & 0.80 & 0.80 & & 0.72 & 0.74 & 0.67 & 0.78 & 0.77 & 0.74 & 0.61 & 0.64 & 0.69 & 0.62 & 0.59 & 0.57 & 0.61 \\
						13 & 0.71 & 0.72 & 0.57 & 0.57 & 0.49 & 0.75 & 0.70 & 0.74 & 0.68 & 0.82 & 0.78 & 0.72 & & 0.84 & 0.76 & 0.85 & 0.85 & 0.82 & 0.72 & 0.75 & 0.77 & 0.65 & 0.70 & 0.63 & 0.67 \\
						14 & 0.71 & 0.70 & 0.54 & 0.58 & 0.48 & 0.72 & 0.67 & 0.70 & 0.66 & 0.77 & 0.78 & 0.74 & 0.84 & & 0.78 & 0.85 & 0.87 & 0.85 & 0.74 & 0.79 & 0.81 & 0.70 & 0.71 & 0.64 & 0.69 \\
						15 & 0.67 & 0.63 & 0.51 & 0.50 & 0.44 & 0.72 & 0.62 & 0.68 & 0.65 & 0.79 & 0.70 & 0.67 & 0.76 & 0.78 & & 0.84 & 0.88 & 0.80 & 0.81 & 0.83 & 0.76 & 0.59 & 0.62 & 0.63 & 0.65 \\
						16 & 0.78 & 0.74 & 0.63 & 0.64 & 0.55 & 0.83 & 0.77 & 0.80 & 0.74 & 0.87 & 0.83 & 0.78 & 0.85 & 0.85 & 0.84 & & 0.91 & 0.88 & 0.80 & 0.80 & 0.79 & 0.66 & 0.70 & 0.68 & 0.71 \\
						17 & 0.77 & 0.75 & 0.62 & 0.61 & 0.53 & 0.83 & 0.75 & 0.80 & 0.75 & 0.86 & 0.83 & 0.77 & 0.85 & 0.87 & 0.88 & 0.91 & & 0.89 & 0.82 & 0.84 & 0.83 & 0.69 & 0.70 & 0.68 & 0.71 \\
						18 & 0.74 & 0.73 & 0.61 & 0.61 & 0.53 & 0.78 & 0.71 & 0.76 & 0.70 & 0.81 & 0.83 & 0.74 & 0.82 & 0.85 & 0.80 & 0.88 & 0.89 & & 0.75 & 0.79 & 0.77 & 0.67 & 0.69 & 0.67 & 0.71 \\
						19 & 0.64 & 0.60 & 0.50 & 0.47 & 0.43 & 0.68 & 0.61 & 0.67 & 0.64 & 0.74 & 0.68 & 0.61 & 0.72 & 0.74 & 0.81 & 0.80 & 0.82 & 0.75 & & 0.79 & 0.72 & 0.53 & 0.58 & 0.59 & 0.61 \\
						20 & 0.62 & 0.64 & 0.48 & 0.46 & 0.40 & 0.66 & 0.57 & 0.63 & 0.62 & 0.74 & 0.70 & 0.64 & 0.75 & 0.79 & 0.83 & 0.80 & 0.84 & 0.79 & 0.79 & & 0.80 & 0.67 & 0.64 & 0.66 & 0.68 \\
						21 & 0.71 & 0.68 & 0.53 & 0.50 & 0.47 & 0.70 & 0.65 & 0.69 & 0.65 & 0.76 & 0.75 & 0.69 & 0.77 & 0.81 & 0.76 & 0.79 & 0.83 & 0.77 & 0.72 & 0.80 & & 0.75 & 0.67 & 0.60 & 0.64 \\
						22 & 0.54 & 0.59 & 0.40 & 0.42 & 0.36 & 0.57 & 0.53 & 0.56 & 0.52 & 0.61 & 0.65 & 0.62 & 0.65 & 0.70 & 0.59 & 0.66 & 0.69 & 0.67 & 0.53 & 0.67 & 0.75 & & 0.61 & 0.56 & 0.58 \\
						23 & 0.56 & 0.59 & 0.45 & 0.46 & 0.37 & 0.61 & 0.56 & 0.58 & 0.56 & 0.65 & 0.65 & 0.59 & 0.70 & 0.71 & 0.62 & 0.70 & 0.70 & 0.69 & 0.58 & 0.64 & 0.67 & 0.61 & & 0.59 & 0.60 \\
						24 & 0.54 & 0.58 & 0.48 & 0.45 & 0.40 & 0.61 & 0.55 & 0.59 & 0.54 & 0.63 & 0.61 & 0.57 & 0.63 & 0.64 & 0.63 & 0.68 & 0.68 & 0.67 & 0.59 & 0.66 & 0.60 & 0.56 & 0.59 & & 0.61 \\
						25 & 0.58 & 0.61 & 0.49 & 0.47 & 0.43 & 0.62 & 0.57 & 0.62 & 0.57 & 0.66 & 0.66 & 0.61 & 0.67 & 0.69 & 0.65 & 0.71 & 0.71 & 0.71 & 0.61 & 0.68 & 0.64 & 0.58 & 0.60 & 0.61 & \\
						\hline \hline
					\end{tabular}
				}
				\vspace{7pt}
				\subcaption*{\footnotesize The table reports the cosine similarity coefficients between the 25 layers of the network. Please refer to Table \ref{tab:layer_list} for the list of layers.}
				\label{fig:similarity}
			\end{table}
		\end{landscape}

		\subsubsection{Network factors}
\begin{figure}[t!]
			\centering
			\caption{Network factors (averages 2001-2019).}
			\begin{tabular}{lll}
				\includegraphics[width=0.3\textwidth, height= 0.3\textwidth]{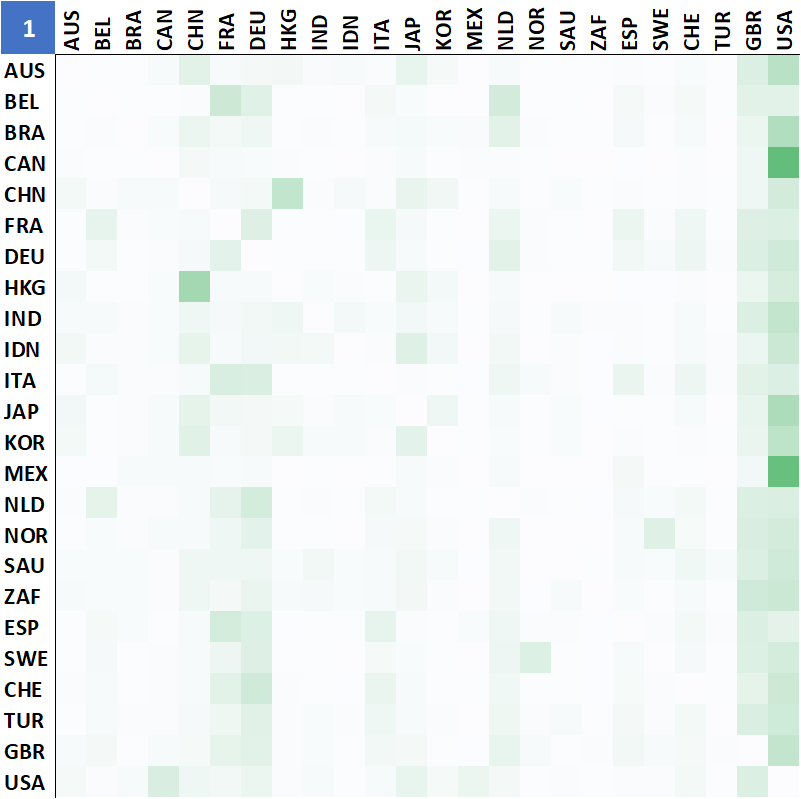}&
				\includegraphics[width=0.3\textwidth, height= 0.3\textwidth]{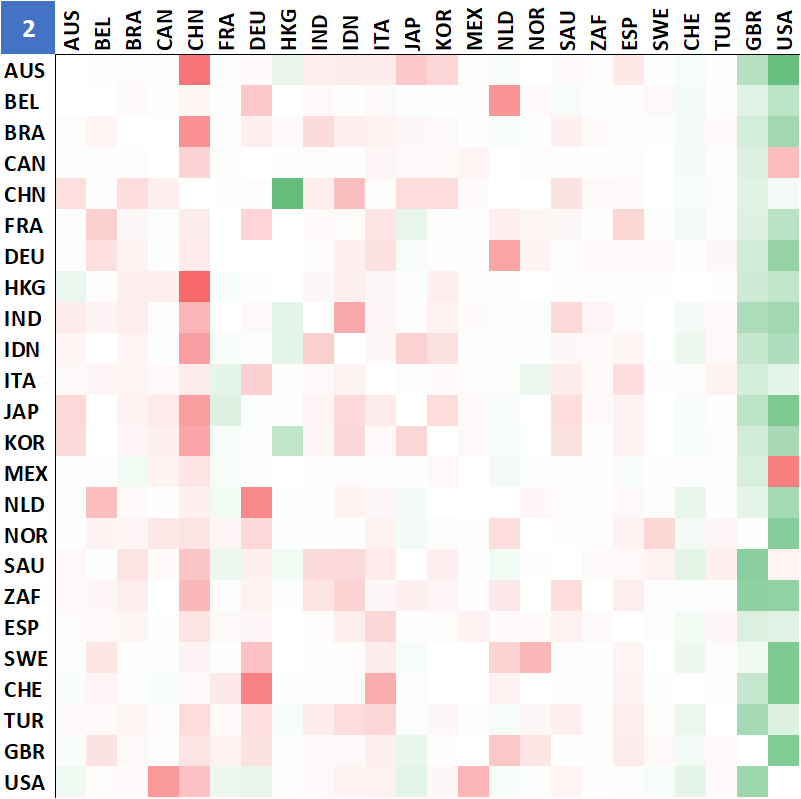}&
				\includegraphics[width=0.3\textwidth, height= 0.3\textwidth]{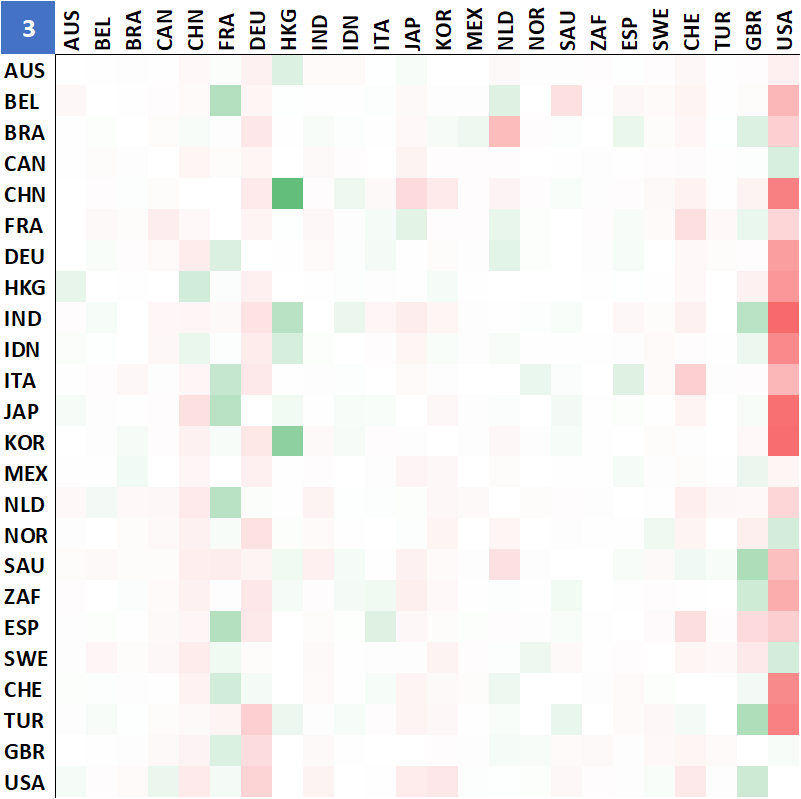}\\
				\includegraphics[width=0.3\textwidth, height= 0.3\textwidth]{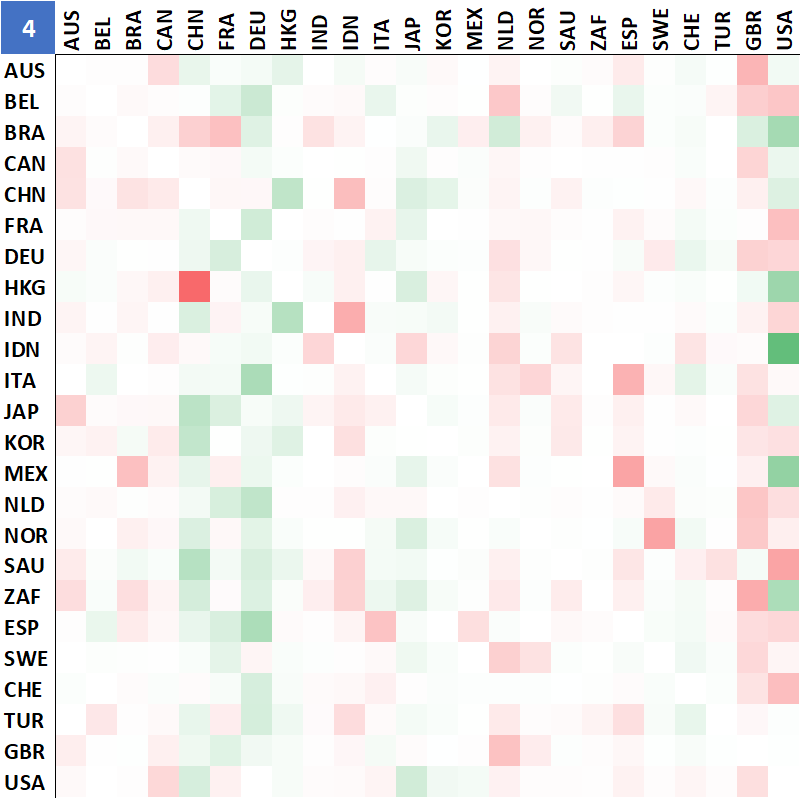}&
				\includegraphics[width=0.3\textwidth, height= 0.3\textwidth]{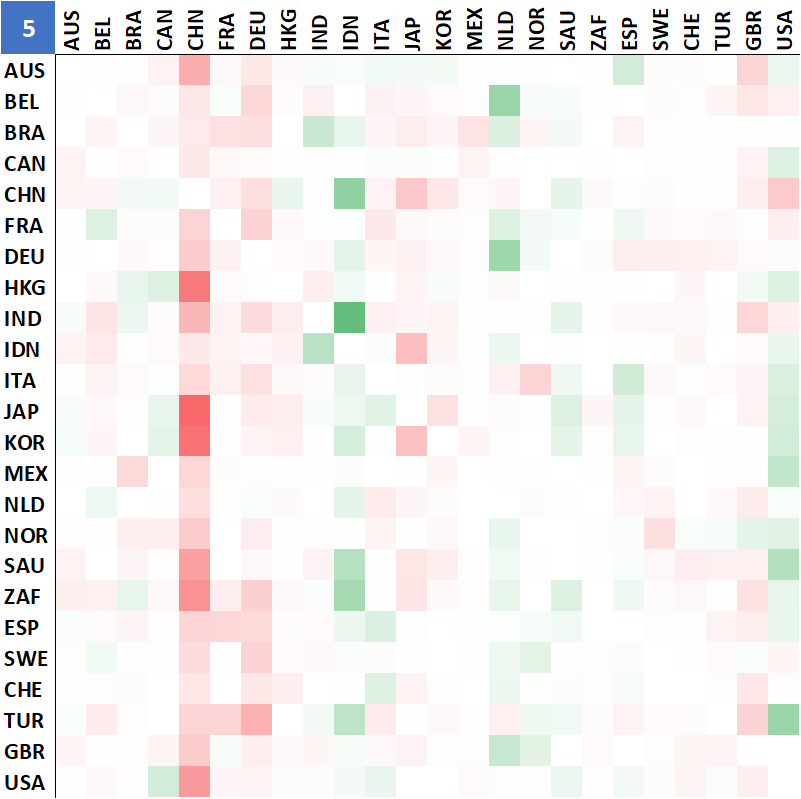}&
				\includegraphics[width=0.3\textwidth, height= 0.3\textwidth]{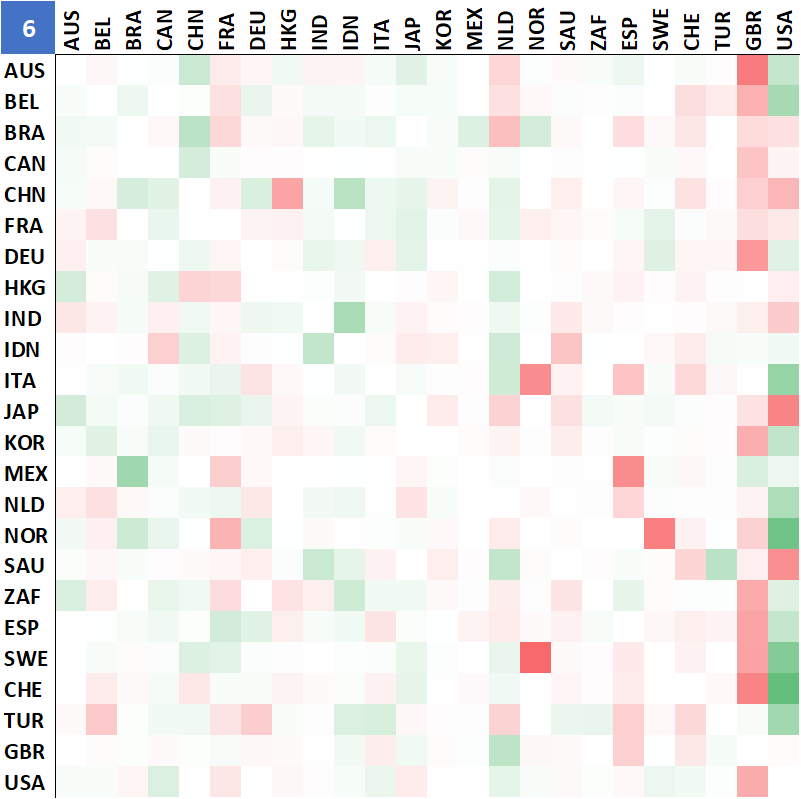}
			\end{tabular}
			\vspace{7pt}
			\subcaption*{\footnotesize Each panel reports  the estimated network factors $\widehat F_{j,t}$, $j=1,\ldots, 6$, averaged over the period 2001-2019, using color scales. Green cells identify positive entries and red cells negative entries, with darker shades of color indicating larger entries in absolute value. }
			\label{fig:Wtilde}
		\end{figure}

\begin{figure}[t!]
			\centering
			\caption{Factor loadings for different network layers.}
			\begin{tabular}{lll}
				\includegraphics[scale=0.072]{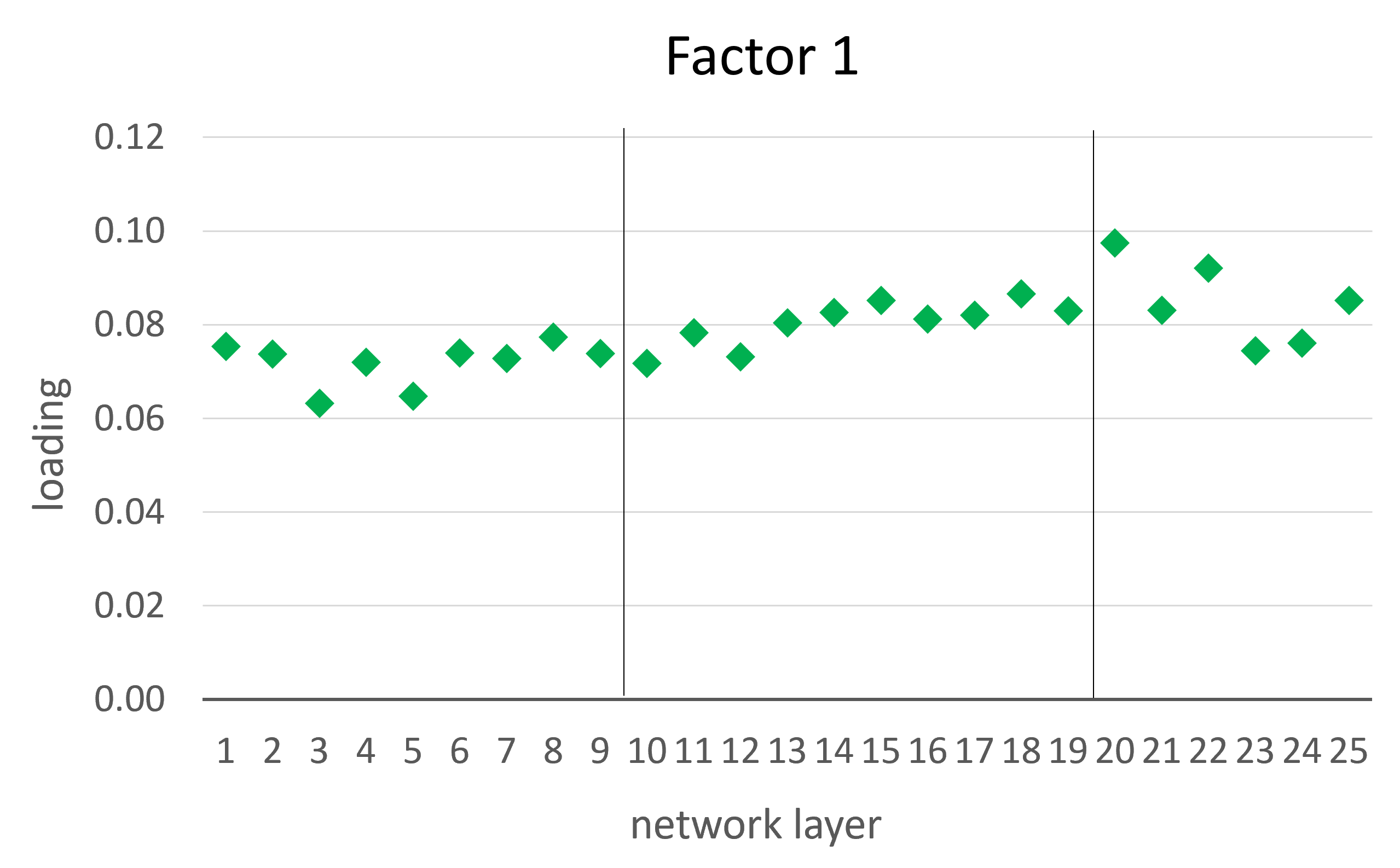} &
				\includegraphics[scale=0.072]{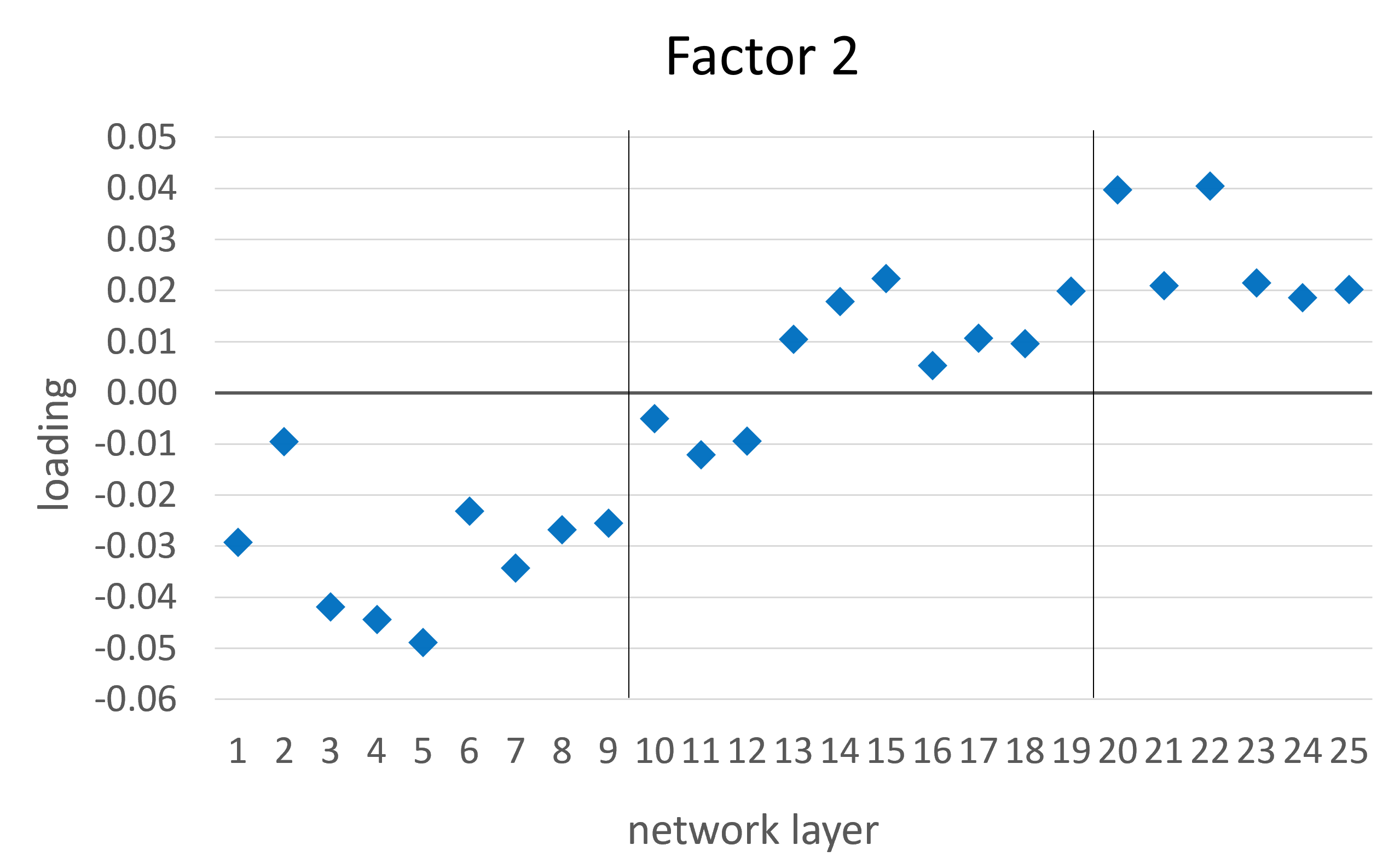} &
				\includegraphics[scale=0.072]{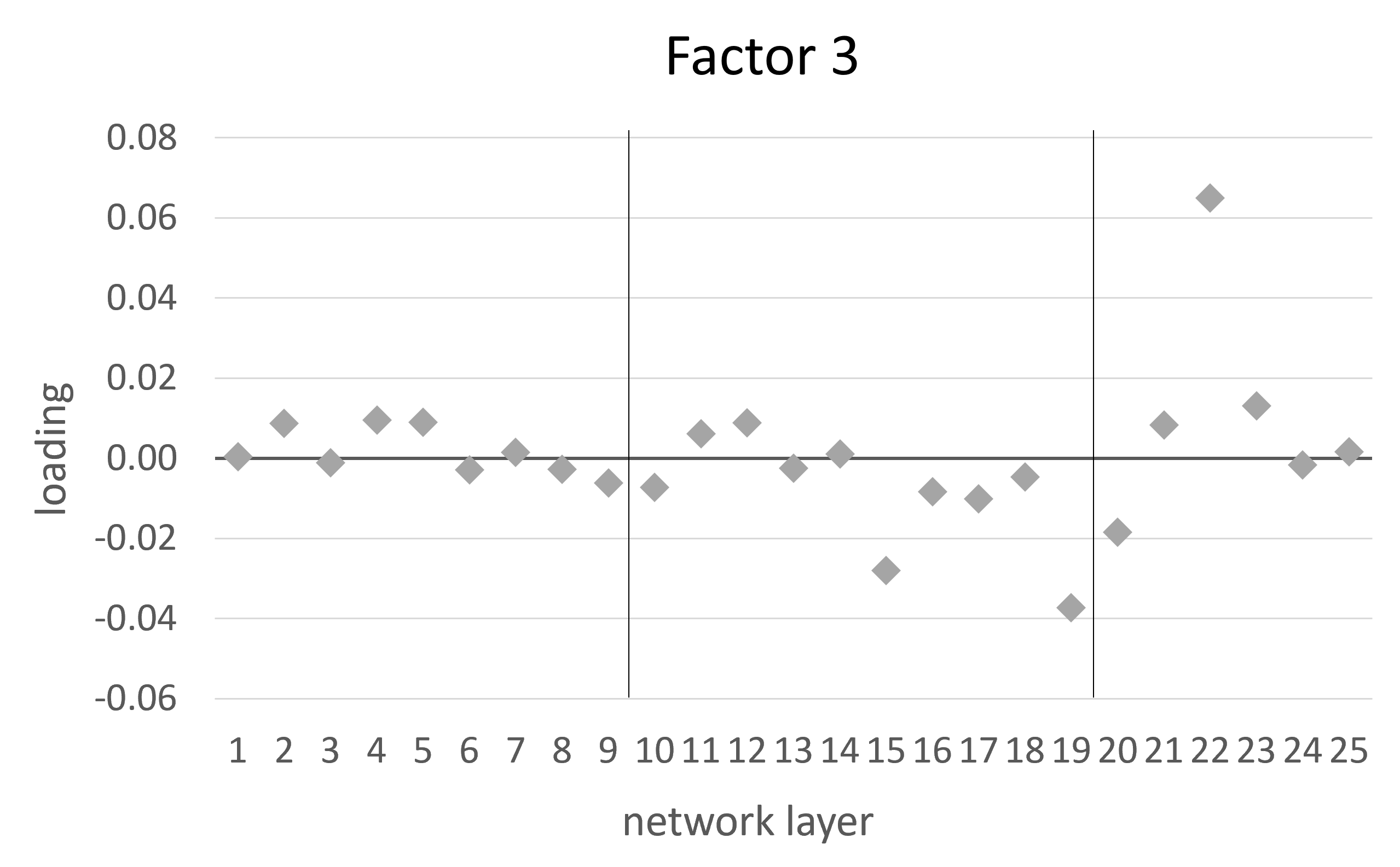} \\
				\includegraphics[scale=0.072]{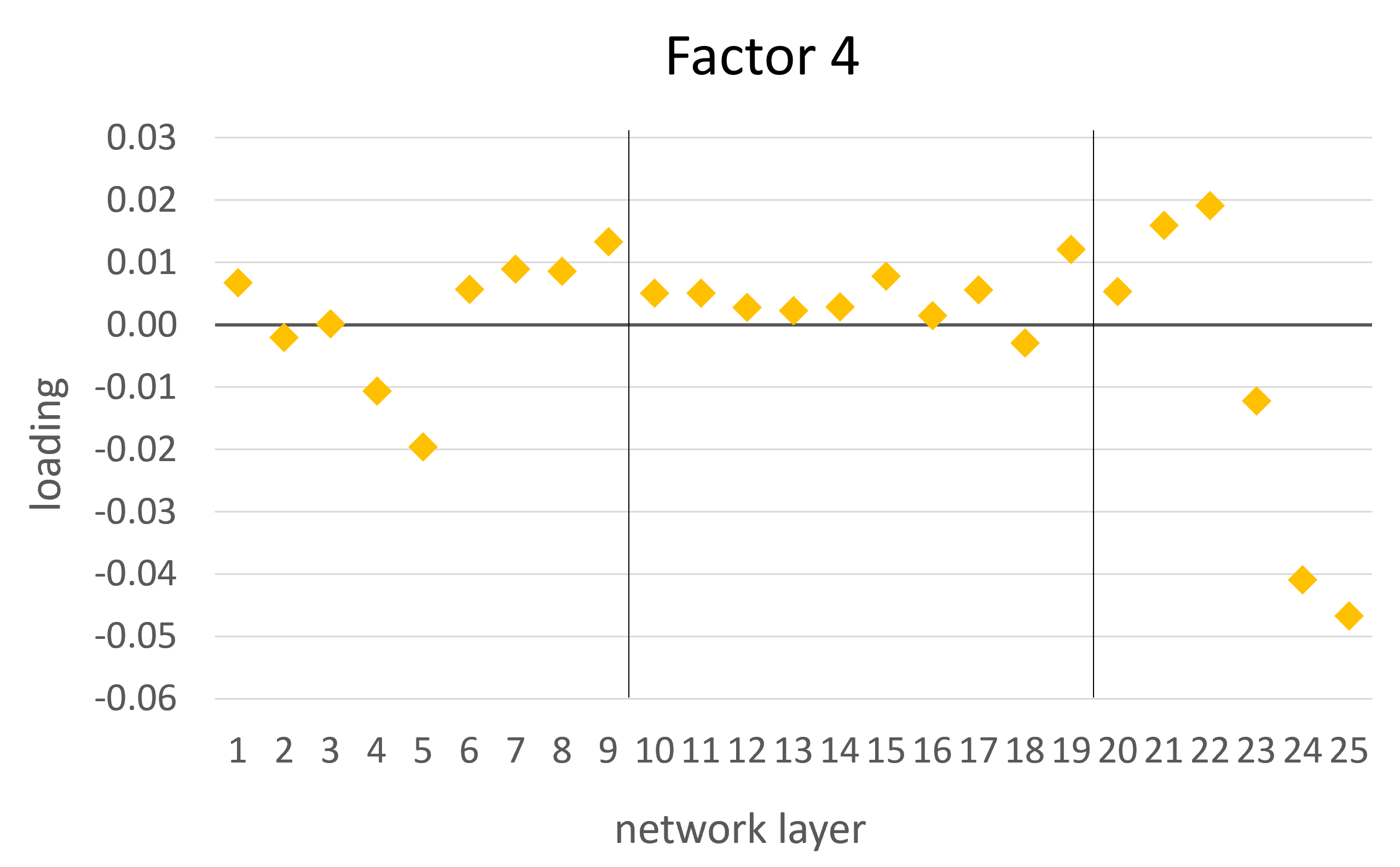} &
				\includegraphics[scale=0.072]{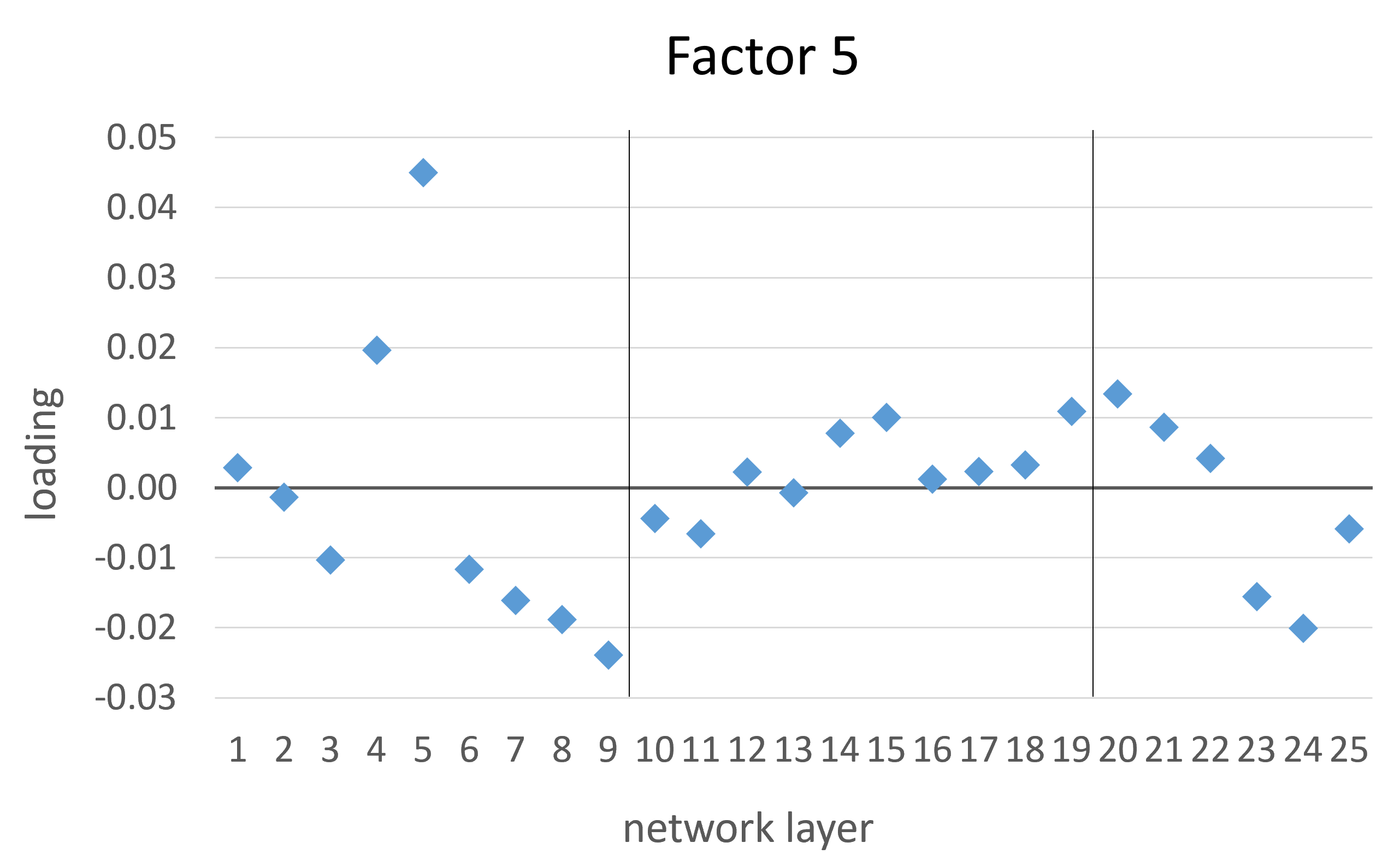} &
				\includegraphics[scale=0.072]{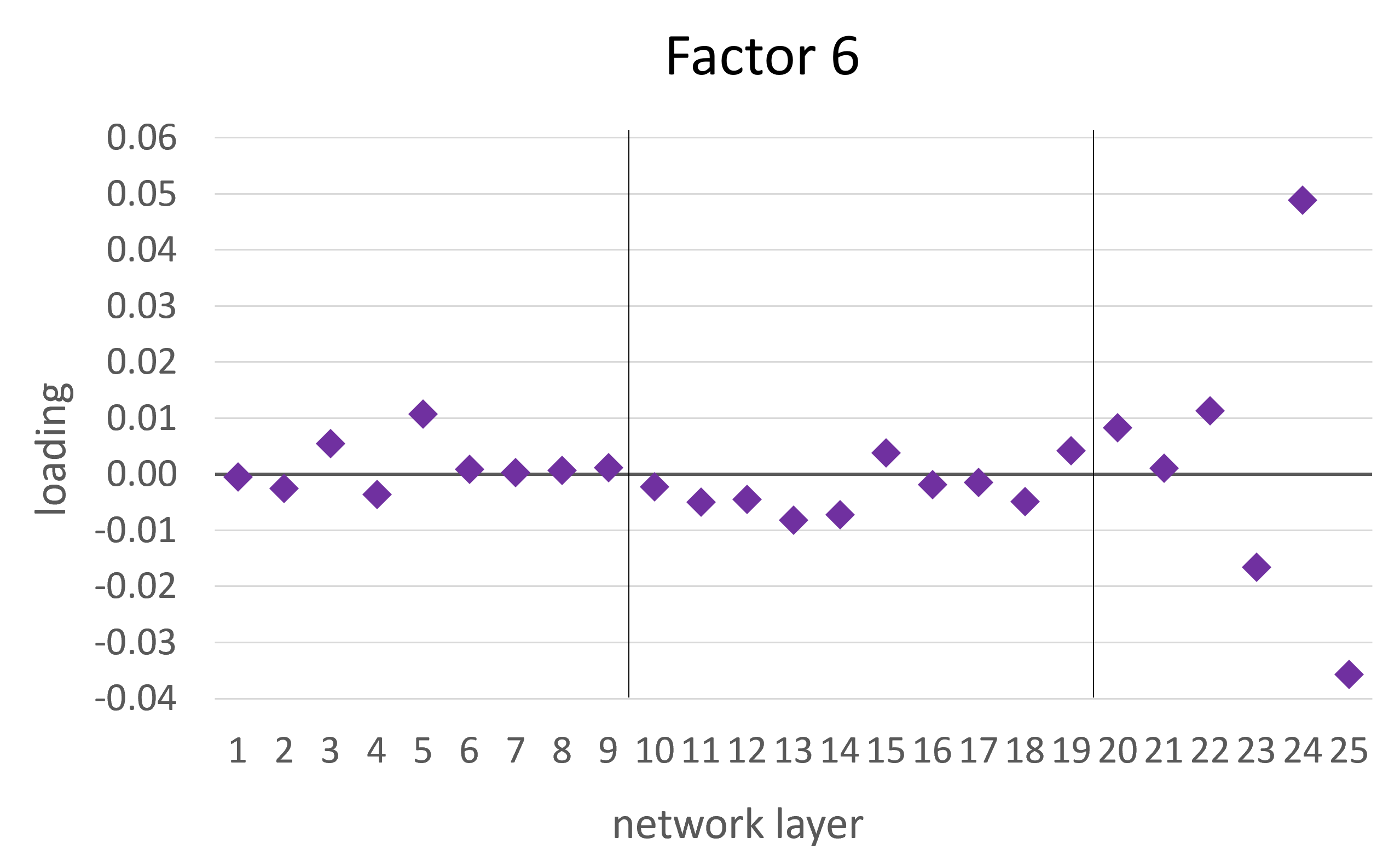} 
			\end{tabular}
			\vspace{7pt}
			\subcaption*{\footnotesize  The figure reports the loading coefficients $\widehat U_{ik}$, $i=1,\ldots,25$, $k=1,\ldots, 6$, of the 6 network factors for the 25 layers of the multinetwork. The vertical lines separate layers related to trade in goods (1-9), layers related to trade in services (10-19) and financial layers (20-25). See Table \ref{tab:layer_list} for the complete list of layers. We report all loadings scaled (divided) by $N$.}
			\label{fig:loadings_u3}
		\end{figure}

		Figure \ref{fig:Wtilde}  displays the average values of the factors over the period 2001-2019 using color scales. Green cells identify positive weights and red cells negative weights, with darker shades of color indicating larger weights in absolute value. Figure \ref{fig:loadings_u3} plots the factor loadings for different layers of the network.
		
\begin{figure}[t!]
			\centering
			\caption{Network factors for the United States.}
			\begin{tabular}{lll}
				\includegraphics[scale=0.068]{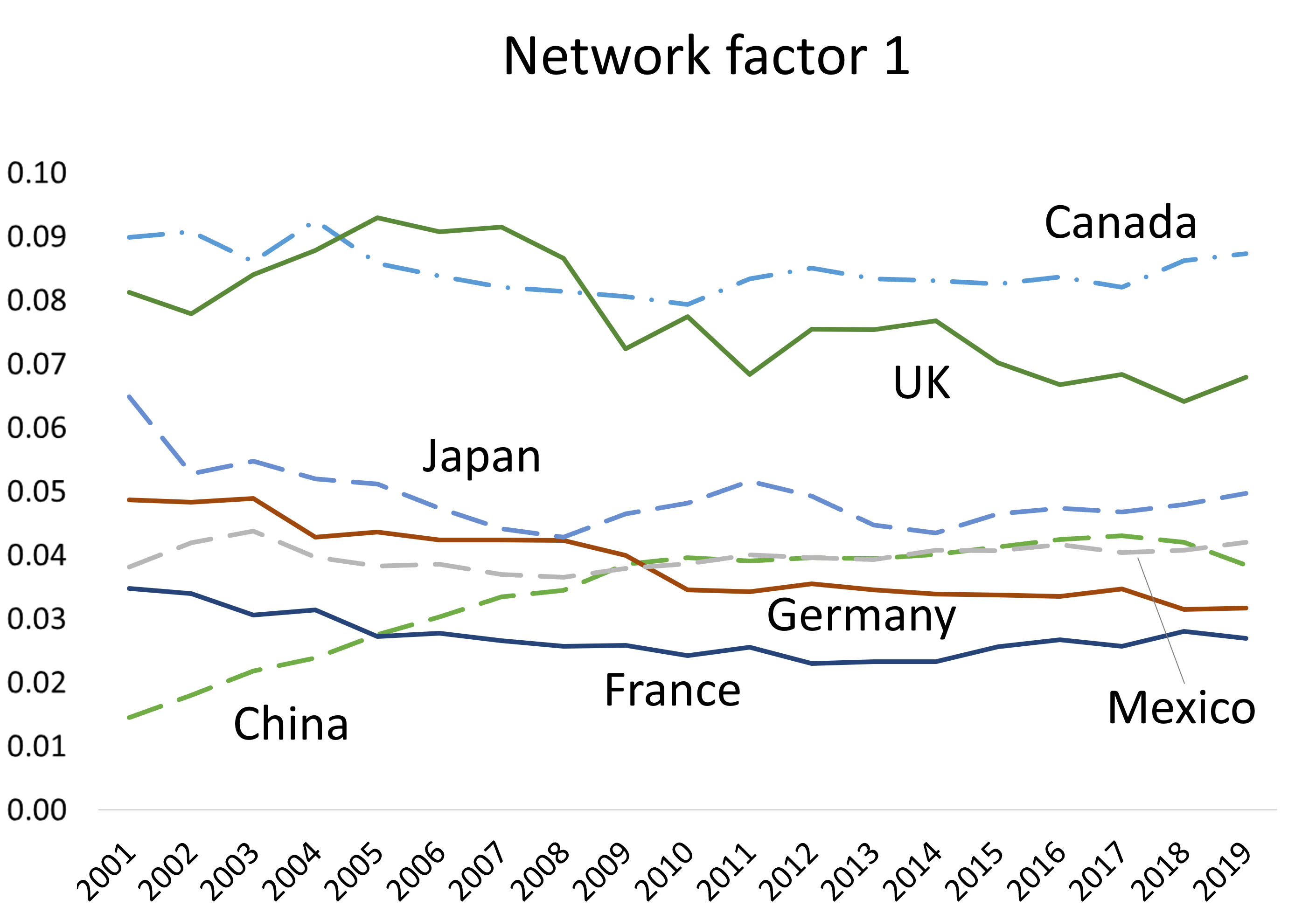}&
				\includegraphics[scale=0.068]{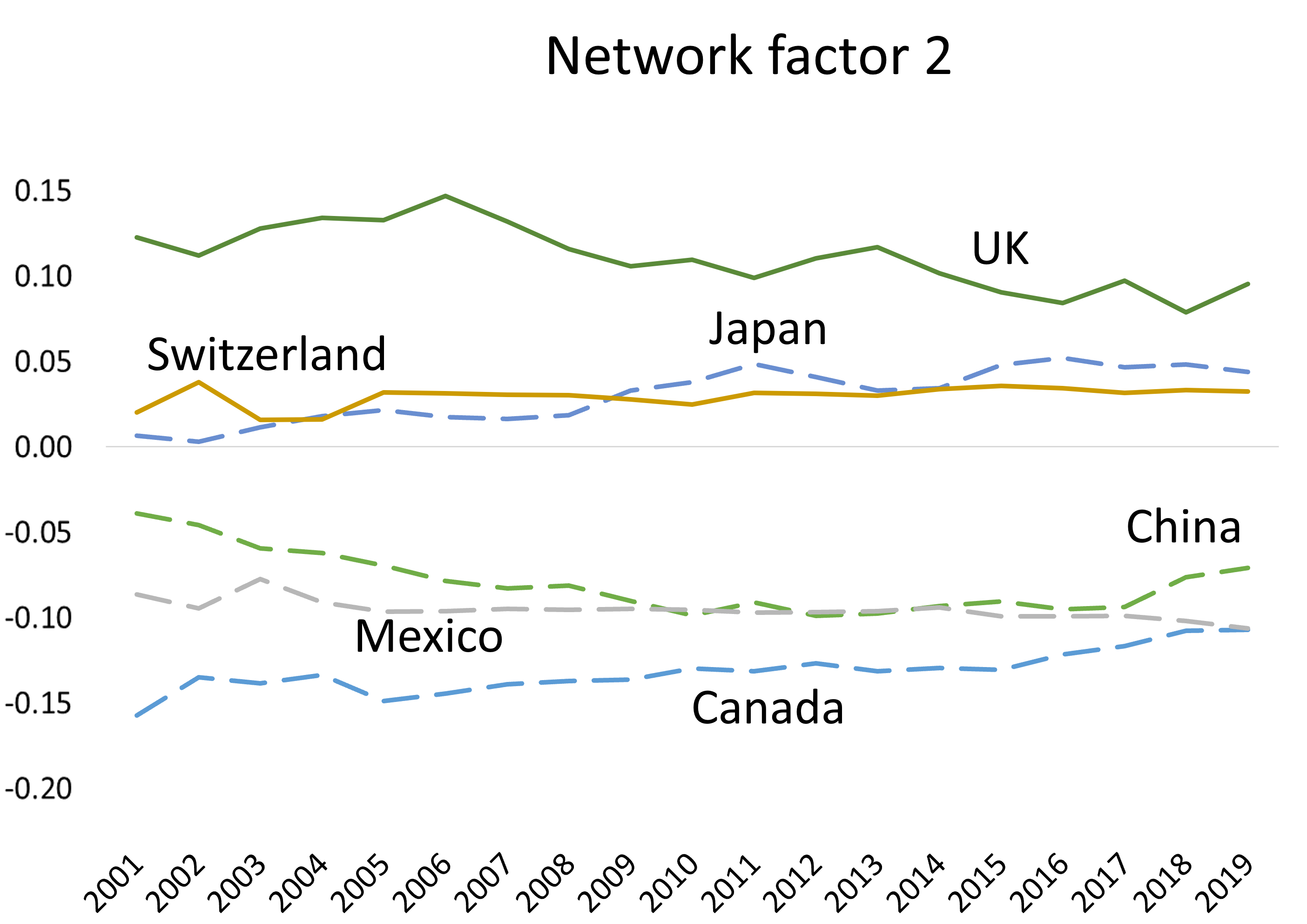}&
				\includegraphics[scale=0.068]{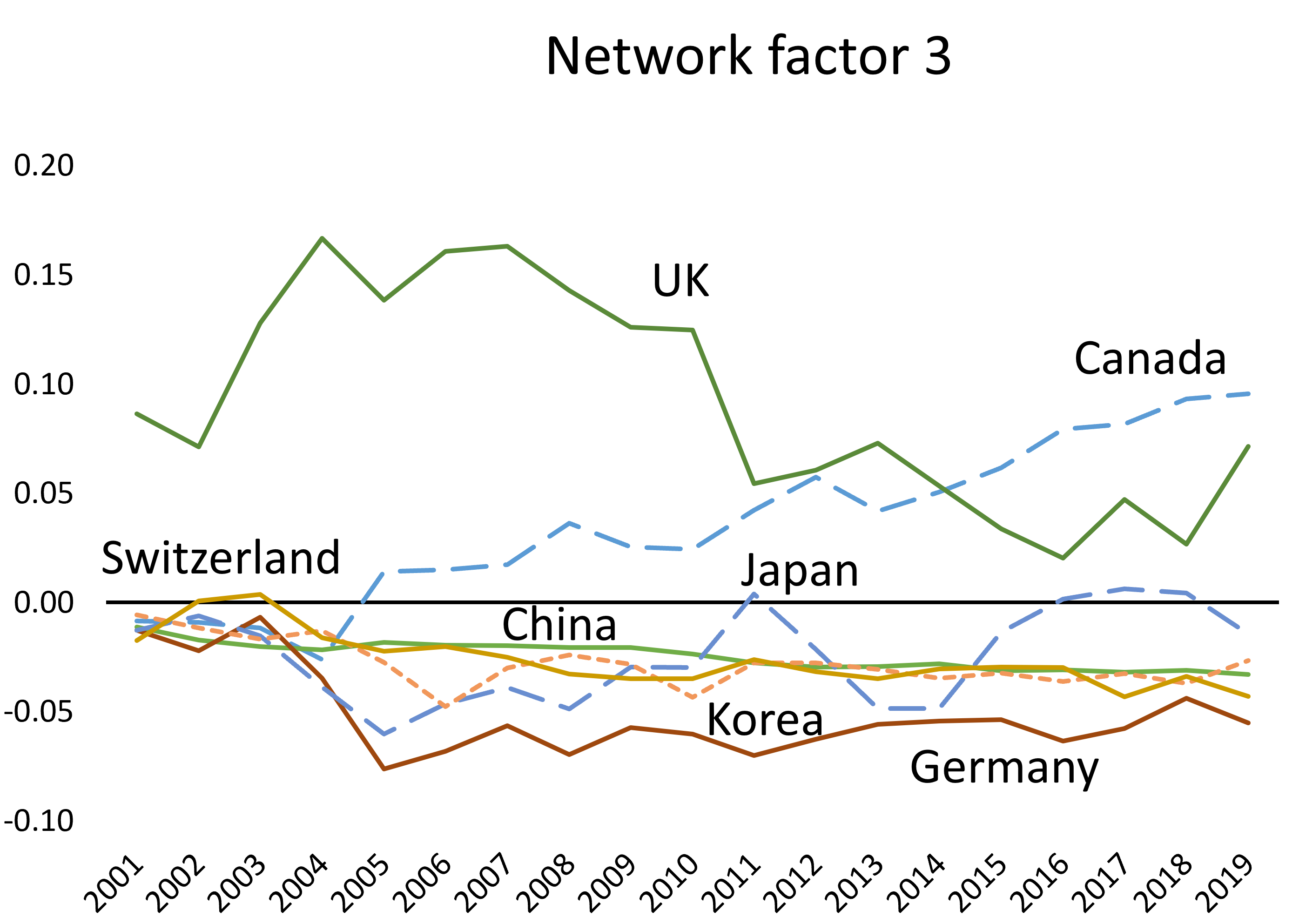}\\
				\includegraphics[scale=0.068]{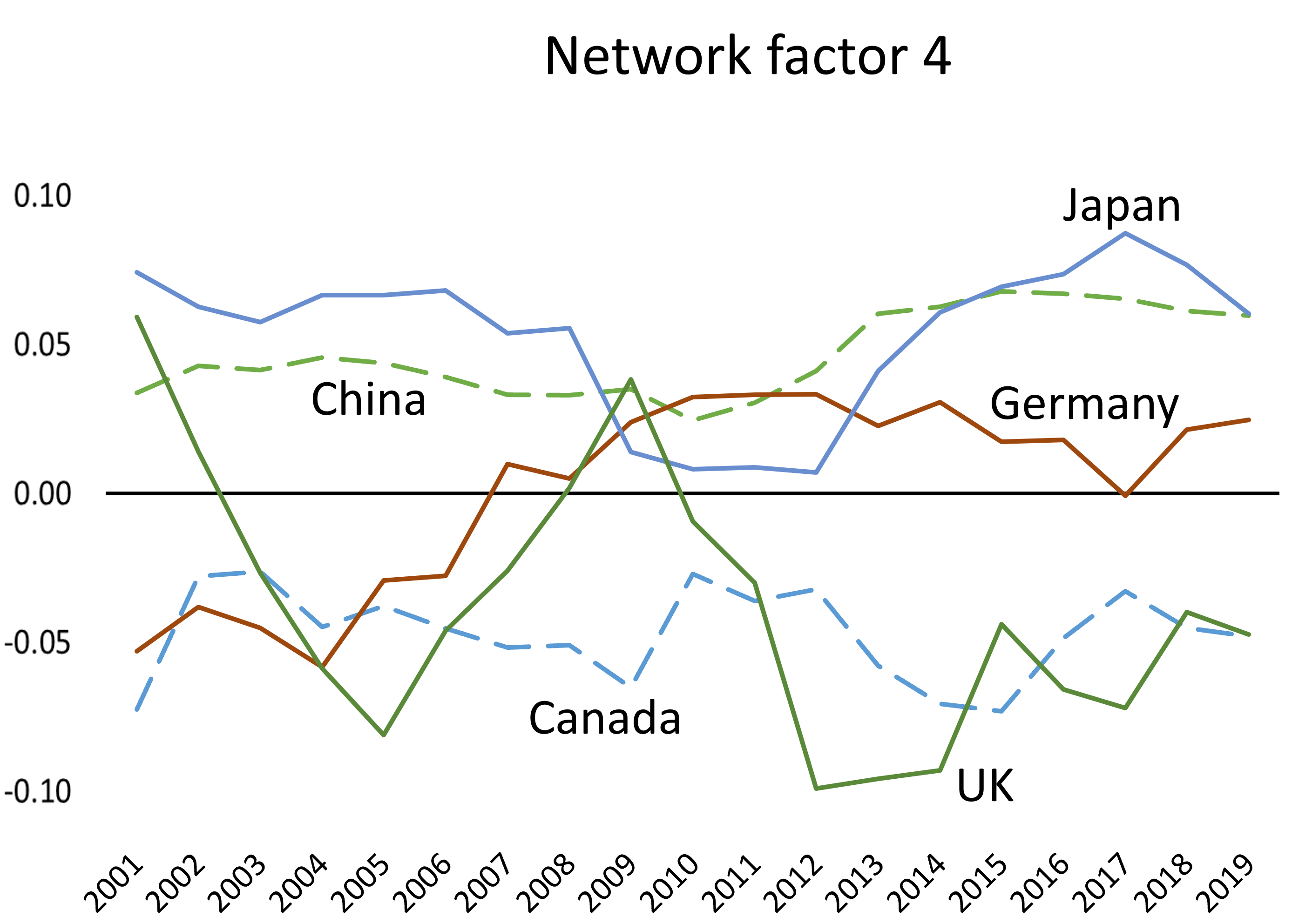}&
				\includegraphics[scale=0.068]{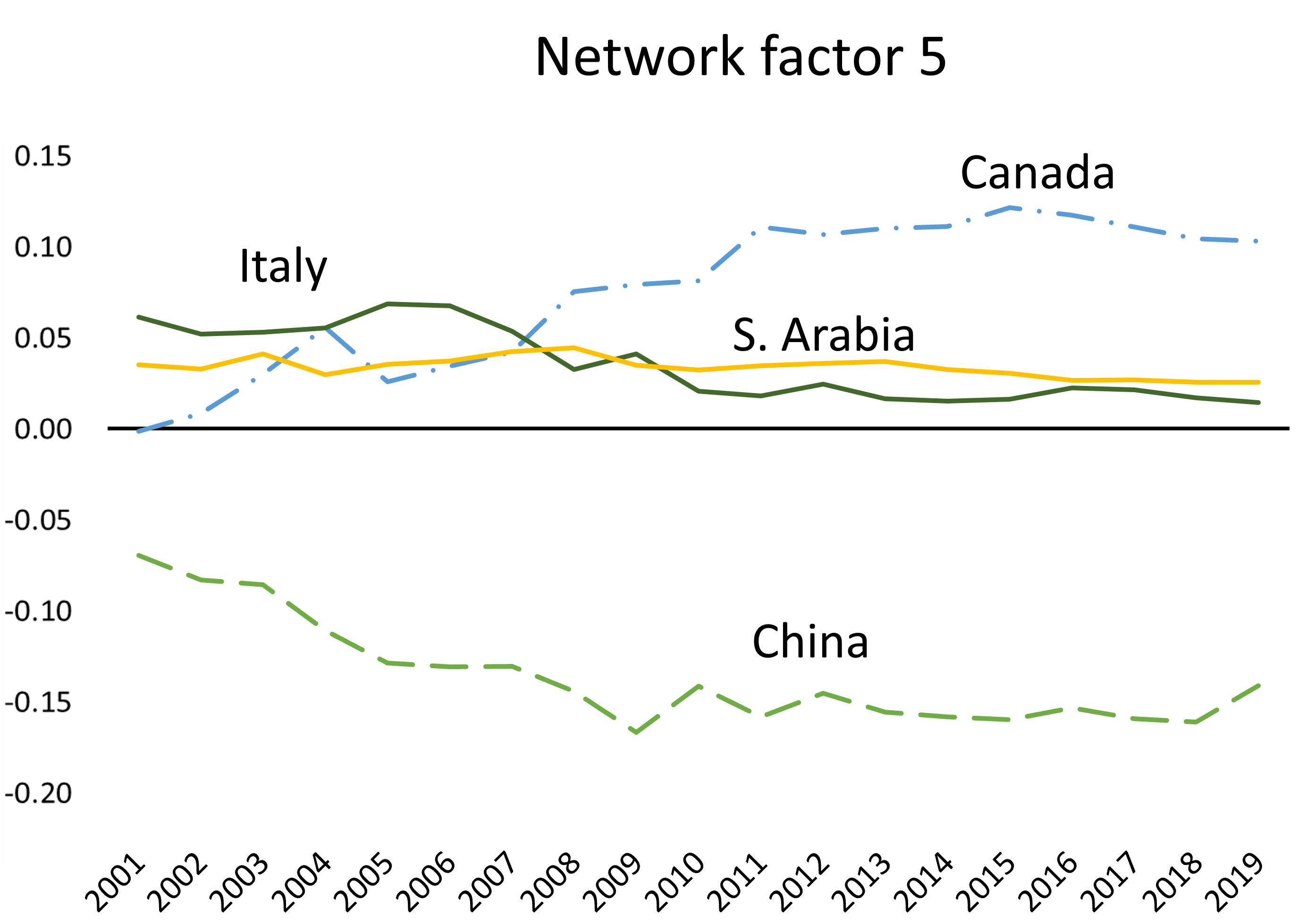}&
				\includegraphics[scale=0.068]{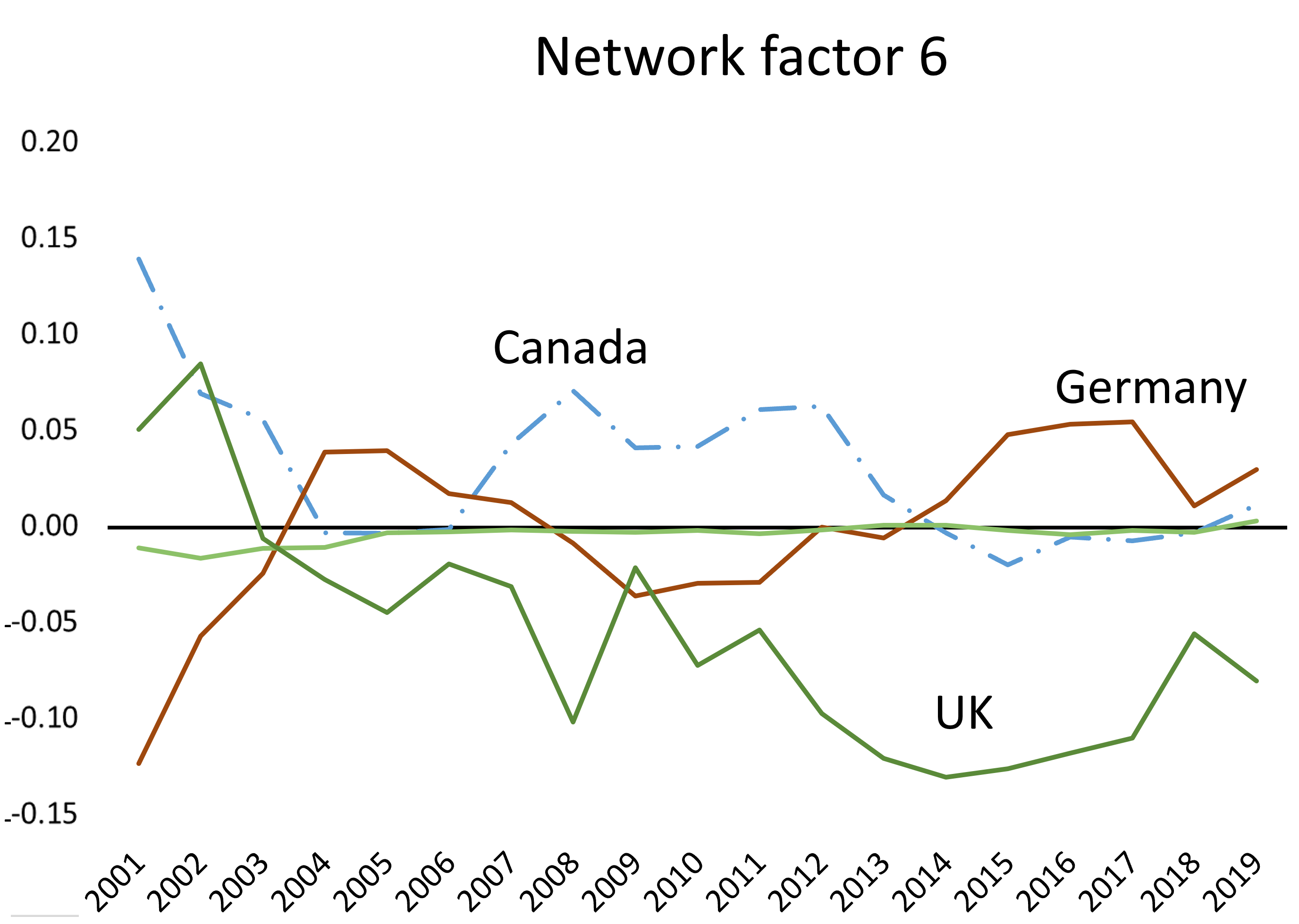}
			\end{tabular}
			\vspace{7pt}
			\subcaption*{\footnotesize  Time-varying weights of different countries in the network of US.}
			\label{fig:Wtilde_usa}
		\end{figure}

\begin{figure}[t!]
			\centering
			\caption{Network factors for China.}
			\begin{tabular}{lll}
				\includegraphics[scale=0.068]{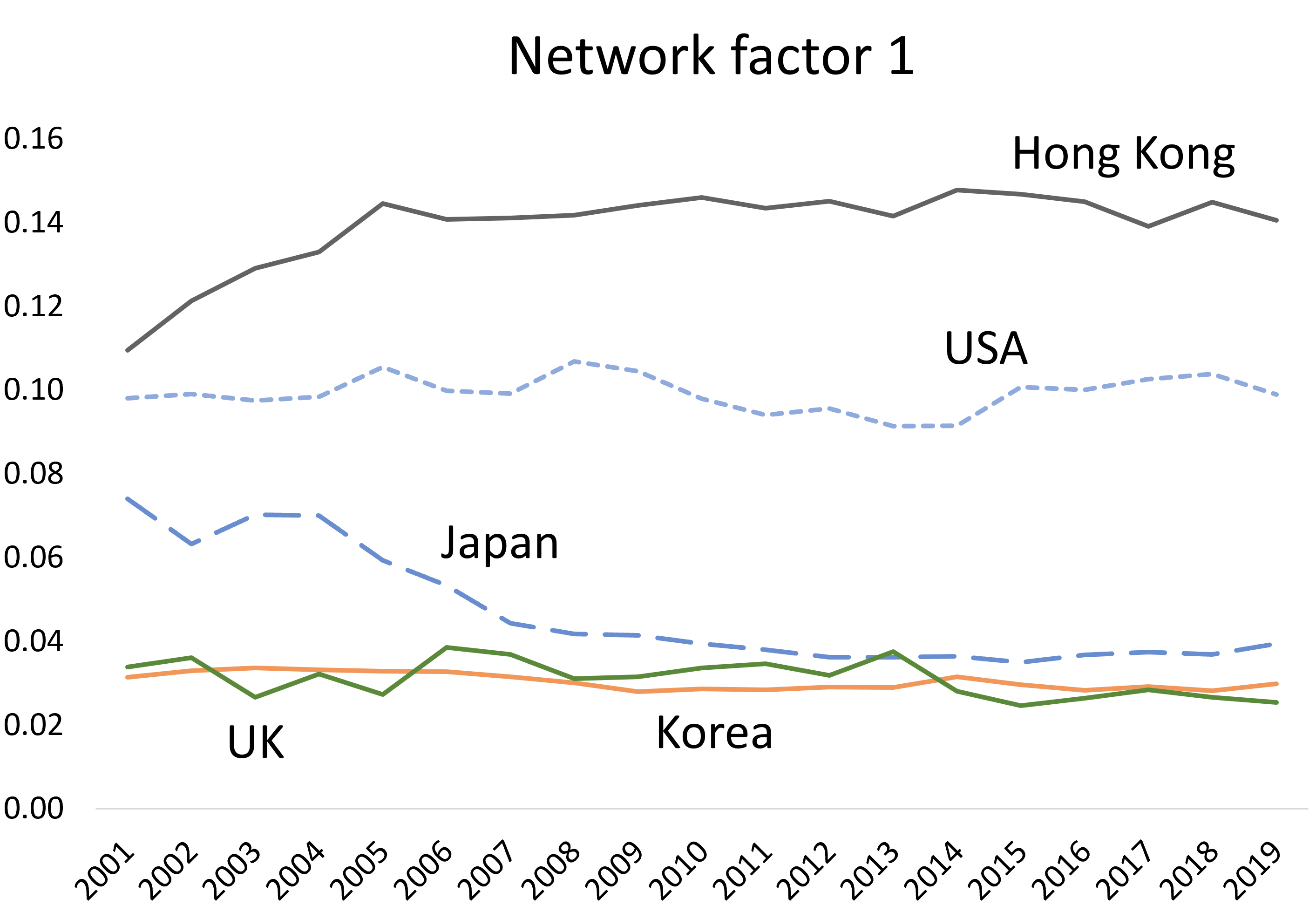}&
				\includegraphics[scale=0.068]{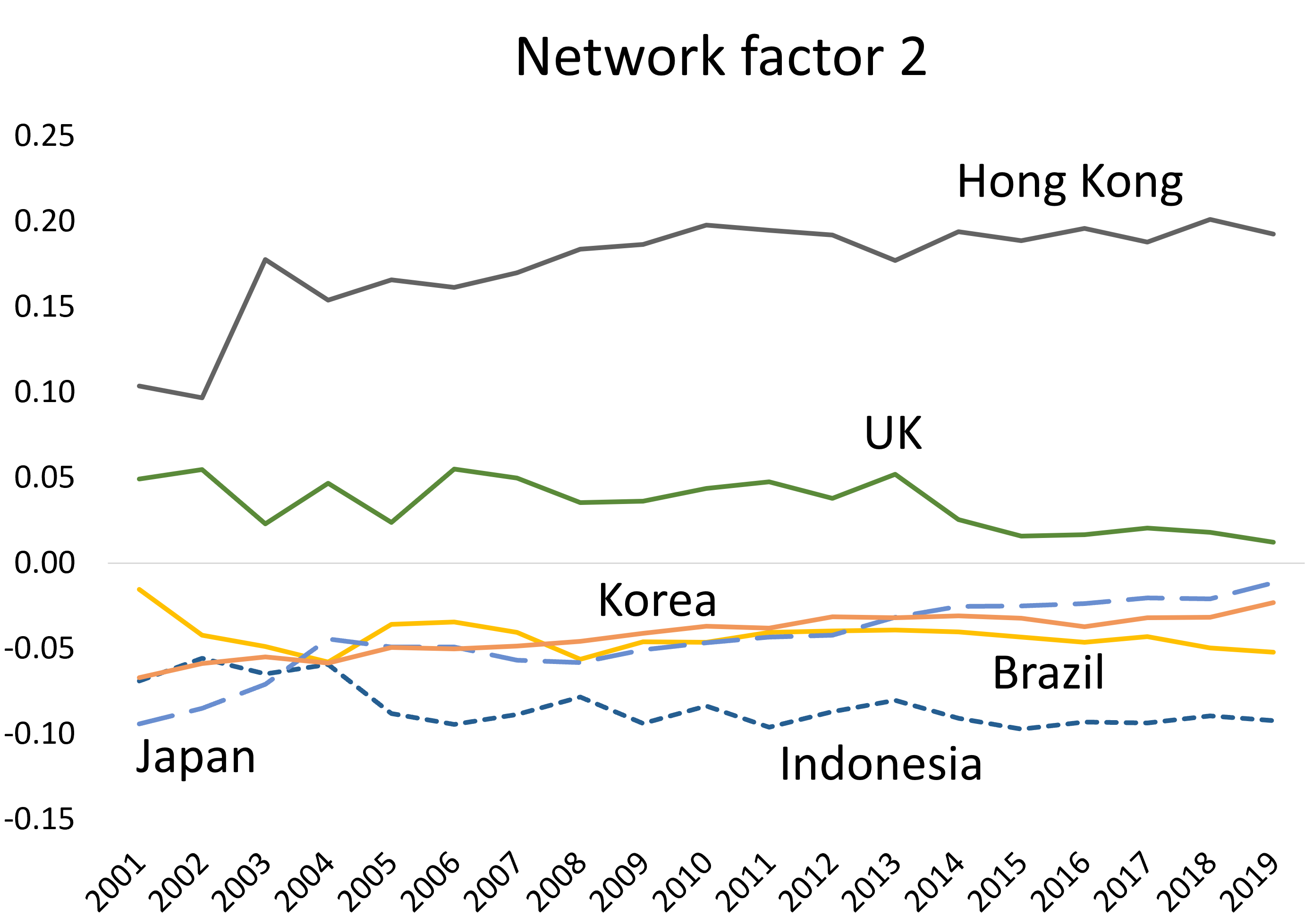}&
				\includegraphics[scale=0.068]{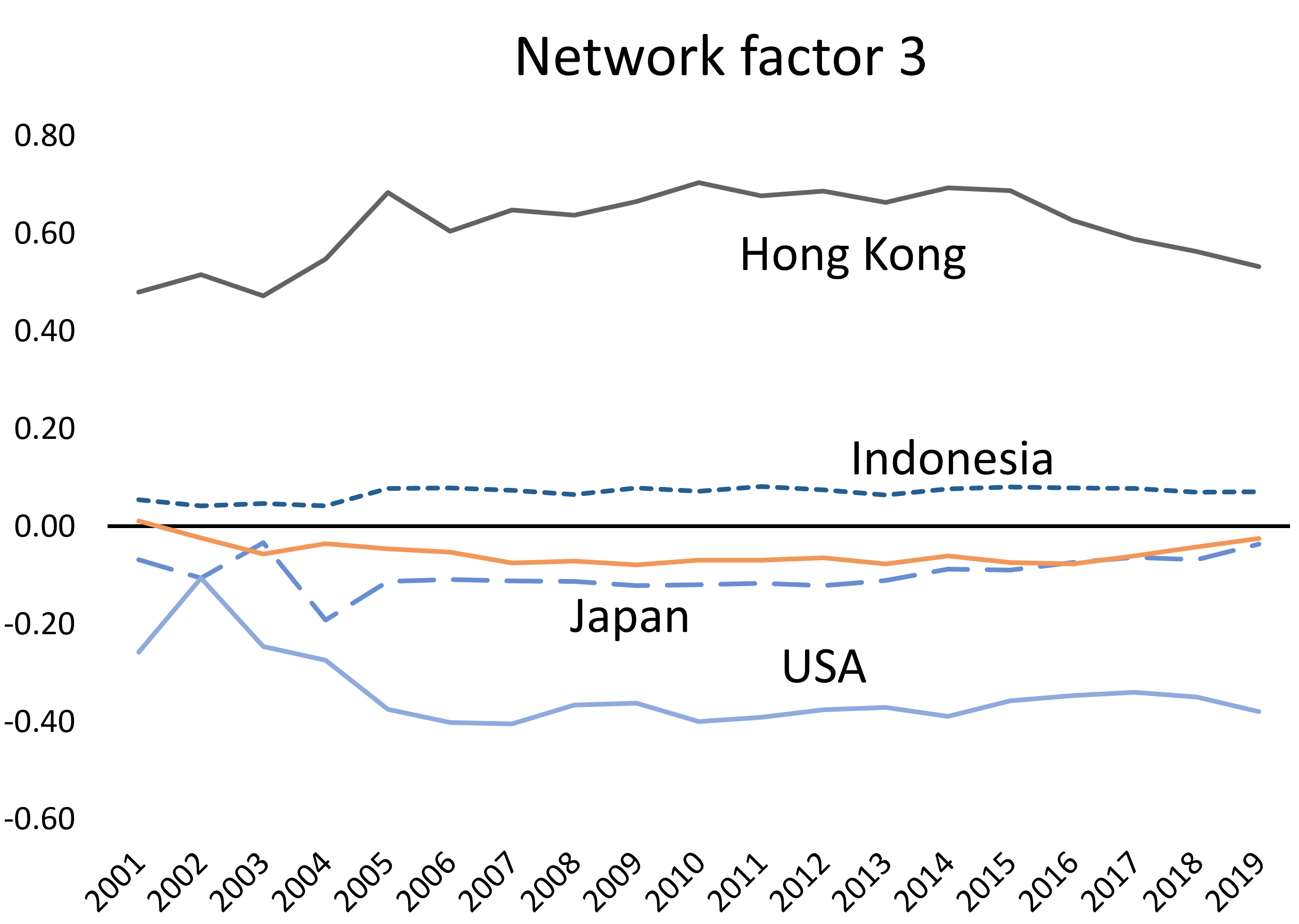}\\
				\includegraphics[scale=0.068]{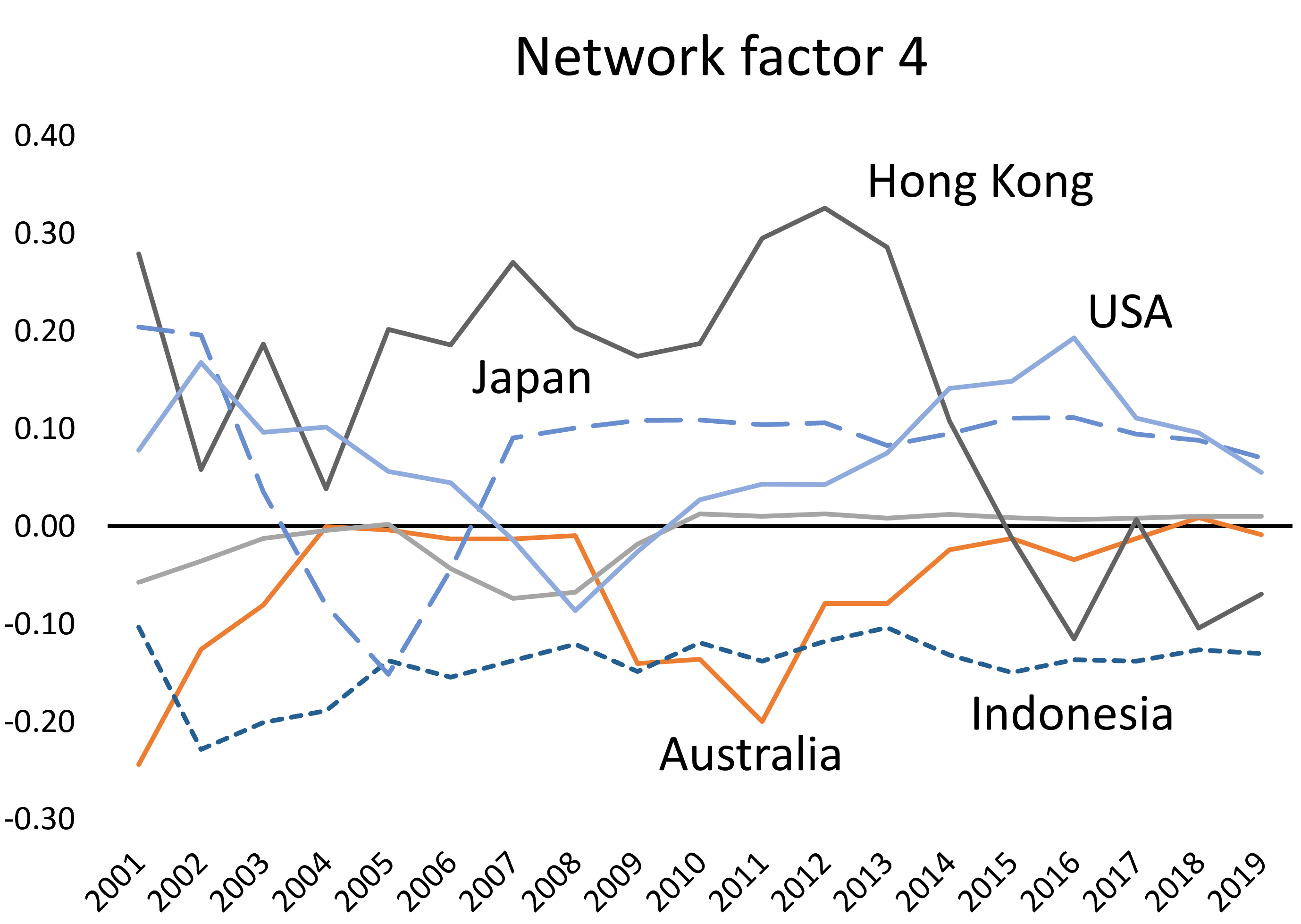}&
				\includegraphics[scale=0.068]{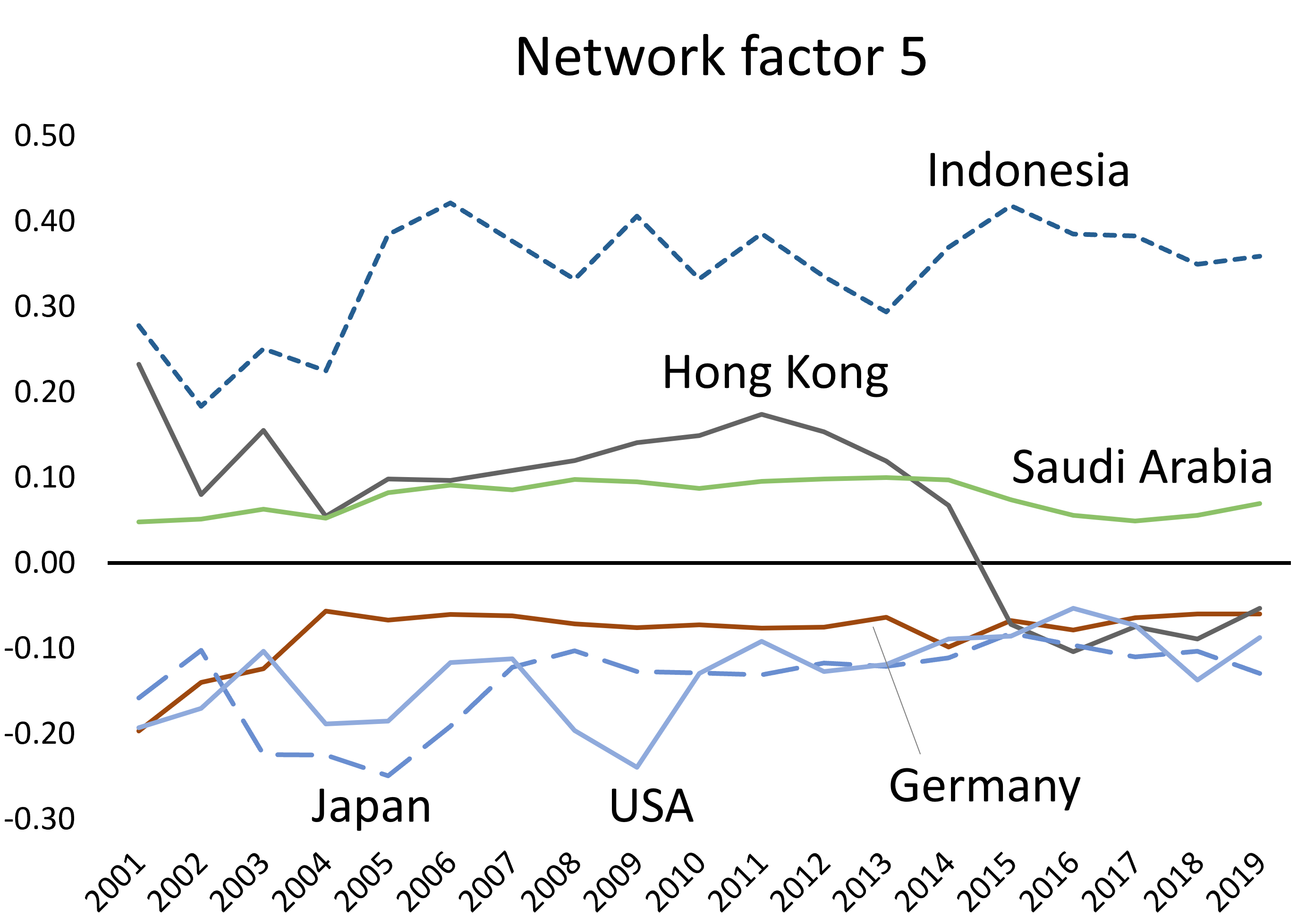}&
				\includegraphics[scale=0.068]{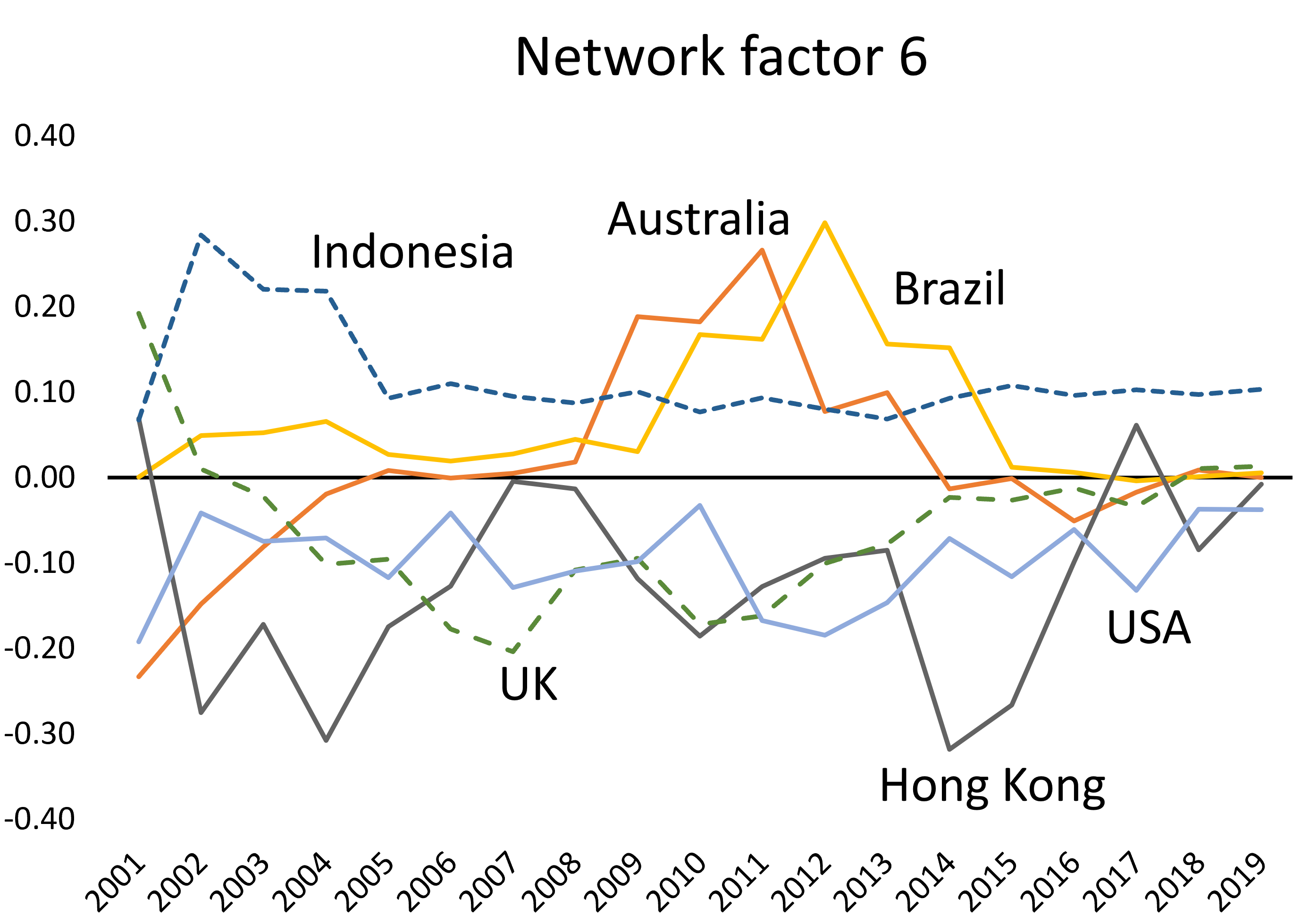}
			\end{tabular}
			\vspace{7pt}
			\subcaption*{\footnotesize  Time-varying weights of different countries in the network of China.}
			\label{fig:Wtilde_china}
		\end{figure}

		Figure \ref{fig:Wtilde_usa} reports the time-varying weights of different countries in the network of US. For each network factor, the countries reported are those that have on average over the sample 2001-2019 the largest weights in absolute values.  Similarly, Figure \ref{fig:Wtilde_china} reports the time-varying weights of different countries in the network of China. Again, for each network factor, the countries reported are those that have on average over the sample 2001-2019 the largest weights in absolute values. The estimated network factors are divided by $N$.

		\subsubsection{Explained variance} 	Consider the order-4 tensor $\mathcal{W}$ of size $ N \times N \times m \times T$ such that $\text{mat}_{(4)}(\mathcal W)$ is $T\times N^2m$ with $t$-th row $\text{vec}(\text{mat}_{(1)} \mathcal{W}_{t})$,
		and, letting $\widehat{\mathcal{W}}_t^{(k)} := \widehat{\mathcal{F}}_{\cdot\cdot,k,t} \times_3 \widehat{U}_{\cdot k}$, $t=1,\ldots, T$, $k=1,\dots,r$, we define 
		the tensor $\widehat{\mathcal{W}}^{(k)}$ of size $N \times N \times m \times T$ in the same way. 
		The fraction of variance in $\mathcal{W}$ explained by the $k$-th factor, denoted as $v^{(k)}$, is then:
		\[
		v^{(k)} = \frac{ \norm{ \widehat{\mathcal{W}}^{(k)}}_F^2}{ \norm{ \mathcal{W} }_F^2}, \quad k=1,\ldots, r.
		\]
		And we have $v^{(1)}=0.68$, $v^{(2)}=0.07$, $v^{(3)}=0.03$, $v^{(4)}=0.03$, $v^{(5)}=0.02$, and $v^{(6)}=0.02$. Thus, overall the 6 factors explain about 85\% of the total variance of $\mathcal W$. However, the importance of different factors varies greatly across countries. Let $v_{ij}^{(k)}$ be the fraction of variance in the weight of country $j$ for country $i$ explained by the $k$-th factor. Then, we have that:
		\[
		v_{ij}^{(k)} = \frac{ \norm{ \widehat{\mathcal{W}}_{ij,\cdot\cdot}^{(k)} }_F^2}{ \norm{ \mathcal{W}_{ij,\cdot\cdot} }_F^2}, \quad i,j=1,\ldots, N, \quad k=1,\ldots, r.
		\] 
		For each factor, Table \ref{tab:vd} reports the ten network links for which the factor explains the largest share of variance. 
		
		\begin{table}[t!]
			\caption{Variance decomposition of network links: share of variance explained by each factor.}
			\begin{minipage}{.5\textwidth}
				\centering
				\subcaption{Factor 1}
				\scriptsize{
					\begin{tabular}{cccc}
						\hline \hline
						ranking & country $i$    & country $j$    & share of var.      \\
						\hline
						1  & CAN & USA & 0.969 \\
						2  & MEX & USA & 0.948 \\
						3  & BRA & USA & 0.851 \\
						4  & JAP & USA & 0.829 \\
						5  & DEU & FRA & 0.828 \\
						6  & GBR & USA & 0.825 \\
						7  & BEL & FRA & 0.813 \\
						8  & KOR & USA & 0.800 \\
						9  & FRA & DEU & 0.797 \\
						10 & FRA & GBR & 0.796 \\
						\hline \hline
					\end{tabular}
				}
			\end{minipage}%
			\begin{minipage}{.5\textwidth}
				\centering
				\subcaption{Factor 2}
				\scriptsize{
					\begin{tabular}{cccc}
						\hline \hline
						ranking & country $i$    & country $j$    & share of var.      \\
						\hline
						1  & AUS & KOR & 0.368 \\
						2  & BRA & CHN & 0.360 \\
						3  & CHE & ITA & 0.338 \\
						4  & AUS & CHN & 0.338 \\
						5  & KOR & IDN & 0.331 \\
						6  & USA & MEX & 0.326 \\
						7  & CHN & IND & 0.289 \\
						8  & IDN & CHN & 0.287 \\
						9  & IND & BRA & 0.282 \\
						10 & BRA & IND & 0.276\\
						\hline \hline
					\end{tabular}
				}
			\end{minipage}
			\begin{minipage}{.5\textwidth}
				\vspace{10pt}
				\centering
				\subcaption{Factor 3}
				\scriptsize{
					\begin{tabular}{cccc}
						\hline \hline
						ranking & country $i$    & country $j$    & share of var.      \\
						\hline
						1  & TUR & HKG & 0.438 \\
						2  & JAP & FRA & 0.417 \\
						3  & KOR & HKG & 0.406 \\
						4  & ZAF & ITA & 0.340 \\
						5  & IND & NOR & 0.313 \\
						6  & IND & HKG & 0.301 \\
						7  & BRA & KOR & 0.272 \\
						8  & AUS & HKG & 0.261 \\
						9  & BRA & MEX & 0.241 \\
						10 & KOR & BRA & 0.231\\
						\hline \hline
					\end{tabular}
				}
			\end{minipage}
			\begin{minipage}{.5\textwidth}
				\vspace{10pt}
				\centering
				\subcaption{Factor 4}
				\scriptsize{
					\begin{tabular}{cccc}
						\hline \hline
						ranking & country $i$    & country $j$    & share of var.      \\
						\hline
						1  & MEX & ESP & 0.358 \\
						2  & CAN & AUS & 0.356 \\
						3  & IDN & CAN & 0.298 \\
						4  & IDN & SAU & 0.281 \\
						5  & SAU & AUS & 0.276 \\
						6  & BEL & TUR & 0.275 \\
						7  & HKG & NLD & 0.260 \\
						8  & BRA & ZAF & 0.259 \\
						9  & MEX & BRA & 0.252 \\
						10 & HKG & ESP & 0.251\\
						\hline \hline
					\end{tabular}
				}
			\end{minipage}
			
			\begin{minipage}{.5\textwidth}
				\vspace{10pt}
				\centering
				\subcaption{Factor 5}
				\scriptsize{
					\begin{tabular}{cccc}
						\hline \hline
						ranking & country $i$    & country $j$    & share of var.      \\
						\hline
						1  & IND & IDN & 0.378 \\
						2  & ZAF & IDN & 0.360 \\
						3  & ITA & IDN & 0.357 \\
						4  & CHN & IDN & 0.357 \\
						5  & TUR & IDN & 0.352 \\
						6  & BRA & IND & 0.349 \\
						7  & AUS & ESP & 0.324 \\
						8  & DEU & IDN & 0.322 \\
						9  & NLD & IDN & 0.321 \\
						10 & SAU & IDN & 0.321\\
						\hline \hline
					\end{tabular}
				}
			\end{minipage}
			\begin{minipage}{.5\textwidth}
				\vspace{10pt}
				\centering
				\subcaption{Factor 6}
				\scriptsize{
					\begin{tabular}{cccc}
						\hline \hline
						ranking & country $i$    & country $j$    & share of var.      \\
						\hline
						1  & KOR & BEL & 0.503 \\
						2  & BEL & KOR & 0.395 \\
						3  & TUR & ZAF & 0.388 \\
						4  & BEL & BRA & 0.382 \\
						5  & HKG & NLD & 0.339 \\
						6  & BEL & IDN & 0.328 \\
						7  & DEU & IND & 0.303 \\
						8  & MEX & BRA & 0.300 \\
						9  & FRA & SWE & 0.294 \\
						10 & BRA & ZAF & 0.284\\
						\hline \hline
					\end{tabular}
				}
			\end{minipage}
			\vspace{5pt}
			\caption*{\footnotesize  For each network factor $\widehat F_{k,t}$, the table reports the 10 bilateral links for which the factor explains the largest share of variance (calculated  both across network layers and across time). Each link represents the weight of country $j$ for country $i$.}
			\label{tab:vd}
		\end{table}

		\clearpage
		
		\subsubsection{Details on the forecasting application}\label{app:pred}
		Forecasts are computed according to the following recursive forecasting scheme. We initially estimate the model on a sample window ending in 2001Q4, then we expand the sample by one-quarter increments up to 2019Q3, and we use the estimates obtained in each window to produce 1-quarter-ahead forecasts of GDP growth, from 2002Q1 to 2019Q4. The smallest window ends in 2001 as this is the first year in which data on financial networks are available, as explained in Section \ref{sec:application} of the paper. To make the exercise feasible, the start date of the estimation sample is moved back in time, specifically to 1980Q1, and we assume that the network tensor until 2000 is constant at its first available value (2001). 
		We consider the following competing estimators.
		
		\begin{enumerate}
			
			\item The tensor-based estimators developed by \citet{wangetal21} to estimate coefficients of high-dimensional VARs. Namely, we consider the multilinear low-rank (MLR) estimator and the sparse higher-order reduced-rank (SHORR) estimator. These estimators apply a Tucker decomposition to the order-3 tensor of unknown VAR coefficients, where the first two dimensions of the tensor are given by the number of variables in the VAR and the third dimension is given by the number of lags. To allow for a proper order-3 tensor, we consider a VAR with 2 lags in this case. In each estimation window, the multilinear ranks (i.e., the dimensions of the core tensor in the Tucker decomposition) are selected using the ridge-type ratio described in \citet{wangetal21}.
			
			\item Two multilayer NARs  \eqref{eq:collinear_r3} and \eqref{eq:FstartuckerFNAR}, based on networks extracted from two alternative full Tucker tensor decompositions, as detailed in Appendix \ref{app:Tucker}.

			\item A multilayer NAR that includes all 25 layers composing the original weight tensor $\mathcal{W}_t$, estimated by means of the LASSO or Ridge estimators.
			%
			%
			
			\item A simple VAR(1) model that does not utilize any information on the underlying networks.  
		\end{enumerate}

		\section{NAR estimation based on full Tucker decompositions}\label{app:Tucker}
		
		
		Consider  the $N\times N\times m$ tensor of observed data $\mathcal W_t$ having as slices the $m$ layers $W_{k,t}$. 
		A full Tucker decomposition of $\mathcal W_t$ reads:
		\begin{align}
			\mathcal W_t = \mathcal G_t \times_1 V_1\times_2 V_2 \times_3 V_3 + \mathcal Z_t, \quad t=1,\ldots, T,\label{eq:tucker_r3}
		\end{align}
		where the tensor of factors is $\mathcal G_t$ of size $p\times q\times r$ and the loadings are $V_1$ of size $N\times p$, $V_2$ of size $N\times q$, and $V_3$ of size $m\times r$. 
		We can get estimated loadings $\widehat V_j$ in \eqref{eq:tucker_r3}, using the TOPUP estimators of \citet{chenyangzhang22}. The estimated factors are then obtained by linear projection as 
		$\widehat{\mathcal G}_t= \mathcal W_t \times_1 (\widehat V_1'\widehat V_1)^{-1} \widehat V_1'\times_2 (\widehat V_2'\widehat V_2)^{-1} \widehat V_2'\times_3 (\widehat V_3'\widehat V_3)^{-1} \widehat V_3'$. We cannot directly use $\widehat{\mathcal G}_t$ for estimating the multilayer NAR as now these do not have the right dimensions and cannot be interpreted as networks. 
		We then consider two possibilities. 
		
		The first one is to use the multilayer common network 
		\begin{align}
			\widehat S_t=\widehat{  \mathcal G}_t \times_1\widehat V_1\times_2\widehat V_2 \times_3\widehat V_3.\label{eq:Stucker}
		\end{align}
		Denote as $\widehat S_{t,k}$, $k=1,\ldots, m$, the $m$ layers  of $\widehat {\mathcal S}_t$, each of size $N\times N$. 
		Thus, we could think of estimating a new multilayer NAR of the form
		\begin{align}
			y_t &=N^{-1} \gamma_1 \widehat S_{1,t-1}  y_{t-1}+ \ldots +N^{-1} \gamma_m \widehat S_{m,t-1}  y_{t-1}+\varsigma y_{t-1}+ a+ \omega_t,  \quad t=1,\ldots, T. \label{eq:collinear_r3}
		\end{align}
		However, this model has by construction collinear regressors. Indeed, write \eqref{eq:collinear_r3} as $y_t=\widehat X_t\theta+\omega_t$ with $\widehat X_t := (\widehat{\mathsf S}_{1,t-1,\cdot}',\cdots, \widehat{\mathsf S}_{m,t-1,\cdot}', y_{t-1}, \iota_N)$ where  $\widehat{\mathsf S}_{k,t-1,\cdot}$ is the $t$-th row of $\widehat{\mathsf S}_k:=\left(\widehat S_{k,0} y_{0},\cdots, \widehat S_{k,T-1} y_{T-1}\right)^\prime$. Then,
		$\sum_{t=1}^T \widehat X_t' \widehat X_t$ is $m+2\times m+2$ but it is not invertible, since it has rank $r+2$. We therefore applied either LASSO or Ridge to estimate \eqref{eq:collinear_r3}.

		The second possibility is to calculate the $N \times N \times r$ tensor 
		\begin{align}\label{eq:Fstartucker}
			\widehat{\mathcal{F}}_t^* = \widehat{ \mathcal G}_t \times_1 \widehat V_1\times_2 \widehat V_2
		\end{align}
		and then to use the matrices $\widehat{F}_{k,t}^*$, with $k=1,\ldots, r$, defined as the slices of $\widehat{\mathcal{F}}_t^*$ of size $N \times N$, in place of our factors $\widehat{F}_{k,t}$, $k=1,\ldots, r$, in the FNAR, i.e.,
		\begin{align}
			y_t &=N^{-1}\beta^*_1 \widehat F^*_{1,t-1}  y_{t-1}+ \ldots +N^{-1} \beta^*_r \widehat F^*_{r,t-1}  y_{t-1}+ \rho^* y_{t-1}+ \alpha^* + \nu^*_t,  \quad t=1,\ldots, T.\label{eq:FstartuckerFNAR}
		\end{align}

		We refer to Table \ref{tab:forecast_comparison} in Section \ref{sec:application} of the paper for an empirical comparison between our approach and these two alternative full Tucker approaches. The results show that our approach delivers in general better predictions. 
		
		We conclude with a comment on consistency rates. First, note that our estimator of the loadings $\widehat U$ (in the paper) has the same rate as $\widehat V_3$, however, both $\widehat{\mathcal S}_t$ and $\widehat{\mathcal F}_t^*$ depend also on the estimated loadings $\widehat V_1$ and $\widehat V_2$, while our network factors $\widehat{\mathcal F}_t$ used in the FNAR do not. Therefore the consistency rates of $\widehat{\mathcal S}_t$ and $\widehat{\mathcal F}_t^*$  might differ from those we derived for $\widehat{\mathcal F}_t$. In particular, $\widehat{\mathcal S}_t$ and $\widehat{\mathcal F}_t^*$ depend on the estimated loadings both directly in their definition and indirectly through the estimated core tensor $\widehat{\mathcal G}_t$.
		Tables \ref{tab:rates1} and \ref{tab:rates} summarize this comparison.

		\renewcommand{\arraystretch}{0.7}
		\begin{table}[H]
			\centering
			\caption{Comparison of consistency rates - Loadings}
			{ \footnotesize
				\scalebox{0.9}[0.8]
				{\begin{tabular}
						{l p{5.75cm} p{2.75cm}p{2.75cm}}
						\hline
						\hline
						&& (A)& (B)\\
						\hline
						Loadings - Theorem \ref{theorem:CLT_loadings}&	$m^{-1/2}\Vert \widehat U-U H_3\Vert$ &  $m$ & $N\sqrt{T}$\\
						\hline
						Loadings - full Tucker decomposition \eqref{eq:tucker_r3}&	$N^{-1/2}\Vert \widehat V_1-V_1 H_1\Vert$ &  $N$ & $\sqrt{NmT}$\\
						&$N^{-1/2}\Vert \widehat V_2-V_2 H_2\Vert$ &  $N$ & $\sqrt{NmT}$\\ 
						&$m^{-1/2}\Vert \widehat V_3-V_3 H_3\Vert$ &  $m$ & $N\sqrt{T}$\\
						\hline
						\hline
					\end{tabular}
				}
				\vspace{7pt}
				\subcaption*{\footnotesize 
					In this table $H_1$ $(p\times p)$, $H_2$ $(q\times q)$, and $H_3$ $(r\times r)$ are invertible matrices which depend on the identification assumptions, e.g., in Theorem \ref{theorem:CLT_loadings} $H_3$ is a diagonal matrix with entries $\pm 1$. 
				}
				\label{tab:rates1}
			}
		\end{table}

		\renewcommand{\arraystretch}{0.7}
		\begin{table}[H]
			\centering
			\caption{Comparison of consistency rates - Factors}
			{ \footnotesize
				\scalebox{0.9}[0.8]
				{\begin{tabular}
						{l p{5.75cm} p{2.75cm}p{2.75cm}}
						\hline
						\hline
						&& (A)& (B)\\
						\hline
						Network factors - Theorem \ref{theorem:CLT_factors}&	$N^{-1}\Vert \widehat{\mathcal F}_t-\mathcal F_t\times_3 H_3^{-1}\Vert$& $N^2 T$ & $\sqrt m$\\
						\hline
						Factors - full Tucker decomposition \eqref{eq:tucker_r3}&$\Vert \widehat{\mathcal G}_t-\mathcal G_t\times_1 H_1^{-1}\times_2 H_2^{-1}\times_3 H_3^{-1}\Vert $& $\min(NmT, N^2 T)$& $\sqrt m N$\\
						\hline
						Networks in \eqref{eq:Stucker}&$N^{-1}\Vert \widehat{\mathcal S}_t-\mathcal S_t\Vert$&$\min(\sqrt{NmT}, N\sqrt T)$& $\sqrt m N$\\
						\hline
						Networks in \eqref{eq:Fstartucker}&	$N^{-1}\Vert \widehat{\mathcal F}_t^*-\mathcal F_t^*\times_3 H_3^{-1}\Vert$&$\sqrt{NmT}$& $\sqrt m N$\\
						\hline
						\hline
					\end{tabular}
				}
				\vspace{7pt}
				\subcaption*{\footnotesize 
					In this table $H_1$ $(p\times p)$, $H_2$ $(q\times q)$, and $H_3$ $(r\times r)$ are invertible matrices which depend on the identification assumptions, e.g., in Theorem \ref{theorem:CLT_loadings} $H_3$ is a diagonal matrix with entries $\pm 1$. 
				}
				\label{tab:rates}
			}
		\end{table}
		
		Consider estimation of the loadings in Table \ref{tab:rates1}, then the rates depend on two terms:
		(A) a term due to the fact that the factors are unknown, and (B) a classical term which would be obtained by linear projection of the data onto known factors.
		Similarly, considering estimation of the factors in Table \ref{tab:rates}, the rates also depend on two terms:
		(A) a term due to the fact that the loadings are unknown, and (B) a classical term which would be obtained by linear projection of the data onto known loadings.
		
		The results in this table follow directly by noticing that the rates of the TOPUP estimator of the loadings by \citet{chenyangzhang22} coincide with those derived by \citet{helitrapani2022}, and from the results in the latter paper the rates for $\widehat{\mathcal G}_t$ can be obtained. Those for $\widehat{\mathcal S}_t$ and $\widehat{\mathcal F}_t^*$ are then simply the worst between those of $\widehat{\mathcal G}_t$ and the estimated loadings.
		
		It is then clear that the alternative approaches based on a full-Tucker decomposition have a term in the consistency rates which is always faster than ours (column B) and have a term which is faster than ours if $N/m\to 0$ (column A). Nevertheless, as we mentioned, these alternative approaches, although interesting, do not deliver networks that are economically interpretable nor are able to outperform our approach in the considered empirical application.

	\clearpage

	\let\section\oldsection   	
	
	\bibliography{fnar}

\end{document}